\documentclass[aps, prd,floats, floatfix, twocolumn,superscriptaddress, nofootinbib, showpacs]{revtex4-1}

\usepackage{graphicx}
\usepackage{hyperref}
\usepackage{amsmath,amssymb}
\usepackage{amsfonts}
\usepackage{xspace} 
\usepackage[usenames]{color}
\usepackage{dcolumn} 
\usepackage{bm} 
\usepackage{epsfig}
\usepackage{aas_macros}

\definecolor {darkgreen}{rgb}{0.2,0.7,0.2}

\newcommand{\be}{\begin{equation}}
\newcommand{\ba}{\begin{eqnarray}}
\newcommand{\ee}{\end{equation}}
\newcommand{\ea}{\end{eqnarray}}

\newcommand{\SMBH}{\bullet}
\newcommand{\CO}{\star}
\newcommand{\GW}{{\mbox{\tiny GW}}}
\newcommand{\SPA}{{\mbox{\tiny SPA}}}
\newcommand{\Tot}{{\mbox{\tiny tot}}}
\newcommand{\Bondi}{{\mbox{\tiny B}}}

\newcommand{\Newt}{{\mbox{\tiny Newt}}}

\newcommand{\Edd}{{\mbox{\tiny Edd}}}

\newcommand{\Msun}{\,{\rm M_\odot}}
\newcommand{\yr}{\,{\rm yr}}
\newcommand{\cm}{\,{\rm cm}}
\newcommand{\gm}{\,{\rm g}}

\newcommand{\D}{\mathrm{d}}

\newcommand{\R}{{\bar{r}}}
\newcommand{\RR}{{\tilde{r}}}
\newcommand{\Ls}{\ell}
\newcommand{\gas}{\mathrm{gas}}

\newcommand{\cs}{c_{\rm s}}

\begin{document}

\title{Observable Signatures of EMRI Black Hole Binaries \\ Embedded in Thin Accretion Disks}

\author{Bence Kocsis}
\affiliation{Harvard-Smithsonian Center for Astrophysics, 60 Garden St., Cambridge, MA 02138, USA.}

\author{Nicol\'as Yunes}
\affiliation{Dept. of Physics and MIT Kavli Institute, Massachusetts Institute of Technology, 
77 Massachusetts Avenue, Cambridge, MA 02139, USA.}
\affiliation{Harvard-Smithsonian Center for Astrophysics, 60 Garden St., Cambridge, MA 02138, USA.}

\author{Abraham Loeb}
\affiliation{Harvard-Smithsonian Center for Astrophysics, 60 Garden St., Cambridge, MA 02138, USA.}

\begin{abstract}

We examine the electromagnetic (EM) and gravitational wave (GW) signatures of stellar-mass compact objects (COs) spiraling into a supermassive black hole (extreme mass-ratio inspirals or EMRIs), embedded in a thin, radiation-pressure dominated, accretion disk. At large separations, the tidal effect of the secondary CO clears a gap.  We show that the gap refills during the late GW-driven phase of the inspiral, leading to a sudden EM brightening of the source. The accretion disk leaves an imprint on the GW through its angular momentum exchange with the binary, the mass increase of the binary members due to accretion, and its gravity. We compute the disk-modified GWs both in an analytical Newtonian approximation and in a numerical effective-one-body approach. We find that disk-induced migration provides the dominant perturbation to the inspiral, with weaker effects from the mass accretion onto the CO and hydrodynamic drag. Depending on whether a gap is present, the perturbation of the GW phase is between 10 and 1000 radians per year, detectable with the future Laser Interferometer Space Antenna (LISA) at high significance. The Fourier transform of the disk-modified GW in the stationary phase approximation is sensitive to disk parameters with a frequency trend different from post-Newtonian vacuum corrections. Our results suggest that observations of EMRIs may place new sensitive constraints on the physics of accretion disks.

\end{abstract}
\date{\today \hspace{0.2truecm}}
\maketitle
\tableofcontents

\section{Introduction}\label{s:intro}

The full exploitation of gravitational wave (GW) signals will hinge on
controlling all systematics associated with their astrophysical
sources. One can classify such systematics into three major groups:
instrumental, theoretical, and astrophysical. Instrumental systematics
are associated with possible issues related to the detector. For
example, the future Laser Interferometer Space Antenna
(LISA)~\cite{Danzmann:2003tv,Danzmann:2003ad,Prince:2003aa,lisa} might
suffer from instrumental glitches ~\cite{Carre:2010ra}.  Such
glitches, and other potential instrumental issues, might lead to a
foreground of noise artifacts that might have to be either removed, or
dealt with via data analysis techniques.

Theoretical systematics are due to incomplete modeling of waveform
templates~\cite{Cutler:2007mi}. The extraction of GWs from noisy data
requires the construction of these optimized filters, which represent
our best guess model for the GWs generated by the source.  Since
approximation schemes (either analytical or numerical) are employed to
solve Einstein's equations of General Relativity (GR) for the source,
the templates used for data analysis are not exact solutions and can
introduce errors in parameter estimation~\cite{Cutler:2007mi}.

Astrophysical systematics arise from modifications to the waveforms
caused by the environment.  For example, when modeling GWs from black
hole (BH) or neutron star (NS) binary coalescences, one usually assumes
the binary is isolated from external perturbers and ambient
electromagnetic (EM) or matter fields.  However, the unresolved GW
foreground of Galactic and extragalactic white dwarfs (WDs),
and possibly extreme mass ratio inspirals, introduces additional astrophysical noise
for LISA sources (see Ref.~\cite{2004PhRvD..70l2002B} and references
therein).  Furthermore, an additional nearby supermassive BH (SMBH) in the
vicinity of a merging binary can lead to detectable Doppler-shifts in
the GW signal~\cite{Yunes:2010sm}.

Astrophysical systematics are expected to be negligible for the final
stages of the inspiral and merger of two SMBHs, because the SMBH's
inertia greatly exceeds that of the environment. However, this is not
the case for extreme mass-ratio inspirals (EMRIs), where a small
compact object (CO) spirals into a SMBH~\cite{2007CQGra..24..113A}.
In this case, the GW inspiral rate and signal amplitude are
decreased by a factor of the mass-ratio, making these systems more sensitive to
astrophysical perturbations as well as theoretical uncertainties.
EMRIs produce millions of GW cycles in the LISA frequency band with
signal-to-noise ratios (SNRs) around 20 with a GW phasing accuracy
better than $1$ radian and logarithmic mass measurement accuracy of $10^{-3}$--$10^{-5}$
for a typical source at 1\,Gpc observed for a year~\cite{Barack:2003fp}.

In this paper we consider the most important effects that a
radiatively-efficient, thin accretion disk might have on an EMRI that
is embedded in it:
\begin{enumerate}
\item[(i)] SMBH mass increase due to accretion;
\item[(ii)] CO mass increase due to accretion;
\item[(iii)] modification of the gravitational potential due to the disk's self-gravity
(e.g.~changing the angular velocity of the orbit as a function of radius, and inducing
additional apsidal and nodal precession);
\item[(iv)] modification in the energy and angular momentum dissipation rate
(e.g.~hydrodynamic drag from winds, torques from spiral arms, and
resonant interactions analogous to planetary migrations).
\end{enumerate}
We examine the conditions necessary for the tidal gravity of the CO to
open a gap in radiation-pressure supported accretion disks, and its
implications on the EM and GW signals.  In particular, we study
whether LISA will have sufficient sensitivity to resolve the presence
and structure of the accretion disk.

\subsection{Relevance of Accretion Disks to EMRIs}

EMRIs are expected to form in dense galactic nuclei of stars, WDs,
NSs, and BHs in orbit around a central SMBH. These dense nuclei are
sometimes called {\emph{galactic cusps}} (see
e.g.~\cite{2007CQGra..24..113A}), because of their sharply peaked
density profile at zero radius. Some of these galactic nuclei are
coincidentally {\emph{active}}, meaning that gas is currently
accreting onto the central SMBH and produces bright EM
radiation. Accretion disk effects on EMRIs are most prominent in
active galactic nuclei (AGN) where gas is actively feeding the
central SMBH.

Plausible arguments have been put forth both against and in favor of
the common existence of EMRIs in AGN disks. EMRIs can only be detected
at relatively low redshift ($z\lesssim 0.5$)~\cite{Gair:2008bx}, but
only a small fraction of galaxies within $z<2$ are active and host a
massive gaseous disk.  AGN activity may be triggered by the inflow of
gas during major galaxy mergers~\cite{1992ARA&A..30..705B}. However,
the SMBHs in the centers of merging galaxies form a binary, which may
deplete the central cusp of stars~\cite{1980Natur.287..307B}, thereby
reducing the probability of EMRI events.  On the other hand, AGN
activity may be fueled by the tidal disruption of stars in dense
central cusps, which have large EMRI
rates~\cite{2005ApJ...619...30M}. Stars may be captured or may form in
accretion disks by fragmentation and/or coagulation of density
enhancements~\cite{2004ApJ...604L..45M,2004ApJ...608..108G,2007MNRAS.374..515L}.
The remnants of these stars would be pushed inwards by the disk and
could provide a reservoir of EMRI events in AGNs.

Astrophysical evidence already exists for tightly bound BH binaries
with accretion disks. OJ287 is believed to be an SMBH-SMBH binary,
with masses of $10^{8} M_{\odot}$ and $\sim 10^{10} M_{\odot}$,
respectively, orbiting in an inclined accretion disk of mass $\sim
10^{2} M_{\odot}$ \cite{2010ApJ...709..725V}. For this system, optical
flashes are observed periodically and interpreted as crossings of the
accretion disk by the smaller object. For EMRIs with stellar mass COs,
similar EM flares will be much harder to detect.

Estimates on the expected EMRI rates are very uncertain, around a few
tens to a few hundreds per year~\cite{Gair:2008bx}. LISA is expected
to be sensitive to EMRIs up to redshifts close to unity \cite{2004PhRvD..70l2002B}, although most
events should be at redshift much smaller than unity.  A few percent
among these might be in AGN environments, where accretion disk effects
are non-negligible.  These sources, if observed, may become the most
interesting EMRI sources for studying astrophysics with LISA.

We examine whether an EMRI, if present in the accretion disk, is capable
of regulating the accretion of the SMBH. If the secondary is
sufficiently massive, its tidal gravity would expel gas from the inner regions,
greatly reducing the amount of gas that would fall into the SMBH, and thus
decreasing the disk's EM luminosity. However, as the inspiral
proceeds, and the relative importance of tidal field changes compared
to the local radiation pressure in the disk, the disk might refill,
reigniting bright AGN activity.

The presence of an accretion disk around EMRIs also leads to interesting possibilities
for future GW detections. A detection of the imprint of an accretion disk on the GW
signal could inform us about accretion disk physics. But by the same
token, the presence of the disk complicates the modeling of the GW
signal, as the rather uncertain accretion disk physics introduces
additional theoretical errors and potentially makes tests of GR with
EMRIs more difficult.

If GWs could inform us about accretion disk physics, they could then provide candidates for
EM counterpart searches. Indeed, accretion disks are very sensitive to the accretion
rate parameter. A LISA measurement of the accretion rate would imply a
plausible range of AGN luminosity. Then, by looking at the LISA source
location box (approximately $1^{\circ}$ angular, $10^{-3}$ distance
measurement accuracy \cite{Barack:2003fp}), EM instruments could
search for AGNs with the predicted luminosity and redshift
\cite{2008ApJ...684..870K}. Since bright AGNs are relatively sparsely
distributed in the universe relative to the LISA error volume, a
search could cut down the number of galaxy candidates to just
one~\cite{2006ApJ...637...27K}. Alternatively, the EM counterpart
could be identified in case it is strongly modulated in time by the
EMRI.

The identification of an EM counterpart would allow the use of
EMRIs as standard sirens: measuring the distance using the EM redshift
and the GW luminosity, which would allow to independently test
cosmological
models~\cite{1986Natur.323..310S,2005ApJ...629...15H,2007ApJ...668L.143D}. Peculiar
velocities and weak lensing are expected to be the main limitation for
studying cosmology with LISA, implying that a large number of low
redshift sources such as the EMRIs considered here, are necessary for
tightening existing cosmological constraints
\cite{2006ApJ...637...27K}.

\subsection{Previous Explorations}
\label{sec:previous}

Accretion disks are common astrophysical systems that have been
studied in depth, but only recently has there been some effort to
discuss their effect on GW sources. To our knowledge, the first study
of accretion disk effects on GWs was by
Giampieri~\cite{Giampieri:1993pt}. He considered an equal-mass binary,
where each component is gaining mass due to accretion, in turn leading
to a modified radial inspiral rate. He then argued that a measurement
of the so-called GW braking index, $k_{\GW} = f_{\GW}
\ddot{f}_{\GW}/\dot{f}^{2}_{\GW}$, where $f_{\GW}$ is the GW
frequency, could lead to information on the rate of
accretion. However, if limited by the Eddington rate\footnote{The
Eddington limit is defined as the mass accretion rate at which the
inward gravitational force is balanced by the radiation-pressure force
produced by the in-falling matter in spherical symmetry; see
Sec.~\ref{s:Edd} for details.}, the accretion timescale is $\sim 10^8$ times
larger than the observation time and the effects on GWs is
insignificant (see Sec.~\ref{s:Edd} below).

Almost simultaneously with Giampieri,
Chakrabarti~\cite{1993ApJ...411..610C,Chakrabarti:1995dw} and then
Giampieri, Gerardi and Molteni~\cite{1994ApJ...436..249M} considered
an EMRI embedded in an accretion disk, where the CO accretes in the
Bondi-Hoyle-Lyttleton (BHL) approximation\footnote{BHL accretion
results when the accreting object is completely embedded in a gaseous
medium and accretes isotropically; see Sec.~\ref{sec:Bondi} for further
details.}
\cite{1939PCPS...35..405H,1944MNRAS.104..273B,1952MNRAS.112..195B,1986bhwd.book.....S,2004NewAR..48..843E}. Assuming
the gas velocity is different from the CO's velocity due to pressure
gradient effects, they found that the CO experiences a head-wind
(accelerating the coalescence) or a tail-wind (delaying the
coalescence), depending on whether the disk is rotating at
sub-Keplerian or super-Keplerian velocities.  For very high accretion
rates, the flow is trans-sonic and super-Keplerian, and the tailwind
could supersede the angular momentum loss via GW emission, leading to
a stalled orbit or even an out-spiral.

Later on, Narayan~\cite{2000ApJ...536..663N} examined the effect of
radiatively-inefficient accretion flows on EMRIs.  In quiescent
nuclei, the accretion rate is often much lower than the limiting
Eddington value and the accretion flow is dominated by advection of
the thermal energy. Accretion disks of this type are commonly referred
to as being ``radiatively inefficient,'' since the thermal energy of
the gas is advected inwards rather than radiated away as 
in thin accretion disks.  Narayan estimated the importance of
hydrodynamic drag by computing the ratio of the timescale on which the
EMRI loses angular momentum due to hydrodynamic torques and GW
emission.  For all reasonable sets of ADAF disk parameters, he found
that the GW phase is changed by order $10^{-2}$ radians, well below
the measurement accuracy of LISA.

Shortly after, {\v S}ubr and Karas
\cite{1999A&A...352..452S,2001A&A...376..686K} investigated the
evolution of the eccentricity and inclination of a CO when crossing a
thin radiatively-efficient Shakura-Sunyaev $\alpha$-disk\footnote{See
Sec.~\ref{sec:acc-disk-models} for further details on Shakura-Sunyaev
$\alpha$-disk models.}~\cite{1973A&A....24..337S}.  Assuming that the
disk is not perturbed by the CO significantly, they found that the
orbit circularizes and aligns with the plane of the disk in its
outskirts. In a follow-up work~\cite{2001A&A...376..686K}, they
extended their study to the inner parts of the Shakura-Sunyaev disks,
where the radiation pressure significantly modifies the density
profile, and examined hydrodynamic drag during disk crossing, and
angular momentum exchange with spiral density waves analogous to Type
I and II planetary migration. They have provided formulas for the
relative timescale of the effects, and concluded that GW emission
drives the evolution interior to $\sim 100\,{\rm G} M/{\rm c}^2$.

Another important study in this field was presented by
Levin~\cite{2007MNRAS.374..515L}. Motivated by simulations of Gammie
\cite{2001ApJ...553..174G}, Levin constructed models of thin
self-gravitating, radiatively-efficient disks, including optically
thick (i.e.~photons scatter several times before escaping the disk)
and thin regimes.  He re-derived order-of-magnitude estimates of
the non-relativistic hydrodynamic drag and planetary migration
timescales, similar to that of Karas and {\v S}ubr
\cite{2001A&A...376..686K}, and also included the effect of azimuthal
winds. Based on these estimates, Levin argued that, although GW
emission drives the evolution inside $\sim 100\,{\rm G} M/{\rm c}^2$, such disk effects
may be important for LISA EMRIs.

More recently, Barausse and Rezzolla~\cite{2008PhRvD..77j4027B}
examined the effects of relativistic hydrodynamic drag on EMRIs
embedded in a thick torus. They found that the hydrodynamic drag
drives the EMRI toward alignment overcoming the GW radiation reaction,
which by itself would drive the orbital plane toward anti-alignment
with the MBHs spin~\cite{1995PhRvD..52.3159R}.  However, stationary
thick massive tori with constant specific angular momentum are
unstable to global non-axisymmetric modes that grow on a dynamical
timescale~\cite{1984MNRAS.208..721P,1985MNRAS.212P..37B,1986PThPh..75..251K},
making their conclusions on EMRIs rather uncertain.

The EM emission of accretion disks around sub-parsec scale BH binaries
have been considered by many authors. However, previous explorations
focused on comparable mass binaries. In this case, the gravitational
effects would clear a gap around the binary, significantly reducing
accretion. As long as the binary separation is large enough that the
gas can follow the GW inspiral rate of the binary, tidal stresses
would act to increase the EM luminosity of the disk
\cite{2010PhRvD..82l3011L}. Eventually, however, the gas is left
behind at radii exceeding $100\,{\rm G} M/{\rm c}^2$ for equal mass systems and the gap
freezes relative to the rapid GW-induced inspiral. Consequently,
bright EM emission would be expected only several years after the
merger, when the gas has had time to diffuse inwards and accrete onto
the remnant SMBH~\cite{2005ApJ...622L..93M}. However, periodic inflow
across the gap may generate EM variability prior to
merger~\cite{2008ApJ...684..835S,2008ApJ...672...83M,2009MNRAS.393.1423C,2011ApJ...731....7B}
or the secondary can shepherd any gas remaining interior to its orbit
into the SMBH after gap freez-out~\cite{2010MNRAS.407.2007C}.  The
conditions for gap opening have not been examined in radiation-pressure
dominated accretion disks, which is most relevant at separations
smaller than $\sim 1000\, {\rm G}M/{\rm c^2}$.  Here, we investigate whether this can lead
to gap refilling prior to merger, which could have important
consequences for EM signatures of EMRIs.

None of the previous studies examined in detail whether LISA has the
sufficient accuracy to resolve the imprint of accretion disk effects
on EMRI GWs and whether any of the accretion disk parameters may be
recovered; this is the main topic of this paper and a companion
paper~\cite{2011arXiv1103.4609Y} (hereafter Paper I).  In Paper I, we
examined the detectability of the angular momentum exchange of EMRIs
with an ambient accretion disk using a large class of torque models
parameterized by two free parameters.  In particular, we pointed out
that planetary migration models are examples that generate a very
significant GW phase shift for LISA.

In this paper, we begin by reviewing the astrophysical models of the
most important accretion-disk effects on EMRIs in AGNs,
many of which were not included in Paper I. We focus on
standard radiatively-efficient thin disks, where the viscosity is
proportional to the total pressure (Shakura-Sunyaev
$\alpha$-disks~\cite{1973A&A....24..337S}) or proportional to gas
pressure only (which we refer to as
$\beta$-disks)~\cite{1981ApJ...247...19S}. These disks constitute the
standard model of luminous AGN accretion
disks (see~Refs. \cite{1986bhwd.book.....S,2002apa..book.....F} and
Sec.~\ref{sec:acc-disk-models} below).  For both of these models, we include
a more detailed analysis of the effects considered previously (SMBH mass
accretion, CO mass accretion, hydrodynamic drag, torques from spiral density waves
and resonant interactions), investigating the detailed circumstances under
which such effects are possible or suppressed, and derive the effects
of axisymmetric disk self-gravity on GWs, which was not considered before.
We provide a detailed study of the consequences on the GW
observable, using a few GW data analysis tools.

The analysis presented here is by no means exhaustive.  For
simplicity, we restrict attention to EMRIs on non-inclined,
quasi-circular orbits. This restriction does not allow us to model
binaries in which eccentricity is excited in the presence of an
accretion
disk~\cite{2005ApJ...634..921A,2008ApJ...672...83M,2009PASJ...61...65H,2009MNRAS.393.1423C,2011MNRAS.tmp..363N}.
Other effects that we do not discuss here include the following:
GWs generated by accretion flows through the excitation of BH quasinormal
modes~\cite{1997PhRvL..78.2894L,Papadopoulos:1998nc,Nagar:2004ns,2005PhRvD..72b4007N,Nagar:2006eu,2007PhRvD..75d4016N};
GW energy flux dissipation by an ambient viscous disk, driving transverse and longitudinal
density waves \cite{1971ApJ...165..165E} that could heat the disk significantly and
result in an observable infrared flare~\cite{2008PhRvL.101d1101K};
EM radiation generated by GWs in a strongly magnetized plasma,
boosting the frequency of
photons~\cite{2000ApJ...536..875M,2001PhRvD..63l4003B}, driving
magnetosonic waves~\cite{2001A&A...377..701P}, focusing of EM
radiation~\cite{2004PhRvD..70d4014K}, generating photons in a static
magnetic field~\cite{2004ApJ...613..492C,2010PhRvD..81f4017M} and
photons back-converting to GWs in a magnetized
plasma~\cite{1974PhLB...49..185J,2006CQGra..23L...7K,2002PhRvD..65f4009M}.
We also do not consider the direct scattering of GWs by the gravity of
the gas, nor the GW radiation of spiral arms in the disk. Clearly,
these effects are interesting and should be studied in more detail,
but they go beyond the scope of this paper.

\subsection{Executive Summary of Results}

\begin{figure*}
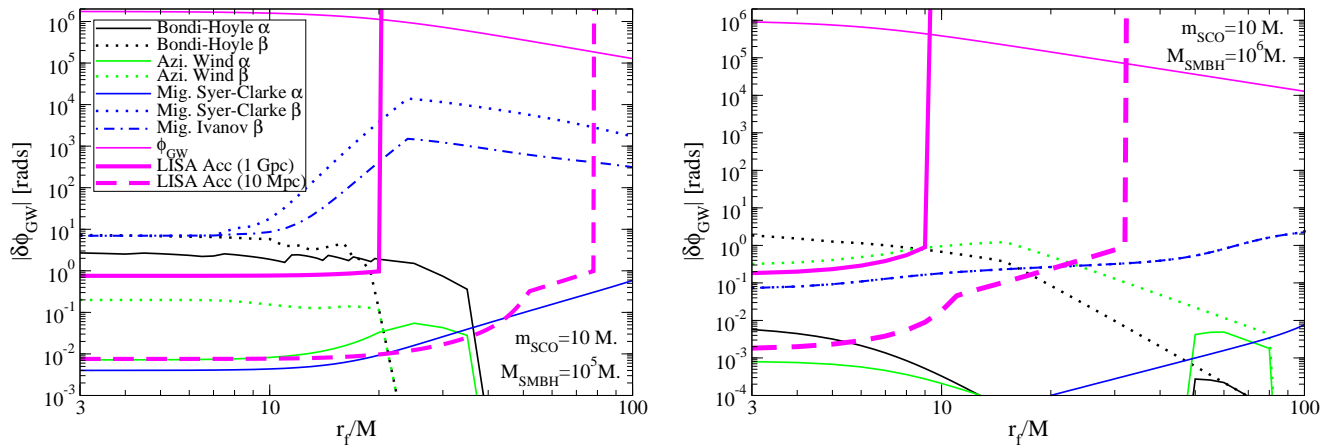

\begin{center}
\begin{tabular}{c}
 \includegraphics[width=8.5cm,clip=true]{analytic-dephasing-mSCO=10-mSMBH=1e5.eps}
 \quad
 \includegraphics[width=8.5cm,clip=true]{analytic-dephasing-mSCO=10-mSMBH=1e6.eps}
\end{tabular}
\end{center}
 \caption{\label{fig:dephasing-analytics-intro} The GW phase shift as
a function of final radius in units of $M_{\SMBH}$ induced by different accretion disk effects relative
to vacuum waveforms (see Sec.~\ref{s:conventions} for the conventions used here).
Solid (dotted) curves correspond to $\alpha$ ($\beta$) disks,
with different colors indicating different disk effects: black corresponds to
Bondi-Hoyle-Lyttleton (BHL) accretion, green to azimuthal wind and
blue to migration. The thin, solid magenta line is the total
accumulated GW phase in vacuum.  The thick, solid (dashed) magenta
line corresponds to a measure of the accuracy to which LISA can
measure the GW phase for a source at $1$ Gpc ($10$ Mpc). Observe that certain
disk effects, like migration, can leave huge imprints on the GW observable, inside
the LISA accuracy bucket.}
\end{figure*}

This sub-section of the introduction is an executive
summary of our main results, intended for non-specialists or for
readers who might not be interested in all the related technical
details of the paper, considering its length.

We derive the necessary conditions for the tidal effect of the CO to
open a gap in radiation-pressure dominated $\alpha$ and $\beta$-disks.
We find that EMRIs can open gaps at large radii in both $\alpha$ and
$\beta$-disks.  Depending on the EMRI masses and accretion disk
parameters, the gap typically closes during the inspiral due to strong
radiation-pressure gradients.  Gap refilling occurs at orbital
separations outside (inside) the LISA frequency band at orbital radii
(in geometric units, see Sec.~\ref{s:conventions})
$r \gtrsim 300$ ($24\,M_{\SMBH}$) for $\alpha$ ($\beta$)
disks, under typical EMRI parameters [SMBH and CO masses of
$(M_{\SMBH},m_{\CO})=(10^{5},10)M_{\odot}$].  Complete gap refilling
occurs within 9 months in $\beta$-disks for these EMRI masses, well
before the inspiral terminates in coalescence.  This implies that
bright AGN activity is coincident with LISA EMRIs in both cases.

We calculate the perturbations of the GW waveforms for quasi-circular,
non-inclined EMRI orbits, due to the effects of mass accretion onto
the SMBH and the CO, the hydrodynamic drag caused by an azimuthal and
radial wind, and the axisymmetric and non-axisymmetric gravitational
effects of the disk. Some of these effects lead to a strong imprint on
the GW phase, while others do not.

We find that the gravitational torque from spiral density waves in the
accretion disk (also known as {\emph{migration}} torque in planetary
dynamics) provides the most significant deviation from vacuum EMRI
dynamics. Migration leads to the strongest GW imprints when the CO
clears a gap, leading to gas accumulation near the outer edge of the
gap, like in a hydroelectric dam.  The CO's mass accretion also leaves
a significant imprint on the GW signal, if described by spherical BHL
accretion \cite{2007MNRAS.374..515L}. However, we find that the CO
accretion rate is significantly reduced by many processes,
predominantly by limited gas supply and radiation pressure.
Accounting for these limitations, the GW signature due to the CO's
mass accretion is typically much smaller than that of migration. If
tidal torques from the CO open a gap, the mass accretion onto the CO
may be greatly reduced.  All other effects are typically less
significant.

We calculate the effect of all of the above mentioned processes on the
GW observables, namely the waveform amplitude and phase, both with a
leading-order Newtonian waveform model as well as with a relativistic
effective-one-body (EOB)
model~\cite{Yunes:2009ef,2009GWN.....2....3Y,Yunes:2010zj}.
Figure~\ref{fig:dephasing-analytics-intro}
summarizes the imprint of such processes on the Newtonian phase for
the dominant GW mode as a function of the final orbital radius for a
one-year observation. Different curve colors correspond to the phase
difference between vacuum GW phases and those that include various
disk effects, while solid and dotted lines correspond to $\alpha$ and
$\beta$-disks respectively. We provide simple asymptotic analytical
formulas describing the phase shift perturbation for arbitrary
accretion disk, EMRI, and observation time parameters in
Eqs.~(\ref{e:long-gen-form}--\ref{e:short-gen-form}) and
Table~\ref{Table-pars4} in Sec.~\ref{sec:comparison}.
Figure~\ref{fig:dephasing-analytics-intro} shows that the GW phase is
modified significantly for EMRIs relevant to LISA (those with masses
$(M_{\SMBH}, m_{\CO})=(10^{5}, 10)$ and final orbital radii
$r_f\lesssim 50 M_{\SMBH}$) at 1\,Gpc and an observing time of one
year. The upper and lower thick magenta lines represent a rough measure of
LISA sensitivity to the phase shift for a source at 1\,Gpc and 10\,Mpc, respectively.

The various curves in Figure~\ref{fig:dephasing-analytics-intro}
exhibit interesting features at different radii, which correspond to
different astrophysical mechanisms that come into play.  Most notably,
the big decrease for the blue migration curves at $r \lesssim
24\,M_{\SMBH}$ correspond to a transition from Type-II to Type-I
migration as the gap refills for $\beta$-disks.  Coincidentally,
roughly interior to that radius, BHL accretion and hydrodynamic drag
from azimuthal winds are activated. The wiggles and the cutoff in the
BHL accretion induced phase shift (black curves) correspond to a
variety of effects.  For $\alpha$-disks (solid black curve), the gas density
and sound speed determine the BHL accretion rate at small
separations. At larger radii, the Bondi accretion radius becomes
larger than the disk thickness, reducing the accretion rate. Even
farther out, differential rotation of the disk reduces the BHL rate
and the decrease in the average background radial gas velocity makes
the amount of gas supply an important limitation. In the innermost
region of the disk, photon diffusion is slow and the radiation is
trapped within the BHL flow leading to super-Eddington accretion
rates.  However, at $r \gtrsim 35 M_{\SMBH}$ this is no longer true for an
$\alpha$-disk and the flow becomes Eddington limited, greatly reducing
BHL accretion effects.

Relativistic waveform models yield similar results to those presented
in Fig.~\ref{fig:dephasing-analytics-intro}. After aligning the
waveforms in time and phase (equivalent to a maximization of the SNR
over the corresponding extrinsic parameters in white noise), we find
changes in the GW phase after a typical one-year inspiral of up to
${\cal{O}}(10^{4})$ radians when modeling migration in
$\beta$-disks. Migration effects for $\alpha$-disks are much smaller,
since these disks are less dense. As the gap is typically expected to
close for EMRIs in the most sensitive LISA frequency band ($r \lesssim
25 M_{\SMBH}$), Type-I migration is the most relevant process.  Supply-limited,
BHL accretion and wind effect lead to a dephasing of ${\cal{O}}(1)$
rads.  Other effects are less significant: ${\cal{O}}(10^{-3})$ radians
for SMBH mass accretion, and ${\cal{O}}(10^{-4})$ radians for
axisymmetric self-gravity effects.

We then proceed with a more careful data analysis study on the
distinguishability of accretion disk effects by computing a data
analysis measure for two representative systems at 1\,Gpc with
component masses $(10, 10^5)\Msun$ and $(10, 10^6)\Msun$,
respectively. We calculate the SNR in the waveform {\it difference}
between signals accounting for accretion disk perturbations and those
that do not, marginalizing over an overall time and phase shift.  We
find that, for these systems, $\rho(\delta h)>10$ after just 4 months
of evolution for $\beta$-disk migration, while it takes one full year
of integration to reach the same SNR for BHL accretion and wind
effects. All other accretion disk effects are less significant.

Finally, we examine possible degeneracies between accretion disk
effects and vacuum EMRI parameters.  We analytically derive the
Fourier transform of the waveforms in the stationary phase
approximation.  We find that the disk-induced perturbation to the
frequency-domain GW phase depends on the GW frequency to a high
negative power relative to the Newtonian term, multiplied by a
function of the initial binary masses, the $\alpha$-disk parameter and
the SMBH accretion rate $\dot{m}_{\SMBH}$.  In contrast, the phase of
the Fourier transform of vacuum waveforms is a positive power of
frequency relative to the Newtonian term, when including
post-Newtonian (PN) corrections.  The difference in the frequency
scaling arises because the accretion disk effects grow with orbital
separation (lower frequency), as opposed to PN corrections which grow
with decreasing separation (higher frequency).  This suggests that
accretion disk effects are not strongly correlated with general
relativistic vacuum terms in the frequency-domain GW phase. Whether
this statement holds in a realistic data analysis implementation
requires a much more detailed analysis of the likelihood surface that
is beyond the scope of this paper.

Our results suggest that if a GW signal is detected from an EMRI in an
accretion disk, then matched filtering with accretion disk templates
could allow for the measurement of certain disk parameters to an
interesting fractional accuracy (early estimates suggest $10\%$ accuracy
for certain parameters~\cite{Yagi:prep}). The precise magnitude of the latter 
requires the detailed mapping  of the likelihood surface with relativistic EMRI
signals, full Fourier transforms, and improved disk modeling (including relativistic
effects and magnetic fields), which is beyond the scope of this paper.   

We caution, however, that the models considered here might not provide
a fully realistic description of the angular momentum exchange between
the binary and the accretion disk, leading to systematic theoretical
uncertainties in interpreting GW measurements.  Since EM observations
of the disk luminosity are also sensitive to $\dot{m}_{\SMBH}$, the
combination of contemporaneous EM and GW observations might hold the
key for constraining the accretion disk physics most reliably.

\subsection{Organization and Conventions}
\label{s:conventions}

This paper is aimed at both the General Relativity and Astrophysics
communities. We present a significant amount of background material to
make the paper self-contained for both communities.
Sec.~\ref{sec:EMRI-basics} reviews the basics of EMRIs as relevant
to GW physics and a rough measure of the accuracy to which the
waveforms need to be computed for LISA parameter estimation.
Sec.~\ref{sec:acc-disk-models} presents the basic elements of the
thin accretion disk models that are considered in this paper, and derives 
the necessary conditions for tidal effects to open a gap in the disk around the secondary.
Sec.~\ref{sec:mass-accretion} studies the effect of binary mass increase
due to gas accretion on the GW signal.
Sec.~\ref{s:wind} focuses on the effect of hydrodynamic drag on the
GW signal, induced by the gas velocity relative to the CO (i.e.~wind).
Sec.~\ref{sec:gravitationaleffects} discusses the effects of the
axisymmetric gravity of the disk.
Sec.~\ref{s:migration} concentrates on gravitational angular
momentum exchange with the disk (i.e.~migration) and its effects on
the GW signal.
Sec.~\ref{sec:comparison} compares and contrasts the effect of the
different accretion disk effects on the GW phase.
Sec.~\ref{sec:GWmodeling} describes the theoretical framework
through which we compute effective-one-body waveforms in the presence
of an accretion disk.
Sec.~\ref{sec:DA} performs a simple data analysis study to infer
the detectability of accretion disk effects.
Finally, Sec.~\ref{sec:conclusions} concludes and suggests future work.

Throughout the paper, we employ the following conventions.  We use
geometric units with $G = c = 1$ unless otherwise noted.  This implies
that masses are in units of length or time, where the mapping is
simply $M_{\odot} = 1.476 \; {\textrm{km}} = 4.92 \; \mu{\rm s}$.  The
EMRI is assumed to be composed of a SMBH with mass $M_{\SMBH}$ and a
CO with mass $m_{\CO}$. The SMBH is assumed to be spinning with spin
angular momentum $S_{\SMBH} = a_{\SMBH} M_{\SMBH}$, aligned with the
orbital one. We do not model here the spin of the CO.  We measure
quantities relative to their typical magnitudes and denote $A_{b} =
A/(10^{b} \Msun)$.  For example, $M_{\SMBH 5} = M_{\SMBH}/(10^5
\Msun)$ and $m_{\CO 1} = m_{\CO}/(10 \Msun)$. The radial orbital
separation is always scaled in terms of the SMBH's mass, such that $\R
\equiv r/M_{\SMBH}$.  The natural scale for the start and end of
observation in the most sensitive part of the LISA band is $20
M_{\SMBH}$ and $10 M_{\SMBH}$, so we use $\bar{r}_{20}=r/(20
M_{\SMBH})$ and $\bar{r}_{10}$ accordingly. We use $r'$ to denote
distance from the CO to a field point.

\section{Review of EMRI GWs}
\label{sec:EMRI-basics}

We start by reviewing some basic facts about EMRI dynamics, focusing
only on leading-order effects. Sec.~\ref{sec:GWmodeling} provides a
more detailed analysis that includes higher-order relativistic
effects.

\subsection{Basics of EMRI Dynamics}

Since the CO orbits very close to the SMBH, GR effects cause
the largest perturbations of Newtonian orbits. In this section we
consider eccentric EMRI dynamics, although in most of what follows we
restrict our attention to quasi-circular EMRIs. In the absence of
spin, the binary's energy is \be\label{e:E} \frac{E}{ m_{\CO}} =
\frac{1}{2} M_{\SMBH}^2 \Omega^{2} \R^{2} - \frac{1}{\R} =
-\frac{1}{2\R} \,, \ee the Keplerian orbital frequency is
\be
\Omega = M_{\SMBH}^{-1} \R^{-3/2},
\label{kep-ang-vel}
\ee
where $\R \equiv r/M_{\SMBH}$ is the semi-major axis of the orbit in units of
$M_{\SMBH}$. Apsidal or pericenter precession for an orbit with eccentricity $e$ is
\be
\Omega_{\rm GR, ap} = 3 M_{\SMBH}^{-1} (1-e^2)^{-1} \R^{-5/2},
\ee
while Lense-Thirring precession of the node of an inclined orbit around a spinning SMBH
is
\be
\Omega_{\rm LT} = 2 a_{\SMBH} M_{\SMBH}^{-1} \R^{-3}.
\ee

The CO inspiral produces GWs that remove binding energy and specific angular momentum  from the system at the rate
\ba
\label{e:Egw}
\dot{E}_{\GW} &=&-\frac{32}{5} \; \frac{m_{\CO}^{2}}{M_{\SMBH}^{2}} \; \frac{g_1(e)}{\R^{5}}
=-6.4\times 10^{-13} \; \frac{m_{\CO 1}^2}{M_{\SMBH 5}^{2}} \;  \frac{g_1(e)}{\R_{10}^{5}},
\\
\dot{\Ls}_{\GW} &=& -\frac{32}{5} \;  \frac{m_{\CO}}{M_{\SMBH}} \;  \frac{g_{2}(e)}{\R^{7/2}}
=-2\times 10^{-7} \;  \frac{m_{\CO 1}}{M_{\SMBH 5}} \;  \frac{g_2(e)}{\R_{10}^{7/2}},
\label{e:Lgw}
\ea
to leading order in $m_{\CO} \ll M_{\SMBH}$, where $\R_{10}=\R/10$ and where
\ba
g_1(e)&=&\frac{1+\frac{73}{24} e^{2}+ \frac{37}{96} e^{4}}{(1-e^2)^{7/2}}\,,
\qquad
g_{2}(e) = \frac{1 + \frac{7 }{8}e^{2}}{(1 - e^{2})^{2}}\,.
\ea
We have here introduced the notation $A_{b} = A/(10^{b} \Msun)$ for any quantity $A$, such that
$M_{\SMBH 5} = M_{\SMBH}/(10^5 \Msun)$ and $m_{\CO 1} = m_{\CO}/(10 \Msun)$.

Such loss of energy and angular momentum leads to the decrease of the semi-major axis
and eccentricity at a rate
\begin{align}\label{e:dotr-inspiral}
\dot{r}_{\GW} &=-\frac{64}{5}  \frac{m_{\CO}}{M_{\SMBH}} g_1(e) \R^{-3}
= -1.3 \times 10^{-6} \; \frac{m_{\CO 1}}{M_{\SMBH 5}} g_{1}(e) \R_{10}^{-3}
\\
\left|\dot{e}_{\GW}\right| &= \frac{304}{15} \frac{m_{\CO}}{M_{\SMBH}^{2}} g_3(e) \R^{-4}
= 2 \times 10^{-12} M_{\odot}^{-1} \frac{m_{\CO 1}}{M_{\SMBH 5}^{2}} g_3(e) \R_{10}^{-4}
\nonumber
\end{align}
where we have defined
\be
g_3(e) = \frac{e(1+\frac{121}{304}e^{2})}{(1-e^{2})^{5/2}}\,.
\ee
Equivalently, we can parameterize the orbit in terms of the change in the orbital frequency
\ba
\dot{\Omega}_{\GW} &=&
\frac{96}{5} \frac{m_{\CO}}{M_{\SMBH}^{3}} g_1(e) \R^{-11/2}
\ea

For quasi-circular orbits, several formula simplify. For example, one
can easily see that $\dot{E}_{\GW} = \Omega \dot{L}_{\GW}$. Similarly,
the inward inspiral velocity $v_{\CO r}$ is simply given by
Eq.~\eqref{e:dotr-inspiral} with $g_{1}(e) \to 1$.  The orbital
evolution of the semi-major axis, Eq.~(\ref{e:dotr-inspiral}), can be
integrated~\cite{1963PhRv..131..435P,1964PhRv..136.1224P} for
quasi-circular orbits, \be\label{e:r0} \R_{0} = \R_{f} \left( 1 +
\frac{\tau}{\R_{f}^{4}} \right)^{1/4},{\rm~~where~} \tau \equiv
\frac{256}{5} \frac{m_{\CO} }{M_{\SMBH}^{2}}T \,.  \ee
and where $\R_0$ and $\R_{f}$ are the initial and final separations in units of $M_{\SMBH}$
for an observation time $T$.
Let us define the critical radius and observation time
where $\R_f$ starts to deviate significantly from $\R_0$ as
\begin{align}\label{e:rf-crit}
\R_{f,{\rm crit}} &\equiv \tau^{1/4} =24\, m_{\CO 1}^{1/4} M_{\SMBH 5}^{-1/2} T_{\yr}^{1/4}\,,\\
T_{\rm crit} &\equiv \frac{5}{256}  \frac{M_{\SMBH}^{2}}{m_{\CO} } \R_f^4
= 0.031\yr\, \frac{M_{\SMBH 5}^{2}}{m_{\CO 1} } \R_{f,10}^4\,.
\label{e:t-crit}
\end{align}

The measured GW phase to leading order is twice the orbital phase for
a quasi-circular orbit $\phi_{\GW}=2\phi_{\rm orb}$. The total
accumulated phase for a quasi-circular inspiral is then
\ba
\phi_{\GW}
&=&
2 \int^{t_f}_{t_f - T_{\rm obs}} \Omega(t) dt
= 2 \int^{\R_f}_{\R_0} \Omega(\bar{r}) \frac{d\bar{r}}{\dot{\bar{r}}}\,
\nonumber \\
&=&
\frac{1}{16}\frac{M_{\SMBH}}{m_{\CO}} \R_{f}^{5/2}
\left[\left(1 + \frac{\tau}{\R_{f}^4} \right)^{5/8}-1\right]\,
\label{e:phiGW}
\ea
where $\tau$ is the dimensionless observation time defined in
Eq.~(\ref{e:r0}), and $\R_{f}$ is the final radius at the end of the
observation. Depending on whether the observation time is short or
long compared to the inspiral timescale $\tau/\R_{f}^4\ll 1$ or
$\tau/\R_{f}^4\gg 1$, Eq.~(\ref{e:phiGW}) becomes
%
\begin{align}\label{e:phiGW-short}
\phi_{\GW}^{\rm short}
&\approx 4 \times 10^6\, {\rm{rads}} \;  \frac{T_{\yr}}{M_{\SMBH 5} \R_{f,10}^{3/2}}
\left(1 - 6 \,\frac{m_{\CO 1}}{M_{\SMBH 5}^2}\frac{T_{\yr}}{\R_{f, 10}^{4}}\right)\\
\phi_{\GW}^{\rm long}
&\approx 2\times 10^6\, {\rm{rads}} \; \frac{T_{\yr}^{5/8}}{M_{\SMBH 5}^{1/4} m_{\CO 1}^{3/8}}
\left(1 - 0.1 \frac{M_{\SMBH 5}^{5/4}}{m_{\CO 1}^{5/8}}\frac{\R_{f,10}^{5/2}}{T_{\yr}^{5/8}} \right)
\label{e:phiGW-long}
\end{align}
for short observations or widely separated binaries ($T\ll T_{\rm
crit}$ and $\R_{f}\gg \R_{f, {\rm crit}}$), and long observations or
close-in orbits ($T\gg T_{\rm crit}$ and $\R_{f}\ll \R_{f, {\rm
crit}}$), respectively.\footnote{ We note that $\phi_{\GW}^{\rm long}$
is the standard PN expression for the phase evolution as a function of
time~\cite{Blanchet:2002av}, when the phase evolution culminates in
merger. Individual EMRIs, however, may outlive LISA observations,
which is why we choose to use Eq.~\eqref{e:phiGW} along with the
asymptotes Eq.~(\ref{e:phiGW-short}--\ref{e:phiGW-long}).}

Equations~(\ref{e:phiGW-short}--\ref{e:phiGW-long}) show that
initially the GW phase accumulates quickly as $c_1 T + c_2 T^2$, but
then saturates to a rate $c_3 T^{5/8}$, where $c_{i}$ are
time-independent constants.  Setting $m_{\CO} = 10 M_{\odot}$ and
$M_{\SMBH} = 10^{5} M_{\odot}$, we find that $\phi_{\GW} \sim
{\cal{O}}(10^{6})$ rads in a one year observation.

\subsection{Measures of LISA Sensitivity to GWs}
\label{LISA-sense}

GW detectors are most sensitive to the phase of the GW signal. Since
EMRIs can accumulate millions of GW cycles in the detector's
sensitivity band, they make for excellent probes of the astrophysical
environment.

\subsubsection{Simple Mass and Time-Scale Measures}\label{s:simple-LISA-measures}
A rough measure of the accuracy to which LISA can extract parameters
can be derived by looking at the mass measurement accuracy. This
quantity is of order~\cite{Barack:2003fp}
\be
\frac{\delta M_{\SMBH}}{M_{\SMBH}} \sim \frac{\delta m_{\CO}}{m_{\CO}} \sim \frac{10
^{-3}}{\rho}\,,
\label{e:dM}
\ee where $\rho$ is the SNR (see below).  The SNR can be roughly as
large as a few tens for a $10\Msun$ CO spiraling into a $10^6 \Msun$
within $\R\lesssim 10 M$ at a distance of $1\,$Gpc. Thus, the relative
mass estimation precision is at best around $10^{-4}-10^{-5}$~\cite{Barack:2003fp}.
Of course, for less distant sources, the SNR can be larger, allowing a
better determination of the masses.  This mass accuracy, compared to
the accretion mass or local disk mass, provides a rough measure of
whether the disk may generate important perturbations for LISA.

Another rough measure to decide whether certain effects are important
for LISA is the following.  If a perturbation has an associated
timescale ${\cal{T}}$ on which it changes the inspiral phase by a
factor of order unity, then the magnitude of the phase correction
corresponding to this process is roughly
\be
\delta \phi_{\GW} \sim \frac{T_{\rm obs}}{{\cal{T}}} \phi_{\GW}^{\Tot}\,
\label{timescale-measure}
\ee
where $\phi_{\GW}^{\Tot}$ is the total, accumulated GW phase in the
observation.  We assume here that this $\delta \phi$ is not a simple,
constant phase shift, but a modification in the phase evolution, such
that at the end of the observation, the template dephases from the
signal by an amount $\delta \phi_{\GW}$.  LISA can detect phase
differences of order a few radians (see below). Thus an effect is
important if ${\cal{T}} \lesssim 4.4 \times 10^{5} \; {\rm yrs}$ for a
single year observation, see Eq.~\eqref{e:phiGW}. In practice,
${\cal{T}} \equiv x/\dot{x}$ may be used where $x(t)$ represents any
of the following physical quantities: the mass of the SMBH or CO, the
induced angular momentum or energy dissipation rate relative to the
GW-driven dissipation rates, the inspiral rate relative to the GW
inspiral rate, or the frequency shift relative to the Keplerian
frequency.

\subsubsection{Dephasing Measure}
\label{deph-section}

A more accurate criterion to decide whether a certain GW modification
is detectable for LISA is:
\be\label{simple-LISA-acc-est}
\delta \phi_{\GW} \geq
\left\{
\begin{array}{lr}
10/\rho & \quad {\rm{if}} \; \rho \geq 10\,,
\\
0  &  \quad  {\rm{if}} \; \rho \leq 10\,,
\end{array}
\right.
\ee
where $\rho^{2}$ is the square of the signal-to-noise ratio (SNR), defined as
\be
\rho^{2}(h) = 4 \int \frac{df}{S_{n}(f)} |\tilde{h}|^{2}\,.
\ee
The quantity $\tilde{h}$ is the Fourier transform of the measured GW strain amplitude
and $S_{n}(f)$ is the spectral noise density curve of the detector. Accounting for white-dwarf confusion noise and not averaging over sky-angles, the noise curve is
\be
S_{n}(f) = {\rm{min}} \left\{S^{\rm NSA} \; e^{4.5 T_{\yr}^{-1} N'},
S^{\rm NSA} + S^{\rm gal}\right\} + S^{\rm ex-gal}\,,
\label{Sn}
\ee
where $(S^{\rm NSA},S^{\rm gal},S^{\rm ex-gal},N')$ are functions of frequency, that can be found,
for example, in~\cite{2004PhRvD..70l2002B}. We account for sky position and binary orientation averaged
response functions and noise curves, by multiplying Eq.~\eqref{Sn} by a prefactor of $20/3$.

Equation~\eqref{simple-LISA-acc-est} is motivated by the following
arguments. Any accretion disk effect is measurable only if the EMRI is
detected in the first place. We have here conservatively chosen $\rho
\geq 10$ as the threshold for detection.  Once the EMRI is detected, the
accuracy to which a phase difference can be measured is roughly
$10/\rho$, where $\rho$ is the total SNR.

Let us now relate this phase shift measure to the SNR of the waveform
difference between signals that include and those that neglect disk
effects: $\delta \tilde{h} \equiv \tilde{h}_{1} - \tilde{h}_{2}$. If
these waveforms differ only in phase by an amount $\delta \phi_{\GW}$,
then
\be
\delta \tilde{h} = \tilde{h}_{1} \left(1 - e^{i \delta \phi_{\GW}} \right)\,.
\ee
The SNR of the difference is then
\ba
\rho^{2}(\delta h) &=& 4 \int \frac{df}{S_{n}(f)} |\delta \tilde{h}|^{2}\,,
\nonumber \\
&=& 8\int \frac{df}{S_{n}(f)} |\tilde{h}_{1}|^{2} \left[1 - \cos(\delta \phi_{\GW}) \right]\,,
\nonumber \\
&\sim& 4\int \frac{df}{S_{n}(f)} |\tilde{h}_{1}|^{2} \delta \phi_{\GW}^{2}\,,
\ea
where in the last line we have assumed that $\delta \phi_{\GW} \ll 1 \; {\rm{rad}}$.

The perturbation of the waveform is significant if $\rho(\delta
h)\gtrsim 10$, which is similar to the accuracy requirements
constructed
in~\cite{Lindblom:2008cm,2009PhRvD..80f4019L,2010arXiv1008.1803L}.  If
the instantaneous SNR, $|\tilde{h}_{1}|^{2} S_{n}^{-1}(f)$ does not
vary greatly while the phase difference accumulates, then $\rho(\delta
h) \sim (\delta \phi_{\GW}) \rho(h_{1})$ for $\delta \phi_{\GW} \ll 1$ rad,
which leads to the simple phase shift criterion, Eq.~(\ref{simple-LISA-acc-est}), above.

\subsubsection{Degeneracies and Template Placement}

One can generalize the above measures by allowing for both amplitude
and phase modifications. The SNR of the waveform difference then
becomes
\be
\rho^2(\delta h) \equiv \min_{\lambda_2} \left[4 \int \frac{df}{S_{n}(f)} \left|\tilde{h}_{1}(f) - \tilde{h}_{2}(f;\lambda_2)\right|^2 \right],
\label{full-SNR-diff}
\ee
where $\tilde{h}_{1}$ and $\tilde{h}_{2}$ are the Fourier transforms
of two waveforms (the ``signal'' and ``template''), normalized such
that $\rho(h_1)=\rho(h_2)=1$. The template may depend on free
parameters $\vec{\lambda}_2$ that may be different from the true
astrophysical values, where the minimum difference corresponds to the
best fit.  Expanding Eq.~\eqref{full-SNR-diff}, we find
$\rho^{2}(\delta h) = 2 MM$, where $MM = 1 - {\cal{O}}$ is the
mismatch and
\be
{\cal{O}}(h_{1},h_{2}) \equiv \max_{\lambda_2} \left[4 \Re \int \frac{df}{S_{n}(f)} \tilde{h}_{1}(f) \tilde{h}_{2}^{*}(f;\lambda_2)\right]\,.
\ee
is the overlap.

For a simple estimate, we minimize $\rho^{2}(\delta h)$ over certain
non-physical {\emph{extrinsic parameters}} only, i.e.~an overall phase
and time shift, as opposed to {\emph{intrinsic parameters}}, such as
the binary's masses or spins, or other extrinsic parameters, such as
the distance to the source, the polarization angles or the sky
position.  When the SNR of the difference is computed in this way,
estimates of measurability are optimistic as they do not account for
possible degeneracies between all parameters. For example, a signal
$h_1(f,\vec{\lambda}_1)$ with true parameters $\lambda_1$ may be
mimicked by a waveform model $h_2(f,\vec{\lambda}_2)$ with different
intrinsic parameters $\vec{\lambda}_2\neq \vec{\lambda}_1$, even if
the extrinsic parameters are the same.  In Sec.~\ref{degen-sect} we
show that the incorrect determination of EMRI parameters in a
fiducial model (e.g.~with no accretion disk) cannot mimic the
waveform of a different model (e.g.~including the effects of an
accretion disk), because of the particular spectral features
introduced by accretion disks in the Fourier transform of the response
function.

\section{Review of Accretion Disk Models}
\label{sec:acc-disk-models}

We proceed with an overview of the accretion disk models under
consideration.  As this paper aims to bridge between the accretion
disk astrophysics community and GW physics community, we have chosen
to provide a complete description of background material.

\subsection{Basics of Accretion-Disk Models}

Despite a long history of observational, theoretical, and numerical
investigations, accretion disks remain one of the most exciting
unsolved problems in astrophysics. The complexity is related to the
modeling of magnetohydrodynamic (MHD) flows, turbulence, radiative
transport, and plasma physics. Here we provide a brief overview of the
formulas used to model accretion disks; for more details see textbooks
by Shapiro~\&~Teukolsky \cite{1986bhwd.book.....S} and Frank
et.~al.~\cite{2002apa..book.....F}.

We restrict our attention to geometrically thin, radiatively
efficient, stationary accretion disks, responsible for the observed
bright emission around AGNs. In this case, as gas orbits around the
central object, it radiates thermal energy away much faster than the
timescale over which the gas particles drift inward, and the disk
maintains a thin configuration. Radiatively efficient disks are the
most massive among accretion disks, as the mass accretion rate is
largest and the inward drift velocity is smallest. For lower radiative
efficiencies around quiescent SMBHs, accretion is described by other
models, such as advection-domination, which we ignore here.

Radiatively-efficient, stationary accretion disks can be described by
the Shakura-Sunyaev $\alpha$-disk
model~\cite{1973A&A....24..337S,1986bhwd.book.....S}.  The $t_{r\phi}$
component of the viscous stress tensor corresponds to an effective
viscosity\footnote{Throughout this section $\rho\equiv \rho(r)$
denotes gas density (i.e.~not the SNR of GWs), $\nu$ is the kinematic
viscosity coefficient in units of ${\rm cm}^2/{\rm s}$, and $v$ is the
gas velocity.} $t_{ij}= \rho \nu \nabla_i v_j$ and it is responsible
for the slow inflow of gas. In the $\alpha$-disk model, the viscous
stress is assumed to be proportional to the total pressure $p_{\Tot}$
in the disk at each radius: $t_{r\phi} = - (3/2)\alpha \;
p_{\Tot}$. The total pressure includes both thermal gas pressure and
radiation pressure, and the dimensionless constant of proportionality
$\alpha$ is a free model parameter.  These disks are viscously,
thermally, and convectively unstable to linear perturbations
\cite{1974ApJ...187L...1L,
1976MNRAS.175..613S,1977A&A....59..111B,1978ApJ...221..652P}.
In the alternative model~\cite{1981ApJ...247...19S}, hereafter denoted
$\beta$-disks, viscous stress is proportional to the gas pressure
only, $t_{r\phi} =- (3/2) \alpha \; p_{\gas}$, and such models are
stable.\footnote{Another popular model that is stable assumes
$t_{r\phi} = -(3/2) \alpha \sqrt{p_{\gas}p_{\Tot}}$.}

The nature of viscosity, however, is not sufficiently well-understood to predict
which of these prescriptions is closer to reality.  Recent MHD
numerical simulations of accretion disks indicate that stresses
correlate with total pressure as in the $\alpha$-disk
model~\cite{2009PASJ...61L...7O} and are thermally
stable~\cite{2009ApJ...691...16H}, though they might be viscously
unstable~\cite{2009ApJ...704..781H}. In cases where the diffusion
scale is larger than the wavelength of the magneto-rotational
instability, other simulations are consistent with the $\beta$-disk
model~\cite{2003ApJ...593..992T,2004ApJ...605L..45T}. The instability
in $\alpha$-disks implies spectral variations which are not observed
in many systems, while the $\beta$-disk model provides a better match
to spectral constraints~\cite{2008ApJ...683..389D}. In this paper, we
remain agnostic about the disk model and carry out calculations for
both of them.

Shakura-Sunyaev $\alpha$ and $\beta$-disks for a BH of fixed
$M_{\SMBH}$ mass are described by two free parameters: the accretion
rate $\dot{M}_{\SMBH}$ and the $\alpha$ parameter in the viscosity
prescription. AGN observations show that the accretion rate relative
to the Eddington rate\footnote{We define the Eddington rate
$\dot{M}_{\SMBH \rm Edd}$ more precisely in Eq.~(\ref{e:MBondi})
below.}, $\dot m_{\SMBH} \equiv \dot{M_{\SMBH}}/\dot{M}_{\SMBH\rm
Edd}$, is typically around 0.1--1 with a statistical increase towards
higher luminosities \cite{2006ApJ...648..128K,2009ApJ...700...49T}.
Theoretical limits based on simulations of MHD turbulence around BHs
are inconclusive, but are consistent with $\alpha$ in the range
0.001--1 \cite{2007ApJ...668L..51P}.  Its estimated value in
protoplanetary accretion disks is lower, $\alpha\lesssim 10^{-3}$
\cite{2009Icar..199..338L}.  Observations of outbursts in binaries
with an accreting WD, NS, or stellar BH imply $\alpha=0.2$--$0.4$
\citep{2001A&A...373..251D,2007MNRAS.376.1740K}. The value of $\alpha$
in AGN accretion disks is uncertain, but might be expected to be
similar (for a review, see Ref.~\cite{2007MNRAS.376.1740K}).  In the
following, we assume
\be
\dot{m}_{\SMBH 1} \equiv \frac{\dot m_{\SMBH}}{0.1} = 1,
\quad
\epsilon_{1} \equiv \frac{\epsilon}{0.1} = 1,
\quad
\alpha_{1} \equiv \frac{\alpha}{0.1} = 1\,,
\label{acc-disk-pars}
\ee
but retain $\dot{m}_{\SMBH 1}$ and $\alpha_{1}$ to be able to describe
different values. Here $\epsilon$ is the radiation efficiency (see
Eq.~(\ref{e:MEdd}) below).
\footnote{Note that in all of our formulas $\dot{m}_{\SMBH 1}$ and
$\epsilon_{1}$ always appear in the combination $\dot{m}_{\SMBH
1}/\epsilon_{1}$. To simplify notation we suppress the $\epsilon_{1}$
scaling.}

The surface density, $\Sigma(\R)$, and the scale height of the disk
from the mid-plane, $H(\R)$, can be calculated as
\cite{2004ApJ...608..108G},
\begin{align}\label{e:Sigma_definition}
 \Sigma(\R) &= \frac{2^{4/5}\sigma_{\rm SB}^{1/5}}{3\pi^{3/5}f_{T}^{2}\kappa^{1/5}}
\left(\frac{\mu_0 m_{\rm p}}{{\rm k_B \alpha}}\right)^{4/5}   \dot M_{\SMBH}^{3/5} \Omega^{2/5} \beta^{4(1-b)/5},\\
 \label{e:H_definition}
H(\R) &= \frac{f_{T} \kappa \dot M_{\SMBH} }{2\pi {\rm c} (1-\beta)}.
\end{align}
where $b=0$ for $\alpha$-disks and $1$  for $\beta$-disks,
and the radial dependence is implicit in the orbital velocity
$\Omega=M_{\SMBH}^{-1} \R^{-3/2}$ and $\beta$.
Here, $\beta(\R) \equiv p_{\gas}/p_{\Tot}$,
where $p_{\Tot} = p_{\gas} + p_{\rm rad}$,
$p_{\gas}$ is the thermal gas pressure, $p_{\rm rad}$ is the radiation pressure,
and $\beta$ satisfies
\begin{align}
  \frac{\beta^{(1/2) + (1/10)(b-1)}}{1-\beta} &=
\frac{2^{3/5}\pi^{4/5}{\rm c}}{ \sigma_{\rm SB}^{1/10} \alpha^{1/10} \kappa^{9/10}}
\left(\frac{{\rm k_B}}{\mu_0 m_{\rm p}}\right)^{2/5}\nonumber\\
&\quad
\times \dot M_{\SMBH}^{-4/5} \Omega^{-7/10} \label{e:beta_definition},
\end{align}
with $\mu_0=2/(3X_H+1)=0.615$ the mean molecular weight, $k_{\rm B}$
the Boltzmann constant, $\sigma_{\rm SB}$ the Stefan-Boltzmann
constant, and $f_{T} =3/4$ a constant related to the assumption of
optical depth (see \cite{2009ApJ...700.1952H}). The quantity $\kappa =
\mu_e \sigma_T/m_p$ is the electron scattering opacity in the medium,
with $\mu_e$ the number density of electrons relative to the total
density, $m_{\rm p}$ the proton mass, and $\sigma_{T}$ the Thompson
scattering cross-section.  In practice $\mu_e=(1+X_H)/2=0.875$ and
$\kappa=0.348\cm^2\gm^{-1}$ for a fully ionized gas of hydrogen and
helium, where the mass fraction of hydrogen is
$X_H=0.75$. Equation~(\ref{e:beta_definition}) can be solved
numerically for $\beta(\R)$ at each radius and substituted in
Eqs.~(\ref{e:Sigma_definition}--\ref{e:H_definition}) to obtain
$\Sigma(\R)$ and $H(\R)$.  The kinematic viscosity coefficient in the
disk is \be\label{e:nu} \nu = \alpha \beta^b H \cs =
\frac{\dot{M}_{\SMBH}}{3\pi \Sigma} \ee \cite{2003MNRAS.339..937G}.

Equations~(\ref{e:Sigma_definition}--\ref{e:beta_definition}) give a
self-consistent non-relativistic description of a geometrically thin,
radiatively efficient, stationary accretion disk, provided the
following conditions are satisfied: the disk is optically thick to
electron scattering, opacity is dominated by electron scattering, the
disk is hot enough to be fully ionized, the self-gravity of the gas is
negligible relative to the gravity of the accreting object, and the
modifications near the inner boundary condition are neglected (see
below).  These conditions are all satisfied within $6\ll \R<10^3$ for
$M_{\SMBH}<10^7\Msun$, $\dot m > 0.1$, $\alpha>0.1$
\cite{2009ApJ...700.1952H}.  Equation~(\ref{e:beta_definition}) shows
that well within \be\label{e:rmax} \R_{\rm rad}=600 \; \dot{m}_{\SMBH
1}^{16/21} \alpha_{1}^{2/21} M_{\SMBH 5}^{2/21}\,, \ee $\beta\ll 1$
holds, so that radiation pressure dominates over thermal gas pressure.
In this case, Eq.~(\ref{e:beta_definition}) can be inverted
analytically \be\label{e:beta} \beta(\R) \approx 4.3\times 10^{-4}
\alpha_1^{-1/5} \dot{m}_{\SMBH 1}^{-8/5} M_{\SMBH 5}^{-1/5}
\R_{10}^{21/10}\quad\text{for~$\beta$-disks} \ee and
Eqs.~(\ref{e:Sigma_definition}--\ref{e:H_definition}) simplify to
%
\begin{align}\label{e:Sigma}
\Sigma(\R) &\approx \left\{
\begin{array}{l}
538.5  \gm\cm^{-2}\,\alpha_{1}^{-1} \dot{m}_{\SMBH 1}^{-1} \R_{10}^{3/2}  \;\;\text{for~$\alpha$-disks},\\
1.262 \times 10^6 \gm\cm^{-2}\,\alpha_{1}^{-4/5} \dot{m}_{\SMBH 1}^{3/5} M_{\SMBH 5}^{1/5} \R_{10}^{-3/5}\\  \;\quad\text{for~$\beta$-disks},
\end{array}\right.\nonumber\\
&=
\left\{
\begin{array}{l}
5.907\times 10^{-21} \Msun^{-1}\,\alpha_{1}^{-1} \dot{m}_{\SMBH 1}^{-1} \R_{10}^{3/2}  \;\;\text{for~$\alpha$-disks},\\
1.384 \times 10^{-17} \Msun^{-1}\,\alpha_{1}^{-4/5} \dot{m}_{\SMBH 1}^{3/5} M_{\SMBH 5}^{1/5} \R_{10}^{-3/5} \\ \;\quad\text{for~$\beta$-disks},
\end{array}\right.
\\
H(\R) &\approx 1.5  \; {\dot m}_{\SMBH 1} \; M_{\SMBH} =  1.5\times 10^5  \Msun {\dot m}_{\SMBH 1} M_{\SMBH 5}\,.
\label{e:H}
\end{align}
The disk scale height is the same for the two models in the radiation
pressure dominated regime, approximately constant in radius. Note that
the thin disk assumption $2H \ll r$ breaks down within $\R \lesssim 3
{\dot m}_{\SMBH 1}$. Since radiatively efficient thick disks have no
widely accepted analytical models to date, we extrapolate
Eqs.~(\ref{e:Sigma_definition}--\ref{e:H_definition}) to this regime,
as well.

Equation~(\ref{e:Sigma}) shows that $\beta$-disks are much more
massive within $\R\ll 1000$.  This is to be expected, as the effective
viscosity is much smaller for $\beta$-disks relative to $\alpha$-disks
by a factor of order $\beta$.  Since the orbital velocity of gas is
different for different radii, viscosity leads to energy dissipation,
and a slow radial inflow. A smaller viscosity implies a smaller
radial inflow velocity, which for a fixed accretion rate corresponds
to a larger mass.

The local disk mass near the binary in a logarithmic radius bin is defined as
\begin{align}
m_{\rm d}(\R) &= 4\pi r^2 \Sigma\nonumber\\
&=
\left\{
\begin{array}{l}
7.4\times 10^{-8} \Msun\,\alpha_{1}^{-1} \dot{m}_{\SMBH 1}^{-1}  M_{\SMBH 5}^{2} \R_{10}^{7/2}
\\ \; \quad \text{for~$\alpha$-disks},\\
1.7\times 10^{-4} \Msun\,\alpha_{1}^{-4/5} \dot{m}_{\SMBH 1}^{3/5} M_{\SMBH 5}^{11/5} \R_{10}^{7/5} \\ \;\quad\text{for~$\beta$-disks}\,.
\end{array}\right.
\label{e:md}
\end{align}
The gas density in the disk, the mean radial inflow velocity, and the
isothermal sound speed are
%
\begin{align}
\rho(\R) &\equiv \frac{\Sigma}{2 H} \label{e:rho}\\
&\approx
\left\{
\begin{array}{l}
2.0 \times 10^{-26}\Msun^{-2}\,\alpha_{1}^{-1} \dot{m}_{\SMBH 1}^{-2} M_{\SMBH}^{-1}\R_{10}^{3/2} \\
\quad\;\text{for~$\alpha$-disks}\\
4.6\times 10^{-23}\Msun^{-2}\, \alpha_{1}^{-4/5}\dot{m}_{\SMBH 1}^{-2/5} M_{\SMBH}^{-4/5} \R_{10}^{-3/5}\\
\quad\;\text{for~$\beta$-disks}
\end{array}\right.\nonumber\\
v_{r}^{\gas}(\R) &= \frac{3}{2} \frac{\nu}{r}
= -\frac{\dot{M}_{\SMBH}}{2\pi r \Sigma } =  -\frac{\dot{M}_{\SMBH} r}{m_{\rm d} } \label{e:vr}\\
&\approx
\left\{
\begin{array}{l}
-3.8\times 10^{-4} \alpha_{1}\dot{m}_{\SMBH 1}^2 \R_{10}^{-5/2} \;\text{for~$\alpha$-disks}\\
-6.9\times 10^{-7} \alpha_{1}^{4/5}\dot{m}_{\SMBH 1}^{2/5} M_{\SMBH 5}^{-1/5}  \R_{10}^{-2/5}
\text{for\,$\beta$-disks}
\end{array}\right.\nonumber\\
\cs(\R) &\equiv \sqrt{\frac{p_{\Tot}}{\rho}} = H \Omega \approx 0.047\, \dot{m}_{\SMBH 1} \R_{10}^{-3/2}.
\label{e:cs}
\end{align}
where $\dot{M}_{\SMBH}$ is the SMBH accretion rate, defined more precisely in Sec.~\ref{s:Edd}.

\begin{figure}
\centering
\mbox{\includegraphics[width=87mm,clip=true]{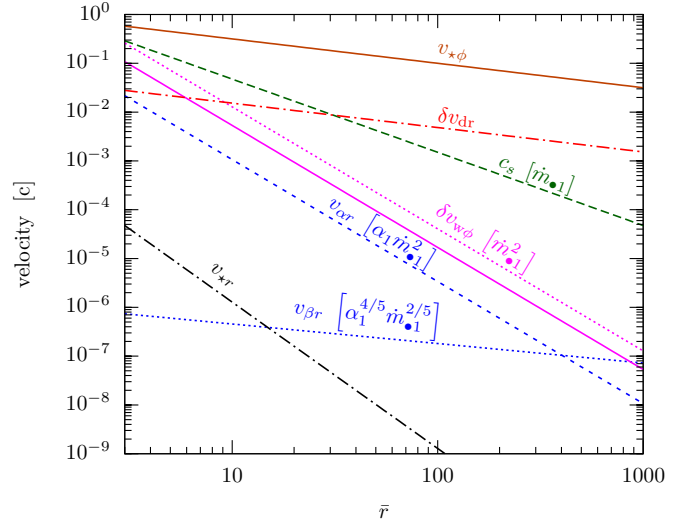}}
\caption{The velocity scales in the problem as a function of radius.
From top to bottom (with corresponding equations): Keplerian orbital
velocity (\ref{kep-ang-vel}), differential rotation at the Hill's
radius for $\beta$- and $\alpha$-disks (\ref{e:vr}), isothermal sound
speed (\ref{e:cs}), azimuthal wind (\ref{e:dvphi}), radial inflow
velocity for $\alpha$-- and $\beta$-disks (\ref{e:vr}), and the GW
radiation reaction inspiral velocity (\ref{e:dotr-inspiral}).  The
scalings with accretion disk parameters $(\alpha_1,\dot{m}_{\SMBH 1})$
are as labelled. Fiducial parameters used: $\alpha_1 = \dot{m}_{\SMBH
1} = M_{\SMBH 5} = m_{\CO 1} = 1$.  }
\label{f:velocity}
\end{figure}

Figure~\ref{f:velocity} compares the typical velocity scales in the
problem for quasi-circular EMRIs.  Typically, $|v_{r \CO}| \ll
|v_{r,\beta}| \ll |v_{r,\alpha}| \ll \cs $ for $\R\gtrsim 10$, where
$v_{r,\alpha}$ and $v_{r,\beta}$ corresponds to $v_{r}^{\gas}$ for $\alpha$ and $\beta$-disks. The
figure also depicts other relevant velocity scales which we shall
derive in Secs.~\ref{sec:mass-accretion} and~\ref{s:wind} below.
\begin{figure}
\centering
\mbox{\includegraphics[width=87mm,clip=true]{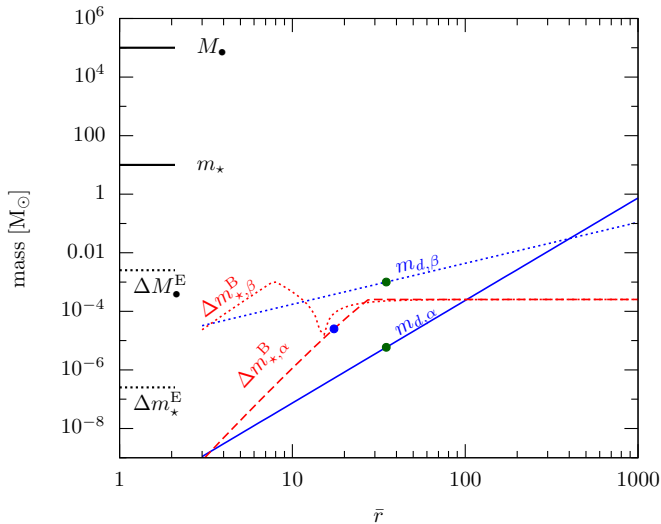}}
\caption{The mass scales in the problem. Plotted are
the MBH mass, $M_{\SMBH}$,
the EMRI mass, $m_{\CO}$,
the local disk mass, $m_{\rm d}$,
the accreted mass at Eddington rate after $1 \yr$ (efficiency $\epsilon=0.1$),
 $\Delta M_{\SMBH}^{\rm E},\Delta m_{\CO}^{\rm E}$ (Sec.~\ref{s:Edd}),
and the accreted mass at BHL rate onto the CO for the $\alpha$ and $\beta$-disks
per year, $\Delta m_{\CO \alpha,\beta}^{\rm B}$ (Sec.~\ref{sec:Bondi}).
A gap opens outside the radius marked by green dots (Sec.~\ref{s:gap}).}
\label{f:mass}
\end{figure}
Similarly, to get a feel of the typical disk mass scales,
Figure~\ref{f:mass} compares the MBH and EMRI mass, and the local disk
masses for $\alpha$ and $\beta$-disks, as well as the CO accretion
rate per year derived in Sec.~\ref{sec:Bondi} below.  The local
$\beta$-disk mass and $\Delta m_{\CO}$ are close to the LISA detection
uncertainty of the EMRI mass, $\delta m_{\CO}$ [Eq.~(\ref{e:dM})]
suggesting that the disk gravity and accretion may lead to detectable
effects for LISA observations.

Note that the above mentioned simple formulas describing accretion
disks are non-relativistic and neglect modifications related to the
inner boundary condition of the accretion disk. If assuming zero
torque at the inner boundary of the disk, $r_{0}$, this introduces
additional factors of $1-(r_{0}/r)^{1/2}$ for isothermal disks
\cite{1986bhwd.book.....S}, making the surface density profile no
longer a simple power of $r$. General relativistic corrections
introduce additional similar factors near the innermost stable
circular orbit (ISCO), light-ring, and horizon
\cite{1973blho.conf..343N,1974ApJ...191..499P}.  Among these, the
ISCO radius is the outermost one, at $1 \leq \R_{\CO}\leq 9$ in the
equatorial plane for spinning BHs.  If the shear stress generated by
the CO heats the disk, then this may further affect the scale height
and the density profile. Non-axisymmetric inflow across the CO orbit,
and the inward migration of the CO leads to a more complicated
time-dependent density profile
\cite{1996ApJ...467L..77A,1999MNRAS.307...79I,2002ApJ...567L...9A,2009MNRAS.393.1423C,2010MNRAS.407.2007C,2010ApJ...714..404T}.
We neglect these additional factors for simple order-of-magnitude
estimates and extrapolate the disk down to $\R=3$ in many of our
Figures (which is close to the ISCO for a spinning BH with $a/M\sim
0.9$).

\subsection{Gap opening}
\label{s:gap}
Up to this point, we neglected the effects of the CO. If the CO is
sufficiently massive so that its gravitational torque moves gas away
faster than viscosity can replenish it, then an annular gap opens in
the disk around the CO. The gap width can be obtained from the balance
between these two competing effects and it is given by\footnote{Here,
and throughout the paper, primed distances correspond to radial
distances measured from the CO, i.e.~$r'=|r-r_{\CO}|$.}
\be\label{e:Delta} \R'_{\Delta} = \left( \frac{f_g}{3\pi}
\frac{r_{\CO}^2 \Omega}{\nu} q^2 \right)^{1/3} \R_{\CO}\,, \ee where
$\R_{\CO}=r_{\CO}/M_{\SMBH}$ is the dimensionless orbital radius of
the CO, $q=m_{\CO}/M_{\SMBH}$ is the mass-ratio, and $f_g$ is a
geometrical factor for which $3\pi/f_g \approx 40-50$ according to
numerical simulations
\cite{1986ApJ...309..846L,1999ApJ...514..344B,2006Icar..181..587C}.
Typically, the gap width is much larger than the horizon radius of the CO.

Gap opening requires that the equilibrium width $\R'_{\Delta}$ be
larger than (i) the torque cutoff scale around the CO, and (ii) the
scale on which the tidal field of the CO dominates over the SMBH
\cite{1980ApJ...241..425G,1986ApJ...309..846L,1994ApJ...421..651A,1999ApJ...514..344B}.
The CO's tidal torque is shifted out of resonance by the mid-plane
radial pressure gradient and saturates interior to the torque cutoff
scale, $r'_{\rm cutoff}$.  This is roughly equal to the disk scale
height [see Eq.~(\ref{e:H})], $r'_{\rm cutoff}\sim H$
\cite{1993ApJ...419..155A}.  The tidal field of the CO dominates
inside the Hill radius or Roche lobe, ${r}'_{\rm H}$,
\be\label{e:rHill} \bar{r}'_{\rm H} = (q/3)^{1/3} \R_{\CO} = 0.32\,
m_{\CO 1}^{1/3} M_{\SMBH 5}^{-1/3} \R_{\CO 10}.  \ee Gas entering
within ${r}'_{\rm H}$ gets either accreted by the CO or it may flow
around the CO toward the SMBH.

Thus, gap opening requires\footnote{In the planetary context, the gap
opening condition is sometimes written as $H\lesssim r'_{\rm H}
\lesssim r'_{\Delta}$ where $H\lesssim r'_{\rm H}$ guarantees that
pressure effects are less important than the gravity of the CO and
nonlinearities become significant. However, the validity and
interpretation of this condition is disputed
\cite{1997Icar..126..261W}, and we shall not require it here over the
two criteria in Eq.~(\ref{e:gapopening}).}  \be\label{e:gapopening}
H\lesssim r'_{\Delta}\;{\rm and~} r'_{\rm H} \lesssim r'_{\Delta}\,.
\ee

Combining Eqs.~(\ref{e:Delta}--\ref{e:gapopening}), we get that a gap
opens if the mass-ratio satisfies
\begin{align}
q &> \max\left\{
 \sqrt{\frac{3\pi}{f_g} \frac{\nu}{r^2 \Omega}} \left(\frac{H}{r_{\CO}}\right)^{3/2},
\frac{3\pi}{f_g} \frac{\nu}{r_{\CO}^2 \Omega^2}
\right\}
\nonumber\\
 &= \max\left\{
\sqrt{\frac{3\pi}{f_g} \alpha \beta^b}\left(\frac{H}{r_{\CO}}\right)^{5/2},
\frac{3\pi}{f_g} \alpha\beta^b \left(\frac{H}{r_{\CO}}\right)^{2} \right\}\,,
\label{e:qgap0}
\end{align}
where in the last line we have utilized Eq.~(\ref{e:nu}) for  $\nu$, and the terms in the brackets
correspond respectively to $ H \leq  r'_{\Delta}$, and  $ r'_{\rm H} \leq r'_{\Delta}$.
Eq.~(\ref{e:qgap2}) is general for both $\alpha$ and $\beta$-disks. For $\alpha$-disks
$b=0$, but for $\beta$-disks $b=1$ and the RHS depends implicitly on $q$ and $r$ through
$\beta$.
Substituting $H$ from Eq.~(\ref{e:H}), and $\beta$ from Eq.~(\ref{e:beta}) and
solving for $q$ gives the mass-ratios that lead to gap opening
\begin{align}\label{e:qgap1}
q_{\rm gap,\alpha} &> \max\left\{
\,0.018\, \alpha_1^{1/2} \frac{\dot{m}_{\SMBH 1}^{5/2}}{\R_{\CO 10}^{5/2}},
\,0.092\, \alpha_1 \frac{\dot{m}_{\SMBH 1}^{2}}{\R_{\CO 10}^{2}}
\right\}\,,\\
q_{\rm gap,\beta} &> \max\left\{
3.6\times 10^{-4}\,  \alpha_1^{2/5} \dot{m}_{\SMBH 1}^{17/10}M_{\SMBH 5}^{-1/10} \R_{\CO 10}^{-29/20},
\right.\nonumber\\&\qquad \left.
3.9\times 10^{-5}\, \alpha_1^{4/5} \dot{m}_{\SMBH 1}^{2/5} M_{\SMBH 5}^{-1/5}\R_{\CO 10}^{1/10}
\right\}\,,
\label{e:qgap2}
\end{align}
for $\alpha$ and $\beta$-disks, respectively, and we assumed $f_{g}=0.23$.

Equations~(\ref{e:qgap1}--\ref{e:qgap2}) can be used to find the CO
orbital radius $\R_{\CO}$ at which a gap opens.  For $\alpha$-disks
both terms decrease quickly with radius, thus the gap may eventually
close if the CO orbital radius $\R_{\CO}$ is sufficiently
small. However, for $\beta$-disks, the second term depends very mildly
on radius. Hence, this term for $\beta$-disks is best viewed as a
radius independent necessary condition for the CO mass for gap
formation.  The CO radius where a gap opens, satisfies
\begin{align}
\label{e:rgap1}
\R_{\CO \rm gap,\alpha}&\geq \max
\left\{
\begin{array}{l}
79\, \alpha_{1}^{1/5} \dot{m}_{\SMBH 1} M_{\SMBH 5}^{2/5} m_{\CO 1}^{-2/5}\\
300\, \alpha_{1}^{1/2} \dot{m}_{\SMBH 1} M_{\SMBH 5}^{1/2} m_{\CO 1}^{-1/2}
\end{array}
\right\}\,,\\\label{e:rgap2}
\R_{\CO \rm gap,\beta}&\geq 24\, \alpha_{1}^{8/29} \dot{m}_{\SMBH 1}^{34/29} M_{\SMBH 5}^{18/29} m_{\CO 1}^{-20/29}{~\rm and}\\
m_{\CO \rm gap,\beta} &\gtrsim 3.9 \Msun\, \alpha_1^{4/5}\dot{m}_{\SMBH 1}^{2/5} M_{\SMBH 5}^{4/5}\,.\label{e:mgap}
\end{align}

Therefore, EMRIs in radiation-pressure dominated $\alpha$-disks in the
LISA frequency band ($\R_{\CO}\lesssim 50$) typically do not open gaps
unless $\alpha\lesssim 10^{-3}$.  However, a gap typically opens
around the CO in a radiation-pressure dominated $\beta$-disk, provided
the CO mass exceeds a value given by Eq.~(\ref{e:mgap}) and it is
captured by the accretion disk at a large radius $\R_{\CO} > 100$.  As
the CO travels inwards, however, it eventually crosses the radius
[given by Eq.~(\ref{e:rgap2})] where the gap starts to refill for
$\beta$-disks too.

If the CO gets first captured in the accretion disk around the SMBH at
some large radius $\R_{\CO}\gg 100$, a gap is expected to be cleared
quickly, and the gas interior to the orbit slowly drains down the SMBH
on the viscous timescale. Depending on disk parameters and less
understood non-axisymmetric inflows \cite{1996ApJ...467L..77A}, the
inner disk may be completely or partially cleared by the time the CO
reaches the LISA frequency band at separation $\R_{\CO}\lesssim
50$. If there is still residual gas interior to $r_{\CO}$, the EMRI
may eventually catch up with the inner disk, shepherding it into the
SMBH and causing an EM brightening of the AGN
\cite{2010MNRAS.407.2007C}.  We estimate the radius at which this
first happens in Eq.~(\ref{e:rtr}) below.  Eventually, close to the
merger, the gap would refill interior to the radii given by
Eqs.~(\ref{e:rgap1}--\ref{e:rgap2}), reigniting the AGN activity.

The fact that in AGN disks, gaps open around EMRIs for large $r_{\CO}$
but then eventually close for smaller separations may seem surprising,
because it has the opposite behavior in protoplanetary disks. The
reason for the difference is the large radiation pressure in AGN
disks, which makes $H$ to be a constant, so that $H/r_{\CO}$ decreases
outwards in Eq.~(\ref{e:qgap0}). In contrast, $H/r_{\CO}$ is nearly
constant, slowly increasing outwards for gas pressure dominated or
self-gravitating disks
\cite{2003MNRAS.339..937G,2009ApJ...700.1952H}. The other unusual
feature in Eq.~(\ref{e:qgap0}) is the $\beta^b$ factor, where we
recall $\beta=p_{\rm gas}/(p_{\rm gas} + p_{\rm rad})$. This factor
approaches 1 and becomes unimportant in the gas pressure dominated
regime for $\beta$-disks where $b=1$, while $b=0$ makes it identically
1 for $\alpha$-disks. However, this factor makes a big difference in
the radiation-pressure dominated regime for $\beta$-disks, where
$\beta\ll 1$, see Eq.~(\ref{e:beta}).\footnote{The $\beta^b$ factors
were incorrectly missing from the gap opening criteria in
Ref.~\cite{2009ApJ...700.1952H} for $\beta$-disks.}

We note that the above conditions for gap opening, based on Eq.~(\ref{e:gapopening}) are probably necessary but perhaps
insufficient in realistic disks. While these conditions have been well tested for protoplanetary disks using
simulations \cite{1999ApJ...514..344B,2006Icar..181..587C},
we are not aware of any studies discussing their applicability in 3D radiation-pressure dominated,
turbulent MHD flows for the typical EMRI and accretion disk parameters in AGNs.
MHD simulations of turbulent protostellar disks show that in some cases an annular gap may form with an ``anti-gap'' interior
to that region where the gas density is increased compared to the unperturbed case \cite{2003ApJ...589..543W}.

\subsubsection{Gap decoupling}\label{s:gap:longrange}

As explained above, a gap is expected to form for $\beta$-disks in the LISA band for a wide range of parameters. 
The outer edge of the gap at
$\lambda \R$ can initially follow the secondary as long as the GW inspiral rate is smaller than the viscous gas
inflow rate. As the binary separation shrinks, the GW inspiral velocity eventually overtakes the viscous inflow rate,
$v_{\CO r}(\R_{\CO}) \geq v_{\gas, r}(\lambda\R_{\CO})$ and the gas outside the gap cannot keep up
with the CO: the outer disk decouples \cite{2005ApJ...622L..93M}. Coincidentally, the CO can catch up
with the disk interior to the gap, if present, causing an EM flare \cite{2010MNRAS.407.2007C}.
The evolution of the gap and binary decouple at $\R_{\CO}\leq \R_{\rm d}$, which
using Eqs.~(\ref{e:dotr-inspiral}) and (\ref{e:vr}), is given by
%
\be\label{e:rtr}
\R_{\rm d} = \left\{
\begin{array}{l}
1.4 \times 10^{-5} \;  \alpha_{1}^{-2} \dot{m}_{\SMBH 1}^{-4} M_{\SMBH 5}^{-2} m_{\CO 1}^{2} \lambda^{5}
\;\text{for~$\alpha$-disks}\,, \\
15\; m_{\CO 1}^{5/13} \alpha_{1}^{-4/13} \dot{m}_{\SMBH 1}^{-2/13}  M_{\SMBH 5}^{-4/13} m_{\CO 1}^{5/13}
\lambda^{2/13}\\ \quad\;  \text{for~$\beta$-disks}\,,
\end{array}
\right.
\ee
where in the following we adopt $\lambda = 1.7$ \cite{1994ApJ...421..651A}.

The criterion given by Eq.~(\ref{e:rtr}) is not satisfied anywhere where a gap has been opened
for our nominal set of parameters in either $\alpha$ or $\beta$-disks, see Eqs.~(\ref{e:rgap1}--\ref{e:rgap2}).
However, the gap can decouple in $\beta$-disks for $m_{\CO}\gtrsim 15 \Msun$  or $\alpha\lesssim 0.05$ 
[see Fig.~\ref{fig:radii} below].

\subsubsection{Density enhancement outside the gap}
As inflowing gas is repelled by
the secondary at $r_{\CO}$, gas accumulates outside the gap, and the surface density is
modified relative to its original value without the perturber $\Sigma(\R)$ [Eq.~(\ref{e:Sigma})]
as
\be
\label{density-gap-eq}
\Sigma_{\rm gap}(\R) = \Sigma(\R)\times
\left\{
\begin{array}{ll}
1 &{\rm~if~} \R > \R_{\rm n}\\
\left[m_{\CO}/m_d(r_{\CO}\right]^{k} &{\rm~if~} \R_{\rm n}>\R>\R_{\rm g}\\
0 &{\rm~if~} \R<\R_{\rm g}
\end{array}
\right.
\ee
where $\R_{\rm g}=\lambda \R_{\CO}$ is the outer boundary of the gap\footnote{We assume that the
inflow across the gap interior to the CO is significantly reduced by the CO. This is a
conservative estimate since, non-axisymmetric inflow is expected to occur across the inner
edge of the annulus at a reduced rate \cite{1996ApJ...467L..77A}.
}, and
$k=3/8$ for $\beta$-disks\footnote{In the radiation-pressure
dominated regime, $k$ has not been determined for $\alpha$-disks.
We conservatively adopt $k=0$ in this case.} \cite{1995MNRAS.277..758S},
while $r_{\rm n}$ is the radius at which
the density enhancement disappears. In practice $r_{\rm n}$ is time dependent; it moves outward with velocity
of order $|v_{\gas, r}|$ \cite{1995MNRAS.277..758S}. By the time the CO enters the LISA band $r_{\rm n}\gg 100$.

\bigskip
In Sec.~\ref{sec:mass-accretion} through~\ref{s:migration} we discuss various effects the disk has on EMRIs,
and show that the opening of a gap has a serious impact.
Depending on whether a gap is opened or not, the CO is subject to Type-II or Type-I migration,
respectively. Moreover, if a gap is opened, then the mass density
near the secondary is significantly reduced, accretion and hydrodynamic drag effects are quenched. In that case,
only the gravitational effects play a role (i.e.~axisymmetric disk gravity and Type-II migration).

\subsection{Twists, warps, and disk alignment}
In general, MBHs are expected to have a non-negligible spin and dominate
the angular momentum of the EMRI and the disk within $\R\lesssim 10^3$.
In the absence of an accretion disk, radiation-reaction tends to
circularize the EMRI. Due to GW emission, the orbital angular momentum
evolves slowly toward anti-alignment with the MBH spin, although the total perturbation of
the orbital inclination is very small \cite{1995PhRvD..52.3159R,2000PhRvD..61h4004H,*2001PhRvD..63d9902H,*2002PhRvD..65f9902H,*2003PhRvD..67h9901H,*2008PhRvD..78j9902H}.
If the CO is on a misaligned orbit, collisions with the disk will cause it to align or counteralign
with the disk \cite{1999A&A...352..452S,2005ApJ...619...30M,2008PhRvD..77j4027B}.
In the absence of a CO, if the disk is initially misaligned
with the MBH spin axis, Lense-Thirring precession will cause the disk to warp and twist,
and viscous dissipation and radiative cooling leads to an aligned or anti-aligned
configuration with the MBHs spin (Bardeen--Petterson effect,
\cite{1975ApJ...195L..65B,2005MNRAS.363...49K,2006MNRAS.368.1196L}).
Similarly, a CO can cause warps and twists in the disk if on an inclined orbit
leading to alignment (or anti-alignment) of the disk plane with the EMRI \cite{1999MNRAS.307...79I}.
Finally, the accretion disk may be warped by the star cluster surrounding the
MBH in the outskirts of the disk \cite{2009ApJ...700L.192B}, by an intermediate mass
BH \cite{2007ApJ...666..919Y}, or by its own self gravity \cite{2009MNRAS.398..535U}.
Henceforth, we neglect such complications, and consider the MBH, EMRI and disk
angular momenta to be aligned or anti-aligned in the relative radial range of LISA
observations for simplicity.

\section{Mass Increase via Gas Accretion}
\label{sec:mass-accretion}

Next, we concentrate on the mass increase effect due to accretion. Our
goal is to make order of magnitude estimates and compare them to the
detectability measures described in Sec.~\ref{LISA-sense}.

\subsection{Primary Mass Increase}
\label{s:Edd}

The accretion disk feeds matter to the SMBH, but such process is bounded by the Eddington limit.
This limit corresponds to the balance between radiation pressure and the gravitational force in
spherical symmetry\footnote{This limit may sometimes be violated as shown in Sec.~\ref{sec:Bondi} below.}.
The corresponding luminosity due to accretion is
\be
L^{\Edd} = \frac{4 \pi {\rm G} {\rm c} }{\kappa} M_{\SMBH}\,,
\label{e:LEdd}
\ee
where $M_{\SMBH}$ is the SMBH's mass, which is here the accreting object,
and $\kappa$ is the opacity (see discussion below Eq.~\eqref{e:beta_definition}).

If a fraction $\epsilon$ of the rest mass energy is converted into radiation, then the corresponding accretion rate is
\be
\frac{\dot{M}^{\Edd}_{\SMBH}}{M_{\SMBH}} = \frac{L^{\Edd}}{\epsilon M_{\SMBH} {\rm c}^{2} } = 2.536\times 10^{-8} \epsilon_{1}^{-1} {\yr}^{-1},
\label{e:MEdd}
\ee
where $\epsilon_{0.1}=\epsilon/0.1$ is the normalized efficiency. Notice that the right-hand side of this equation is mass-independent. If the SMBH is accreting at a {\emph{constant}} rate $\dot{m}_{\SMBH} = \dot{M}_{\SMBH}/\dot{M}^{\Edd}_{\SMBH}$, then
\be
M_{\SMBH}(t) \approx M_{\SMBH,0} + \dot{M}_{\SMBH}^{\Edd} \; t
=  M_{\SMBH,0} \left(1 + \frac{\dot{m}_{\SMBH 1}}{\epsilon_{1}}  \frac{L^{\Edd}}{{\rm c}^{2}}  \; t \right)\,,
\label{change-in-SMBH-mass-Edd-limited}
\ee
where $M_{\SMBH,0}$ is the initial SMBH mass. In the analysis of Sec.~\ref{sec:acc-disk-models} we dropped most
factors of $\epsilon_{1}^{-1}$, but as is clear from here, every factor of $\dot{m}_{\SMBH}$ is accompanied by a
factor of $\epsilon_{1}^{-1}$.

Is such a change in the mass observable via EMRI GWs?
 Equation~\eqref{e:MEdd} tells us that, during a $1\yr$ observation,
 the SMBH's mass changes by $\Delta M_{\SMBH}/M_{\SMBH} = 2.2\times
 10^{-9} \dot{m}_{\SMBH 1}/\epsilon_{1}$, which is clearly below the
 LISA mass measurement accuracy of Eq.~(\ref{e:dM}). Another way to
 see this is to compute the phase shift
 (Eq.~\eqref{timescale-measure}) for Eddington-limited accretion,
 which yields $\delta \phi_{\GW}^{\Edd} \sim \phi_{\GW}^{\Tot} T_{\rm
 obs}/{\cal{T}}_{\Edd} \approx 10^{-3} \dot{m}_{\SMBH 1} M_{\SMBH
 5}^{-1/4} m_{\CO 1}^{3/8} T_{\yr}^{13/8}$, much smaller than the
 phase measurement accuracy.  We conclude then that the change in the
 SMBH's mass via Eddington-limited accretion has a negligible effect
 on the GW signal, irrespective of the type of disk modeled.  The
 accretion has to be very super-Eddington or the radiation efficiency
 very small ($\dot{m}_{\SMBH,1}/\epsilon_{1} \to 10^{3}$) for it to
 have any impact on the GW signal for typical EMRIs at 1\,Gpc.

\subsection{Secondary Mass Increase}
\label{sec:Bondi}
The CO itself increases in mass too, as it feeds from the ambient gas in the accretion disk.
We consider the case when the CO orbital inclination is aligned with the disk.
The thickness of the accretion disk [Eq.~(\ref{e:H})] is much larger than the horizon diameter by a factor of
$H/m_{\CO}\sim 1.5\, {\dot m}_{\SMBH 1} M_{\SMBH}/m_{\CO} \sim 10^4$, and thus, it completely surrounds the CO.

In such circumstances, accretion can be analyzed within the framework of
Bondi-Hoyle-Lyttleton (BHL)~\cite{1944MNRAS.104..273B,1952MNRAS.112..195B,1986bhwd.book.....S}.
The characteristic radius of accretion can be calculated as the
radius at which the thermal energy of particles is less than the gravitational
potential energy\footnote{We use $r'$ to distinguish orbital distances measured from the CO.
We denote the CO orbital radius by $r$ unless it leads to confusion, otherwise $r_{\CO}$.
Over-bar denotes units of $M_{\SMBH}$. }
\be\label{e:rBondi}
r'_{\Bondi} = \frac{2 m_{\CO}}{v_{\rm rel}^2 + \cs^2}
\approx \frac{2 m_{\CO}}{M_{\SMBH}}\frac{r^3}{H^2}
=8.9 \times 10^{3} \Msun m_{\CO 1} \dot{m}_{\SMBH 1}^{-2}\R_{10}^3\,.
\ee
The second and third equality correspond to corotating quasi-circular COs,
neglecting $v_{\rm rel}$ and using Eqs.~(\ref{e:cs}) and (\ref{e:H}).

Assuming isotropic accretion and an adiabatic equation of state,
the Euler and continuity equations can be integrated to give
%
\begin{align}\label{e:MBondi}
\frac{\dot{m}_{\CO}^{\Bondi}}{m_{\CO}} &= 4 \pi \rho \frac{ m_{\CO}}{\left(v_{\rm rel}^2 + \cs^2\right)^{3/2}}\\
&\approx
\left\{
\begin{array}{l}
1.5\times 10^{-7}\yr^{-1}\,\alpha_{1}^{-1}\dot{m}_{\SMBH 1}^{-5} M_{\SMBH 5}^{-1} m_{\CO 1} \R_{10}^{6}\\
\;\text{~for~$\alpha$-disks,~corotating,~circular~CO},\\
3.5\times 10^{-4}\yr^{-1}\,\alpha_{1}^{-4/5}\dot{m}_{\SMBH 1}^{-17/5} M_{\SMBH 5}^{-4/5} m_{\CO 1} \R_{10}^{39/10}\\
\;\text{~for~$\beta$-disks,~corotating,~circular~CO}.
\end{array}
\right.
\nonumber
\end{align}
where $\rho$ is the ambient density, $\cs$ is the sound speed, and
$v_{\rm rel}$ is the relative velocity of the gas with
respect to the medium, Eqs.~(\ref{e:dotr-inspiral}) and (\ref{e:rho}--\ref{e:cs}).
The numerical values shown in the second line are only representative, they
assume $v_{\rm rel} = |\bm{v}_{\gas} - \bm{v}_{\star}|\ll \cs$.
This is approximately satisfied near the SMBH, but in the numerical calculations we substitute the
estimated value of $v_{\rm rel}$ (see Eq.~(\ref{e:vrel}) below).
Figure~\ref{f:mass} shows the corresponding mass accretion rate per year for $\alpha$ and
$\beta$-disks, including the quenching processes discussed next.

\subsection{Quenching of BHL Accretion}

Accretion can be ``quenched'' or suppressed by several different
astrophysical processes.  In this subsection, we summarize all such
quenching effects that severely modify the accretion rates quoted in
Eq.~\eqref{e:MBondi}.

\subsubsection{Quenching by wind and tidal effects}
When $v_{\rm rel}$ is not neglected, the estimates of
Eq.~\eqref{e:MBondi} are reduced.  The relative velocity between the
CO and the gas contains contributions from the relative bulk motion of
the gas, differential rotation of the disk, and turbulence.

The effect of differential rotation can be estimated as follows.
Since the gas velocity is different at the edge of the accretion range
relative to the bulk velocity at $r$ at orbital radii $r \pm \delta r'$, then
\be\label{e:vdr}
\delta v_{\rm dr} = \left|\delta {r}'^i \nabla_i v_{\gas}^{j}\right|
\approx \left|\partial_r v_{\phi} + \Gamma^{\phi}_{\phi r} v_{\phi}\right| r_{\rm H}=
\frac{3}{2} \R'_{\rm H} \R^{-3/2}\,,
\ee
where in the second equality we have set $(\delta r'^r,\delta r'^{\phi})
= (r'_{\rm H},0)$, and in the third used $v_{\phi}=\sqrt{M_{\SMBH}/r}$
and $\Gamma^{\phi}_{\phi r}=-1/r$ for the Christophel symbol in flat space.
Substituting in for the Hill radius [Eq.~(\ref{e:rHill})], this becomes
\be
\delta v_{\rm dr} = \frac{3}{2} \frac{r'_{\rm H}}{H} \cs
= 0.015 \left(\frac{m_{\CO 1}}{M_{\SMBH 5}}\right)^{1/3} \R_{10}^{1/2}\,.
\ee

We estimate the radial wind using $\delta v_{r} = |v_{\gas, r}-v_{\CO r}|$,
where $v_{\gas, r}$ and $v_{\CO r}$ can be found in
Eqs.~(\ref{e:dotr-inspiral}) and (\ref{e:vr}). The relative velocity induced by an azimuthal
wind $\delta v_{\phi}$ is
\be\label{e:dvphi0}
\delta v_{\phi} = \frac{3-\gamma}{2}\frac{H}{r} \cs
=
\left\{
\begin{array}{ll}
0.0053\;\dot{m}_{\SMBH 1}^{2} \R_{10}^{-5/2} &\text{for~$\alpha$-disks}\\
0.013\;\dot{m}_{\SMBH 1}^{2} \R_{10}^{-5/2} &\text{for~$\beta$-disks}\\
\end{array}
\right.\,,
\ee
which we derive in Sec.~\ref{s:wind} below.

With all of this in mind, the relative velocity is then
\be\label{e:vrel} v_{\rm rel}^2 = \left(\delta v_{\phi} + \delta
v_{\rm dr}\right)^2 + \delta v_{r}^2 \,.  \ee Figure~\ref{f:velocity}
shows that $\delta v_{\rm dr}$ dominates the relative gas velocity
relative to the bulk azimuthal and radial wind velocities.  Indeed,
$\delta v_{\rm dr}>\cs$ when the CO's mass is larger than $m_{\CO}
\gtrsim 300\, M_{\SMBH 5} \dot{m}_{\SMBH 1}^{3} \R_{10}^{-3}$ or if
$\R > \R_{\rm dr} \equiv 31\, \dot{m}_{\SMBH 1} M_{\SMBH 5}^{1/3}
m_{\CO 1}^{-1/3}$. In that case, one might expect large deviations in
Eq.~\eqref{e:MBondi}. As we show in Sec.~\ref{sec:Bondi:limited-gas},
however, this quenching mechanism is typically superseded by limited gas
supply. Nevertheless, we include $v_{\rm rel}$ in our calculations
below.

The accretion is very anisotropic due to differential rotation and turbulence in the accretion disk.
The accretion rate in such an advection dominated accretion flow may be significantly less than the
BHL rate \cite{2005ApJ...618..757K,2006ApJ...638..369K}.
We consider our simple estimates to be accurate only to the order of magnitude.

\subsubsection{Quenching by thin disk geometry}

Spherical BHL accretion is valid in the region where the Bondi radius is
less than (i) the scale height of the disk $H$ [Eq.~(\ref{e:H})],
and (ii) the Hill's radius or Roche lobe where tidal effects from the SMBH are negligible [Eq.~\eqref{e:rHill}].
These constraints are satisfied interior to
\begin{align}
\label{e:thin}
\R_{\rm thin} &= 19\, \dot{m}_{\SMBH 1} M_{\SMBH 5}^{1/3} m_{\CO 1}^{-1/3}\,,
\\\label{e:tidal}
\R_{\rm tidal} &= 26\, \dot{m}_{\SMBH 1} M_{\SMBH 5}^{1/3} m_{\CO}^{-1/3}\,.
\end{align}
Thus, the thin disk requirement is always more restrictive for radiation-pressure dominated disks.

Beyond orbital radius $\R_{\rm thin}$
[Eq.~(\ref{e:thin})],
the accretion cross section is reduced from $4\pi {r}'^2_{\Bondi}$ to
$4 \pi {r}'_{\Bondi} H$, because of cylindrical symmetry. Consequently, the accretion rate
is modified as
\be\label{e:MBondi'}
\dot{m}_{\CO}'^{\Bondi} \approx
\min\left[1,\; \frac{H}{r'_{\Bondi}} \right] \dot{m}^{\Bondi}_{\CO}\,.
\ee

\subsubsection{Quenching by limited gas supply}
\label{sec:Bondi:limited-gas}

The BHL accretion rate might also be limited by the amount of gas supply near the CO.
First, note that the radial inspiral velocity $v_{\CO r}$ is typically much slower than the
radial inflow velocity of the gas $v_{\alpha r}$ and $v_{\beta r}$ for $\alpha$ and $\beta$-disks
(Fig.~\ref{f:velocity}).
This implies that there is a constant gas flux across the CO orbit from the outer
regions. However, if the Bondi rate in Eq.~(\ref{e:MBondi}) was greater than the radial gas
flux toward $M_{\SMBH}$, then the accretion onto $m_{\CO}$ would be limited by the rate
at which gas flows in from the outer regions. The mass flux across the CO's orbit is
\be
\dot{M}_{\rm flux, \CO} = 2 \pi r \Sigma | v_{\gas, r} - v_{\CO r} |
 =\dot{M}_{\SMBH} \left| 1 - \frac{v_{\CO r}}{v_{\gas, r}} \right|
 \label{e:M-flux}
\ee
where $v_{\CO r}$ and $v_{\gas, r}$ are given by Eqs.~(\ref{e:dotr-inspiral}) and (\ref{e:vr}).
Thus, the accretion rate onto the CO becomes
\be\label{e:dmSCO''}
\dot{m}_{\CO}'' = \min\left[\dot{M}_{\rm flux, \CO} , \dot{m}_{\CO}'^{\Bondi}\right]\,.
\ee
At large separations, $|v_{\CO r}| \ll |v_{\gas, r}|$, the CO accretes inflowing gas
from the outside and the accretion rate $\dot{m}_{\CO}$ becomes independent of radius
(as long as a gap is not opened, see Sec.~\ref{s:gap}).
At very small separations, given by Eq.~(\ref{e:rtr}), $|v_{\CO r}| > |v_{\gas, r}|$,
the CO sweeps up the disk interior to its orbit, and the accretion rate becomes sensitive
to the assumptions on the available gas supply interior to the orbit.

Is Eq.~(\ref{e:dmSCO''}) an important constraint for EMRI systems observable by LISA? If the mass-ratio is extreme
$\eta\lesssim 10^{-4}$, then $|v_{\CO r}|\ll |v_{\gas, r}|$ (Fig.~\ref{f:velocity}),
so Eq.~(\ref{e:dmSCO''}) imposes a constraint if and only if
$\dot{m}_{\CO}^{\Bondi} \leq \dot{M}_{\SMBH}$ is
violated for the radial separations covered by the CO during the observation.
This is the case outside $\R\geq \R_{\rm q}$, where
%
\be\label{e:rq}
\R_{\rm q} =
\left\{
\begin{array}{l}
 24\, \alpha_{1}^{1/6}  \dot{m}_{\SMBH 1} M_{\SMBH 5}^{1/3}m_{\CO 1}^{-1/3} \quad\text{for~$\alpha$-disks}\,, \\
   5.1\,\alpha_{1}^{8/39} \dot{m}_{\SMBH 1}^{44/39} M_{\SMBH 5}^{6/13}m_{\CO 1}^{-20/39} \;\,\text{for~$\beta$-disks}\,.
\end{array}
\right.
\ee
Figure~\ref{f:velocity} shows that the accretion rate of the CO saturates
at $\dot{M}_{\SMBH}$ at large $\R$. BHL accretion is supply limited for at
least part of the observation for the particular EMRI systems we consider in
Sec.~\ref{sys-invest} below.

\subsubsection{Quenching by radiation pressure}

Since the BHL accretion rate is typically super-Eddington for a mass $m_{\CO}$, does
radiation pressure quench such large rates?
One has to be careful about this point since the derivation of the BHL
accretion rate restricts to adiabatic flows, neglecting the effects of
radiation pressure, heat transport, and cooling.
Super-Eddington mass accretion onto the CO is possible if the radiation is transported inward
with the inflow faster than how it can diffuse out
\cite{1978PhyS...17..193R,1979MNRAS.187..237B,1986ApJ...308..755B}.
We compare the diffusion time with the infall time below.

The infall time of the fluid element from the Bondi radius to
a distance $r'$ from the CO is approximately
\begin{align}
t_{\rm in} &\approx r_{\Bondi}' \frac{1 - \RR'}{\sqrt{v_{\rm rel}^2 + \cs^2}}\,,
\label{e:tin}
\end{align}
where $\RR'=r'/r_{\Bondi}'$ is the dimensionless radial distance from the
CO, and the denominator is the RMS gas velocity.

The diffusion time from a distance $r'$ from the CO to $r_{\Bondi}'$ is
\be
t_{\rm diff}(r') = \frac{1}{2} \int_{r'}^{r_{\Bondi}'} \kappa \rho_{\CO}(\xi)  \xi  d\xi\,,
\label{e:tdiff0}
\ee
where we recall that $\kappa$ is the gas opacity (see discussion after Eq.~\eqref{e:beta_definition}).
To avoid confusion, we label the local gas density in the close vicinity of the CO as $\rho_{\CO}(r')$,
which at the Bondi radius is equal to the mean disk density near the location of the CO,
$\rho_{\CO}(r_{\Bondi}')\equiv\rho(r_{\CO})$ in Eq.~\eqref{e:rho}.
For BHL accretion, the density increases toward the CO as
$\rho_{\CO}(r') = \rho_{\CO}(r_{\Bondi}') \RR'^{-3/2}$~\cite{1944MNRAS.104..273B,1952MNRAS.112..195B,1986bhwd.book.....S}.
Thus, Eq.~(\ref{e:tdiff0}) simplifies to
\begin{align}
t_{\rm diff} &=
\kappa \rho(r_{\CO}) r_{\Bondi}'^{2}\left( 1 -  \RR'{}^{1/2}\right)
\nonumber\\\label{e:tdiff} &= \frac{\kappa}{\pi} \frac{\dot{m}_{\CO}^{\Bondi}}{\sqrt{v_{\rm rel}^2 + \cs^2}} \left( 1 -  \RR'{}^{1/2}\right)\,,
\end{align}
where in the second line we have used Eq.~(\ref{e:MBondi}) to relate $t_{\rm diff}$ to the BHL accretion rate.
Comparing Eqs.~(\ref{e:tin}) and (\ref{e:tdiff}), we find that the
diffusion time is larger than the infall time, precisely if $\dot{m}_{\CO}^{\Bondi}\geq \dot{m}_{\CO}^{\rm crit}$
where
\begin{align}\label{e:MBondicrit}
\frac{\dot{m}_{\CO}^{\rm crit}}{m_{\CO}}
&\equiv \frac{\pi}{\kappa} \frac{r_{\Bondi}'}{m_{\CO}} \left( 1 +  \RR'{}^{1/2}\right)
=\frac{2\pi}{\kappa} \;\frac{1 +  \RR'{}^{1/2}}{v_{\rm rel}^2 + \cs^2}
\\\nonumber
&\approx5.6\times 10^{-7}\yr^{-1}\, \dot{m}_{\SMBH 1}^{-2} \R_{10}^{3}\left( 1 +  \RR'{}^{1/2}\right)\,.
\end{align}
The second line corresponds to $v_{\rm rel}\ll \cs$, but in the numerical calculations we
substitute $v_{\rm rel}$ from Eq.~(\ref{e:vrel}).

Interestingly, radiation pressure does not have any impact if the BHL accretion rate exceeds the
limit, $\dot{m}_{\CO}^{\rm crit}$. For the nominal parameter values for a $\beta$-disk,
Eqs.~(\ref{e:MBondi}) and (\ref{e:MBondicrit}) show that $\dot{m}_{\CO}^{\Bondi}\geq \dot{m}_{\CO}^{\rm crit}$
is satisfied for all $0 \leq \RR' \leq 1$. Therefore, we are reassured that radiation is trapped and advected inward
in this case. However, $\alpha$-disks are much less dense, and this condition is violated interior to
$20\, \alpha_{1}^{1/3} \dot{m}_{\SMBH 1} M_{\SMBH 5}^{1/3} m_{\CO 1}^{-1/3}$.

More generally, BHL accretion may be quenched by the various other effects discussed above,
modifying $r_{\Bondi}'$ and decreasing the gas density, and thereby the diffusion and infall times.
In this case, radiation pressure may further suppress the
accretion rate onto the CO if $\dot{m}_{\CO} \leq \dot{m}_{\CO}^{\rm crit}$. This criterion can be
fulfilled by both $\alpha$ and $\beta$-disks. The accretion rate then becomes
\begin{equation}\label{e:dmSCO3}
\dot{m}_{\CO}''' =
\left\{
\begin{array}{l}
\dot{m}_{\CO}'' {~\rm if~}\dot{m}_{\CO}''\geq \dot{m}_{\CO}^{\rm crit}\,,\\
\dot{m}_{\CO}^{\Edd} {~\rm otherwise}\,.
\end{array}\right.
\end{equation}
where $\dot{m}_{\CO}''$ is given by Eq.~(\ref{e:MBondi'})
and we model the radiation-pressure quenched accretion as Eddington limited,
replacing $M_{\SMBH}$ with $m_{\CO}$ in Eq.~(\ref{e:MEdd}). We here choose $\RR' = 1$,
as this gives the most conservative (smallest) estimate for $\dot{m}_{\CO}'''$.

\subsubsection{Quenching by gap formation}
If the tidal torques of the CO are sufficiently strong to dominate over the viscous inflow,
an annular gap forms around the CO, where the gas density is significantly reduced
(see Sec.~\ref{s:gap}). Gap formation requires $m_{\CO}$ and $r_{\CO}$ to be sufficiently large,
[Eqs.~(\ref{e:rgap1}--\ref{e:mgap})]. These conditions can be satisfied for $\beta$-disks during the
final year of the inspiral, but not for typical $\alpha$-disks. If a gap forms, the accretion onto the CO ceases.

For large CO masses $m_{\CO}\gtrsim 15\Msun$ or $\alpha\lesssim 0.05$, the inspiral rate
becomes faster than the viscous inflow rate of gas outside the annular gap if
$r_{\rm gap} < r_{\CO} < r_{\rm d}$ (see Eq.~\ref{e:rtr}).
In this case, the CO may ``catch up'' with the gas interior to the orbit \cite{2010MNRAS.407.2007C}.
The inner disk may be filled by non-axisymmetric or three dimensional overflow \cite{1996ApJ...467L..77A}.
In fact, in turbulent MHD disks, the region interior to the annular gap
may have an over-density (``anti-gap'') relative to the case without an EMRI \cite{2003ApJ...589..543W}.
In this case, $\dot{m}_{\CO}$ may be restarted interior to $r_{\rm d}$, and may exceed the BHL rate
of the original unperturbed surface density of the disk [Eq~(\ref{e:MBondi})]. However, it is also possible that
the inner disk drains away before $r_{\rm d}$ is reached, implying no accretion.
We conservatively assume no accretion onto the CO if a gap is present,
\begin{equation}\label{e:dmSCO4}
 \dot{m}_{\CO} = \left\{
\begin{array}{l}
\dot{m}_{\CO}''' {~\rm if~}r \leq r_{\rm gap}\,,\\
0 {~\rm otherwise}\,.
\end{array}\right.
\end{equation}

\subsubsection{Summary of quenching processes}\label{s:BondiSummary}
\begin{figure}
 \includegraphics[width=8.5cm,clip=true]{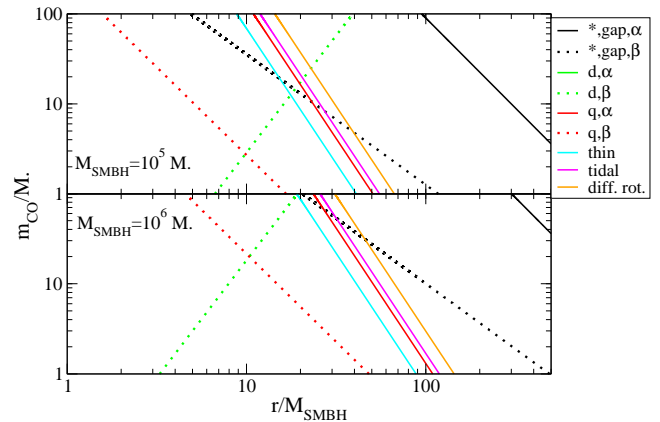}
 \caption{\label{fig:radii} Critical CO mass as a function of CO orbital radius for various mechanisms to quench
 BHL accretion onto the CO. The accretion rate is reduced for larger $m_{\CO}$ or larger $r_{\CO}$.
 Top and bottom panels correspond to $M_{\SMBH}=10^{5}$ and $10^{6} M_{\odot}$, respectively.
Gap-decoupling occurs interior to the green curves.}
\end{figure}
The mass increase of the CO is very sensitive to the complicated
details of accretion disk astrophysics. Most of these processes act to decrease the accretion rate from
$\dot{m}_{\CO}^{\Bondi}$. We summarize the EMRI parameters where
various quenching mechanisms are in play in Figure~\ref{fig:radii}. This figure depicts the minimum CO mass $m_{\CO}$
and orbital radii where particular processes become significant to quench the BHL accretion rate
onto the CO for $\alpha_{1}=\dot{m}_{\SMBH 1}=1$ for different $M_{\SMBH}=10^5 \Msun$ (top panel) and
$10^6\Msun$ (bottom panel). For these parameters, accretion is first completely quenched by gap formation
for $\beta$-disks, but gaps do not form for $\alpha$-disks for EMRIs in the LISA range. Then the gap refills, and
accretion is limited by the amount of inflowing gas, radiation pressure, differential rotation, and the thin disk geometry.
Closer to the SMBH, these processes become less and less significant and the accretion rate increases to the BHL rate
$\dot{m}_{\CO}^{\Bondi}$.  This decreases inward with the decrease of gas density
and the increase in the sound speed. The corresponding CO mass increase is shown in Figure~\ref{f:velocity} above.

\subsection{Implications of BHL Accretion}

Let us now describe the implications of BHL accretion on EMRI formation and the
GW phase. The former is relevant to understand whether EMRIs can remain extreme mass-ratio systems as they inspiral
in the accretion disk toward the LISA band, or if they grow in mass to form an intermediate mass BH.
Then, we estimate the corresponding perturbation to the GW phase and discuss its detectability with LISA.

Equations~(\ref{e:MBondi}), (\ref{e:MBondi'}), (\ref{e:dmSCO''}), (\ref{e:dmSCO3}), and (\ref{e:dmSCO4}) show that the CO mass growth rate
is sensitive to the location of the CO, i.e.~$\dot{m}_{\CO}$ is not a constant.
The mass at radius $\R$ can be estimated as
\be
m_{\CO} =
m_{\CO,0}  + \int_{t_0}^{t} \dot{m}_{\CO} \D t =
m_{\CO,0}  + \int_{\R_0}^{\R} \frac{\dot{m}_{\CO}}{\dot{\R}} \D \R\,,
\label{mSCO-expansion-num}
\ee
where $m_{\CO,0}$ is the initial CO mass, and $\dot{m}_{\CO}$ is the accretion rate and $\dot{\R}$
is the inspiral rate (\ref{e:dotr-inspiral}).

\subsubsection{EMRI formation scenarios}

Does the CO mass grow significantly in the disk prior to the LISA observation? If the CO migrated through the
disk from very large radii, $m_{\CO} \gg 10 M_{\SMBH}$ could be expected by the time
the CO reaches detectable separations for LISA observations \cite{2004ApJ...608..108G,2007MNRAS.374..515L}.
At large orbital radii the accretion rate is supply limited. Assuming that a gap does not form and
the CO consumes the inflowing gas completely and that
$|\dot{r}_{\CO}| \ll |v_{\gas, r}|$
\begin{align}\label{e:Deltam}
\Delta m_{\CO}
&\leq \dot{M}_{\SMBH} \,t_{\rm merger}
\leq 1.2\Msun \frac{\dot{m}_{\SMBH 1} M_{\SMBH 5}^3}{m_{\CO 1}} \R_{200}^4\,,
\end{align}
where $\R_{200}=\R_0/200$ and we have substituted the inspiral time to coalescence $t_{\rm merger}$ given by
Eq.~(\ref{e:t-crit}).\footnote{In the opposite extreme $\dot{r}_{\CO} \gg v_{\gas, r}$,
the CO growth is limited by the
interior disk mass, which is typically less than $10 M_{\SMBH 5}^2 \Msun$ within $\R \sim 10^{3}$, see Fig.~\ref{f:mass}.}
The merger time is typically less than the GW driven inspiral time
due to angular momentum exchange with the disk discussed in Sec.~\ref{s:migration} \cite{2001A&A...376..686K,2007MNRAS.374..515L},
so that Eq.~(\ref{e:Deltam}) is only an upper limit on $\Delta m_{\CO}$.
Note that in this case, the mass increase is larger than the instantaneous amount of disk mass within a few Hill's radii [Eq.~(\ref{e:rHill})],
called the isolation mass, due to the viscous inflow of gas from the outer regions across the orbit.
This situation is most relevant for $\alpha$-disks where a gap does not form easily [see Eq.~(\ref{e:rgap1})].

Once the CO mass has grown sufficiently, the tidal torque of the CO
eventually opens a radial gap in the disk, halting further
growth. This leads to a limit called the starvation mass\footnote{If a
gap opens and the CO is transported inward by Type-II migration
together with the flow (see Sec.~\ref{s:migration} below), then $\Delta
m_{\CO}$ is limited by the local disk mass in a few Hill's radii,
i.e.~the isolation mass \cite{2007astro.ph..1485A}. However, typical
CO masses exceed the local disk mass within $\R\lesssim 10^3$ (see
Fig.~\ref{f:mass}), and the inward migration rate is slower than the
gas inflow rate. In this case, gas can build up near the edge of the
gap and cause the object to grow to the starvation mass.}
\cite{2010ApJ...710.1375K}, which from
Eq.~(\ref{e:qgap1}--\ref{e:qgap2}) is
\begin{align}
\Delta m_{\CO \alpha} &\leq \max\left\{
1800\Msun\, \alpha_1^{1/2} \dot{m}_{\SMBH 1}^{5/2} M_{\SMBH 5} \R_{\CO 10}^{-5/2},\,
\right.\nonumber\\&\qquad \left.
9200\Msun\, \alpha_1 \dot{m}_{\SMBH 1}^{2} M_{\SMBH 5}\R_{10}^{-2}
\right\}\,,
\label{e:Deltam2a}\\
\Delta m_{\CO \beta} &\leq \max\left\{
36\Msun\,  \alpha_1^{2/5} \dot{m}_{\SMBH 1}^{17/10} M_{\SMBH 5}^{9/10} \R_{10}^{-29/20},
\right.\nonumber\\&\qquad \left.
3.9\Msun\, \alpha_1^{4/5} \dot{m}_{\SMBH 1}^{2/5} M_{\SMBH 5}^{4/5}\R_{\CO 10}^{1/10}
\right\}\,.
\label{e:Deltam2b}
\end{align}

Note that both Eq.~(\ref{e:Deltam}) and (\ref{e:Deltam2a}) must be satisfied for $\alpha$-disks, and
Eq.~(\ref{e:Deltam2b}) for $\beta$-disks.
In both cases, the CO mass remains small $m_{\CO}\ll 100\Msun$
for $M_{\SMBH}=10^5\Msun$.
It grows to at most $\sim 76\Msun$ until reaching $\R=25$ for $M_{\SMBH}\sim 10^6\Msun$ in a $\beta$-disk,
but can grow beyond $100\Msun$ for $M_{\SMBH}\gtrsim 10^6\Msun$ in an $\alpha$-disk,
if captured in the disk outside $\R \gtrsim 50$.
If so, an initial EMRI would morph into an intermediate mass-ratio inspiral before entering the LISA band.
Growth beyond a mass given by Eq.~(\ref{e:Deltam2b}) is halted by gap formation \cite{2007MNRAS.374..515L}.

We conclude that EMRIs can remain extreme in mass-ratio on their journey to the LISA band ($\R\lesssim 50$)
for a non-negligible set of disk parameters.
Conversely, the mass measurement with LISA could have interesting implications on the structure of the
accretion disk. Suppose LISA measures $m_{\CO}$ to be large, consistent with either
Eq.~(\ref{e:Deltam2a}) or (\ref{e:Deltam2b}). This information alone would suggest that the CO has grown by accretion
in an $\alpha$ or $\beta$-disk, and suggest the possible presence of a disk, even without a direct GW phase shift
measurement. In the opposite case, if LISA measures $m_{\CO}$ to be larger than what is expected from the
growth arguments given by Eq.~(\ref{e:Deltam2a}--\ref{e:Deltam2b}), then this could point to the common existence
of intermediate mass BHs, which is  debated at the time of the writing of this manuscript.

\subsubsection{GW Observations}
BHL accretion changes the GW inspiral rate of the EMRI due to the increase in the radiating mass
quadrupole. This leads to a GW phase shift relative to a constant $m_{\CO}$. Is BHL accretion measurable for EMRIs
with LISA observations?

For a crude first estimate let us consider the corresponding limits on the mass and timescales of Sec.~\ref{s:simple-LISA-measures}.
Figure~\ref{f:mass} shows the BHL mass accretion rate as a function of radius (red lines). For these masses ($10\Msun, 10^5\Msun$),
$\Delta m_{\CO}/m_{\CO}\sim 10^{-5}$ to $10^{-4}$ for $\alpha$ and $\beta$-disks. This is comparable to the mass
measurement precision of LISA for a source at $\sim 1\,$Gpc, suggesting that the perturbation caused by the CO mass growth
may be marginally significant.
The timescale argument with ${\cal{T}}_{\Bondi} \sim m_{\CO}/\dot{m}_{\CO}$ is not directly applicable as $\dot{m}_{\CO}$
varies significantly during a one-year LISA measurement.

A more accurate analytical estimate is computed in Appendix~\ref{B-H-accurate-estimate}, where we integrate the total perturbation to the GW phase assuming that $\dot{m}_{\CO}= A \R^{B}$, where $A$ and $B$ are constants.  The phase shift will be presented in Sec.~\ref{sec:comparison}, Eqs.~(\ref{e:long-gen-form}--\ref{e:short-gen-form}) and Table~\ref{Table-pars4}; it shows that the phase shift accumulates with time initially as $\delta\phi \approx a_1 T^3 + a_2 T^4$, for short observations relative to the inspiral rate, and eventually asymptotes to $a_3 T^{c_5}$, where $(a_1, a_2, a_3,c_5)$ are constant coefficients that depend on the EMRI and the accretion disk parameters.

We find that the quenching by gas supply has a major effect on the GW phase shift. Without quenching, the effect would be of order
$\delta \phi_{\GW}^{\Bondi}\sim 7$ and $3000$ rads for $\alpha$ and $\beta$-disks respectively, and even larger for larger $m_{\CO}$, assuming $T_{\rm obs}=1\yr$ and $M_{\SMBH 5}=m_{\CO 1}=\dot{m}_{\SMBH}=1$.
However, at most separations, the BHL rate is significantly suppressed by gas supply for $\beta$-disks, but the reduced
dephasing is still about $13\,$rad per a year, and larger for larger $M_{\SMBH}$.
The phase shift is proportional to the following combination of accretion disk parameters: $\alpha_1^{-1}\dot{m}_{\SMBH 1}^{-5}$ and
$\alpha_1^{-4/5}\dot{m}_{\SMBH 1}^{-17/5}$ for the unquenched BHL rate for $\alpha$ and $\beta$-disks,
and $\dot{m}_{\SMBH 1}$ for the supply limited rate. These combinations may be marginally measurable by LISA observations, given sufficiently strong signals.

\section{Hydrodynamic Drag}
\label{s:wind}

Next, we consider the drag induced by a difference between the gas and
CO velocities, sometimes called {\emph{wind}}.  This relative velocity
is a consequence of a pressure gradient in the disk and results in a
force that pushes the CO both azimuthally and
radially~\cite{1993ApJ...411..610C,Chakrabarti:1995dw}. As in
Sec.~\ref{sec:mass-accretion}, the goal is to make order of magnitude
estimates on the corresponding GW phase shift and compare them to the
detectability measures of Sec.~\ref{LISA-sense}.

\subsection{Azimuthal wind}\label{s:wind-azimuthal}

If the orbital velocity of the gas is different from the CO's orbital
velocity, the latter will experience an azimuthal headwind or backwind
(relative to its unperturbed azimuthal motion).  To estimate the
orbital velocity of the gas, let us write the radial equation of
motion assuming that $M_{\SMBH}\gg m_{\CO}$, the orbital velocity is
much larger than the radial velocity, and that the flow is subsonic:
\be\label{e:vorb1} -\frac{v_{\phi}^2}{r} + v_{r} \nabla_r{ v_r} =
-\frac{M_{\SMBH}}{r^2} - \nabla_r \Phi_{\CO} - \nabla_r \Phi_{\rm
disk} - \frac{\nabla_r p_{\Tot}}{\rho}\,, \ee where $v_{\phi}$ and
$v_{r}$ are the azimuthal and radial velocities of the gas, $p_{\Tot}$
is the total pressure of the gas in a comoving frame, and $\rho$ is
the gas density. For standard thin disks, $p_{\Tot}=\rho \cs^2 = \rho
H^2 \Omega^2$, where the scale height $H$ is independent of radius in
the radiation-pressure dominated regime [see e.g.~Eqs.~(\ref{e:H}) and
(\ref{e:cs})].  The orbital average gravitational potential of the CO,
$\Phi_{\rm disk}$, acting on the fluid element at radius r is \be
\Phi_{\CO} = -\frac{2 }{\pi} \frac{ m_{\CO}
}{r+r_{\CO}}K\hspace{-3pt}\left(\frac{2 \sqrt{r
r_{\CO}}}{r+r_{\CO}}\right)\,,
\label{SCOpotential}
\ee
where $K(k)=\int_0^{\pi/2} (1-k^2 \sin^2\theta)^{-1/2} d\theta$ is the complete elliptic integral of the first kind.
The gravitational potential of the disk, $\Phi_{\rm disk}$, will be given in Sec.~\ref{sec:gravitationaleffects} [see e.g.~Eq.~(\ref{e:phi-disk0})].

Let us parameterize the radial density profile via
$\rho(r) \propto r^{\gamma}$, where the exponent $\gamma = 3/2$ and $\gamma = -3/5$
for $\alpha$ and $\beta$-disks respectively [Eq.~(\ref{e:Sigma})].
Note that $v_{r} \nabla_r{ v_r}$ is negligible in Eq.~(\ref{e:vorb1}) since
$v_r \ll \cs$ and $v_r \propto 1/(r\Sigma)$, see Eq.~(\ref{e:vr}), and
Fig.~\ref{f:velocity}.
Equation~(\ref{e:vorb1}) then becomes
\be
-\frac{v_{\phi}^2}{r} =  -\frac{M_{\SMBH}}{r^2} - \nabla_r \Phi_{\CO} - \nabla_r \Phi_{\rm disk}+ (3-\gamma) \frac{H^2\Omega^2}{r}\,,
\ee
which one can solve to obtain
\be
v_{\phi}^2 =  \frac{ M_{\SMBH}}{r}\left[ 1  + (\gamma - 3) \frac{H^2}{r^2} + \frac{r}{M_{\SMBH}} \nabla_r (\Phi_{\CO}+\Phi_{\rm disk})\right]\,.
\label{v-phi-corr}
\ee
In the following we neglect the effects of the potential due to the secondary, an approximation valid
if the gas accretes onto the CO from outside the Hill's sphere, i.e.~the $r'_{\Bondi}<r'_{\rm H}$, see Eqs.~(\ref{e:tidal}),
as well as the disk gravity.

We then find that a corotating CO always experiences an azimuthal headwind with velocity
(i.e.~orbital velocity of gas with respect to the CO),
\begin{align}
\delta v_{\phi}&\equiv \Omega_{\rm vac} r-v_{\phi}\nonumber \\\label{e:dvphi}
&\approx \frac{3-\gamma}{2}\frac{H^2}{r^2} \sqrt{\frac{ M_{\SMBH}}{r}}
= \frac{3-\gamma}{2} (1.5\dot{m}_{\SMBH 1})^2 \R^{-5/2}\,,
\end{align}
where we have used Eqs.~(\ref{e:H}) and (\ref{v-phi-corr}).
This equation agrees with Tanaka et.~al.~\cite{2002ApJ...565.1257T} or the approximate equation of
Levin~\cite{2007MNRAS.374..515L}. This estimate, however,
does not hold for transsonic flows, as in this case the CO experiences a backwind,
as found by Chakrabarti~\cite{Chakrabarti:1995dw}. Since such flow requires very high
accretion rates where the thin disk approximation may not hold, we ignore this possibility here.

An azimuthal headwind leads to additional dissipation of the CO's specific angular momentum,
$\dot{\Ls}_{\rm wind} = r \dot{P}/m_{\CO}$,
where $\dot{P}$ is the rate of change of the linear momentum, so that
\be
 \dot{\Ls}_{\rm wind} = - \frac{r \,\dot{m}_{\CO} \delta v_{\phi}}{m_{\CO}}
 = -\frac{3-\gamma}{2} \frac{\dot{m}_{\CO} M_{\SMBH}}{m_{\CO}} (1.5\dot{m}_{\SMBH 1})^2 \R^{-3/2}\,,
 \label{Ldotwind-with-gamma}
\ee
where we have used Eq.~(\ref{e:dvphi}).
Clearly, this is typically a small perturbation relative even to the loss of angular
momentum through GW emission. For unsaturated BHL accretion
$\dot{m}_{\CO}=\dot{m}_{\CO}^{\Bondi}$, substituting Eq.~(\ref{e:MBondi}) yields
\be
 \frac{\dot{\Ls}^{\Bondi}_{\rm wind}}{\dot{\Ls}_{\GW}}
\approx
\left\{
\begin{array}{l}
6.1\times 10^{-10} \alpha_{1}^{-1} \dot{m}_{\SMBH 1}^{-3}M_{\SMBH 5} \R_{10}^{8}
\;\text{~for~$\alpha$-disks},\\
3.4\times 10^{-6} \alpha_{1}^{-4/5} \dot{m}_{\SMBH 1}^{-7/5} M_{\SMBH 5}^{6/5} \R_{10}^{59/10}
\text{\,for~$\beta$-disks}.
\end{array}
\right.
\label{e:L'wind}
\ee
For supply limited BHL accretion $\dot{m}_{\CO}=\dot{M}_{\SMBH}$, Eq.~(\ref{e:M-flux}), yields
\be
 \frac{\dot{\Ls}^{\rm sup.Bondi}_{\rm wind}}{\dot{\Ls}_{\GW}}
\approx
\left(\begin{array}{c}
1.0\times 10^{-7} \\
2.5\times 10^{-7}
\end{array}\right)
\times
\dot{m}_{\SMBH 1}^{3} \frac{M_{\SMBH 5}^3}{m_{\CO 1}^2} \R_{10}^{2}\,,
\label{e:L'wind-supply-limited}
\ee
where the top and bottom rows correspond to $\alpha$ and $\beta$-disks respectively.
The change in the angular momentum dissipation rate leads to a modified inspiral rate
that we discuss in Sec.~\ref{wind-GWimp}.

\subsection{Radial wind}

In addition to the azimuthal headwind,
the CO also experiences a wind in the radial direction. The corresponding force is
\begin{align}
\dot{P}_{\rm wind} &= \dot{m}_{\CO} \left|v_{\CO r} - v_{\gas, r}\right|\\
&\approx
 \frac{  \dot{M}_{\SMBH} m_{\CO}^{2}}{H \cs^3 r}
=
2.5\times 10^{-22} \dot{m}_{\SMBH 1}^{-3} M_{\SMBH 5}^{-1} m_{\CO 1}^2 \R_{10}^{7/2}\,,
\end{align}
where we have assumed $|v_{r \CO}| \ll |v_{\gas, r}|$, see Fig.~\ref{f:velocity}.
The radial equation of motion is thus
\be
m_{\CO} \Omega^2 r = \frac{m_{\CO} M_{\SMBH}}{r^2} + \dot{P}_{\rm wind}\,.
\ee
The first term on the RHS is the gravitational force, which satisfies
$m_{\CO} M_{\SMBH} /r^2 = m_{\CO} \Omega_{\rm vac}^2 r$, by the definition of $\Omega_{\rm vac}$.
Therefore, unlike the azimuthal wind, the radial wind does not dissipate angular momentum,
but it modifies the orbital velocity $\Omega_{\rm vac}$ relative to Keplerian:
\begin{align}
\frac{\Omega^2 - \Omega_{\rm vac}^2}{\Omega_{\rm vac}^2}
&= \frac{\dot{P}_{\rm wind}}{m_{\CO} \Omega_{\rm vac}^2 r}
= 2.5 \times 10^{-16} \frac{m_{\CO 1} \R_{10}^{11/2}}{\dot{m}_{\SMBH 1}^{3}}\,.
\end{align}
The impact of this modification on the GW phase will be discussed in Sec.~\ref{wind-GWimp}.

\subsection{Dynamical friction}
\label{s:dynamical-friction}
Dynamical friction is generated by the gravitational interaction of a perturber traveling at some relative velocity
in an ambient medium \cite{1943ApJ....97..255C}. The gravity of the perturber deflects the particles of the medium and generates a density wake trailing
the perturber. In turn, the gravitational pull of the density wake acts like friction, decreasing the speed of the perturber.
This process is analogous to Landau-damping in plasma physics and is also important in
galactic dynamics for objects moving through a population of stars \cite{1962MNRAS.124..279L,2008gady.book.....B}.

The standard treatment of dynamical friction in a gaseous medium usually assumes that the medium has a spatially
uniform initial velocity distribution relative to the perturber.
If the perturber moves on a linear trajectory with a subsonic relative velocity  in a gaseous medium, then the density
wake in front and behind the perturber approximately cancel, leading to a small drag force \cite{1999ApJ...513..252O}.
However, dynamical friction is more significant for a supersonic perturber.
Dynamical friction has also been investigated for perturbers moving on quasi-circular orbits in an initially static (i.e.~non-rotating)
medium \cite{2007ApJ...665..432K}. In this case the density wake has a spiral structure and the drag force is enhanced.
A fully relativistic treatment was presented in Refs.~\cite{2007MNRAS.382..826B,2008PhRvD..77j4027B}.

Equations~(\ref{e:vr}), (\ref{e:cs}), and (\ref{e:dvphi}) and Figure~\ref{f:velocity} show that the relative wind velocity at corotating orbits
is typically non-relativistic and subsonic in standard $\alpha$ and $\beta$-disks. The arguments mentioned above
then imply that the standard dynamical friction effect
is greatly suppressed.\footnote{Regular dynamical friction may be significant for standard $\alpha$ and $\beta$-disks for retrograde orbits
\cite{2011MNRAS.tmp..363N}.}

The relative gas velocity at different radii outside $r_{\CO}\pm \frac{2}{3}H$, however, is supersonic [see Eq.~(\ref{e:vdr})] and thus, dynamical friction with respect to the gas in this region may be significant. In this case, the velocity of the medium is mostly due to
differential rotation of the accretion disk and not the wind generated by the pressure gradient effects of Sec.~\ref{s:wind-azimuthal}.
Differential rotation causes the density waves to wind up and standard dynamical friction formulas are not applicable. This
regime has been well studied in planetary dynamics,
which leads to the phenomenon called planetary migration. In this paper, we explicitly distinguish between such migration, described separately in Sec.~\ref{s:migration} below, and standard dynamical friction, which as argued above has a negligible effect on LISA EMRIs.

\subsection{GW Implications}
\label{wind-GWimp}

Are such changes in the orbital dynamics measurable with EMRI GWs?
Let us first consider the effect of an azimuthal wind, for which the inspiral rate is increased with the timescale
${\cal{T}}_{\rm{a.wind}} \equiv \Ls/\dot{\Ls}_{\rm wind}$:
%
\be
{\cal{T}}_{\rm aw}^{\alpha} \approx
- 4.0 \times 10^{8} \; \yr \; \alpha_{1} \dot{m}_{\SMBH 1}^{3} M_{\SMBH 5}^{} m_{\CO 1}^{-1} \R_{10}^{4}\,,
\ee
for $\alpha$-disks and
\be
{\cal{T}}_{\rm aw}^{\beta} \approx
-7.1\times 10^{4} \; \yr \; \alpha_{1}^{4/5} \dot{m}_{\SMBH 1}^{7/5} M_{\SMBH 5}^{4/5} m_{\CO 1}^{-1} \R_{10}^{19/10}\,,
\ee
for $\beta$-disks. We have here used the fact that the specific angular momentum is
$\Ls = r^2 \Omega$ where $\Omega$ is given in Eq.~\eqref{kep-ang-vel}.

These timescales can now be used to estimate the GW dephasing. The effect of the headwind is to change the
orbital angular velocity of the EMRI system, i.e.~it induces a $\delta{\Omega}_{\rm{wind}}$. This change in frequency
can be approximated as $\delta \Omega_{\rm wind} \sim (\dot{\Ls}_{\rm wind}/\Ls) \Omega \delta t$, so the phase shift
is proportional to $\cal{T}_{\rm aw}=\dot{\Ls}_{\rm wind}/\Ls$.
Such a change in orbital frequency induces a change in orbital phase, which we can approximate as
$\delta \phi_{\rm aw} \sim \delta \Omega_{\rm wind} \delta t \sim (\delta t/{\cal{T}}_{\rm aw}) \phi_{\GW}^{\Tot}$.
For $\alpha$-disks we find
$\delta \phi_{\rm aw}^{\alpha} \approx 0.03 \; {\rm{rads}}$, while for $\beta$-disks we find
$\delta \phi_{\rm aw}^{\beta} \approx 33 \; {\rm{rads}}$, using a typical set of EMRI and disk parameters and and initial separation of $\R_{0} = 20$.

A more accurate measure of the GW phase shift can be obtained by integrating the perturbation to the inspiral
rate due to the hydrodynamic drag. The results of a similar calculation was presented in Paper I~\cite{2011arXiv1103.4609Y}
for a general model, of which the azimuthal wind is a special case. We provide the details of the derivation in
Appendix~\ref{azi-wind-estimate} below. The asymptotic analytical solution is presented in
Eqs.~\eqref{e:long-gen-form}-\eqref{e:short-gen-form}, where the coefficient and exponents are given in Table~\ref{Table-pars4}.
For the default parameters, we get that the perturbation to the GW phase is
$\delta \phi_{\rm aw}^{\alpha} \sim 0.03\, {\rm rad}$ for $\alpha$-disks,
$\delta \phi_{\rm aw}^{\beta} \sim 47\,{\rm rad}$ for $\beta$-disks,
assuming BHL accretion is not saturated for the final year of inspiral. Including the limitation by the gas supply does limit the process leading to
$\delta \phi_{\rm aw}^{\alpha} \sim 0.15\, {\rm rad}$ for $\alpha$-disks and $\delta \phi_{\rm aw}^{\beta}\sim 0.35\, {\rm rads}$ for $\beta$-disks.
The result of the full calculation including all other saturation effects is shown in Figures~\ref{fig:dephasing-analytics-intro} and
\ref{fig:dephasing-analytics} for different EMRI masses and final inspiral radii for a one year observation.
We conclude then that the azimuthal wind affects the EMRI signal only marginally for typical parameters.

Let us now consider the effect of a radial wind on the GW phase. Since the GW phase is proportional to twice the orbital phase for quasi-circular orbits, the total GW phase shift induced by the radial wind is roughly
\begin{align}
\delta {\phi}_{\GW}^{\rm rad.~wind}  &= 2(\Omega - \Omega_{\rm vac}) T_{\rm obs}
\approx \frac{  \dot{M}_{\SMBH} m_{\CO}}{ \Omega_{\rm vac} H \cs^3 r^2} T_{\rm obs}
\\  &= 5.0\times 10^{-10} \; {\rm rads} \;M_{\SMBH 5}^{-1} m_{\CO 1} \dot{m}_{\SMBH 1}^{-3} \R_{10}^{4}\,.
\label{e:dphi-rpgr}
\end{align}
This radial wind is thus completely negligible for the LISA measurement unless
$\R \gtrsim 1000$.\,\footnote{Note that $\dot{m}_{\SMBH 1}\ll 1$ is unrealistic for radiatively efficient, thin disk models, but even so, BHL accretion would be quenched by the limited gas supply, as explained in Sec.~\ref{sec:Bondi:limited-gas}.}

\section{Axisymmetric Gravitational Effects}
\label{sec:gravitationaleffects}

The axisymmetric component of the disk gravity induces several effects in the orbital evolution of an EMRI:
it modifies the angular velocity of
the orbit and the inspiral rate and induces additional apsidal precession for eccentric orbits.
We examine these effects here in turn, making order of magnitude estimates of the corresponding GW phase shifts.

\subsection{Accretion Disk Potential}

The gravitational potential of a thin disk may be much stronger than the isotropic component
of the enclosed mass, $M_{\rm disk}(\R)/r$. This is to be expected since a thin ring exerts a much stronger
force than a spherical shell of the same mass, which can point both in or out for
exterior and interior test particles, respectively. Here we estimate the Newtonian
gravitational potential of a stationary planar disk.

The total potential of the disk is a superposition of the
contributions of infinitesimal concentric rings of mass $\D m = 2\pi
r\, \D r\Sigma $. Using dimensionless radius variables,
\be\label{e:Phidisk} \Phi_{\rm disk}(\R) = -M_{\SMBH}
\int_{\R_{\min}}^{\R_{\max}} \Sigma(\R_0) \frac{4 \R_0}{\R+\R_0}
K\hspace{-3pt}\left(\frac{2 \sqrt{\R \R_0}}{\R+\R_0}\right) \D \R_0.
\ee where the surface density $\Sigma(\R)$ is to be substituted from
Eq.~(\ref{e:Sigma_definition}) or (\ref{e:Sigma}). We here used the
fact that a circular ring of mass $\D m$ and radius $r_0$ generates a
potential at $r$ as given in Eq.~\eqref{SCOpotential}.  Here
$\R_{\min}$ and $\R_{\max}$ are the inner and outer radii defining the
radial extent of the disk, we assume $\R_{\min}\sim 0$ and
$\R_{\max}\sim \R_{\rm rad}$ in practice [Eq.~(\ref{e:rmax})].

For inclined orbits, with the CO angular momentum vector $\bm{L}_{\CO}$
at inclination $\iota$ relative to the total angular momentum vector of the CO and the disk,
the potential generated by the disk can be expressed more conveniently with the Legendre
polynomials $P_{\ell}(x)$
\begin{align}
\Phi_{\rm disk}(\R)
 &= -\sum_{\ell=0}^{\infty} 2\pi [P_{2\ell}(0)]^2 P_{2\ell}(\cos \iota)\nonumber\\
&\quad\times M_{\SMBH}\int_{\R_{\min}}^{\R_{\max}} \Sigma(\R_0) \R_0 \frac{\R_<^{2\ell}}{\R_>^{2\ell + 1}}
\D \R_0\,,
\label{e:Phidisk2}
\end{align}
where $\R_< = \min(\R,\R_0)$ and $\R_> = \max(\R,\R_0)$ \cite{2010arXiv1006.0001K}.
The integrals in Eq.~(\ref{e:Phidisk2}) can be simply evaluated analytically for the particular
form of $\Sigma(\R)$ given by Eq.~(\ref{e:Sigma}). In Appendix~\ref{s:potential-app}, we carry out this
exercise and show that in the limit $\R_{\min} \ll \R \ll \R_{\max}$,
\begin{align}\label{e:phi-alpha}
\Phi_{\alpha\text{-disk}}(\R)
&\approx
 - \pi \Sigma_{\alpha 0} \sqrt{\R_{\max}}\,
\R^2\,, \\\label{e:phi-beta}
\Phi_{\beta\text{-disk}}(\R) &\approx
2\pi c_{0} \Sigma_{\beta 0}\, \R^{2/5}\,,
\end{align}
where $c_0=1.38$, and Eq.~(\ref{e:Sigma}) was used to define the dimensionless density scales
$\Sigma_{\alpha 0}$ and $\Sigma_{\beta 0}$ as
$M_{\SMBH}\Sigma(\R) = \Sigma_{\alpha 0} \R ^{3/2}$ and $\Sigma_{\beta 0} \R ^{-3/5}$, for $\alpha$ and $\beta$-disks, respectively.
Substituting $\R_{\max} = \R_{\rm rad}$ from Eq.~(\ref{e:rmax}), we get
\be\label{e:phi-disk0}
\frac{\Phi_{\rm disk}(\R)}{\Phi(\R)}
\approx
\left\{
\begin{array}{l}
-1.4\times 10^{-12} \alpha_1^{-20/21} \dot{m}_{\SMBH 1}^{-13/21} M_{\SMBH 5}^{22/21} \R_{10}^{3}\\
\quad\text{~for~$\alpha$-disks}\,,\\
1.2\times 10^{-9} \alpha_1^{-4/5} \dot{m}_{\SMBH 1}^{3/5} M_{\SMBH 5}^{6/5} \R_{10}^{7/5} \\
\quad\text{~for~$\beta$-disks}\,,
\end{array}
\right.
\ee
where we have used $\Phi(\R)=\R^{-1}$ for the SMBH potential.

Equation~(\ref{e:phi-disk0}) shows that the disk potential decreases outwards and inwards for $\alpha$ and
$\beta$-disks. Thus interestingly, the disk exerts an outward force on the CO for $\alpha$-disks.
In Appendix~\ref{s:potential-app}, we show that this is due to  a strong quadrupolar field generated by the outskirts of
a radiation-pressure dominated $\alpha$-disk.

\subsection{Change in the orbital frequency}

The orbital angular frequency $\Omega(\R)$ is modified due to the axisymmetric gravitational field of the disk
from its value without the disk $\Omega_{\rm vac}$ given by Eq.~\eqref{kep-ang-vel}.
Equating the centripetal acceleration to the gradient of the gravitational potential we find
\be
r \Omega^{2} = \frac{M_{\SMBH}}{r^{2}} + \frac{d \Phi_{\rm disk}}{dr}
 = r \Omega_{\rm vac}^{2} + \frac{d \Phi_{\rm disk}}{dr}\,.
\ee
Therefore the orbital frequency is
\be
\Omega \approx \Omega_{\rm vac} \left(1 + \frac{\R^{2}}{2} \frac{d \Phi_{\rm disk}}{d\R}\,\right)\,.
\label{e:Omega-SG}
\ee

\subsection{Change in the inspiral rate}

The binding energy in the Newtonian approximation is
\be\label{e:ESCO}
E = \frac{1}{2} \Omega^{2} r^{2} \eta M - \frac{\eta M^{2}}{r} + m_{\CO} \Phi_{\rm disk}\,,
\ee
where $M$ is the total mass and $\eta = M_{\SMBH} m_{\CO}/M^{2}$ is the symmetric mass-ratio.
For EMRIs, $\eta M \approx m_{\CO}$ and $M \approx M_{\SMBH}$.
Assuming no other source of energy loss, other than the GW emission, and quasi-circular orbits,
\be\label{e:dotE-general}
\frac{dE}{dt} = - \frac{32}{5} \eta^{2} M^{2} r^{4} \Omega^{6}\,.
\ee

We can use Eq.~(\ref{e:Omega-SG}) to express the radial and time derivatives of the energy explicitly as a
function of radius alone:
\ba
\frac{dE}{d\R} &=& \frac{\eta M}{2 \R^{2}} \left( 1 + 3 \R^{2} \frac{d \Phi_{\rm disk}}{d\R}
+ \R^{3} \frac{d^{2} \Phi_{\rm disk}}{d \R^{2}} \right)\,,
\label{dEdr-SG}
\\
\frac{dE}{dt} &=& - \frac{32}{5} \eta^{2} \R^{-5} \left(1 + 3 \R^{2} \frac{d \Phi_{\rm disk}}{d\R} \right)\,.
\label{dEdt-SG}
\ea
Therefore the radial velocity is modified as
\be\label{e:dr-sG}
\frac{d \R}{dt} = \left(\frac{dE}{d\R}\right)^{-1} \frac{dE}{dt}
= \dot{\R}_{\rm vac} \left(1 - \R^{3} \frac{d^{2} \Phi_{\rm disk}}{d\R^{2}} \right)\,,
\ee
where $\dot{\R}_{\rm vac} $ is the GW inspiral rate neglecting the effects of the disk [Eq.~(\ref{e:dotr-inspiral})]. Notice that the first derivative of the disk potential does not enter the radial velocity.

\subsection{Apsidal precession}
While clearly unimportant for quasi-circular EMRIs, we briefly show that
the axisymmetric disk gravity causes a negligible amount of apsidal precession in
the eccentric case too.

The radial oscillation frequency for a nearly circular orbit in a perturbed potential is
\be\label{e:kappa}
\kappa^2(r) = \frac{M_{\SMBH}}{r^3} + \frac{d^2 \Phi_{\rm disk}}{dr^2} + \frac{3}{r}\frac{d \Phi_{\rm disk}}{dr}\,,
\ee
(see, e.g.~\cite{2008gady.book.....B}). If $\kappa \neq \Omega$,
the ellipse precesses in its plane at a rate $\Omega - \kappa$.
The dimensionless apsidal precession rate relative to the Keplerian frequency from Eqs.~(\ref{e:Omega-SG})
and (\ref{e:kappa}) is thus
\be\label{e:Omega-ap2}
\frac{\Omega_{\rm ap}}{\Omega_{\rm vac}}  \equiv \frac{\Omega - \kappa}{\Omega_{\rm vac}} =
-\frac{\R^3 }{2}\left(\frac{d^2 \Phi_{\rm disk}}{d \R^2} + \frac{2}{\R}\frac{d \Phi_{\rm disk}}{d\R}\right)\,.
\ee
We relate $\Omega_{\rm ap}/\Omega_{\rm vac}$ to the induced phase shift below; 
this quantity is proportional to $\Phi_{\rm disk}/\Phi \ll 1$.

\subsection{GW Implications}

The disk potential changes the GW frequency and the inspiral rate at
fixed orbital radii. The dimensionless change in the frequency $\delta
\Omega = (\Omega - \Omega_{\rm vac})/\Omega_{\rm vac}$, inspiral rate
$\delta \dot{\R} = (\dot{\R} - \dot{\R}_{\rm vac})/\dot{\R}_{\rm
vac}$, and the dimensionless apsidal precession rate $\Omega_{\rm
ap}/\Omega_{\rm vac}$ are all proportional to derivatives of
$\Phi_{\rm disk}(\R)/\Phi(\R)$, up to a factor of order unity [see
Eqs.~(\ref{e:Omega-SG},\ref{e:dr-sG},\ref{e:Omega-ap2})].
Equation~(\ref{e:phi-disk0}) shows that this ratio is typically very
small, at the level of $\Phi_{\rm disk}/\Phi \sim (2 \times 10^{-6},
10^{-5})$ for $(\alpha,\beta)$--disks with very small $\alpha$ and
large $M_{\SMBH}$ and $\R$, $(\alpha,M_{\SMBH},\R)\sim
(10^{-3},10^6\Msun,100)$, and even smaller for more typical
values. This leads to a small change in the GW phase of
Eq.~\eqref{e:phiGW}. In Appendix~\ref{SG-estimate}, we carry out a
detailed calculation that shows that for both $\alpha$ and
$\beta$-disks, self-gravity effects induce a dephasing of
approximately $10^{-4}-10^{-7}$\,rads for our nominal set of
parameters in a one year observation, which is clearly below LISA's
observational threshold.

These estimates are modified if the gravitational torques from the CO quenches gas inflow onto the
SMBH, and clears a gap (Sec.~\ref{s:gap}). In this case, the gas density is greatly reduced interior to the CO, but
extra gas accumulates outside its orbit, changing the potential. In Appendix~\ref{SG-estimate}
we show that this effect increases the dephasing by roughly an order of magnitude for $\beta$-disks, but such
an increase is still well below the threshold of detectability.
Axisymmetric gravitational effects may be more important for eccentric orbits with large semi-major axis $\R\sim 1000$.

\section{Migration}\label{s:migration}

Let us now examine the role of non-axisymmetric gravitational effects
induced by the disk, which lead to a phenomenon known as ``migration''
in planetary dynamics. As before, the goal is to make order of
magnitude estimates of the migration effect on the GW phase and
compare them with the detectability measures of Sec.~\ref{LISA-sense}.

\subsection{General Properties}

Consider the non-axisymmetric gravitational effects of the disk,
leading to angular momentum dissipation, analogous to planetary
migration (see \cite{1997Icar..126..261W,2007astro.ph..1485A} for
reviews).  The orbiting CO exerts a nonzero average gravitational
torque on the gaseous disk, creating a spiral density wave. The total
angular momentum budget is dissipated through viscosity and the
outward angular momentum transport of the spiral density wave. The
gravitational torque of the spiral density wave exchanges angular
momentum resonantly with the CO causing it to migrate
\cite{1980ApJ...241..425G}.  This effect is analogous to dynamical
friction (see Sec.~\ref{s:dynamical-friction}), but accounts for the
inhomogeneous velocity field of the gaseous medium.

In planetary dynamics, this phenomenon in different regimes is called
Type-I, Type-II, and Type-III migration.
The distinction is whether a gap is opened (Type-II) or not (Type-I) and whether
strong corotation torques related to horseshoe orbits are taken into account (Type-III).
In the following we neglect Type-III migration because it is most relevant only
if the disk mass near corotating orbits is comparable or larger than the secondary mass.
Figure~\ref{f:mass} shows that this is not the case for EMRIs in the LISA frequency band.

In a pioneer paper, Goldreich and Tremaine presented the first study on Type I migration,
using 2 dimensional (2D) linear perturbation theory \cite{1980ApJ...241..425G}.
Their results were later improved by Tanaka et.~al.~to account for corotation resonances
and 3D effects in isothermal disks \cite{2002ApJ...565.1257T}, which are also consistent with 3D
hydrodynamic simulations of laminar disks \cite{2003ApJ...586..540D}.
Further generalizations exist for locally adiabatic 2D disks \cite{2010MNRAS.401.1950P}.
However, recent  MHD shearing box simulations of turbulent protoplanetary disks with a low mass planet
show that the torques exhibit stochastic fluctuations for Type I migration, where
even the sign of the torque changes rapidly \cite{2004MNRAS.350..849N,2004ApJ...608..489L}.
Stochastic migration is not expected if the satellite
is more massive relative to the disk.
The migration is also very sensitive to the value of opacity and radiation processes \cite{2006A&A...459L..17P}.
Recently, Hirata generalized the original study of Goldreich and Tremaine for laminar 2D relativistic
disks and found the angular momentum transport to be larger by a factor $\sim 4$
close to the central BH \cite{2010arXiv1010.0758H,2010arXiv1010.0759H}.
Finally, higher order PN corrections may resonantly excite persistent spiral density waves
close to the SMBH even without an $m_{\CO}$, which could modify the torque estimates for EMRIs \cite{2006MNRAS.370L..42A}.

To our knowledge, Type-I migration is unexplored for the expected environments for LISA EMRI sources:
for radiation-pressure-dominated, optically-thick, geometrically-thin, relativistic, magnetized and turbulent disks,
where the mass of the perturbing body $m_{\CO}$ exceeds the disk's mass.
For a simple estimate below, we explore the GW phase shift in an EMRI system due to Type I migration,
using the isothermal non-relativistic laminar formulas of Tanaka et.~al \cite{2002ApJ...565.1257T},
where we include the effects of radiation pressure by using the corresponding expressions for the
sound speed and the scale height.

If the CO is sufficiently massive and/or far out, a central region is cleared out and a
circular gap develops (see Sec.~\ref{s:gap}).
Once a gap opens, the angular momentum exchange becomes much more regular than Type-I and it is determined by the angular momentum transport through the more distant part of the
disk~\cite{2003ApJ...586..540D,2004MNRAS.350..849N,2004ApJ...608..489L}. This type of
angular momentum exchange is referred to as Type II migration. If the local mass of
the disk is greater than the mass of the perturbing object, then the object migrates inward
on a viscous timescale with the velocity of the accreting gas.
However, for LISA EMRI sources, the local disk mass is smaller than the CO's (see Fig.~\ref{f:mass}).
In this case, the migration slows down, and becomes ``secondary dominated Type-II migration''
\cite{2007astro.ph..1485A}.

Simple steady-state estimates, based on angular momentum balance,
were presented by Syer and Clarke~\cite{1995MNRAS.277..758S}.
These estimates account for the increase of the gas density relative to an isolated disk, assuming that the angular momentum
exchange is dominated by that near the inner-edge of the disk.
Later, Ivanov et.~al.~\cite{1999MNRAS.307...79I} relaxed the steady-state assumption and estimated the
quasi-stationary, time-dependent evolution of the disk and satellite using a zero stress boundary condition
at the location of the binary.

Both of these studies focused on thin, one-zone, gas pressure dominated, Shakura-Sunyaev
disks, where the density is an increasing function of stress. This
assumption is satisfied for $\beta$-disks but not for radiation-pressure dominated
$\alpha$-disks. In the later case, we are not aware of literature that is applicable to Type II migration.
Recent studies have shown that migration with an annular gap is affected by global edge-modes for massive
disks, and can cause stochastic migration of planets either inward or outward (similar to the Type I case with MHD)
\cite{2011arXiv1103.5036L}, and the migration rate is also sensitive to vortex forming instabilities \cite{2011arXiv1103.5025L}.
However, these phenomena have not yet been explored in AGN accretion disks for EMRIs.
For an order of magnitude analysis of Type II migration, we use both Syer and Clarke and the asymptotic Ivanov et.~al.~equations to
estimate the corresponding GW phase shifts for EMRIs in $\beta$-disks. We
make a conservative estimate for radiation-pressure dominated $\alpha$-disks, based on angular momentum
balance between $m_{\CO}$ and the local disk, neglecting the accumulation of gas outside the gap.

\subsection{Type-I migration}

For an isothermal 3D disk, Type I migration removes or increases angular momentum at a rate
\begin{align}
\dot{\Ls}_{\rm mig,I} = \pm c_1
\frac{m_{\CO} \Sigma r^6 \Omega^4 }{M_{\SMBH}^2 \cs^2} =
\pm c_1 \frac{m_{\CO}}{M_{\SMBH}} \Sigma  \frac{r^3}{H^2}\,,
\end{align}
where $c_1=(1.4-0.5\gamma)$ and  $\gamma=3/2$ or $-3/5$ for $\alpha$ and $\beta$-disks, respectively,
and the $\pm$ signs highlight the stochastic nature of migration in a turbulent disk
\cite{2002ApJ...565.1257T}.
The magnitude of $c_1$ is different for locally adiabatic 2D disks by a factor $\sim 2$, but the
scaling with other parameters remains the same~\cite{2010MNRAS.401.1950P}.
Relative to the GW rate of angular momentum loss,
\begin{align}
\label{type-1-ldot}
\frac{\dot{\Ls}_{\rm mig,I}}{\dot{\Ls}_{\GW}} =
\left\{
\begin{array}{ll}
\pm 7.2\times 10^{-11}
\alpha_1^{-1} \dot{m}_{\SMBH 1}^{-3} M_{\SMBH 5} \R_{10}^{8}\;\text{~for~$\alpha$-disks}\,, \\
\pm 5.1\times 10^{-7}
\alpha_1^{-4/5} \dot{m}_{\SMBH 1}^{-7/5} M_{\SMBH 5}^{6/5} \R_{10}^{59/10}
&\\\quad \text{~for~$\beta$-disks}\,. \\
\end{array}\right.
\end{align}
Notice that the Type I migration dominates over even the GW loss of angular momentum at
sufficiently large radii, beyond $\R\approx150$ for $\alpha$-disks and
$\R\approx112$ for $\beta$-disks.

\subsection{Type-II migration}
If the CO is sufficiently massive and/or far out that a gap opens (see Sec.~\ref{s:gap})
the CO is subject to Type II migration.
The angular momentum exchange depends on the local disk mass near the binary, $m_{\rm d}$
given by Eq.~(\ref{e:md}).
Typically, $m_{\rm d}< m_{\CO}$ (see Fig.~\ref{f:mass}), and the specific angular momentum dissipation rate,
in the quasi-stationary approximation of Syer \& Clarke \cite{1995MNRAS.277..758S},
is
\be
\dot{\Ls}_{\rm mig,II,SC} =
 \left(\frac{m_{\rm d}}{m_{\CO}}\right)^{k} \dot{\Ls}_{\rm gas} = - \left(\frac{m_{\rm d}}{m_{\CO}}\right)^{k}
\frac{v_{\rm orb} v_{\gas, r}}{2}\,,
\label{e:Syer-Clarke}
\ee
where $\dot{\Ls}_{\rm gas}$ is the angular momentum loss in the gas due to viscosity,
$v_{\gas, r} = 2\dot{M}_{\SMBH} r/m_{\rm d}(r)$ is the radial velocity of gas given by Eq.~(\ref{e:vr}), $v_{\rm orb}$ is the orbital velocity
Eq.~(\ref{kep-ang-vel}), and
$k=3/8$ for electron-scattering opacity and $\beta$-disks. If neglecting
the banking up of gas near the outer edge of the gap, then angular momentum
balance implies $k=1$, which we adopt conservatively for $\alpha$-disks [see eq.~(42) in Ref.~\cite{1999MNRAS.307...79I}].
We note that this assumption is different than those used in Refs.~\cite{2005ApJ...634..921A,2009ApJ...700.1952H}.
Substituting Eqs.~(\ref{e:Lgw}), (\ref{e:md}), and (\ref{e:vr}),
%
\be
\frac{\dot{\Ls}_{\rm mig,II,SC}}{\dot{\Ls}_{\GW}} =
\left\{
\begin{array}{l}
6.2 \times 10^{-6} \; \dot{m}_{\SMBH 1} M_{\SMBH 5}^{3} m_{\CO 1}^{-2} \R_{10}^{4}
\\ \; \quad \text{for~$\alpha$-disks}\,,
\\
5.8\times 10^{-3}\, \alpha_{1}^{1/2} \dot{m}_{\SMBH 1}^{5/8} M_{\SMBH 5}^{13/8} m_{\CO}^{-11/8} \R_{10}^{25/8}
\\ \; \quad \text{for~$\beta$-disks}\,.
\end{array}
\right.
\label{ldot-TypeII,SC}
\ee
Notice that for $\alpha$-disks, this is independent of $\alpha$; the dissipation of the CO's angular
momentum is proportional to $\dot{M}_{\SMBH}$. For $\beta$-disks, Eq.~(\ref{ldot-TypeII,SC}) also accounts for the
accumulation of mass near the CO, which leads to additional angular momentum dissipation, sensitive to $\alpha$.

We also consider the time-dependent solution of Ivanov et.~al.~\cite{1999MNRAS.307...79I} for $\beta$-disks
that accounts for the accumulation of gas and a zero stress boundary condition at the edge of the gap
(see also Eq.~(24) in Ref.~\cite{2011MNRAS.411.1467K}):
\be\label{e:Ivanov}
\dot{\Ls}_{\rm mig,II,IPP}^{\beta} =  -c_2 v_{\rm orb}  \frac{\dot{M}_{\SMBH}}{m_{\CO}}(r \; r_{\rm dd})^{1/2}\,,
\ee
where $r_{\rm dd}$ is the ``radius of disk dominance'', which satisfies $m_{\rm d}(r_{\rm dd}) = m_{\CO}$ [see Eq.~(\ref{e:md})],
$c_2\equiv \{1+\delta[1-(r/r_{\rm dd})^{1/2}]\}^{k_2}$, $\delta = 6.1$, and $k_2=0.26$.
Note that $c_2$ is only mildly $r$ dependent. For $r=r_{\rm dd}$, $c_2=1$ so that both estimates
[Eqs.~(\ref{e:Syer-Clarke}) and (\ref{e:Ivanov})] imply a migration rate tracking the radial velocity of gas, so that
$\dot{\Ls}_{\rm mig,II,IPP}^{\beta}=\dot{\Ls}_{\rm mig,II,SC}^{\beta}=\dot{\Ls}_{\rm gas}=-v_{\rm orb}v_{\gas, r} /2$,
while for $r\ll r_{\rm dd}$, $c_2\approx 1.66$, asymptotically independent of $r$.
Similar to Eq.~(\ref{e:Syer-Clarke}), Eq.~(\ref{e:Ivanov}) is valid for electron scattering
opacity $\beta$-disks, but is not applicable for radiation-pressure dominated $\alpha$-disks.
Substituting Eq.~(\ref{e:md}) for $r_{\rm dd}$, we find for $r\ll r_{\rm dd}$,
\be
\frac{\dot{\Ls}_{\rm mig,II,IPP}^{\beta}}{\dot{\Ls}_{\GW}} =  5.1 \times 10^{-4}\, \alpha_{1}^{2/7} \dot{m}_{\SMBH 1}^{11/14} M_{\SMBH 5}^{31/14} m_{\CO 1}^{-23/14} \R_{10}^{7/2}\,.
\label{ldot-TypeII,IPP}
\ee

\subsection{Quenching of migration}
\label{s:migration:quench}

In the following we assume that inside and outside the minimum gap
opening radius $r_{\rm gap}$ [Eq.~(\ref{e:rgap1}--\ref{e:rgap2})]
Type-I and Type-II migration operate, where we use
Eq.~(\ref{type-1-ldot}) and Eqs.~(\ref{ldot-TypeII,SC}---\ref{ldot-TypeII,IPP}),
respectively.  Once the CO has crossed inside
$r_{\rm gap}$, the gas is no longer expelled efficiently by the tidal
field of the CO, and the gas is free to flow in on the accretion
timescale with a radial velocity $v_{{\rm gas}, r}$
[Eq.~(\ref{e:vr})]. If this is faster than the GW inspiral rate of the
CO $v_{\CO r}$ [Eq.~(\ref{e:dotr-inspiral})], then the gap can
refill. In Sec.~\ref{s:gap:longrange}, we have shown that $|v_{\CO
r}| \leq |v_{{\rm gas}, r}|$ is satisfied if $\R\geq \R_{\rm d}$, given by
Eq.~(\ref{e:rtr}).  Thus, if $\R=\R_{\rm gap}\geq \R_{\rm d}$ is met
[see Eqs.~(\ref{e:rgap1}--\ref{e:rgap2})], we assume that the gap
refills, and switch from the disk generated torque from Type-II to
Type-I.  Otherwise, if a gap has formed and cannot follow the inspiral
rate of the CO, disk torques are expected to shift out of resonance,
and become too distant for efficient angular momentum exchange. Then
the interaction is greatly suppressed, and we assume $\dot{\Ls}_{\rm
mig}(\R)=0$ if $\R_{\rm gap}\leq\R\leq\R_{\rm d}$. Once the gap has
formed and decoupled, it can no longer refill, so we further assume
$\dot{\Ls}_{\rm mig} = 0$ if $\R\leq \R_{\rm gap}\leq\R_{\rm d}$.
Finally, if the gap refills (i.e.~$\R_{\rm gap}\geq \R_{\rm d}$) then
Type-I migration can operate efficiently even if the inspiral rate is
faster than the viscous inflow rate ($\R\leq \R_{\rm d}$), since the
spiral patterns can form much faster on a dynamical timescale $t_{\rm
dyn}\sim \Omega^{-1}$.  The interaction may be significantly different
only much later when the inspiral time $t_{\rm GW}\equiv
r/\dot{r}_{\GW}$ becomes faster than the cooling time of the disk
\cite{1999agnc.book.....K}, i.e.~
\be
\label{e:tcool}
t_{\rm cool } \equiv \frac{1}{\alpha\Omega}\,,
\ee
where $\dot{r}_{\GW}$ is the GW
driven inspiral rate, $\Omega$ is the orbital angular velocity and
$\alpha$ is the parameter in the viscosity prescription.  From
Eqs.~(\ref{kep-ang-vel}), (\ref{e:dotr-inspiral}), and (\ref{e:tcool})
we get $t_{\rm GW}<t_{\rm cool}$ inside
\be
\R_{\rm c} \equiv 0.055\,\alpha_{1}^{-2/3} M_{\SMBH 5}^{-2/3} m_{\CO 1}^{2/3}\,.
\ee
Typically the disk cooling does not impose a limitation outside the ISCO for EMRIs unless $\alpha$ is
very small and the mass-ratio is not extreme.

To summarize we assume,
\begin{align}
\dot{\Ls}_{\rm mig}' =
&
\left\{
\begin{array}{ll}
\dot{\Ls}_{\rm mig, I}\,, & {\rm~if~} \R<\R_{\rm gap}\,,\\
\dot{\Ls}_{\rm mig, II}\,,& {\rm~if~} \R>\R_{\rm gap}\,,
\end{array}
\right.
\label{e:mig-decoup1}
\end{align}
and
\begin{align}
\dot{\Ls}_{\rm mig} =
&
\left\{
\begin{array}{lc}
0 &{\rm~if~}\R<\R_{c} {~\rm~or~} \R_{\rm gap}<\R<\R_{\rm d} \\
  &{\rm~or~}\R<\R_{\rm gap}<\R_{\rm d}\,,\\
\dot{\Ls}_{\rm mig}' & {\rm otherwise}\,,
\end{array}
\right.
\label{e:mig-decoup}
\end{align}
where for $\dot{\Ls}_{\rm mig, I}$ we use Eq.~(\ref{type-1-ldot}), while
for $\dot{\Ls}_{\rm mig, II}$ either the Syer-Clarke model Eq.~(\ref{ldot-TypeII,SC})
or the Ivanov et.~al.~model (\ref{ldot-TypeII,IPP}). The later is only available for $\beta$-disks in the
radiation-pressure dominated regime, while the Syer-Clarke is applicable both for $\alpha$ or $\beta$-disks.
In summary, we consider three cases for migration, which utilize the same Type-I model, but differ
in the Type-II regime
\begin{enumerate}
\item M $\alpha_{\rm SC}$: $\alpha$-disks with Syer-Clarke Eq.~\eqref{ldot-TypeII,SC};
\item M $\beta_{\rm SC}$: $\beta$-disks with Syer-Clarke Eq.~\eqref{ldot-TypeII,SC};
\item M $\beta_{\rm IPP}$: $\beta$-disks with Ivanov et.~al.~Eq.~\eqref{e:Ivanov}.
\end{enumerate}

\subsection{GW Implications}

The change in the angular momentum dissipation rate modifies the GW driven inspiral rate, which modifies the GW
phase evolution. The corresponding phase shift $\delta\phi_{\rm GW}$ can be calculated in a similar way as for an azimuthal
wind (see Sec.~\ref{wind-GWimp} above, and Paper I~\cite{2011arXiv1103.4609Y}).
The interested reader can find the details of the derivation in Appendix~\ref{azi-wind-estimate}.

For the default parameters, we find that Type-I migration produces a typical GW dephasing on the
order of $10^{-2}$ and $10$ radians during the final year of observation in $\alpha$ and $\beta$-disks, respectively.
The large increase in dephasing for the $\beta$-disk case is due to the much larger surface density.

Even more interesting is the case where a gap opens; Type-II migration is clearly the most dominant
perturbation among all the ones considered in this paper.
However, as showed in Sec.~\ref{s:gap}, a gap is expected to close for most EMRIs in the LISA band.
For $\alpha$-disks, Type-II migration can generate GW phase shifts of order $10-100$ radians in one year,
and up to $10^{3}-10^{4}\,$rad for $\beta$-disks. The modifications to the GW spectrum are so large that if this
effect is in play, that vacuum EMRI templates might be ineffective to extract LISA GWs.
The blue lines in Figure~\ref{fig:dephasing-analytics-intro} and \ref{fig:dephasing-analytics} show the phase shift
due to migration as a function of final orbital radius including quenching effects [Eq.~(\ref{e:mig-decoup})]
for both Type-II migration models (Syer \& Clarke steady state and the
Ivanov et.~al.~quasi-stationary models, Eq.~(\ref{ldot-TypeII,SC}) and (\ref{ldot-TypeII,IPP}), respectively), assuming
$\alpha_1=\dot{m}_{\SMBH 1}=1$. For other system parameters and observation times, we provide asymptotic
analytical expressions in Eqs.~(\ref{e:long-gen-form}--\ref{e:short-gen-form}) and Table~\ref{Table-pars4} below.

\section{Comparison of Accretion Disk Effects}
\label{sec:comparison}

\begin{table}
\begin{ruledtabular}
\begin{tabular}{lcccc}
 & $(10,10^{5})$ & $(10^{2},10^{5})$ & $(10,10^{6})$ & $(10^{2},10^{6})$\\
\hline
Primary acc. & $1.0\,(-3)$ & $1.0\,(-3)$ & $1.0\,(-3)$ & $1.0\,(-3)$\\
\hline
BHL $\alpha$ & $1.9\,(+0)$ & $4.6\,(-3)$ & $5.7\,(-3)$ & $2.9\,(-3)$\\
BHL $\beta$ & $4.6\,(+0)$ & $0.0\,(+0)$ & $1.8 \,(+1)$ & $7.0\,(+0)$\\
\hline
W $\alpha$ & $1.5\,(-2)$ & $8.0\,(-5)$ & $8.0\,(-4)$ & $1.1\,(-4)$\\
W $\beta$ & $1.4\,(-1)$ & $0.0\,(+0)$ & $3.2\,(-1)$ & $8.3\,(-1)$\\
\hline
SG $\alpha$  & $2.5\,(-5)$ & $4.9\,(-5)$ & $4.5\,(-6)$ & $8.7\,(-6)$\\
SG  $\beta$ & $6.8\,(-4)$ & $5.5\,(-4)$ & $1.1\,(-3)$ & $8.9\,(-4)$ \\
\hline
M\,$\alpha_{\rm SC}$& $6.2\,(-3)$ & $1.8\,(-1)$ & $2.4\,(-6)$ & $9.6\,(-5)$ \\
M\,$\beta_{\rm SC}$ & $6.9\,(+2)$ & $1.8\,(+2)$ & $7.3\,(-2)$ & $8.8\,(-1)$ \\
M\,$\beta_{\rm IPP}$ & $8.2\,(+1)$ & $1.4\,(+1)$ & $7.3\,(-2)$ & $8.8\,(-1)$ \\
\end{tabular}
\caption{\label{Table-acc-disk-effec}Summary of accretion disk effects on the GW phase shift
induced by different accretion disk effects relative to vacuum waveforms. Rows correspond to different
accretion disk effects: Primary Eddington-limited accretion (Primary acc.~), secondary Bondi-Hoyle-Lyttleton (BHL) accretion, azimuthal winds (W), disk's self-gravity (SG) and migration (M). Columns correspond to different EMRI systems assuming a 1 year
observation. The entries $x\,(y)$ denote $x\times 10^y$ in radians. The phase shift is
negative for all effects except for SG $\alpha$. Observe that $\beta$-disk migration is the dominant effect for the first
two columns, while $\beta$-disk BHL accretion is dominant for the last two columns.}
\end{ruledtabular}
\end{table}
\begin{figure*}
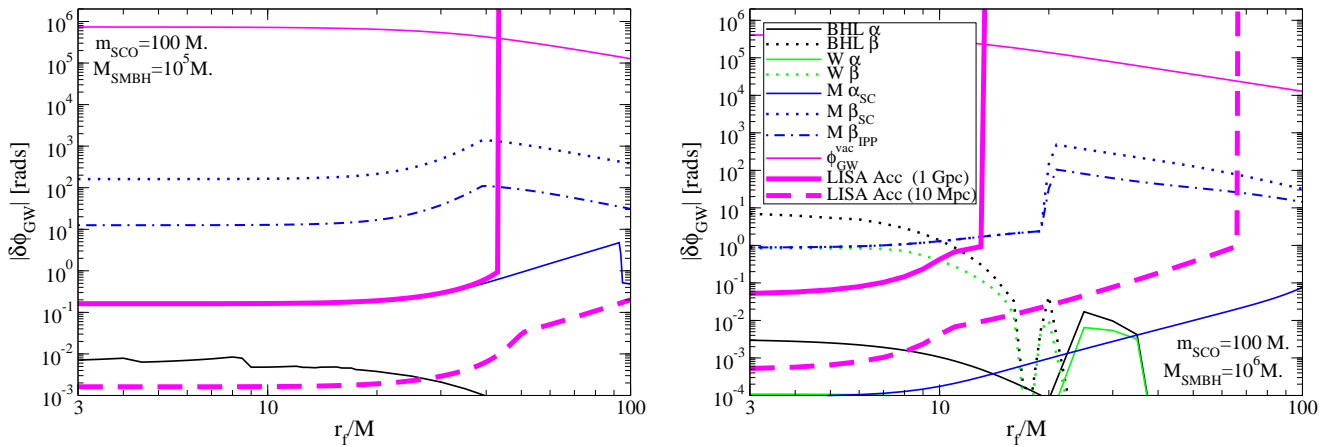

\begin{center}
\begin{tabular}{c}
 \includegraphics[width=8.5cm,clip=true]{analytic-dephasing-mSCO=100-mSMBH=1e5.eps}
 \quad
 \includegraphics[width=8.5cm,clip=true]{analytic-dephasing-mSCO=100-mSMBH=1e6.eps}
\end{tabular}
\end{center}
 \caption{\label{fig:dephasing-analytics} The GW phase shift as a function of final radius in units of $M_{\SMBH}$ induced by different accretion disk effects relative to vacuum waveforms. As in Fig.~\ref{fig:dephasing-analytics-intro}, solid curves correspond to $\alpha$-disks, while dotted ones to $\beta$-disks. Different color curves represent different accretion disk effects ($\alpha_1=\dot{m}_{\SMBH 1}=1$): black is for secondary Bondi-Hoyle-Lyttleton (BHL) accretion, green is for azimuthal winds (W) and blue is for migration (M), with dotted and dot-dashed curves for $\beta$-disks in the Syer-Clarke and Ivanov et.~al.~model respectively. The thin, solid magenta line is the total accumulated GW phase in vacuum, while the thick, solid (dashed) magenta line is a measure of the sensitivity to which LISA can measure the GW phase for a source at $1$ Gpc ($10$ Mpc). This figure presents similar conclusions to those found in Fig.~\ref{fig:dephasing-analytics-intro}, but for a different set of EMRI parameters.}
\end{figure*}

We summarize the GW phase shift generated by the dominant accretion
disk effects in Table~\ref{Table-acc-disk-effec}.  Different rows show
different accretion disk effects with the parameters of
Eq.~\eqref{acc-disk-pars}, while columns show different EMRI systems
with component masses given by all four combinations of
$M_{\SMBH}=(10^{5},10^{6}) \Msun$ and $m_{\CO} = (10, 100) \Msun$. The
entries represent the GW phase shift between the standard vacuum
waveform and those including the effects of the accretion disk in the
Newtonian approximation.  The phase shift is between one-year long GW
waveforms with the same final radius in the most sensitive LISA
frequency band.\footnote{Here we don't marginalize over an arbitrary
phase shift between the two waveform templates for simplicity. These
estimates are in good agreement with the more complicated calculations
presented in Sec.~\ref{sec:DA} below.}  $(\R_{f},\R_{0})\approx(16,25)$
for $(M_{\SMBH},m_{\CO})=(10^{5},10)M_{\odot}$;
$(\R_{f},\R_{0})\approx(16,42)$ for
$(M_{\SMBH},m_{\CO})=(10^{5},10^{2})M_{\odot}$;
$(\R_{f},\R_{0})\approx(3,7.6)$ for
$(M_{\SMBH},m_{\CO})=(10^{6},10)M_{\odot}$;
$(\R_{f},\R_{0})\approx(3,13)$ for
$(M_{\SMBH},m_{\CO})=(10^{6},10^{2})M_{\odot}$.  The phase shift
estimates are derived in Appendix \ref{B-H-accurate-estimate},
\ref{azi-wind-estimate}, and \ref{SG-estimate}, using
Eqs.~(\ref{e:dphi_dmSCO}), (\ref{e:dphi_dL}), and
(\ref{e:axisymmetric-gravity}), where the underlying quantities
$\dot{m}_{\CO}$, $\dot{\Ls}$ and $\Phi_{\rm disk}$ are substituted
from Sec.~\ref{sec:mass-accretion}--\ref{sec:gravitationaleffects},
taking into account all of the quenching mechanisms that are in
play. In particular, for migration we use Type I and Type II in the
appropriate radial ranges without and with gaps, respectively, where
we utilize either the Syer-Clarke~(SC) or the Ivanov et.~al.~(IPP)
model for Type II migration, see
Sec.~\ref{s:migration:quench}.\footnote{To understand which quenching
mechanisms are in effect for the particular cases, see
Figure~\ref{fig:radii} in Sec.~\ref{s:BondiSummary}.}

Table~\ref{Table-acc-disk-effec} shows that the migration estimate incorporating the SC model in $\beta$-disks is by far dominant,
followed by migration using IPP in $\beta$-disks, and BHL accretion. The effects of the disk's
self-gravity, wind effects, migration in $\alpha$-disks and Eddington-limited accretion onto the SMBH are all completely negligible for LISA EMRIs. All effects studied lead to a reduction of phase cycles in a fixed observation time for fixed final radius,
except for the effect of disk gravity in $\alpha$-disks. Other detectable EMRIs not shown in this table (different masses and/or
orbital radii) might be more sensitive to disk effects.

\begin{table*}[ht]
\begin{ruledtabular}
\begin{tabular}{ccccccccccccccc}
			& $C_{1}$ 	& $C_{2}$ 	& $c_1$ 	& $c_2$ & $c_3$  & $c_4$ 	& $c_5$ 	 & $c_6$ 	& $D_1$ 		 & $D_{2}$ 	& $d_3$ 	& $d_4$ & $d_5$ 	& $d_6$ 	\\
			& [$\delta \phi_{\rm GW}^{\rm long}$] & [$\delta^2 \phi_{\GW}^{\rm long}$] & [$\alpha_{1}$] & [$\dot{m}_{\SMBH 1}$] & [$M_{\SMBH 5}$]  & [$m_{\CO 1}$] & [$T_{\rm yr}$] 	& [$\R_{f,10}$] 	& [$\delta \phi_{\rm GW}^{\rm short}$] 		& [$\delta^2 \phi_{\rm GW}^{\rm short}$] &	 [$M_{\SMBH 5}$]  & [$m_{\CO 1}$] & [$T_{\rm yr}$] 	& [$\R_{f,10}$]	\\
\hline
BHL\,$\alpha$ & $6.9\,(+0)$ 	& $3.2\,(-1)$ 	& $-1$ 	& $-5$ 	& $-17/4$ & $17/8$ & $25/8$ & $5/2$  	& $2.5\,(+0)$ 	& $-2.6\,(+0)$& $-4$ 	& $2$ 	 & $3$ 	& $1/2$	\\
BHL\,$\beta$ 	& $3.2\,(+3)$ 	& $3.2\,(-1)$ 	& $-4/5$ 	& $-17/5$ & $-3$ 	& $8/5$ 	 & $13/5$ 	& $5/2$ 	 & $5.7\,(+3)$ 	& $8.2\,(+0)$ 	 & $-19/5$ & $2$ 	& $3$ 	& $-8/5$ 	\\
SB  		& $1.3\,(+1)$ 	& $3.7\,(-1)$ 	& $0$ 	& $1$ 	& $3/4$ 	& $-11/8$& $13/8$ 	& $5/2$ 	& $4.2\,(+2)$ 	& $2.8\,(+1)$  	& $-2$ 	& $0$ 	& $3$ 	& $-11/2$\\
\hline
Wind\,$\alpha$ 	& $3.4\,(-2)$ 	& $8.4\,(-4)$ 	& $-1$ 	& $-3$ 	& $-13/4$ & $13/8$	& $21/8$ 	 & $21/2$ 	& $1.5\,(-2)$ 	& $8.2\,(+0)$ 	& $-2$ 	& $1$ 	& $2$ 	& $5/2$ 	\\
Wind\,$\beta$ 	& $4.7\,(+1)$ & $4.3\,(-3)$ 	& $-4/5$ 	& $-7/5$ 	& $-2$ 	& $11/10$& $21/10$& $42/5$& $8.6\,(+1)$ & $1.4\,(+1)$ 	& $-9/5$ 	 & $1$ 	& $2$ 	& $2/5$ 	 \\
SW\,$\alpha$ 	& $1.5\,(-1)$ 	& $7.9\,(-2)$ 	& $0$ 	& $3$ 	& $7/4$ 	& $-15/8$ & $9/8$ 	& $9/2$ 	& $2.6\,(+0)$ 	& $2.5\,(+1)$ 	& $0$ 	 & $-1$ 	& $2$ 	& $-7/2$ 	 \\
SW\,$\beta$ 	& $3.5\,(-1)$ 	& $7.9\,(-2)$ 	& $0$ 	& $3$ 	& $7/4$ 	& $-15/8$ & $9/8$ 	& $9/2$ 	& $6.2\,(+0)$ 	& $2.5\,(+1)$ 	& $0$ 	 & $-1$ 	& $2$ 	& $-7/2$  \\
\hline
M1\,$\alpha$ & $4.0\,(-3)$ 	& $8.4\,(-4)$ 	& $-1$ 	 & $-3$ 	& $-13/4$& $13/8$  & $21/8$ & $21/2$ & $1.8\,(-3)$ 	& $8.2\,(+0)$  & $-2$ 	& $1$ 	 & $2$ 	& $5/2$	\\
M1\,$\beta$ 	& $7.0\,(+0)$ 	& $4.3\,(-3)$ 	& $-4/5$ & $-7/5$ 	& $-2$ 	& $11/10$& $21/10$& $42/5$& $1.3\,(+1)$ & $1.4\,(+1)$ 	& $-9/5$ 	& $1$ 	 & $2$  	& $2/5$ 	\\
M2\,$\alpha_{\rm SC}$ 	& $2.5\,(+1)$	& $1.8\,(-2)$ 	& $0$ 	& $1$ 	& $3/4$     & $-11/8$  & $13/8$    & $13/2$ & $1.5\,(+2)$ & $1.9\,(+1)$  & $0$ 	 & $-1$   & $2$ 	& $-3/2$	 \\
M2\,$\beta_{\rm SC}$ 	& $1.5\,(+4)$ 	& $3.5\,(-2)$ 	& $1/2$ & $5/8$ & $-3/16$   & $-31/32$ & $45/32$   & $45/8$ & $1.5\,(+5)$ & $2.2\,(+1)$  & $-11/8$ & $-3/8$ & $2$ 	& $-19/8$\\
M2$\beta_{\rm IPP}$ 	& $1.6\,(+3)$ 	& $2.7\,(-2)$ 	& $2/7$ & $11/4$& $3/14$    & $-8/7$   & $3/2$     & $6$ 	 & $1.3\,(+4)$ & $2.0\,(+1)$   & $-11/14$& $-9/14$& $2$ 	& $-2$ 	 \\
\hline
SG\,$\alpha$ 	& $2.2\,(-5)$ 	& $1.7\,(-2)$ 	& $-20/21$&$-13/21$& $-59/84$& $3/8$& $11/8$ & $11/2$  & $5.8\,(-6)$ 	& $-1.8\,(+1)$ 	& $1/21$& $0$ & $1$  	& $3/2$ 	\\
SG\,$\beta$ 	& $-6.1\,(-4)$ 	& $-1.0\,(-2)$ 	& $-4/5$ 	& $3/5$ 	& $1/4$ 	& $-1/40$& $39/40$ & $39/10$& $-9.7\,(-4)$ 	& $7.8\,(+0)$ 	& $1/5$ & $0$ 	& $1$ 	& $-1/10$\\
\end{tabular}
\caption{\label{Table-pars4} Constant coefficients and exponents in Eqs.~(\ref{e:long-gen-form}--\ref{e:short-gen-form}) for long and short observations (columns), for different effects (rows). We use the notation $x\,(y) = x\times 10^y$ in radians. Observe again that the dominant
dephasing is due to Type-II migration (M2) both in the short- and long-observation limits. Also notice that all dephasing scales with positive powers
of the radius (i.e.~$c_{6}>0$ and $d_{6}>0$), implying accretion disk effects become stronger for EMRIs orbiting at larger separations.
}
\end{ruledtabular}
\end{table*}

Figures~\ref{fig:dephasing-analytics-intro} and
\ref{fig:dephasing-analytics} show GW phase difference induced by
various disk effects as a function of final orbital radius.  Solid curves correspond to
$\alpha$-disks, and dotted curves to $\beta$-disks. Curve colors
represent different accretion disk effects: black curves
correspond to BHL accretion, green curves to azimuthal wind effects,
and blue curves to migration (dotted for SC-$\beta$, dot-dashed for
IPP-$\beta$). Left and right panels correspond to different mass
ratios and SMBH masses. Additionally,
Figs.~\ref{fig:dephasing-analytics-intro} and
\ref{fig:dephasing-analytics} include three disk-independent reference
curves. The thin, solid magenta line is the total accumulated GW phase
without disk effects. The thick magenta curves represent the
approximate accuracy to which LISA can measure the GW phase as a
function of radius, using the simple estimates given by
Eq.~\eqref{simple-LISA-acc-est} for a typical source located at $1$
Gpc (solid) and 10 Mpc (dashed).  We present more detailed estimates
of measurement accuracy in Sec.~\ref{sec:DA} below.

Typically migration in $\beta$-disks generates the dominant phase shift for all final
radii.  The large drop in the right panel occurs because the gap
closes around $\R \sim 20.4$ for $\beta$-disks for these masses, changing Type-II into
Type-I migration\footnote{Gap decoupling does not occur since
$\R_{\rm d} \sim 19.2$.}.  BHL accretion follows migration in
importance, and in certain cases (as in the right panel of the
figure), the former can be the most important effect if the gap refills.  
The sudden quenching features in
the BHL accretion curves are due to EM radiation pressure, where
photon diffusion becomes sufficiently short for EM radiation to escape
the flow.  Observe that LISA is sensitive to EMRIs that are close to
the SMBH (e.g.~$\R \lesssim 50$); the sharp rise of the LISA accuracy
estimate is because EMRIs at larger radii are not detected with SNR of
10 or bigger. All disk effects above the thick magenta line are
significant and possibly measurable, while wind effects and
self-gravity effects (not shown in the figure) always lead to
dephasings of order $1$ radian or much less. These figures suggest
that GW observations may be used to test the predictions of different accretion
disk models.

We note that for the masses in the left panel of Fig.~\ref{fig:dephasing-analytics},
the CO is located at $\R_{\CO}=43$ and $\R_{\CO}=51$ one and two years prior to merger.
A gap is open beyond $\R_{\CO}>\R_{\rm gap} = 96$ and $5$ for $\alpha$ and $\beta$-disks.
The EMRI evolution becomes much faster than the viscous inflow and the gap decouples
at a radius of $\R_{\rm d} = 39$ for a $\beta$-disk.

To describe the phase shift for different observation times or accretion disk parameters than
those shown in Figures~\ref{fig:dephasing-analytics-intro} and \ref{fig:dephasing-analytics},
we provide analytical expressions for the asymptotic phase shift for various processes in two limits, namely
\begin{align}
\delta \phi_{\GW}^{\rm long} &= -C_{1}
 \alpha_{1}^{c_{1}}
 \dot{m}_{\SMBH 1}^{c_{2}}
M_{\SMBH 5}^{c_{3}}
m_{\CO n}^{c_{4}}
 T_{\yr}^{c_{5}}
 \left(1 - C_{2}
 \frac{M_{\SMBH 5}^{c_{6}/2}}{m_{\CO 1}^{c_{6}/4}}
 \frac{\R_{f,10}^{c_{6}}}{T_{\yr}^{c_{6}/4}} \right)\,
\label{e:long-gen-form}\\
\delta \phi_{\GW}^{\rm short} &= - D_{1}\alpha_{1}^{d_{1}} \dot{m}_{\SMBH 1}^{d_{2}}
 M_{\SMBH 5}^{d_{3}} m_{\CO 1}^{d_{4}} T_{\yr}^{d_{5}} \R_{f, 10}^{d_{6}}\nonumber\\
&\quad\times
 \left(1 - D_{2}
 \frac{m_{\CO 1}}{M_{\SMBH 5}^{2}}
 \frac{T_{\yr}}{\R_{f,10}^{4 }} \right)\,,
\label{e:short-gen-form}
\end{align}
where the $(C,c,D,d)$ parameters are given by the rows of
Table~\ref{Table-pars4}. Here $d_1\equiv c_1$ and $d_2\equiv c_2$.
The first (second) formula is applicable if the observation time is
much shorter (longer) than the inspiral time [Eq.~(\ref{e:t-crit})].
The particular processes are represented by
rows in order: the full BHL rate (i.e.~assuming no quenching
throughout the observation), the supply limited BHL rate (SB, using
Eq.~\ref{e:M-flux} throughout the observation), the corresponding
hydrodynamic drag from the azimuthal wind for the full BHL accretion
and SB (W and SW, respectively), Type-I migration (i.e.~assuming a
gap is not present throughout the observation), steady-state Type-II
migration, quasi-stationary Type-II migration (i.e.~assuming a gap is
open throughout the observation), and disk self-gravity (SG).  Note
that LISA observations are sensitive to the combinations
$\alpha_1^{c_1}\dot{m}_{\SMBH 1}^{c_2}$, see corresponding columns.

Most accretion disk effects are several orders of magnitude larger for $\beta$-disks relative to $\alpha$-disks.
This is because $\beta$-disks can be much more massive in the regime of interest for EMRIs. In particular,
this suggests the GW phase shift may be used to test the predictions of different accretion disk models.

\section{Relativistic Waveforms and Detection}
\label{sec:GWmodeling}

Next, we consider accretion disk effects in more realistic waveform
models. EMRI GWs are highly relativistic, with velocities close to the
speed of light and sometimes skimming the SMBH horizon. As such,
Newtonian waveform estimates for GW data analysis are inaccurate. Here
we investigate the relativistic correction to the EMRI dynamics using
the extended one body framework, and make simple estimates on the
imprint of accretion disk effects on the GW waveform. This analysis,
however, continues to neglect relativistic corrections to accretion
disk effects.

\subsection{Systems Investigated}
\label{sys-invest}
In the rest of this paper, we restrict our investigations to the following two representative EMRI systems:
\begin{itemize}
\item {\bf{System I}}: Masses $M_{\SMBH} = 10^{5} \, M_{\odot}$, $m_{\CO} = 10 \, M_{\odot}$, spin parameter $a_{\SMBH}/M_{\SMBH} = 0.9$, observation time $T=1\yr$, range of orbital radius $\R \in (16,25)$, orbital velocity $v/c \in (0.2,0.25)$, GW frequency $f_{\rm GW} \in (0.005,0.01) \;  {\rm{Hz}}$, GW phase $\phi_{\rm GW} \sim 1.3 \times 10^{6}$ rad.
\item {\bf{System II}}:
 Masses $M_{\SMBH} = 10^{6} \, M_{\odot}$, $m_{\CO} = 10 \, M_{\odot}$, spin parameter $a_{\SMBH}/M_{\SMBH} = 0.9$, observation time $T=1\yr$, range of orbital radius $\R \in (3,7)$, orbital velocity $v/c \in (0.37,0.54)$, GW frequency $f_{\rm GW} \in (0.003,0.01) \;  {\rm{Hz}}$, GW phase $\phi_{\rm GW} \sim 9 \times 10^{5}$ rad. The ISCO is located at $\R_{\rm ISCO} \approx 2.32$.
\end{itemize}
Figures~\ref{fig:dephasing-analytics-intro} and \ref{fig:dephasing-analytics} shows that accretion disk effects are expected to be significant for these systems.

We make the following simplifying assumptions. First, we consider only quasi-circular EMRIs on the equatorial plane, such that the orbital angular momentum is perpendicular to the SMBH's spin angular momentum. We have also investigated Systems with spin anti-aligned or zero and found similar results. The accretion disk is also assumed to be on the equatorial plane, such that the EMRI is completely embedded in the disk. We ignore the CO's spin angular momentum, as well as sub-leading mass-ratio terms in the radiation-reaction fluxes and in the Hamiltonian.

Such simplifying assumptions make the problem analytically tractable within the EOB framework, employed here for waveform modeling. As of the writing of this paper, the EOB framework for EMRIs has not been sufficiently developed for non-equatorial orbits; it has, however, been satisfactorily tested for equatorial orbits with extreme or comparable mass-ratios, $q \lesssim 10^{-4}$ and $q\gtrsim 10^{-2}$.

We expect that many accretion disk effects (migration, BHL, and wind effects)
will be maximal for the equatorial EMRIs studied here. Other effects,
however, are substantially different for EMRIs inclined with respect to the accretion disk.
Similarly, disk effects may excite eccentricity, which
may dramatically increase the impact of accretion disk effects on the GW observables
\cite{2005ApJ...634..921A,2008ApJ...672...83M,2009PASJ...61...65H,2009MNRAS.393.1423C}.
A study of non-equatorial or eccentric EMRIs with an accretion disk is beyond the scope of this paper.

\subsection{Basics of the EOB Framework}
\label{sec:EOB}

We employ the adiabatic EOB framework of~\cite{Yunes:2009ef,2009GWN.....2....3Y,Yunes:2010zj} to model EMRI waveforms. The EOB scheme was first proposed in~\cite{Buonanno99,Buonanno00} to model the coalescence of comparable-mass BH binaries. Since then, this scheme has been greatly enhanced and extended to other type of systems~\cite{Damour00,Damour01,Buonanno06,Damour:2008qf,Barausse:2009xi,Nagar:2006xv,Damour2007,Bernuzzi:2010ty,Damour:1997ub,Damour2007,Damour:2008gu,Pan:2010hz,Fujita:2010xj}. Waveforms constructed in this way have been successfully compared to a set of numerical relativity results~\cite{Damour:2009kr,Buonanno:2009qa,Pan:2009wj} and to self-force calculations~\cite{Barack:2009ey,Damour:2009sm}. Recently,~\cite{Yunes:2009ef,2009GWN.....2....3Y,Yunes:2010zj} proposed the combination of EOB and BH perturbation
theory techniques to model EMRI waveforms for LISA data-analysis purposes. This is the scheme we adopt in this paper.

In the adiabatic, EOB framework for quasi-circular EMRIs, the GW phase can be obtained by solving the adiabatic equation
\begin{equation}
\label{omegaadiab}
\dot{\Omega} = - \left(\frac{d E}{d \Omega}\right)^{-1} \, {\cal F}_{\GW}(\Omega)\,,
\end{equation}
where $\Omega \equiv \dot{\phi}$ is the orbital angular frequency (Eq.~\eqref{kep-ang-vel} but with relativistic corrections), overhead dots stand for time derivatives, $E$ is the binary system's total binding energy and ${\cal F}_{\GW} \equiv dE_{\GW}/dt$ is the GW energy flux. We have here implicitly assumed a balance law: all loss of gravitational binding energy is removed only by GW radiation, $dE/dt = -dE_{\GW}/dt$.

The binary's binding energy in a Kerr background for a quasi-circular orbit is simply given by~\cite{Bardeen:1972fi}
\begin{equation}
E= \mu \,\frac{1-2 \R^{-1} + \chi_{\SMBH}\,\R^{-3/2}}{\sqrt{1-3 \R^{-1}+ 2\chi_{\SMBH}\,\R^{-3/2}}}\,.
\label{E-binding}
\end{equation}
where $\mu \equiv M_{\SMBH} m_{\CO}/M$ is the reduced mass, with $M \equiv M_{\SMBH} + m_{\CO}$ the total mass, $\R \equiv r/M$, and $\chi_{\SMBH} \equiv a_{\SMBH}/M_{\SMBH}$ is the reduced Kerr spin parameter. Notice that it is the binding energy that drives the orbital evolution, and not the total energy of the system, which would also account for the rest-mass energy.

We employ here the factorized form of the GW flux, considered in~\cite{Damour2007,Damour:2008gu,Pan:2010hz}, with the assumption of adiabaticity:
\begin{eqnarray}
{\cal F}_{\GW}(\Omega)= \frac{1}{8 \pi}\,\sum_{\ell=2}^{8}\sum_{m=0}^{\ell} (m\,\Omega)^2\,\left|R \; h_{\ell m} \right|^{2}\,,
\label{pd-flux-adiab}
\end{eqnarray}
where $R$ is the distance to the observer,
$m$ is the azimuthal quantum number of the multipolar-decomposed waveform, and
\begin{equation}
h_{\ell m} = h_{\ell m}^{\Newt,\epsilon_p}\;S^{\epsilon_p}_{\ell m} \; T_{\ell m}\; e^{i \delta_{\ell m}}\; (\rho_{\ell m})^\ell\,,
\label{full-h}
\end{equation}
and where $\epsilon_p$ is the multipolar waveform parity
(i.e.~$\epsilon_p=0$ if $\ell+ m$ is even, $\epsilon_p=1$ if
$\ell+ m$ is odd). All the terms in Eq.~\eqref{full-h}
($S^{\epsilon_p}_{\ell m}$, $T_{\ell m}$, $\delta_{\ell m}$ and
$\rho_{\ell m}$) are functions of $(r,\phi,\Omega)$ that can be found
in~\cite{Damour2007,Damour:2008gu,Pan:2010hz}.  The Newtonian part of
the waveform is given by
\begin{equation}
h_{\ell m}^{\Newt,\epsilon_p} \equiv \frac{M_{\SMBH}}{R}\, n^{(\epsilon_p)}_{\ell m}\, c_{\ell +\epsilon_p}\, v^{\ell +\epsilon_p}\,
Y_{\ell - \epsilon_p,-m}(\pi/2,\phi).
\end{equation}
where $Y_{\ell,m}(\theta,\phi)$ are the standard spherical harmonics, $n_{\ell m}^{(\epsilon_p)}$ and $c_{\ell +\epsilon_p}$ are numerical coefficients that depend on the mass-ratio~\cite{Damour:2008gu}. The orbital velocity $v$ is related to the orbital frequency via $v = (M \Omega)^{1/3}$, which then implies the binary orbital separation is
\be
\R = \frac{\left[1 - \chi_{\SMBH} \bar{\Omega})\right]^{2/3}}{\bar{\Omega}^{2/3}}\,,
\label{kepler-Kerr}
\ee
where recall that the overhead bar stands for normalization with respect to total mass: $\bar{\Omega} = M \Omega \approx M_{\SMBH} \Omega$.

At this stage one might be slightly confused, as the right-hand side
of the evolution equation one wishes to solve (Eq.~\eqref{omegaadiab})
depends on a variety of quantities, including the orbital separation
and the orbital phase.  The assumption of adiabatic quasi-circularity
allows us to replace the orbital separation in terms of the orbital
frequency via Eq.~\eqref{kepler-Kerr}. By definition, the orbital
phase is related to its frequency via the differential equation
$\dot{\phi} = \Omega$. This equation, together with
Eq.~\eqref{omegaadiab} forms a closed system of coupled, first-order
partial differential equations that can be consistently solved.

The flux in Eq.~\eqref{pd-flux-adiab}, however, is not sufficiently
accurate to model EMRIs. First, it neglects the loss of energy due to
the absorption of GWs by the MBH. Second, it is built from a PN
expansion, which is in principle valid only for slowly-moving sources,
which EMRIs are not. This flux can be improved by linearly adding BH
absorption terms and by adding calibration coefficients to
Eq.~\eqref{pd-flux-adiab} that are fitted to a more accurate,
numerical flux. This is the procedure proposed
in~\cite{Yunes:2009ef,2009GWN.....2....3Y,Yunes:2010zj}, which we
follow here. We include up to 8 calibration coefficients, obtained by
fitting to a more accurate Teukolsky evolution in the point-particle
limit, as given in Eqs.~$(26)$-$(29)$ of~\cite{Yunes:2010zj}, as well
as BH absorption terms as given in Appendix~B of~\cite{Yunes:2010zj}.

Initial data for the evolution of the system of differential equations
is obtained through a mock evolution, started at $r=100 M_{\SMBH}$ and
ended at $f_{\rm GW}=0.01$ Hz (see
e.g.~\cite{Yunes:2009ef,2009GWN.....2....3Y,Yunes:2010zj}). The mock
evolution is initialized with the post-circular data
of~\cite{Buonanno00}. Once the evolution terminates, one can read
initial data one-year prior to that point directly from the numerical
evolution of the mock simulation. We obtain initial data of such form
separately both in the case of vacuum EMRIs and for EMRIs in an
accretion disk, as the evolutions are generically different. Once the
orbital phase is obtained by solving Eq.~\eqref{omegaadiab} with
this initial data, the waveforms are readily obtained through
Eq.~\eqref{full-h}.

Before proceeding, let us comment on the differential system one must
solve numerically. As already mentioned, since the source of
Eq.~\eqref{omegaadiab} depends both on orbital phase and frequency,
there are truly two coupled, first-order differential equations that
must be solved. But the source term of this equation is incredibly
more complicated than implied here. Even though $S_{\ell m}$, $T_{\ell
m}$, $\delta_{\ell m}$ and $\rho_{\ell m}$ are known analytically as
functions of $\phi$ and $\Omega$, each term contains very long and
complicated series expansions with fractional exponents that include
special functions, such as the polygamma function. For this reason,
the EOB evolution is not a simple integration, as it naively appears
to be in this section. Instead, the coupled set of first order
equations must be solved simultaneously via numerical methods, where
here we employ a partially optimized, Mathematica routine.

\subsection{Disk Modifications to EOB GWs}
\label{acc-disk-mod}

GW modeling in the adiabatic EOB framework depends sensitively on the energy $E$ and the flux ${\cal{F}}_{\GW}$. Modifications to the MBH or the CO mass naturally change all mass scales that depend on the total mass $M$, such as the symmetric mass-ratio $\eta$. Radiation pressure and migration modifies the rate of change of the angular momentum and thus the flux. In what follows, we explain how we modify the EOB scheme to account for such disk effects.

\subsubsection{Effective Hamiltonian}

The effective Hamiltonian controls the conservative evolution of the EOB model.
We have considered several accretion disk effects that directly modify the Hamiltonian, such as the self-gravity of the disk and the increase in mass of the SMBH and the CO. Of these effects, the latter has been found to be the largest. We concentrate on this effect here.

The increase in the CO's mass can be modeled by solving for the time
evolution of $m_{\CO}$. The differential equation that controls this
evolution is Eq.~\eqref{e:dmSCO''}, where $\dot{M}_{\SMBH}^{\rm
flux}$ is given in Eq.~\eqref{e:M-flux}, while
$\dot{m}_{\CO}'^{\Bondi}$ is given by Eq.~\eqref{e:MBondi'}. Notice
that $\dot{m}_{\CO}$ depends on radius, or on orbital frequency by
Eq.~\eqref{kepler-Kerr}, which itself is a function of time. Since the
accretion rate is not constant, one must solve the system of
differential equations on $\Omega$ and $m_{\CO}$ consistently, which
we do numerically and perturbatively as follows. First, we solve the
frequency evolution equation, setting $m_{\CO}$ to a constant and
neglecting accretion. Second, we use Eq.~\eqref{kepler-Kerr} to
rewrite Eq.~\eqref{e:dmSCO''} in terms of orbital frequency. Third, we
replace this orbital frequency by the time-evolution obtained in
vacuum. Fourth, we numerically solve the evolution equation for
$\dot{m}_{\CO}$, as its source now depends only on time. In step two,
we are implicitly discarding non-linear terms that scale as
$(\dot{m}_{\CO} T_{\rm obs})^{2}$, i.e.~the square of the accretion
rate times the observation time. This quantity is approximately
$10^{-8}$ or much smaller (clearly much smaller than unity) for
typical LISA observation times.

Once the time evolution for the CO's mass has been obtained, one must
then make sure that all quantities that depend on it are properly
promoted to time functions. For example, all quantities that depend on
the total mass, such as the symmetric mass-ratio $\eta$ or the reduced
mass $\mu$ are modified. In particular, the numerical code used to
solve the differential equation (Eq.~\eqref{omegaadiab}) is naturally
written in dimensions of the total mass of the system, which now
becomes a time-function. A simple trick to deal with this is to
rescale all mass-quantities by the factor $M(t)/M(0)$, which is equal
to unity initially, but deviates from unity with time. In particular,
this implies that $\bar{\Omega} \to [M(t)/M(0)] \bar{\Omega}$ and $\R
\to [M(t)/M(0)]^{-1} \R$.

Once these substitutions have been made, one can solve for the frequency and phase evolutions, with $m_{\CO}(t)$ a function of time, by providing appropriate initial data. When considering BHL accretion, we choose the same initial data as in the vacuum case (as explained in the end of Sec.~\ref{sec:EOB}) one year prior to reaching a GW frequency of $10^{-2}$ Hz. The frequency and phase evolution can be compared when $m_{\CO}$ is a constant and when it is not, which provides a measure of the effect of BHL accretion on GWs.

\subsubsection{Radiation-Reaction Force}

The radiation-reaction force controls the rate at which orbits inspiral. This force can be expressed in terms of the rate of change of orbital elements, such as binding energy, angular momentum and the Carter constant. Since we restrict attention to an equatorial, quasi-circular EMRI geometry, we need to consider only the energy flux.

Modifying the EOB model to account for a different radiation-reaction force amounts to the rule
\be
{\cal F}_{\GW} \to {\cal F}_{\GW} \left(1 + \frac{\delta \dot{\Ls}}{\dot{\Ls}_{\GW}}\right)
\label{newflux}
\ee
in Eq.~\eqref{omegaadiab}. When modeling an azimuthal wind, $\delta
\dot{\Ls} = \dot{\Ls}^{\rm wind}$ via Eq.~\eqref{Ldotwind-with-gamma},
while when modeling migration, then $\delta \dot{\Ls}$ is given by
Eqs.~\eqref{e:mig-decoup}; all throughout $\dot{\Ls}_{\GW}$ is given
by Eq.~\eqref{e:Lgw} with $e = 0$. When substituting in for $\delta
\dot{\Ls}$ one must be careful to use the properly quenched
$\dot{m}_{\CO}$ (Eq.~\eqref{e:dmSCO4}), if the accretion disk effect
depends on the rate of BHL accretion. All other aspects of the
framework can be left unchanged, as Eq.~\eqref{newflux} automatically
induces deviations from the Kepler relation.

The system of EOB differential equations can now be solved using appropriate initial data. Initial conditions for the vacuum and accretion-disk case are prescribed via mock evolutions as explained in detail in Sec.~\ref{sec:EOB}, with $\delta \dot{\Ls} = 0$ and $\delta \dot{\Ls} \neq 0$ respectively. As explained above, this guarantees that both simulations will terminate at the same orbital separation. Due to different radiation-reaction force laws, however, the starting radii or frequencies are different in each case for a fixed observation time. To account for this, we will later maximize comparison measures over a time and phase shift between vacuum and accretion disk waveforms, as we explain in Sec.~\ref{Deph-analysis}.

\section{Data Analysis Considerations}
\label{sec:DA}

We are now ready to perform a more detailed data analysis study of the
accretion disk effects on waveforms. We begin by investigating the
dephasing of the EOB waveforms constructed in the previous section. We
then continue with an overlap study and end with a discussion of
degeneracies between accretion disk parameters and EMRI system
parameters.

\subsection{Dephasing Analysis}
\label{Deph-analysis}

As explained in Sec.~\ref{deph-section}, a dephasing study is useful to roughly determine whether two waveforms can be distinguished from each other given a GW detection; if the phase difference or {\emph{dephasing}} between waveforms is large enough, then they are distinguishable (see Eq.~\eqref{simple-LISA-acc-est}). Here, we compare the dephasing of the dominant, $(\ell,m)=(2,2)$ vacuum and non-vacuum GW modes, initialized with the data of Sec.~\ref{sec:EOB}.

If we take one waveform to be ``the signal'' and the other to be ``the template,'' then the dephasing depends on two extrinsic parameters contained in the template: an overall phase $\delta \phi$ and time shift $\delta t$. We here study the dephasing after minimizing it with respect to these extrinsic parameters. The template also depends on other parameters, such as the masses and spins, but we here set these to be equal to the signal's parameters, i.e.~we do not minimize the dephasing over such parameters. In a realistic data analysis implementation, one would maximize the SNR (or minimize the dephasing) over all parameters, at the cost of introducing error into parameter estimation due to the use of incorrect templates.

Before minimizing the dephasing over a time and a phase shift, it is worth pointing out that its magnitude (computed with relativistic EOB waveforms) roughly agrees with the Newtonian results presented in Table~\ref{Table-acc-disk-effec}. For example, the final dephasing after 12 months of evolution (initializing the simulations with the same final frequency and the same initial phase) between vacuum and $\beta$-disk migration is $\sim 217$ rads with the EOB model and $\sim 670$ rads with the Newtonian estimates for System I. In this case, the relativistic model leads to a larger dephasing than the Newtonian estimates because, in the latter, Sys.~I evolves more rapidly due to relativistic corrections and the CO spends less time in the Type-II dominated region. We have verified that for weakly-relativistic EMRIs, where only Type-I or Type-II migration is in play, Newtonian and EOB dephasings agree.

A more appropriate measure of distinguishability, however, requires that one minimizes the dephasing with respect to $\delta t$ and $\delta \phi$. Following the prescription of Eq.~(23) in~\cite{Buonanno:2009qa}, we search for a $\delta t$ and $\delta \phi$ such that $|| f_{1}(t + \delta t) - f_{2}(t) || \leq \delta_{f}$ and $||\phi_{1}(t + \delta t) - \phi_{2}(t) - \delta \phi|| \leq \delta_{t}$, where $\phi_{1,2}$ and $f_{1,2}$ are the dominant GW phase and frequencies for waveforms $h_{1,2}(t)$. The $||A||$ notation stands for the integral of $A$ over a time window of length $64 \lambda_{\GW}$, where $\lambda_{\GW}$ is the GW wavelength (see e.g.~\cite{Yunes:2010zj} for details). The signal $h_{1}(t)$ is assumed to be a vacuum template, while $h_{2}$ is a non-vacuum template for a specific accretion disk effect. We choose the tolerances $\delta_{f} = 10^{-11}$ and $\delta_{t} = 10^{-6}$; decreasing these magnitudes does not visibly change the dephasing results shown below. The value of $\delta t$ and $\delta \phi$ are unique for a specific set of $h_{1,2}$, i.e.~for a given accretion disk effect. This {\emph{alignment}} procedure has been shown to be equivalent to maximizing the fitting factor over time and phase of coalescence in a matched filtering calculation with white noise~\cite{Buonanno:2009qa}.

\begin{figure*}
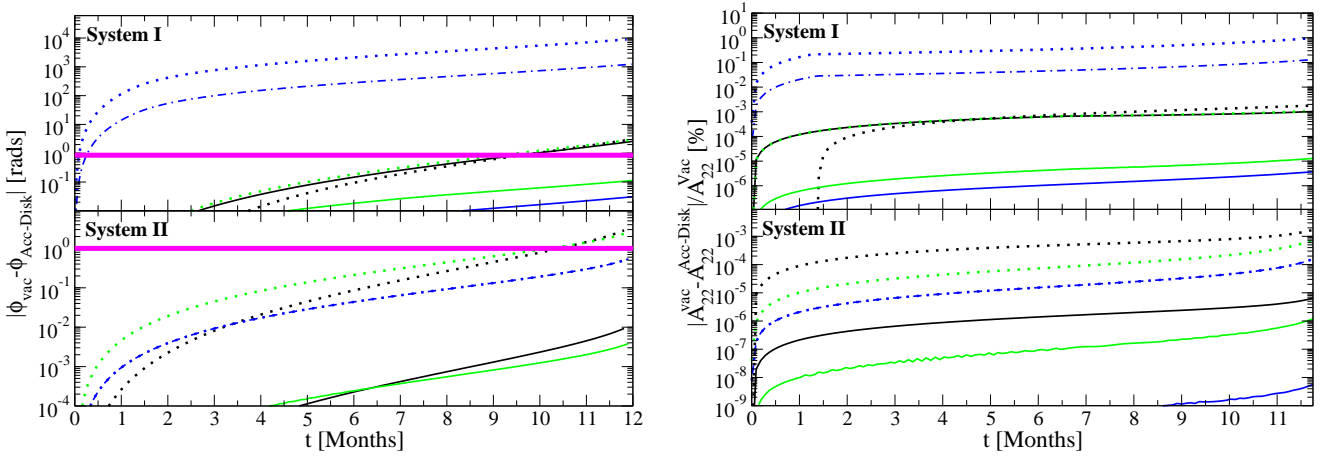

\begin{center}
\begin{tabular}{c}
 \includegraphics[width=8.5cm,clip=true]{phase-comp.eps}
 \quad
 \includegraphics[width=8.5cm,clip=true]{amp-comp.eps}
\end{tabular}
\end{center}
 \caption{\label{fig:dephasing-EOB-aligned} Aligned dephasing (left) and aligned fractional amplitude difference (right) as a function of time in units of months for the dominant GW mode. Line style and color follow the same notation as in Fig.~\ref{fig:dephasing-analytics}. The thick solid lines signal a 1 radian dephasing. The top and bottom panels correspond to System I and II respectively. Observe that the minimized dephasing exceeds unity in a short observation time for $\beta$-disk migration and System~I, while for System~II, only BHL accretion and $\beta$-disk wind effects do so after a full-year of observation.}
\end{figure*}

Figure~\ref{fig:dephasing-EOB-aligned} shows the dominant dephasing (left-panel) and fractional amplitude difference (right-panel) after such alignment. As before, we plot these quantities for the dominant GW mode as a function of time in units of months, using different color curves for different accretion disk effects and different curve styles for different types of disks. Observe that after alignment, the dephasing increases much less rapidly than in the previous case. This implies that the loss of overlap will also grow much more slowly. Dephasings after alignment thus correspond to the least difference between vacuum and non-vacuum waveform phases, without maximizing over intrinsic EOB parameters (such as the SMBH's and CO's mass) or other extrinsic parameters (such as those associated with the observation angles, detector motion, etc.).

Figure~\ref{fig:dephasing-EOB-aligned} confirms that certain accretion disk effects become significant very early during a LISA observation.  As discussed in Sec.~\ref{LISA-sense}, a rough measure of whether the dephasing is ``distinguishable'' for an event with SNR $\sim 10$ is whether $\delta \phi_{\GW} \gtrsim 1\,\rm rad$ (see Sec.~\ref{overlap-analysis} for a more accurate measure). The imprint of migration becomes almost immediately distinguishable for Sys.~I and $\beta$-disks, while it takes at least one full year of observation before one can observe the same type of migration for Sys.~II or BHL accretion for Sys.~I. Wind effects also become important within one year of observation but for $\beta$-disks only.

Whether an effect is distinguishable is naturally sensitive to the EMRI parameters and orbital radii. Indeed, the bottom panel of Figure~\ref{fig:dephasing-EOB-aligned}, corresponding to System-II, shows that in this case most effects are much smaller. This is mainly due to the assumption that System II's orbit is much closer to the SMBH where disk effects are less relevant. The bottom panel further suggests that BHL and azimuthal wind effects might be barely distinguishable after 1 year for $\beta$-disks only. Notice that all of these findings
are consistent with the Newtonian estimates and figures presented in the previous sections.

The right panel of Figure~\ref{fig:dephasing-EOB-aligned} shows the fractional amplitude difference between the dominant mode of vacuum and non-vacuum waveforms after the alignment procedure described above. The amplitude difference follows closely the trend of the dephasing: Type-II migration is clearly visible in amplitude changes, while other effects are greatly suppressed. The amplitude difference plays an important role in the calculation of the overlap and the SNR of the difference that we show below.

\subsection{Overlap Analysis}
\label{overlap-analysis}

Dephasing studies are convenient as rough measure of distinguishability for a fixed SNR. However, the SNR changes during the GW observation, as signal accumulates and its frequency enters the detector's more sensitive domain. A more accurate measure of the detection significance of the particular disk effect in the data stream is the SNR of the waveform {\it difference} between the data streams with and without the effect, and the so called overlap/mismatch. In Sec.~\ref{LISA-sense}, we defined all these quantities in terms of cross-correlation integrals weighted by the spectral noise of the LISA detector. Note that these quantities account for the difference in both the phase and amplitude evolution of the GW signal.

\begin{figure*}
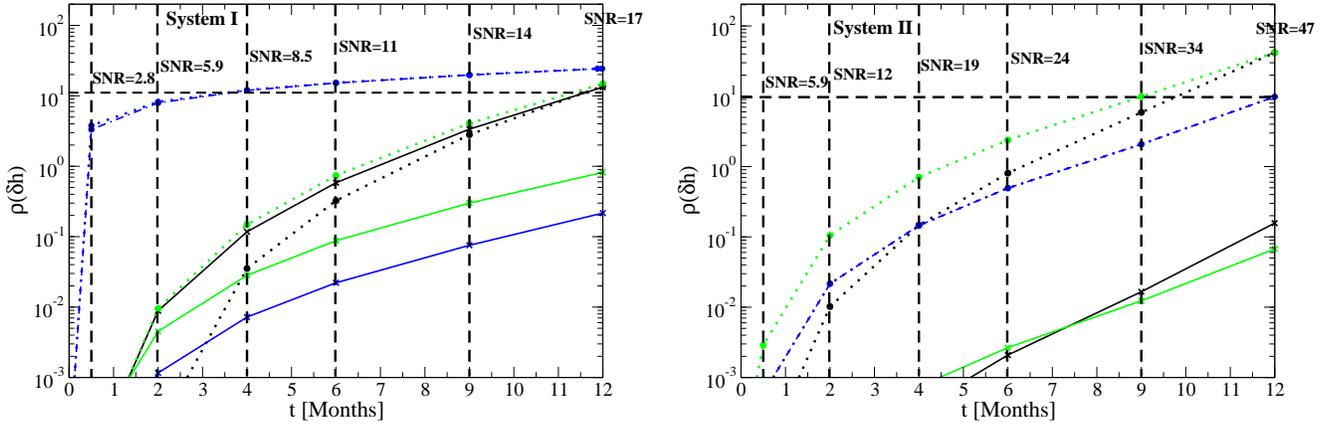

\begin{center}
\begin{tabular}{c}
 \includegraphics[width=8.5cm,clip=true]{SNRofError-SysI.eps}
 \quad
 \includegraphics[width=8.5cm,clip=true]{SNRofError-SysII.eps}
\end{tabular}
\end{center}
 \caption{\label{fig:SNRofError} SNR of the aligned, dominant waveform difference as a function of time in units of months. The left panel corresponds to System I and the right one to System II. Different curve colors and styles correspond to different disk effects and disk models, respectively, as in Fig.~\ref{fig:dephasing-analytics}. The thick horizontal dashed line correspond to an SNR of 10. Observe that the SNR of the difference exceeds the SNR threshold in four months for $\beta$ disk migration and Sys.~I, while it take a full year of observation for the same to occur when modeling BHL accretion and $\beta$-disk wind effects.}
\end{figure*}

Figure~\ref{fig:SNRofError} shows the SNR of the difference $\rho(\delta h)$ in the $(\ell,m)=(2,2)$ GW harmonic after minimization over time and phase shift in units of months. As before, different curve styles and colors correspond to different accretion disk models and effects, as defined in Fig.~\ref{fig:dephasing-analytics-intro}. The thick horizontal line corresponds to an SNR of 10, just about the threshold for detection. The numbers at the top of this figure show how the vacuum waveform SNR builds up as a function of observation time. This figure confirms that $\beta$-disk migration is the dominant perturbation to the measured GW signal and suggests that it is distinguishable within 4 months of observation. Migration is followed in significance by BHL accretion (for either $\alpha$ or $\beta$-disks) and $\beta$-disk azimuthal winds. These effects become distinguishable only after a full year of observation. All other accretion disk effects are insignificant within a 1 year evolution for System I. For System II, only $\beta$-disk BHL accretion and azimuthal disks are significant and only after 1 year of observation. Other effects may become significant for binaries that are closer to Earth than $1\,$Gpc, or if the observation is longer, or if the binary orientation relative to the detector is better than average.

This figure is a more realistic estimate of distinguishability than the dephasing study presented in the
previous section. The increase in realism comes at the cost of a small drop in distinguishability; e.g.~although
Fig.~\ref{fig:SNRofError} suggests that BHL accretion might be measurable after a 1 year observation,
this is only marginal in the figures above. Such a drop is mostly due to the inclusion of detector noise in this
 subsection. Measurability of accretion disk effect would of course improve if the source is closer to Earth,
such that the SNR of the signal is larger. Irrespective of this, all calculations suggest that Type II migration is
such a strong effect that it is likely to be measurable with LISA.

\subsection{Degeneracies}
\label{degen-sect}

Degeneracies between EMRI system parameters, such as the SMBH and CO's mass,
and accretion disk parameters could deteriorate the extraction of accretion disk parameters
from EMRI GW observations. If one were to maximize the overlap function over all
parameters (instead of just a time and phase offset, as done in the
previous section), one might find mismatches much closer to zero, at
the cost of biasing parameter extraction. In this subsection, we
investigate this issue and the spectral signature left by disk-induced
effects.

The effect of possible degeneracies can be assessed by investigating
the Fourier transform of the GW response function, as this is the main
ingredient in matched filtering. We restrict our study of degeneracies
to a simple analytical estimate of the Fourier transform using the
Newtonian stationary phase approximation (SPA) (see
e.g.~\cite{1995PhRvD..52..848P,Yunes:2009yz}).  First, let us review
this approximation in vacuum GR, and then consider the modifications
introduced by leading-order disk effects.

The Fourier transform of the response function $h(t) = A(t) e^{i \phi_{\GW}(t)}$ as
\be
\tilde{h}(f) \equiv
\int_{-\infty}^{\infty} h(t) e^{2\pi i f t} dt\,.
\ee
This generalized Fourier integral can be solved via the method of steepest descent, assuming the amplitude changes slowly relative to the phase and noting that the complex phase $\psi = 2 \pi f t - \phi_{\GW}$ has a stationary point at $d{\psi}(f,t_0)/dt = 2\pi f -d\phi_{\GW}(t_0)/dt = 0$. In this approximation, the Fourier transform becomes (see e.g.~Eq.~$(4.5)$ in~\cite{Yunes:2009yz})
\begin{equation}
\tilde{h}(f) = \frac{8}{5} \frac{A(f)}{2} \; \sqrt{\frac{1}{2 \dot{F}}} e^{i (2 \pi f t_{0} - \phi_{0})} \,,
\label{SPA-wf}
\end{equation}
where the factor of $8/5$ accounts for sky-averaging over beam pattern functions.  The quantities $[t_0, \psi(f,t_0), \ddot\psi(f,t_0)]$ for a fixed $f$ can be found by assuming that the phase and time of merger are fixed $(t_c,\phi_c)$:
\begin{align}
\label{t0-eq}
t_0 - t_{c} &= \int^{\Omega_{0}}_{0} \frac{dE}{d\Omega'} \left(\frac{dE}{dt}\right)^{-1} d\Omega'
= \int^{\R_{0}}_{0} \,\frac{d\R}{\dot{\R}}\,,\\
\phi_{0} - \phi_c &=  \int_{0}^{\Omega_{0}} \frac{dE}{d\Omega'} \left(\frac{dE}{dt}\right)^{-1} \Omega' d\Omega'
= 2 \int_0^{\R_{0}} \Omega \,\frac{d\R}{\dot{\R}}\,,
\label{phi-eq}
\end{align}
where $\Omega_{0} \equiv f/2$ is the stationary point and $\R_{0} \equiv \R(f/2)$. The quantity $\dot\R$ can be constructed from $(dE/d\R)^{-1} \dot{E}$ and $\dot{E}=\mathcal{F}_{\GW}(\R,\Omega)$ is the GW energy loss rate, given in Eqs.~(\ref{e:E},\ref{e:Egw}) in the Newtonian approximation or Eqs.~(\ref{E-binding},\ref{pd-flux-adiab}) in the EOB approximation, if we neglect accretion disk effects.

Neglecting disk effects, we can readily evaluate each of these terms
to leading (Newtonian) order. Using that the orbital frequency is
given by Eq.~\eqref{kep-ang-vel} and the rate of change of the
orbital separation by Eq.~\eqref{e:dotr-inspiral}, Eq.~\eqref{t0-eq}
becomes $t_0 = t_c - (5/256) \mathcal{M} u^{-8/3}$, while
Eq.~\eqref{phi-eq} becomes $\phi(t_0) = \phi_c - (1/16) u^{-5/3}$,
where we have defined the reduced frequency $u\equiv \pi \mathcal{M}
f$, with the chirp mass ${\cal{M}} = q^{3/5} M_{\SMBH}$and $q \equiv
m_{\CO}/M_{\SMBH}$ the mass-ratio. Using the Newtonian expressions for
$A(t)={\cal{M}}/D_{L} ({\cal{M}} \Omega)^{2/3}$ (see e.g.~Eq.$(3.5)$
in~\cite{Yunes:2009yz}), the sky-averaged Fourier amplitude becomes
\be |\tilde{h}|_{\rm vac} = \frac{{\cal{M}}^{5/6}}{\pi^{2/3}
\sqrt{30}\; D_{L}} f^{-7/6}\,.
\label{SPA-amp}
\ee
where $D_L$ is the luminosity distance of the source and $f$ is the observed GW frequency, while the Fourier phase is
\begin{equation}
\psi_{\rm vac}(t_0, f) = 2 \pi f t_0 - \phi(f) = \frac{3}{128} u^{-5/3} + \rm const\,,
\label{phase-SPA}
\end{equation}

Let us now repeat this calculation with accretion disk
modifications. We modify the above algorithm by replacing $\dot{E} =
\dot{E}_{\rm vac} (1 + \delta \dot{\Ls}/\dot{\Ls}_{\GW})$, where
$\delta \dot{\Ls}$ is given by
Eqs.~\eqref{type-1-ldot},~\eqref{ldot-TypeII,SC}
or~\eqref{ldot-TypeII,IPP} for Type I, II-SC and II-IPP migration
respectively, or Eq.~\eqref{Ldotwind-with-gamma} for azimuthal
winds. We replaced $m_{\CO}$ by Eq.~\eqref{mSCO-expansion-num} when
modeling unquenched, BHL accretion. The resulting frequency-domain
phase and amplitude can be parameterized as
\begin{equation}
\label{FFT-phase}
\psi/\psi_{\rm vac} = 1 - \tilde{A}_{1} \alpha_{1}^{{c}_{1}} \dot{m}_{\SMBH 1}^{{c}_{2}} M_{\SMBH 5}^{\tilde{a}_{3}} q_{0}^{\tilde{a}_{4}} u_{0}^{\tilde{a}_{5}}\,,
\end{equation}
and
\begin{equation}
\label{FFT-amp}
|\tilde{h}|/|\tilde{h}|_{\rm vac} = 1 - \tilde{B}_{1} \alpha_{1}^{{c}_{1}} \dot{m}_{\SMBH 1}^{{c}_{2}} M_{\SMBH 5}^{\tilde{a}_{3}} q_{0}^{\tilde{a}_{4}} u_{0}^{\tilde{a}_{5}}\,,
\end{equation}
where $q_{0} \equiv q/10^{-4}$ is the normalized mass-ratio $q = M_{\SMBH}/m_{\CO}$ and $u_{0} \equiv (\pi {\cal{M}} f)/(6.15 \times 10^{-5})$ is a normalized reduced frequency and a GW frequency of $10^{-2}$ Hz. The parameters $(\tilde{A}_{1},\tilde{B}_{1}, \tilde{a}_{i})$ are given in Table~\ref{Table-pars2}, while notice that $(c_{1},c_{2})$ are the same as those in Table~\ref{Table-pars4}. Note that the expressions in Eqs.~\eqref{FFT-phase} and~\eqref{FFT-amp} are valid only in the regime of frequency space where the accretion disk effects are small perturbations away from the vacuum evolution (ie.~at sufficiently small separations).
\begin{table}[ht]
\begin{ruledtabular}
\begin{tabular}{ccccccc}
 & $\tilde{A}_{1}$ & $\tilde{B}_{1}$ & $\tilde{a}_{3}$ & $\tilde{a}_{4}$ & $\tilde{a}_{5}$ \\
\hline
BHL \, $\alpha$ & $3\,(-8)$ & $2\,(-7)$  & $1$ & $4$ & $-20/3$ \\
BHL \, $\beta$ & $1\,(-5)$ & $1\,(-4)$ & $6/5$ & $79/25$ & $-79/15$ \\
\hline
W \, $\alpha$ & $6\,(-17)$ & $1\,(-16)$ & $1$ & $16/5$ & $-16/3$ \\
W \, $\beta$ & $6\,(-12)$ & $4\,(-11)$ & $6/5$ & $59/25$ & $-59/15$ \\
\hline
M1\, $\alpha$ & $3\,(-10)$ & $4\,(-9)$ & $1$ & $16/5$ & $-16/3$ \\
M1\, $\beta$   & $1\,(-6)$ & $3\,(-6)$ & $6/5$ & $59/25$ & $-59/15$ \\
M2\, $\alpha_{\rm SC}$ & $8\,(-6)$ & $2\,(-5)$ & $1$ & $-2/5$ & $-8/3$ \\
M2 \, $\beta_{\rm SC}$ & $6\,(-3)$ & $2\,(-2)$ & $1/4$ & $-1/8$ & $-25/12$ \\
M2 \, $\beta_{\rm IPP}$ & $6\,(-4)$ & $2\,(-3)$ & $4/7$ & $-17/70$ & $-7/3$ \\
\end{tabular}
\caption{\label{Table-pars2} Columns are parameters in
Eq.~\eqref{FFT-phase} and rows are migration effects. As in
Table~\ref{Table-acc-disk-effec}, the notation $x\,(y) = x
\times 10^{y}$ in radians for $\tilde{A}_{1}$ and dimensionless for
$\tilde{B}_{1}$. Observe that the frequency exponent $\tilde{a}_{5} < 0$, implying
that these accretion disk effects are dominant at small frequencies (large radii).}
\end{ruledtabular}
\end{table}

Let us discuss these results further. First, notice that corrections to $\psi(t_{0},f)$ due to Type II migration are orders of magnitude larger than all other effects, as shown by the magnitude of $\tilde{A}_{1}$. Second, notice that all disk-induced corrections depend on {\emph{negative}} powers of frequency (or reduced frequency $u$ in this case). This is because such accretion disk corrections are largest for large radii, equivalent to weak-field GR effects. In fact, they are dominant over the leading-order vacuum term (the factor of $u^{-5/3}$ in $\psi_{\rm vac}$) at low frequency. This suggests that migration effects are not strongly correlated with GR vacuum terms in the PN approximation\footnote{However, the detected signal is also modulated during LISA's orbit around the Sun which we neglect in this paper. This can introduce correlations between parameters that change the GW phase slowly, such as migration effects, source direction, and orientation angles~\cite{2008kocsis-hmd}.}, as the latter depend on positive powers of $u$ relative 
to $u^{-5/3}$. 

\begin{figure}
 \includegraphics[width=8.5cm,clip=true]{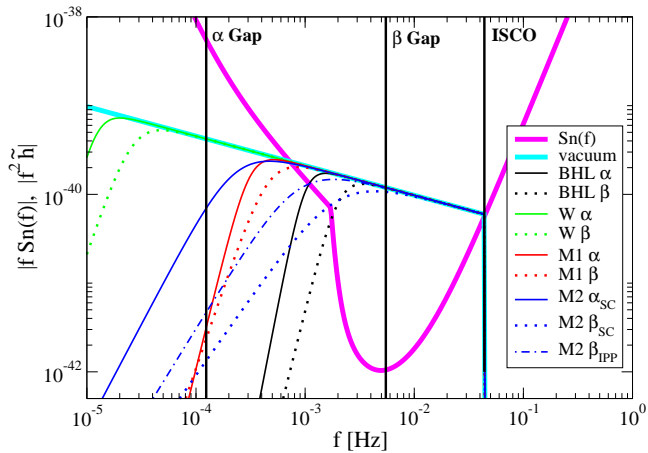}
 \caption{\label{fig:amplitudes} Absolute value squared of the Fourier amplitudes with different accretion disk effects (see label) as a function of frequency in units of Hertz. We also plot this amplitude in vacuum and LISA's spectral noise density. Vertical lines from right to left correspond to the ISCO frequency, the frequency at which a gap forms for $\beta$-disks and for $\alpha$-disks.}
\end{figure}

One might wonder how the accretion disk effects modify the Fourier phase and amplitude when they are not necessary a small perturbation away from the vacuum evolution. In general, the accretion disk correction changes the functional form of the phase or amplitude as follows:
\be
y_{\rm vac} \to \frac{y_{\rm vac}}{1 + \Delta \; \alpha_{1}^{\bar{a}_{1}} \dot{m}_{\SMBH 1}^{\tilde{a}_{2}} M_{\SMBH 5}^{\tilde{a}_{3}} q_{0}^{\tilde{a}_{4}} u_{0}^{\tilde{a}_{5}}}\,
\label{y-func}
\ee
where $y_{\rm vac}=(\psi,|\tilde{h}|)$ when
$\Delta=(\tilde{A}_{1},\tilde{B}_{1})$. This means that, unlike what
Eqs.~\eqref{FFT-phase} and~\eqref{FFT-amp} suggest, the accretion disk
effects always {\emph{suppress}} the vacuum evolution as $\Delta >
0$. Figure~\ref{fig:amplitudes} shows the absolute value squared of
the Fourier amplitudes as a function of frequency for an EMRI with
$M_{\SMBH}=10^{5} M_{\odot}$ and $m_{\CO}=10 M_{\odot}$ and different
accretion disk effects (neglecting all quenching). For comparison, we
also plot the vacuum amplitude and the spectral noise density curve.
Observe that below $f \lesssim 10^{-3}$ Hz, accretion disk induced
migration becomes dominant over GW emission, and the Fourier
amplitude is significantly different.  This effect was also
demonstrated to decrease the GW background for pulsar timing
arrays~\cite{2011MNRAS.411.1467K}. A gap is expected to be opened at
small frequencies, but to close at the radii indicated with vertical
lines (see Sec.~\ref{s:gap}). At these frequencies Type-II migration
transitions to Type-I. Coincidentally, hydrodynamic drag due to an
azimuthal wind and BHL accretion cease at smaller frequencies.

The precise accuracies to which disk parameters can be estimated is difficult
to ascertain. A crude Fisher analysis that neglects degeneracies (a diagonal
approximation) suggests extraction accuracies of up to $1\%$. We expect, 
however, that correlations will deteriorate the accuracy of extraction down
to $10\%$~\cite{Yagi:prep}. Ultimately, a proper assessment of the accuracy
to which disk parameters could be extracted from EMRI observations would
require the detailed mapping  of the likelihood surface with relativistic EMRI
signals, full Fourier transforms, and improved disk modeling (including relativistic
effects and magnetic fields). This, however, is beyond the scope of this paper. 

\section{Discussion}
\label{sec:conclusions}

We have explored the effect of accretion disks on the GWs emitted
during the inspiral of a small BH into a much more massive one. We
have found that disk migration has the biggest influence on the EMRI
dynamics. If EMRIs are detected with LISA, our study suggests that
migration could be measurable within 4 months of observation for a
$\beta$--disk. Depending on the particular EMRI considered, a gap
could open in the accretion disk, leaving an imprint in the GW
observable. We have studied possible degeneracies between accretion
disk and EMRI parameters and found that they are weakly
correlated. This is because disk effects are dominant at large
separations, thus introducing ``negative'' PN terms in the
frequency-domain waveform phase. The LISA detection of a GW from an
EMRI embedded in an accretion disk could therefore allow us to measure
a combination of the $\alpha$ and $\dot{m}_{\SMBH}$ parameters of the disk.

The prospect of probing accretion disks with GWs can be improved
in the presence of an EM counterpart. If the EM luminosity of the SMBH's
accretion disk is observed, one might be able to separately measure
both $\alpha$ and $\dot{m}_{\SMBH}$ by adding the information from a GW
detection. If the GW detection does not show evidence of accretion
disk effects but an EM signal is present, then one might be able to distinguish
between $\alpha$ and $\beta$-disks.

Several caveats should be kept in mind about elements of our analysis
that could be improved on in the future. First, we have considered a
very specific type of EMRI, consisting of quasi-circular orbits on the
equatorial plane. In principle, the CO could be in an inclined and
eccentric orbit. If so, certain accretion disk effects that were
negligible in our analysis, such as the self-gravity of the disk,
could become more important. Such effects would induce additional
apsidal and nodal precession, which could leave a detectable signature
on both the EM and GW signals.

Another important caveat is that we did not explore possible degeneracies 
with other EMRI parameters, like source direction. It is possible that
the orbital modulation of LISA can induces a time dependence mimicing the 
effects of accretion disks. However, since these modulations are periodic, we
expect no strong degeneracies after a multi-year observation.

Another important issue involves the accuracy of the EMRI model
used. Although recently an EOB-inspired EMRI waveform model was
developed~\cite{Yunes:2009ef,2009GWN.....2....3Y,Yunes:2010zj}, no
such model has been thoroughly tested for generic EMRIs. Even in the
case of quasi-circular inspirals on the equatorial plane, the EOB
model is still only accurate up to dephasings of order $10$
radians. This implies that accretion disk effects associated with
dephasing signatures of this magnitude, such as the effect of BHL
accretion, might be difficult to disentangle. However, accretion disk
effects are also important at large separations, whereas mismodeling
errors are of concern only close to the ISCO. Thus, the detection of
EMRIs at moderate binary separations might allow the extraction of
small accretion disk effects. Moreover, the effects of migration are
much larger than any possible waveform mismodeling and so they can be
easily isolated.  Ultimately, a more detailed Markov-Chain Monte-Carlo
study is required to determine the accuracy to which disk effects
can be measured.

Finally, let us highlight the large uncertainty that currently exists in
accretion disk modeling. Even when considering radiatively efficient thin
Newtonian disks, there are several viable models, here parametrized as $\alpha$ and
$\beta$-disks, that can have drastically different effects on EMRI
inspirals. Unfortunately, these Newtonian models are expected to be
highly inaccurate precisely in the regime where EMRIs are most easily
seen, i.e.~in the relativistic regime close to the SMBH. A natural
extension of this work would be to consider EMRIs in the context of
relativistic accretion
disks~\cite{1973blho.conf..343N,1974ApJ...191..499P}, for example
through the inclusion of relativistic BHL
accretion~\cite{2008PhRvD..77j4027B} and
migration~\cite{2010arXiv1010.0758H,2010arXiv1010.0759H}.

\acknowledgments We are grateful to Cole Miller, Ramesh Narayan, Shane
Davis, Scott Hughes, and Zolt\'an Haiman for useful comments and suggestions.  BK and
NY acknowledge support from the NASA through Einstein Postdoctoral
Fellowship Award Number PF9-00063 and PF0-110080 issued by the Chandra
X-ray Observatory Center, which is operated by the Smithsonian
Astrophysical Observatory for and on behalf of the National
Aeronautics Space Administration under contract NAS8-03060.  BK
acknowledges partial support by OTKA grant 68228.  AL acknowledges
support from NSF grant AST-0907890 and NASA grants NNX08AL43G and
NNA09DB30A.

\appendix

\section{Phase Correction due to CO Accretion}
\label{B-H-accurate-estimate}

Let us compute here the GW phase evolution when the CO's mass is varying due to BHL accretion.
The GW phase is given by the first line of Eq.~\eqref{e:phiGW}.
\ba\label{e:phiGW-app}
\phi_{\GW}
&=& 2 \int^{\R_f}_{\R_{0'}} \Omega(\R) \frac{d\R}{\dot{\R}}
= - \frac{5}{32} M_{\SMBH} \int_{\R_{0'}}^{\R_{f}} \frac{\R^{3/2}}{m_{\CO}(\R)} d\R\,.
\ea
Two important corrections are induced by the CO's variable mass: a change in the limits of integration
$\R_{0}\rightarrow \R_{0'}$ and a change in the denominator of the integrand in the phase evolution,
$m_{\CO}\rightarrow m_{\CO}(t)$. We expand both in a Taylor
series around $\R_{0}$ and the initial CO mass $m_{\CO, 0}$ at $\R=\R_{0'}$.
We evaluate the change in the total GW phase induced by accretion while keeping
$\R_f$, the observation time $T=t_f - t_0$, and $m_{\CO, 0}$ fixed.

The limits of integration can be computed by integrating the radial inspiral evolution equation
(\ref{e:dotr-inspiral}) with the time dependent $m_{\CO}(t)$,
\be\label{e:RfR0}
\R^{4}_{f} - \R^{4}_{0'}  =-\frac{256}{5}  \frac{1}{M_{\SMBH}^{2}} \int_{t_{0'}}^{t_f} m_{\CO}(t) dt\,.
\ee
where
\be
m_{\CO}(t) = m_{\CO,0} + \int_{t_{0'}}^{t} \dot{m}_{\CO}(t') dt'\,.
\ee
Note that $\dot{m}_{\CO}\equiv\dot{m}_{\CO}(\R)$ given
by Eqs.~(\ref{e:MBondi},\ref{e:dmSCO4}). Changing integration variable from $t$ to $\R$
using $dt = d\R/\dot{\R}$ using Eq.~(\ref{e:dotr-inspiral}) for $\dot{\R}(\R)$, we get
\be
 \int_{t_0'}^{t_f} m_{\CO}(t) dt = m_{\CO,0} \left( T + \frac{25}{4096} \frac{M_{\SMBH}^{4}}{m_{\CO,0}^{2}} \left< \delta m_{\CO} \right>_{3,3} \right)\,,
\ee
where\footnote{Since $\left<\delta m_{\CO} \right>_{A,B}\ll m_{\CO,0} T$, we approximate the lower integration
bounds in Eq.~(\ref{e:dmab}) with $\R_0$.}
\ba
\left<\delta m_{\CO} \right>_{A,B} &\equiv&
\int_{\R_{f}}^{\R_{0}} d \R \; \R^{A} \int_{\R}^{\R_{0}} d \R'  \; \R'^{B} \frac{\dot{m}_{\CO}(\R')}{{m}_{\CO,0}}\,.
\label{e:dmab}
\ea
Substituting in Eq.~(\ref{e:RfR0}), the initial separation becomes
\be
\R_{0'}^{\dot{m}} = \R_{f} \left[ 1 + \frac{\tau_{\SPA} }{ \R_{f}^{4}}
\left( 1 + \frac{25}{4096} \frac{M_{\SMBH}^{4}}{m_{\CO,0}^{2} T} \left< \delta m_{\CO} \right>_{3,3}  \right) \right]^{1/4}\,.
\ee
where $\tau_{\SPA}$ is the dimensionless observation time given by Eq.~(\ref{e:r0}),
the ${\dot{m}}$ index was introduced to distinguish from other modifications below,
c.f. Eq.~(\ref{e:r0}) for $\R_0$.
Thus, the change in the lower integration bound
in Eq.~(\ref{e:phiGW-app}) is
\be
\delta\R_{0'}^{\dot{m}_{\CO}} = \R_{0'}^{\dot{m}} - \R_{0} \approx
 \frac{5}{64}  \frac{M_{\SMBH}^{2}}{m_{\CO,0} \R_{0}^{3}}\left< \delta m_{\CO} \right>_{3,3}\,.
\ee

Since the total relative change in $m_{\CO}$ is very small during the observation,
we may approximate $1/m_{\CO}(t)$ in Eq.~(\ref{e:phiGW-app})
as
\be
\frac{1}{m_{\CO}(t)} = \frac{1}{m_{\CO,0} + \int_{t_{0}}^{t} \dot{m}_{\CO}(t') dt' }
\approx \frac{1}{m_{\CO, 0}} - \frac{1}{m_{\CO, 0}^2} \int_{t_{0}}^{t} \dot{m}_{\CO}(t') dt'
\label{mSCO-exp}
\ee
so that Eq.~(\ref{e:phiGW-app}) becomes
\begin{align}
\phi_{\GW}^{\dot{m}_{\CO}}
&\approx - \frac{5}{32} M_{\SMBH} \int_{\R_{0'}}^{\R_{f}} d\R \frac{\R^{3/2}}{m_{\CO,0}}
\left[1 - \int_{t(\R_{0'})}^{t(\R)}  dt' \frac{\dot{m}_{\CO}(t')}{m_{\CO,0}} \right]
\nonumber \\
&= \frac{1}{16} \frac{M_{\SMBH}}{m_{\CO,0}}  \left(\R_{0'}^{5/2} - \R_{f}^{5/2} \right)
-\frac{25}{2048} \frac{M_{\SMBH}^{3}}{m_{\CO,0}^{2}} \left<\delta {m}_{\CO} \right>_{3/2,3}\,,
\end{align}
where the ${\dot{m}_{\CO}}$ label denotes that the CO is accreting,
in the second line we have changed integration variables using $dt'  = d\R'/\dot{\R}'$
and used Eq.~(\ref{e:dmab}).

Relative to the GW phase without accretion,
\begin{align}
\delta \phi_{\GW}^{\dot{m}_{\CO}}
&= \frac{5}{32} \frac{M_{\SMBH}}{m_{\CO,0}}   \R_{0}^{3/2}\delta\R_{0'}^{\dot{m}_{\CO}}
-\frac{25}{2048} \frac{M_{\SMBH}^{3}}{m_{\CO,0}^{2}} \left<\delta {m}_{\CO} \right>_{3/2,3}\nonumber\\
&\approx
\frac{25}{2048} \frac{M_{\SMBH}^{3}}{m_{\CO,0}^{2}}
\left( \frac{\left<\delta m_{\CO} \right>_{3,3}}{\R_{0}^{3/2}}
- \left<\delta {m}_{\CO} \right>_{3/2,3} \right).
\label{e:dphi_dmSCO}
\end{align}
Equation~(\ref{e:dphi_dmSCO}) and (\ref{e:dmab}) can be used the calculate the GW
phase shift for arbitrary $\dot{m}_{\CO}$.
In the next two subsections we consider BHL accretion $\dot{m}_{\CO}^{\Bondi}$
and gas supply limited BHL accretion $\dot{m}_{\CO}=\dot{M}_{\SMBH}$.

\subsection{Unsaturated BHL accretion}\label{s:appendix-unsat-Bondi}
First consider the case where BHL accretion is not limited by the amount of local gas supply,
$\dot{m}_{\CO}^{\Bondi}(r)\leq \dot{M}_{\SMBH}$.
BHL accretion is not quenched by local gas supply if
the observation is limited to orbital radii $\R\leq \R_{\rm q}$, or equivalently,
if the dimensionless accretion rate onto the SMBH $\dot{m}_{\SMBH}\geq\dot{m}_{\SMBH, \rm q}$.
In practice, this is the case during a year of observation approaching ISCO,
if the SMBH accretion rate is moderate to high $\dot{m}_{\SMBH}\gtrsim 0.3$
or if the mass-ratio is very small $m_{\CO}/M_{\SMBH}\lesssim 10^{-6}$ [see
Eq.~(\ref{e:rq})].

Substituting the BHL accretion rate, Eqs.~(\ref{e:MBondi},\ref{e:dmab}) for $\alpha$ and $\beta$-disks
into Eq.~(\ref{e:dphi_dmSCO}) yields
%
\begin{align}
\delta \phi_{\GW}^{\alpha,\Bondi} &=
-0.733\, \alpha_{1}^{-1}\frac{M_{\SMBH 5}^2}{\dot{m}_{\SMBH 1}^{5} m_{\CO 1}} \R_{0,20}^{25/2}
\left(1- \frac{280}{99}x^{5/2}\right.\nonumber\\
&\quad\left. + \frac{175}{99}x^{4} + \frac{56}{99}x^{25/2} - \frac{50}{99} x^{14}\right)\,,
\label{e:dphi-Bondia}\\
\delta \phi_{\GW}^{\beta,\Bondi} &=
-495\, \alpha_{1}^{-4/5}\frac{M_{\SMBH 5}^{11/5}}{\dot{m}_{\SMBH 1}^{17/5}m_{\CO 1}}\R_{0,20}^{52/5}
\left(1- \frac{6188}{2133}x^{5/2}\right.\nonumber\\
&\quad\left. + \frac{7735}{4266}x^{4} + \frac{2975}{4266}x^{52/5} - \frac{1300}{2133} x^{119/10}\right)\,,
\label{e:dphi-Bondib}
\end{align}
where $m_{\CO 1}$ refers to the initial CO mass at $\R_{0}$, $x=\R_f/\R_0$,
$\R_{0}\equiv \R_{0}(\R_{f},T)$ is given by Eq.~(\ref{e:r0}), and $\R_{0,20}=\R_{0}/20$.
The phase evolution given by Eq.~(\ref{e:dphi-Bondia}--\ref{e:dphi-Bondib}) depends on only two sets
of parameters: the time-independent coefficient preceding the parentheses, and the time dependent
quantity $\tau/\R_{f}^4$. The later also appears in the standard inspiral phase expression, Eq.~(\ref{e:phiGW}),
which we used to distinguish two cases: when the observation time is long or short relative to the inspiral
timescale at the given radius [see Eqs.~(\ref{e:phiGW-short}--\ref{e:phiGW-long})].

We can similarly distinguish here between two asymptotic cases.
For long observations ($T\gg T_{\rm crit}$) or small separations ($\R_{f}\ll \R_{f,{\rm crit}}$), $\R_{0} \approx \tau^{1/4}$, $x\approx \R_f/\tau^{1/4} \ll 1$, we can approximate the dephasing with the general formula
\begin{equation}
\delta \phi_{\GW, \rm long} = -C_{1}
 \alpha_{1}^{-c_{1}}
 \dot{m}_{\SMBH 1}^{-c_{2}}
 \frac{m_{\CO n}^{c_{3}}}{M_{\SMBH 5}^{c_{4}}}
 T_{\yr}^{c_{5}}
 \left(1 - C_{2}
 \frac{M_{\SMBH 5}^{2 c_{6}}}{m_{\CO 1}^{c_{6}}}
 \frac{\R_{f,10}^{4 c_{6}}}{T_{\yr}^{c_{6}}} \right)\,
\label{long-gen-form}
\end{equation}
where the coefficients $(C_{i},c_{j})$ are given by the first two rows of Table~\ref{Table-pars4} and
$n=2$ for Type II migration, while $n=1$ for all other migration disk effects.
For short observations ($T\ll T_{\rm crit}$) or small separations ($\R_{f}\gg \R_{f,{\rm crit}}$),
$\tau/\R_{f}^4\ll 1$, $\R_{0} \approx \R_f  + \tau/(4\R_f^3)$, $x\approx 1-\tau/(4\R_f^4)$, and we can
approximate the dephasing with the general formula
\begin{equation}
\delta \phi_{\GW, \rm short} = - \frac{D_{1}}{\alpha_{1}^{c_{1}} \dot{m}_{\SMBH 1}^{c_{2}}}
 \frac{m_{\CO n}^{d_{1}}}{M_{\SMBH 5}^{d_{2}}}
 T_{\yr}^{d_{3}} \R_{f 10}^{d_{4}}
 \left(1 - D_{2}
 \frac{m_{\CO 1}}{M_{\SMBH 5}^{2}}
 \frac{T_{\yr}}{\R_{f,10}^{4 }} \right)\,
\label{short-gen-form}
\end{equation}
where the coefficients $(D_{i},d_{j})$ are given by the first two rows of Table~\ref{Table-pars4}
and $n=2$ for Type II migration, while $n=1$ for all other migration disk effects.

\subsection{Gas supply limited BHL accretion}
If $\dot{m}_{\CO}^{\Bondi}(\R)\geq \dot{M}_{\SMBH}$, the accretion onto the CO is
limited by the amount of local gas supply. This is the case
outside $\R\geq \R_{\rm q}$, or if the SMBH accretion rate satisfies
$\dot{m}_{\SMBH}\geq\dot{m}_{\SMBH, \rm q}$, see Eq.~(\ref{e:rq})
for $(\R_{\rm q},\dot{m}_{\SMBH, \rm q})$.
In practice, this is the case for $\dot{m}_{\SMBH}\lesssim 0.1$ for beta disks
if the final orbital radius is not very close to the ISCO (e.g.~$\R_f\gtrsim 10$).
For larger $\dot{m}_{\SMBH}$ accretion rates approaching the ISCO
or intermediate mass-ratio inspirals, BHL accretion starts supply limited
and becomes unsaturated near the ISCO.

Assuming $\dot{m}_{\CO} = \dot{M}_{\SMBH}$
in Eqs.~(\ref{e:dmab},\ref{e:dphi_dmSCO}), we get
%
\begin{align}
\delta \phi_{\GW}^{\rm sup.~BHL} &\approx
-4.2 \frac{\dot{m}_{\SMBH 1} M_{\SMBH 5}^{4} }{m_{\CO 1}^{3}} \R_{0,20}^{13/2}
\left( 1 - \frac{208}{63} x^{5/2}
\right.\nonumber\\&\quad\left.
+ \frac{130}{63} x^4  + \frac{80}{63} x^{13/2} - \frac{65}{63} x^8 \right)
\label{e:dphi-quenched-Bondi}
\end{align}
for both $\alpha$ and $\beta$-disks, where again $\R_{0}\equiv \R_{0}(\R_{f},T)$ given by Eq.~(\ref{e:r0}) and $x=\R_{f}/\R_{0}$.
Relative to the BHL accretion case with unlimited gas supply
Eq.~(\ref{e:dphi-Bondia}--\ref{e:dphi-Bondib}), the phase is a less steep function of radius.
The evolution is again determined by two combination of parameters, the constant coefficient
(which is now different) and the time dependent quantity $\tau/\R_{f}^4$.

In the two limiting cases, where the observation is long or short
relative to the inspiral timescale, we again can parameterize the
dephasing as in Eq.~\eqref{long-gen-form} and~\eqref{short-gen-form}
respectively, where the coefficients $(C_{i},c_{j})$ and
$(D_{i},d_{j})$ are given by the third row of Table~\ref{Table-pars4}.

\section{Phase Correction due to Wind and Migration}
\label{azi-wind-estimate}
In this appendix we derive the correction to the GW phase due to a modification in angular momentum dissipation.
The result of a similar, but more general, angular momentum dissipation rate $\delta\Ls/\Ls_{\rm GW} = A r^B$
is presented in Paper I~\cite{2011arXiv1103.4609Y}.
We focus on quasi-circular orbits only.  From Eq.~(\ref{e:phiGW}),
\begin{align}
\phi_{\GW}
&= 2 \int^{\R_f}_{\R_{0'}} \Omega(\R) \frac{d\R}{\dot{\R}}
= 2 \int^{\R_f}_{\R_{0'}} \Omega(\R) \left(\frac{d\Ls}{dt}\right)^{-1} \left(\frac{d \Ls}{d \R}\right)  d\R\nonumber\\
&= \int^{\R_f}_{\R_{0'}} \R^{-2} \frac{d\R}{\dot{\Ls}(\R)}\,,
\label{e:phiGW-app2}
\end{align}
where we have expressed the GW phase as a function of radial and temporal derivatives of the specific angular
momentum of the CO. Similar to Appendix~\ref{B-H-accurate-estimate},
we calculate the total total change in the GW phase relative to the unperturbed GW inspiral phase,
by keeping the final separation $\R_f$ and the observation time $T$ fixed. Relative to the
vacuum inspiral Eq.~(\ref{e:phiGW}), a modified angular momentum dissipation causes a phase shift by changing the lower integration bound and $\dot{\Ls}$ in Eq.~(\ref{e:phiGW-app2}).
Since the additional specific angular momentum loss $\delta\dot{\Ls}$ is small
relative to $\dot{\Ls}_{\GW}$, we may approximate the result by expanding in a
series in the small quantity $\delta\dot{\Ls}/\dot{\Ls}_{\GW}$.

First to estimate $\R_{0'}$, note that the accelerated dissipation of angular momentum causes
an accelerated inspiral rate.  For
circular orbits, $\Ls=M_{\SMBH} \R^{1/2}$, $E = M_{\CO}/(2\R)$, so that
\be
\dot{\R} = \frac{2\R^{1/2} }{ M_{\SMBH} }\dot{\Ls} =
-\frac{64}{5} \frac{m_{\CO}}{M_{\SMBH}^{2}} \R^{-3} \left(1 + \frac{\delta\dot{\Ls}}{\dot{\Ls}_{\GW}} \right)\,.
\ee
This equation can be integrated to give
\be
\R_{0'}^{\dot{\Ls}} = \R_{f} \left[ 1 + \frac{\tau }{ \R_{f}^{4}}
\left( 1 +  \frac{4}{\tau} \left< \delta \Ls \right>_{3}  \right) \right]^{1/4}\,.
\ee
where we have introduce the index $\dot{\Ls}$ to distinguish from other effects, and we define\footnote{Since
$\dot{\Ls}_{\rm wind}/\dot{\Ls}_{\GW}\ll 1$, we approximate the lower integration bound with $\R_0$ in Eq.~(\ref{e:dLa}).}
\be\label{e:dLa}
\left< \delta \Ls \right>_{A} \equiv \int_{\R_{f}}^{\R_{0}} d\R \; \R^{A} \left(\frac{\delta\dot{\Ls}}{\dot{\Ls}_{\GW}}\right) d\R'\,.
\ee
Comparing to Eq.~(\ref{e:r0}), the change in the lower integration bound
\be
\delta\R_{0'}^{\dot{\Ls}}
= \R_{0'}^{\dot{\Ls}} - \R_{0} \approx
 \frac{\left< \delta \Ls \right>_{3}}{\R_0^3}
\ee

Now using
\begin{align}
\frac{1}{\dot{\Ls}} = \frac{1}{\dot{\Ls}_{\GW} + \delta\dot{\Ls}}
\approx \frac{1}{\dot{\Ls}_{\GW}}\left(1 - \frac{\delta\dot{\Ls}}{\dot{\Ls}_{\GW}}\right)
\end{align}
for the integrand in Eq.~(\ref{e:phiGW-app2}), the GW phase is
\begin{align}
\phi_{\GW}^{\dot{\Ls}} &\approx \frac{5}{32}\frac{M_{\SMBH}}{m_{\CO}}\int_{\R_f}^{\R_{0'}} d\R\, \R^{3/2}
\left(1 - \frac{\delta\dot{\Ls}}{\dot{\Ls}_{\GW}}\right)\nonumber\\
&= \frac{5}{32}\frac{M_{\SMBH}}{m_{\CO}} \left[ \frac{2}{5} \left(\R_{0'}^{5/2} - \R_{f}^{5/2} \right)
- \left< \delta \Ls \right>_{3/2}\right]
\end{align}
so that relative to the vacuum inspiral phase
\begin{align}
\delta\phi_{\GW}^{\dot{\Ls}}
&\approx \frac{5}{32} \frac{M_{\SMBH}}{m_{\CO}} \left( \R_0^{3/2} \delta\R_{0'}^{\dot{\Ls}}  - \left< \delta \Ls \right>_{3/2}\right)
\nonumber\\
&=
\frac{5}{32} \frac{M_{\SMBH}}{m_{\CO}}
\left(
 \frac{\left< \delta \Ls \right>_{3}}{\R_0^{3/2}}
- \left< \delta \Ls \right>_{3/2}\right).
\label{e:dphi_dL}
\end{align}
Equations~(\ref{e:dphi_dL}) and (\ref{e:dLa}) are the analogues of
Eqs~(\ref{e:dphi_dmSCO}) and (\ref{e:dmab}) for the GW
phase shift caused by an additional source of angular momentum dissipation $\delta \dot{\Ls}$.
In the next three subsections we consider the azimuthal wind, Type-I migration, and
Type-II migration. In these cases $\delta \dot{\Ls}/\dot{\Ls}_{\rm GW}=A r^B$ where $A$ and $B$
are constants, so the integral in Eq.~(\ref{e:dLa}) can be evaluated analytically.

\subsection{Wind -- unsaturated BHL accretion}
If the additional angular momentum dissipation is caused by an azimuthal wind
where the accretion onto the CO is the unsaturated BHL rate,
$\delta\dot{\Ls}(\R) = \dot{\Ls}_{\rm wind}^{\Bondi}(\R)$ given by Eq.~(\ref{e:L'wind}).
Substituting in Eqs.~(\ref{e:dphi_dL}) and (\ref{e:dLa}) we get
%
\begin{align}
\delta\phi_{\GW}^{\alpha,\rm Bw}
&\approx
 -5.2 \times 10^{-3} \; \alpha_1^{-1} \frac{M_{\SMBH 5}^2}{\dot{m}_{\SMBH 1}^{3} m_{\CO 1}} \R_{0,20}^{21/2}
 \nonumber \\
&\times  \left(1 - 8 x^{21/2} + 7 x^{12} \right)\,,\\
\delta\phi_{\GW}^{\beta,\rm Bw}
&\approx
 -10 \; \alpha_1^{-4/5} \frac{M_{\SMBH 5}^{11/5}}{\dot{m}_{\SMBH 1}^{7/5}m_{\CO 1}}  \R_{0,20}^{42/5}
\nonumber\\&\quad
  \times \left(1 - \frac{33}{5} x^{42/5}+ \frac{28}{5} x^{99/10} \right)\,,
\end{align}
where $x=\R_f/\R_0$ and $\R_0\equiv\R_0(\R_f,T)$, see discussion
following Eqs.~(\ref{e:dphi-Bondib},\ref{e:dphi-quenched-Bondi}).
Following the derivation presented in
Appendix~\ref{s:appendix-unsat-Bondi}, we can expand in long and short
observation times relative to the inspiral timescale to recover a
dephasing as in Eq.~\eqref{long-gen-form} and~\eqref{short-gen-form}
with the coefficients $(C_{i},c_{j})$ and $(D_{i},d_{j})$ given by the
fourth and fifth rows of Table~\ref{Table-pars4}.

\subsection{Wind -- quenched BHL accretion}
If the additional angular momentum dissipation is caused by an azimuthal wind
where the accretion onto the CO is limited by the amount of gas supply
then we substitute $\delta\dot{\Ls}(\R) = \dot{\Ls}_{\rm wind}^{\rm sup.~BHL}(\R)$
Eq.~(\ref{e:L'wind-supply-limited}) into Eq.~(\ref{e:dphi_dL}) to obtain the corresponding
total GW phase shift
%
\be
\delta \phi_{\GW}^{\rm sBw} =
\left(\begin{array}{c}
-0.065 \\
-0.16
\end{array}\right)
\frac{\dot{m}_{\SMBH 1}^{3}  M_{\SMBH 5}^4}{m_{\CO 1}^3} \R_{0,20}^{9/2}
\left(1+ 3x^{6} - 4x^{9/2} \right)\,,
\label{e:dphi-wind-supply-limited-Bondi}
\ee
where throughout this section the top and bottom rows correspond to
$\alpha$ and $\beta$-disks, respectively, and the parameters $x\equiv
x(\R_f,T)$ and $\R_0\equiv \R_0(\R_f,T)$, see discussion following
Eqs.~(\ref{e:dphi-Bondib},\ref{e:dphi-quenched-Bondi}). Notice that
although we are here setting $\dot{m}_{\CO} = \dot{M}_{\SMBH}$, there
is still a dependance on the type of accretion disk, due to the factor
of $\gamma$ in Eq.~\eqref{Ldotwind-with-gamma}.

Again, as in Appendix~\ref{s:appendix-unsat-Bondi}, we expand in long and short observation times to obtain the dephasing
of Eqs.~\eqref{long-gen-form} and~\eqref{short-gen-form} with the coefficients $(C_{i},c_{j})$ and $(D_{i},d_{j})$ given by
the sixth and seventh rows of Table~\ref{Table-pars4}.

\subsection{Type-I migration}
\label{type-I}

Let us now compute the dephasing introduced by the angular momentum dissipation due to Type-I migration.
Substituting Eqs.~\eqref{type-1-ldot} into Eqs.~(\ref{e:dphi_dL}) and (\ref{e:dLa}) we get
%
\begin{align}
\delta \phi_{\GW}^{\alpha, \rm TI}
&= -6.1 \times 10^{-4} \; \alpha_1^{-1} \frac{\dot{m}_{\SMBH 1}^{-3} M_{\SMBH 5}^{2} }{m_{\CO 1}}
\R_{0,20}^{21/2}
\nonumber \\
&\quad\times
\left(1 - 8 x^{21/2} + 7 x^{12} \right)\,,
\\
\delta \phi_{\GW}^{\beta, \rm TI} &=
-1.5 \; \alpha_1^{-4/5} \frac{\dot{m}_{\SMBH 1}^{-7/5} M_{\SMBH 5}^{11/5}}{m_{\CO 1}}
\R_{0,20}^{42/5}
\nonumber \\
&\quad\times
\left(1 - \frac{33}{5} x^{42/5} + \frac{28}{5} x^{99/10} \right)\,.
\end{align}
where again $x\equiv x(\R_f,T)$ and $\R_0\equiv \R_0(\R_f,T)$

We now take the two limiting cases of long and short observations. For
long and short observations, we find dephasings as in
Eq.~\eqref{long-gen-form} and~\eqref{short-gen-form} with
coefficients $(C_{i},c_{j})$ and $(D_{i},d_{j})$ given by the eighth
and ninth rows of Table~\ref{Table-pars4}.

\subsection{Type-II migration}
\label{type-II}

Consider now the dephasing corresponding to the angular momentum dissipation
due to Type-II migration. Substituting Eqs.~\eqref{ldot-TypeII,SC} and~\eqref{ldot-TypeII,IPP} into
Eqs.~(\ref{e:dphi_dL}) and (\ref{e:dLa}) we get
%
\begin{align}
\delta \phi_{\GW}^{\alpha, \rm TII,SC} &=
-8 \times 10^{-3} \; \frac{\dot{m}_{\SMBH 1} M_{\SMBH 5}^{4} }{m_{\CO 2}^{3}} \R_{0,20}^{13/2}
\nonumber\\&\quad\times
\left(1 - \frac{16}{3} x^{13/2} + \frac{13}{3} x^{8} \right)\,,
\\
\delta \phi_{\GW}^{\beta, \rm TII,SC} &=
-22 \; \alpha_1^{1/2} \frac{\dot{m}_{\SMBH 1}^{5/8} M_{\SMBH 5}^{21/8} }{m_{\CO 2}^{19/8}}\R_{0,20}^{45/8}
\nonumber\\&\quad\times
\left(1 - \frac{19}{4} x^{45/8} + \frac{15}{4} x^{57/8} \right)\,,
\\
\delta \phi_{\GW}^{\beta, \rm TII,IPP} &=
-1.2 \; \alpha_1^{2/7} \frac{\dot{m}_{\SMBH 1}^{11/14} M_{\SMBH 5}^{45/14} }{m_{\CO 2}^{37/14}}\R_{0,20}^{6}
\nonumber\\&\quad\times
\left(1 -5 x^{6} + 4 x^{15/2} \right)\,.
\end{align}
where again $x\equiv x(\R_f,T)$ and $\R_0\equiv \R_0(\R_f,T)$. Notice that we have normalized the CO's mass to
$m_{\CO 2} = m_{\CO}/(100 M_{\SMBH})$, where a gap opens and Type-II migration occurs.

Let us now take the two limiting cases of long and short
observations. For long and short observations, we find a dephasing as
in Eq.~\eqref{long-gen-form} and~\eqref{short-gen-form} respectively,
with $(C_{i},c_{j})$ and $(D_{i},d_{j})$ given by the tenth, eleventh
and twelfth rows of Table~\ref{Table-pars4}.

\section{Phase Correction due to the Disk Gravity}
\label{SG-estimate}

Let us study how the gravitational potential generated by the disk, $\Phi_{\rm disk}$, affects
the GW phase. The latter is given by
\be
\phi_{\GW} = 2 \int_{t_{0}'}^{t_{f}} \Omega dt = 2 \int_{\R_{0}'}^{\R_{f}} \Omega(\R) \frac{d\R}{\dot{\R}}\,,
\label{GWphase-SG}
\ee
where now both the angular velocity $\Omega(\R)$ and the inspiral rate $\dot{\R}$ are modified by the disk potential,
see Eq.~(\ref{e:Omega-SG}) and (\ref{e:dr-sG}). The later also implies a change in the integration bound for a fixed $\R_f$
and observation time.

We can integrate Eq.~(\ref{e:dr-sG}) perturbatively to obtain
\be
\R_{0'} = \R_{f} \left[1 + \frac{\tau}{\R_{f}^{4}} \left(1 + \frac{4}{\tau} \left<\Phi_{\rm disk}\right>_{6,2} \right) \right]^{1/4}\,.
\ee
where we have defined
\be
\left< \delta \Phi \right>_{m,n} \equiv \int_{\R_{f}}^{\R_{0}} \R^{m} \frac{d^{n} \Phi_{\rm disk}}{d\R^{n}} d\R\,.
\ee

Substituting Eqs.~(\ref{e:Omega-SG}) and (\ref{e:dr-sG}) in Eq.~(\ref{GWphase-SG}) gives
\begin{align}
\phi_{\GW} &= \frac{5}{32 \,\eta} \int_{\R_{f}}^{\R_{0}'} \R^{3/2}
\left(1 + \frac{\R^{2}}{2} \frac{d \Phi_{\rm disk}}{d\R}+ \R^{3} \frac{d^{2} \Phi_{\rm disk}}{d\R^{2}} \right) d\R \nonumber\\
&= \frac{5}{32 \,\eta} \left[
\frac{2}{5}\left(\R_{0'}^{5/2} - \R_{f}^{5/2}\right) + \frac{1}{2}\langle \delta\Phi \rangle_{7/2,1} + \left< \delta\Phi \right>_{9/2,2}\right].
\end{align}
Subtracting from this expression the vacuum expression for the GW phase, we find
\be\label{e:axisymmetric-gravity}
\delta \phi_{\GW}^{\rm disk} = -\frac{5}{32} \frac{M}{m_{\CO}} \left( \frac{\left<\delta\Phi\right>_{6,2}}{\R_{0}^{3/2}}
- \frac{1}{2}\left< \delta\Phi \right>_{7/2,1}
- \left< \delta\Phi \right>_{9/2,2} \right)\,.
\ee
The result is algebraically similar to the phase shift due to a modified angular momentum loss rate
Eq.~(\ref{e:dphi_dL}) or a modified CO mass Eq.~(\ref{e:dphi_dmSCO}).
Note that the overall minus sign is compensated for by the sign of the potential
$\Phi_{\rm disk} < 0$.

\subsection{Disk Potential}\label{s:potential-app}
Let us compute a more convenient form of the disk potential $\Phi_{\rm disk}$.
Assuming that $\Sigma(r) = \Sigma_0 r^{\gamma}$, the accretion disk parameters
$(\alpha, \dot{m}_{\SMBH}, M_{\SMBH})$, carried by $\Sigma_0$, can be taken out of the integrals in
Eqs.~(\ref{e:Phidisk},\ref{e:Phidisk2}). The integral only depends on $(\gamma,\R,\R_{\min},\R_{\max})$.
We may evaluate the integrals in the Legendre expansion (\ref{e:Phidisk2}).
\begin{align}\label{e:phidisk-app}
 \Phi_{\rm disk} &= 2\pi\Sigma_0
 \sum_{\ell=0,2}^{\infty} [P_{\ell}(0)]^2
 \left\{
 \frac{\R^{1+\gamma}}{\ell + 2 + \gamma}
\left[1 - \left(\frac{\R_{\min}}{\R}\right)^{\ell+2+\gamma}\right]
 \right.
 \nonumber\\
 &\quad\left.+ \frac{\R^{1+\gamma}}{\ell - 1 - \gamma}
\left[1 - \left(\frac{\R}{\R_{\max}}\right)^{\ell-1-\gamma}\right]
\right\}
\end{align}
Next we exercise the gauge freedom to set
\be
\Phi^{\rm new}_{\rm disk}(\R) \equiv \Phi_{\rm disk}(\R)  + \frac{2\pi r_{\rm max}\Sigma(r_{\max}) }{1+\gamma}
\ee
and in the following drop the ``new'' specifier.
Note that for both disk models $\gamma > -1$; it is $3/2$ and $-3/5$ for $\alpha$ and $\beta$-disks.
The disk potential in Eq.~(\ref{e:phidisk-app}) is expressed as a sum of two terms in the curly
brackets, which correspond to the potential of the disk interior and exterior to the orbit, respectively.

Let us discuss the contribution of various multipolar harmonics, $\ell$.
\begin{align}
\Phi^{\ell=0}_{\rm disk} &= \frac{2 \pi\, \Sigma_0 \;\R^{\gamma+1} }{(1+\gamma)(2+\gamma)}
\left[ 1 +  (1+\gamma) \left(\frac{\R_{\min}}{\R}\right)^{2+\gamma} \right] \nonumber\\
 \Phi_{\rm disk}^{\ell=2} &=
 -\frac{\pi}{2}\frac{ \Sigma_0 \; \R^2}{(\gamma - 1)}
\left(\R_{\max}^{\gamma -1} -  \frac{5}{4+\gamma}\R^{\gamma -1}
-\frac{\gamma-1}{4+\gamma}\frac{\R_{\min}^{4+\gamma}}{\R^5}
\right)
\nonumber\\
 \Phi_{\rm disk}^{\ell=4} &=
 -\frac{81\pi}{32} \frac{ \Sigma_0 \; r^{\gamma + 1} }{(3 - \gamma)(6+\gamma)}
\left[ 1 -  \frac{6+\gamma}{9}\left(\frac{\R}{\R_{\max}}\right)^{3-\gamma}
\right.\nonumber\\&\quad\left.
-\frac{3-\gamma}{9}\left(\frac{\R_{\min}}{\R}\right)^{6+\gamma}\right]\,.
\label{e:phidisk-app-L}
\end{align}
In each case we have arranged the terms in increasing order for $\gamma=3/2$.
Equation~(\ref{e:phidisk-app-L}) shows that for $\gamma>1$ and $\R \ll \R_{\max}$ the potential is dominated by
the quadrupolar harmonic $\ell=2$,
\be\label{e:gamma>}
\Phi_{\gamma > 1}(\R) \approx\Phi_{\ell=2}(\R) \approx
 -\frac{\pi}{2}\frac{ \Sigma_0\; \R_{\max}^{\gamma -1} }{(\gamma - 1)}
\R^2.
\ee
When $-1<\gamma<1$, the disk potential
is asymptotically independent of the inner and outer boundaries for $\R_{\min} \ll \R \ll \R_{\max}$.
We can then analytically evaluate the infinite sum, and get
\be\label{e:gamma<}
\Phi_{\gamma < 1}(\R) \approx
2\pi c_{0} \Sigma_0\, \R^{\gamma+1}
\ee
where
\begin{align}
c_0 =
\frac{1}{\gamma + 1}
\frac{ \Gamma\hspace{-2pt}\left(1+ \frac{\gamma}{2}\right)  \Gamma\hspace{-2pt}\left(\frac{1-\gamma}{2}\right) }{
 \Gamma\hspace{-2pt}\left(\frac{3+\gamma}{2}\right)  \Gamma\hspace{-2pt}\left(\frac{-\gamma}{2}\right) }\,.
\end{align}
so that $c_0=1.38$ for $\beta$--disks where $\gamma=-3/5$.
Equations~(\ref{e:gamma>}) and (\ref{e:gamma<}) represent the asymptotic solutions for $\alpha$ and $\beta$-disks, respectively.
Substituting the particular surface density profiles for the two disk models leads to the potential given by Eq.~(\ref{e:phi-disk0}).

\subsection{Axisymmetric disk gravity without a gap}
We can now evaluate the GW phase shift, Eq.~(\ref{e:axisymmetric-gravity}), for the potential generated
by $\alpha$ and $\beta$-disks
without a gap. We restrict to $\R_{\min}\ll \R \ll \R_{\max}$ and substitute Eq.~(\ref{e:phi-disk0})
%
\begin{align}
\delta\phi_{\GW}^{{\rm adg},\alpha}
&= -8.3\times 10^{-5}\, \alpha_1^{-20/21} \frac{ M_{\SMBH 5}^{43/21} }{\dot{m}_{\SMBH 1}^{13/21} m_{\CO 1} }\R_{0,20}^{11/2}
\nonumber\\&\quad
\times\left(1 - \frac{21}{10}x^{11/2} + \frac{11}{10}x^{7} \right)\,,\\
\delta\phi_{\GW}^{{\rm adg},\beta}
&= 3.0\times 10^{-4}\, \alpha_1^{-4/5}
\frac{ \dot{m}_{\SMBH 1}^{3/5} M_{\SMBH 5}^{11/5}}{ m_{\CO 1} }\R_{0,20}^{39/10}
\nonumber\\&\quad
\times\left(1 + \frac{3}{10}x^{39/10} - \frac{13}{10}x^{27/5} \right)\,,
\end{align}
where again $x=\R_{f}/\R_0$. The sign difference is due to the fact
that the disk exerts an outward pull for $\alpha$ and an inward push
for $\beta$-disks.  In the long and short observation limits, we find
dephasing as in Eq.~\eqref{long-gen-form} and~\eqref{short-gen-form}
with $(C_{i},c_{j})$ and $(D_{i},d_{j})$ given by the thirteenth and
fourteenth rows of Table~\ref{Table-pars4}.

\bibliography{review}

\begin{thebibliography}{176}%
\makeatletter
\providecommand \@ifxundefined [1]{%
 \@ifx{#1\undefined}
}%
\providecommand \@ifnum [1]{%
 \ifnum #1\expandafter \@firstoftwo
 \else \expandafter \@secondoftwo
 \fi
}%
\providecommand \@ifx [1]{%
 \ifx #1\expandafter \@firstoftwo
 \else \expandafter \@secondoftwo
 \fi
}%
\providecommand \natexlab [1]{#1}%
\providecommand \enquote  [1]{``#1''}%
\providecommand \bibnamefont  [1]{#1}%
\providecommand \bibfnamefont [1]{#1}%
\providecommand \citenamefont [1]{#1}%
\providecommand \href@noop [0]{\@secondoftwo}%
\providecommand \href [0]{\begingroup \@sanitize@url \@href}%
\providecommand \@href[1]{\@@startlink{#1}\@@href}%
\providecommand \@@href[1]{\endgroup#1\@@endlink}%
\providecommand \@sanitize@url [0]{\catcode `\\12\catcode `\$12\catcode
  `\&12\catcode `\#12\catcode `\^12\catcode `\_12\catcode `\%12\relax}%
\providecommand \@@startlink[1]{}%
\providecommand \@@endlink[0]{}%
\providecommand \url  [0]{\begingroup\@sanitize@url \@url }%
\providecommand \@url [1]{\endgroup\@href {#1}{\urlprefix }}%
\providecommand \urlprefix  [0]{URL }%
\providecommand \Eprint [0]{\href }%
\providecommand \doibase [0]{http://dx.doi.org/}%
\providecommand \selectlanguage [0]{\@gobble}%
\providecommand \bibinfo  [0]{\@secondoftwo}%
\providecommand \bibfield  [0]{\@secondoftwo}%
\providecommand \translation [1]{[#1]}%
\providecommand \BibitemOpen [0]{}%
\providecommand \bibitemStop [0]{}%
\providecommand \bibitemNoStop [0]{.\EOS\space}%
\providecommand \EOS [0]{\spacefactor3000\relax}%
\providecommand \BibitemShut  [1]{\csname bibitem#1\endcsname}%
\let\auto@bib@innerbib\@empty
\bibitem [{\citenamefont {Danzmann}\ and\ \citenamefont
  {R{\"u}diger}(2003)}]{Danzmann:2003tv}%
  \BibitemOpen
  \bibfield  {author} {\bibinfo {author} {\bibfnamefont {K.}~\bibnamefont
  {Danzmann}}\ and\ \bibinfo {author} {\bibfnamefont {A.}~\bibnamefont
  {R{\"u}diger}},\ }\href@noop {} {\bibfield  {journal} {\bibinfo  {journal}
  {Class. Quantum Grav.}\ }\textbf {\bibinfo {volume} {20}},\ \bibinfo {pages}
  {S1} (\bibinfo {year} {2003})}\BibitemShut {NoStop}%
\bibitem [{\citenamefont {{Danzmann}}(2003)}]{Danzmann:2003ad}%
  \BibitemOpen
  \bibfield  {author} {\bibinfo {author} {\bibfnamefont {K.}~\bibnamefont
  {{Danzmann}}},\ }\href@noop {} {\bibfield  {journal} {\bibinfo  {journal}
  {Advances in Space Research}\ }\textbf {\bibinfo {volume} {32}},\ \bibinfo
  {pages} {1233} (\bibinfo {year} {2003})}\BibitemShut {NoStop}%
\bibitem [{\citenamefont {{Prince}}(2003)}]{Prince:2003aa}%
  \BibitemOpen
  \bibfield  {author} {\bibinfo {author} {\bibfnamefont {T.}~\bibnamefont
  {{Prince}}},\ }\href@noop {} {\bibfield  {journal} {\bibinfo  {journal}
  {American Astronomical Society Meeting}\ }\textbf {\bibinfo {volume} {202}},\
  \bibinfo {pages} {3701} (\bibinfo {year} {2003})}\BibitemShut {NoStop}%
\bibitem [{lis()}]{lisa}%
  \BibitemOpen
  \href@noop {} {\enquote {\bibinfo {title} {{LISA}},}\ }\bibinfo {note} {{\tt
  www.esa.int/science/lisa}, {\tt lisa.jpl.nasa.gov}}\BibitemShut {NoStop}%
\bibitem [{\citenamefont {Carre}\ and\ \citenamefont
  {Porter}(2010)}]{Carre:2010ra}%
  \BibitemOpen
  \bibfield  {author} {\bibinfo {author} {\bibfnamefont {J.}~\bibnamefont
  {Carre}}\ and\ \bibinfo {author} {\bibfnamefont {E.~K.}\ \bibnamefont
  {Porter}},\ }\href@noop {} {\  (\bibinfo {year} {2010})},\ \bibinfo {note} {*
  Temporary entry *},\ \Eprint {http://arxiv.org/abs/arXiv:1010.1641}
  {arXiv:arXiv:1010.1641 [gr-qc]} \BibitemShut {NoStop}%
\bibitem [{\citenamefont {Cutler}\ and\ \citenamefont
  {Vallisneri}(2007)}]{Cutler:2007mi}%
  \BibitemOpen
  \bibfield  {author} {\bibinfo {author} {\bibfnamefont {C.}~\bibnamefont
  {Cutler}}\ and\ \bibinfo {author} {\bibfnamefont {M.}~\bibnamefont
  {Vallisneri}},\ }\href {\doibase 10.1103/PhysRevD.76.104018} {\bibfield
  {journal} {\bibinfo  {journal} {Phys. Rev.}\ }\textbf {\bibinfo {volume}
  {D76}},\ \bibinfo {pages} {104018} (\bibinfo {year} {2007})},\ \Eprint
  {http://arxiv.org/abs/0707.2982} {arXiv:0707.2982 [gr-qc]} \BibitemShut
  {NoStop}%
\bibitem [{\citenamefont {{Barack}}\ and\ \citenamefont
  {{Cutler}}(2004)}]{2004PhRvD..70l2002B}%
  \BibitemOpen
  \bibfield  {author} {\bibinfo {author} {\bibfnamefont {L.}~\bibnamefont
  {{Barack}}}\ and\ \bibinfo {author} {\bibfnamefont {C.}~\bibnamefont
  {{Cutler}}},\ }\href {\doibase 10.1103/PhysRevD.70.122002} {\bibfield
  {journal} {\bibinfo  {journal} {\prd}\ }\textbf {\bibinfo {volume} {70}},\
  \bibinfo {pages} {122002} (\bibinfo {year} {2004})},\ \Eprint
  {http://arxiv.org/abs/arXiv:gr-qc/0409010} {arXiv:gr-qc/0409010} \BibitemShut
  {NoStop}%
\bibitem [{\citenamefont {Yunes}\ \emph
  {et~al.}(2010{\natexlab{a}})\citenamefont {Yunes}, \citenamefont {Miller},\
  and\ \citenamefont {Thornburg}}]{Yunes:2010sm}%
  \BibitemOpen
  \bibfield  {author} {\bibinfo {author} {\bibfnamefont {N.}~\bibnamefont
  {Yunes}}, \bibinfo {author} {\bibfnamefont {M.}~\bibnamefont {Miller}}, \
  and\ \bibinfo {author} {\bibfnamefont {J.}~\bibnamefont {Thornburg}},\
  }\href@noop {} {\  (\bibinfo {year} {2010}{\natexlab{a}})},\ \bibinfo {note}
  {* Temporary entry *},\ \Eprint {http://arxiv.org/abs/arXiv:1010.1721}
  {arXiv:arXiv:1010.1721 [astro-ph.GA]} \BibitemShut {NoStop}%
\bibitem [{\citenamefont {{Amaro-Seoane}}\ \emph {et~al.}(2007)\citenamefont
  {{Amaro-Seoane}}, \citenamefont {{Gair}}, \citenamefont {{Freitag}},
  \citenamefont {{Miller}}, \citenamefont {{Mandel}}, \citenamefont
  {{Cutler}},\ and\ \citenamefont {{Babak}}}]{2007CQGra..24..113A}%
  \BibitemOpen
  \bibfield  {author} {\bibinfo {author} {\bibfnamefont {P.}~\bibnamefont
  {{Amaro-Seoane}}}, \bibinfo {author} {\bibfnamefont {J.~R.}\ \bibnamefont
  {{Gair}}}, \bibinfo {author} {\bibfnamefont {M.}~\bibnamefont {{Freitag}}},
  \bibinfo {author} {\bibfnamefont {M.~C.}\ \bibnamefont {{Miller}}}, \bibinfo
  {author} {\bibfnamefont {I.}~\bibnamefont {{Mandel}}}, \bibinfo {author}
  {\bibfnamefont {C.~J.}\ \bibnamefont {{Cutler}}}, \ and\ \bibinfo {author}
  {\bibfnamefont {S.}~\bibnamefont {{Babak}}},\ }\href {\doibase
  10.1088/0264-9381/24/17/R01} {\bibfield  {journal} {\bibinfo  {journal}
  {Classical and Quantum Gravity}\ }\textbf {\bibinfo {volume} {24}},\ \bibinfo
  {pages} {113} (\bibinfo {year} {2007})},\ \Eprint
  {http://arxiv.org/abs/arXiv:astro-ph/0703495} {arXiv:astro-ph/0703495}
  \BibitemShut {NoStop}%
\bibitem [{\citenamefont {Barack}\ and\ \citenamefont
  {Cutler}(2004)}]{Barack:2003fp}%
  \BibitemOpen
  \bibfield  {author} {\bibinfo {author} {\bibfnamefont {L.}~\bibnamefont
  {Barack}}\ and\ \bibinfo {author} {\bibfnamefont {C.}~\bibnamefont
  {Cutler}},\ }\href {\doibase 10.1103/PhysRevD.69.082005} {\bibfield
  {journal} {\bibinfo  {journal} {Phys. Rev.}\ }\textbf {\bibinfo {volume}
  {D69}},\ \bibinfo {pages} {082005} (\bibinfo {year} {2004})},\ \Eprint
  {http://arxiv.org/abs/gr-qc/0310125} {arXiv:gr-qc/0310125} \BibitemShut
  {NoStop}%
\bibitem [{\citenamefont {Gair}(2009)}]{Gair:2008bx}%
  \BibitemOpen
  \bibfield  {author} {\bibinfo {author} {\bibfnamefont {J.~R.}\ \bibnamefont
  {Gair}},\ }\href {\doibase 10.1088/0264-9381/26/9/094034} {\bibfield
  {journal} {\bibinfo  {journal} {Class. Quant. Grav.}\ }\textbf {\bibinfo
  {volume} {26}},\ \bibinfo {pages} {094034} (\bibinfo {year} {2009})},\
  \Eprint {http://arxiv.org/abs/0811.0188} {arXiv:0811.0188 [gr-qc]}
  \BibitemShut {NoStop}%
\bibitem [{\citenamefont {{Barnes}}\ and\ \citenamefont
  {{Hernquist}}(1992)}]{1992ARA&A..30..705B}%
  \BibitemOpen
  \bibfield  {author} {\bibinfo {author} {\bibfnamefont {J.~E.}\ \bibnamefont
  {{Barnes}}}\ and\ \bibinfo {author} {\bibfnamefont {L.}~\bibnamefont
  {{Hernquist}}},\ }\href {\doibase 10.1146/annurev.aa.30.090192.003421}
  {\bibfield  {journal} {\bibinfo  {journal} {\araa}\ }\textbf {\bibinfo
  {volume} {30}},\ \bibinfo {pages} {705} (\bibinfo {year} {1992})}\BibitemShut
  {NoStop}%
\bibitem [{\citenamefont {{Begelman}}\ \emph {et~al.}(1980)\citenamefont
  {{Begelman}}, \citenamefont {{Blandford}},\ and\ \citenamefont
  {{Rees}}}]{1980Natur.287..307B}%
  \BibitemOpen
  \bibfield  {author} {\bibinfo {author} {\bibfnamefont {M.~C.}\ \bibnamefont
  {{Begelman}}}, \bibinfo {author} {\bibfnamefont {R.~D.}\ \bibnamefont
  {{Blandford}}}, \ and\ \bibinfo {author} {\bibfnamefont {M.~J.}\ \bibnamefont
  {{Rees}}},\ }\href {\doibase 10.1038/287307a0} {\bibfield  {journal}
  {\bibinfo  {journal} {\nat}\ }\textbf {\bibinfo {volume} {287}},\ \bibinfo
  {pages} {307} (\bibinfo {year} {1980})}\BibitemShut {NoStop}%
\bibitem [{\citenamefont {{Miralda-Escud{\'e}}}\ and\ \citenamefont
  {{Kollmeier}}(2005)}]{2005ApJ...619...30M}%
  \BibitemOpen
  \bibfield  {author} {\bibinfo {author} {\bibfnamefont {J.}~\bibnamefont
  {{Miralda-Escud{\'e}}}}\ and\ \bibinfo {author} {\bibfnamefont {J.~A.}\
  \bibnamefont {{Kollmeier}}},\ }\href {\doibase 10.1086/426467} {\bibfield
  {journal} {\bibinfo  {journal} {\apj}\ }\textbf {\bibinfo {volume} {619}},\
  \bibinfo {pages} {30} (\bibinfo {year} {2005})}\BibitemShut {NoStop}%
\bibitem [{\citenamefont {{Milosavljevi{\'c}}}\ and\ \citenamefont
  {{Loeb}}(2004)}]{2004ApJ...604L..45M}%
  \BibitemOpen
  \bibfield  {author} {\bibinfo {author} {\bibfnamefont {M.}~\bibnamefont
  {{Milosavljevi{\'c}}}}\ and\ \bibinfo {author} {\bibfnamefont
  {A.}~\bibnamefont {{Loeb}}},\ }\href {\doibase 10.1086/383467} {\bibfield
  {journal} {\bibinfo  {journal} {\apjl}\ }\textbf {\bibinfo {volume} {604}},\
  \bibinfo {pages} {L45} (\bibinfo {year} {2004})},\ \Eprint
  {http://arxiv.org/abs/arXiv:astro-ph/0401221} {arXiv:astro-ph/0401221}
  \BibitemShut {NoStop}%
\bibitem [{\citenamefont {{Goodman}}\ and\ \citenamefont
  {{Tan}}(2004)}]{2004ApJ...608..108G}%
  \BibitemOpen
  \bibfield  {author} {\bibinfo {author} {\bibfnamefont {J.}~\bibnamefont
  {{Goodman}}}\ and\ \bibinfo {author} {\bibfnamefont {J.~C.}\ \bibnamefont
  {{Tan}}},\ }\href {\doibase 10.1086/386360} {\bibfield  {journal} {\bibinfo
  {journal} {\apj}\ }\textbf {\bibinfo {volume} {608}},\ \bibinfo {pages} {108}
  (\bibinfo {year} {2004})},\ \Eprint
  {http://arxiv.org/abs/arXiv:astro-ph/0307361} {arXiv:astro-ph/0307361}
  \BibitemShut {NoStop}%
\bibitem [{\citenamefont {{Levin}}(2007)}]{2007MNRAS.374..515L}%
  \BibitemOpen
  \bibfield  {author} {\bibinfo {author} {\bibfnamefont {Y.}~\bibnamefont
  {{Levin}}},\ }\href {\doibase 10.1111/j.1365-2966.2006.11155.x} {\bibfield
  {journal} {\bibinfo  {journal} {\mnras}\ }\textbf {\bibinfo {volume} {374}},\
  \bibinfo {pages} {515} (\bibinfo {year} {2007})},\ \Eprint
  {http://arxiv.org/abs/arXiv:astro-ph/0603583} {arXiv:astro-ph/0603583}
  \BibitemShut {NoStop}%
\bibitem [{\citenamefont {{Valtonen}}\ \emph {et~al.}(2010)\citenamefont
  {{Valtonen}}, \citenamefont {{Mikkola}}, \citenamefont {{Merritt}},
  \citenamefont {{Gopakumar}}, \citenamefont {{Lehto}}, \citenamefont
  {{Hyv{\"o}nen}}, \citenamefont {{Rampadarath}}, \citenamefont {{Saunders}},
  \citenamefont {{Basta}},\ and\ \citenamefont
  {{Hudec}}}]{2010ApJ...709..725V}%
  \BibitemOpen
  \bibfield  {author} {\bibinfo {author} {\bibfnamefont {M.~J.}\ \bibnamefont
  {{Valtonen}}}, \bibinfo {author} {\bibfnamefont {S.}~\bibnamefont
  {{Mikkola}}}, \bibinfo {author} {\bibfnamefont {D.}~\bibnamefont
  {{Merritt}}}, \bibinfo {author} {\bibfnamefont {A.}~\bibnamefont
  {{Gopakumar}}}, \bibinfo {author} {\bibfnamefont {H.~J.}\ \bibnamefont
  {{Lehto}}}, \bibinfo {author} {\bibfnamefont {T.}~\bibnamefont
  {{Hyv{\"o}nen}}}, \bibinfo {author} {\bibfnamefont {H.}~\bibnamefont
  {{Rampadarath}}}, \bibinfo {author} {\bibfnamefont {R.}~\bibnamefont
  {{Saunders}}}, \bibinfo {author} {\bibfnamefont {M.}~\bibnamefont {{Basta}}},
  \ and\ \bibinfo {author} {\bibfnamefont {R.}~\bibnamefont {{Hudec}}},\ }\href
  {\doibase 10.1088/0004-637X/709/2/725} {\bibfield  {journal} {\bibinfo
  {journal} {\apj}\ }\textbf {\bibinfo {volume} {709}},\ \bibinfo {pages} {725}
  (\bibinfo {year} {2010})},\ \Eprint {http://arxiv.org/abs/0912.1209}
  {arXiv:0912.1209 [astro-ph.HE]} \BibitemShut {NoStop}%
\bibitem [{\citenamefont {{Kocsis}}\ \emph {et~al.}(2008)\citenamefont
  {{Kocsis}}, \citenamefont {{Haiman}},\ and\ \citenamefont
  {{Menou}}}]{2008ApJ...684..870K}%
  \BibitemOpen
  \bibfield  {author} {\bibinfo {author} {\bibfnamefont {B.}~\bibnamefont
  {{Kocsis}}}, \bibinfo {author} {\bibfnamefont {Z.}~\bibnamefont {{Haiman}}},
  \ and\ \bibinfo {author} {\bibfnamefont {K.}~\bibnamefont {{Menou}}},\ }\href
  {\doibase 10.1086/590230} {\bibfield  {journal} {\bibinfo  {journal} {\apj}\
  }\textbf {\bibinfo {volume} {684}},\ \bibinfo {pages} {870} (\bibinfo {year}
  {2008})},\ \Eprint {http://arxiv.org/abs/0712.1144} {arXiv:0712.1144}
  \BibitemShut {NoStop}%
\bibitem [{\citenamefont {{Kocsis}}\ \emph {et~al.}(2006)\citenamefont
  {{Kocsis}}, \citenamefont {{Frei}}, \citenamefont {{Haiman}},\ and\
  \citenamefont {{Menou}}}]{2006ApJ...637...27K}%
  \BibitemOpen
  \bibfield  {author} {\bibinfo {author} {\bibfnamefont {B.}~\bibnamefont
  {{Kocsis}}}, \bibinfo {author} {\bibfnamefont {Z.}~\bibnamefont {{Frei}}},
  \bibinfo {author} {\bibfnamefont {Z.}~\bibnamefont {{Haiman}}}, \ and\
  \bibinfo {author} {\bibfnamefont {K.}~\bibnamefont {{Menou}}},\ }\href
  {\doibase 10.1086/498236} {\bibfield  {journal} {\bibinfo  {journal} {\apj}\
  }\textbf {\bibinfo {volume} {637}},\ \bibinfo {pages} {27} (\bibinfo {year}
  {2006})},\ \Eprint {http://arxiv.org/abs/arXiv:astro-ph/0505394}
  {arXiv:astro-ph/0505394} \BibitemShut {NoStop}%
\bibitem [{\citenamefont {{Schutz}}(1986)}]{1986Natur.323..310S}%
  \BibitemOpen
  \bibfield  {author} {\bibinfo {author} {\bibfnamefont {B.~F.}\ \bibnamefont
  {{Schutz}}},\ }\href {\doibase 10.1038/323310a0} {\bibfield  {journal}
  {\bibinfo  {journal} {\nat}\ }\textbf {\bibinfo {volume} {323}},\ \bibinfo
  {pages} {310} (\bibinfo {year} {1986})}\BibitemShut {NoStop}%
\bibitem [{\citenamefont {{Holz}}\ and\ \citenamefont
  {{Hughes}}(2005)}]{2005ApJ...629...15H}%
  \BibitemOpen
  \bibfield  {author} {\bibinfo {author} {\bibfnamefont {D.~E.}\ \bibnamefont
  {{Holz}}}\ and\ \bibinfo {author} {\bibfnamefont {S.~A.}\ \bibnamefont
  {{Hughes}}},\ }\href {\doibase 10.1086/431341} {\bibfield  {journal}
  {\bibinfo  {journal} {\apj}\ }\textbf {\bibinfo {volume} {629}},\ \bibinfo
  {pages} {15} (\bibinfo {year} {2005})},\ \Eprint
  {http://arxiv.org/abs/arXiv:astro-ph/0504616} {arXiv:astro-ph/0504616}
  \BibitemShut {NoStop}%
\bibitem [{\citenamefont {{Deffayet}}\ and\ \citenamefont
  {{Menou}}(2007)}]{2007ApJ...668L.143D}%
  \BibitemOpen
  \bibfield  {author} {\bibinfo {author} {\bibfnamefont {C.}~\bibnamefont
  {{Deffayet}}}\ and\ \bibinfo {author} {\bibfnamefont {K.}~\bibnamefont
  {{Menou}}},\ }\href {\doibase 10.1086/522931} {\bibfield  {journal} {\bibinfo
   {journal} {\apjl}\ }\textbf {\bibinfo {volume} {668}},\ \bibinfo {pages}
  {L143} (\bibinfo {year} {2007})},\ \Eprint {http://arxiv.org/abs/0709.0003}
  {arXiv:0709.0003} \BibitemShut {NoStop}%
\bibitem [{\citenamefont {Giampieri}(1993)}]{Giampieri:1993pt}%
  \BibitemOpen
  \bibfield  {author} {\bibinfo {author} {\bibfnamefont {G.}~\bibnamefont
  {Giampieri}},\ }\href@noop {} {\  (\bibinfo {year} {1993})},\ \Eprint
  {http://arxiv.org/abs/astro-ph/9305034} {arXiv:astro-ph/9305034} \BibitemShut
  {NoStop}%
\bibitem [{\citenamefont {{Chakrabarti}}(1993)}]{1993ApJ...411..610C}%
  \BibitemOpen
  \bibfield  {author} {\bibinfo {author} {\bibfnamefont {S.~K.}\ \bibnamefont
  {{Chakrabarti}}},\ }\href {\doibase 10.1086/172863} {\bibfield  {journal}
  {\bibinfo  {journal} {\apj}\ }\textbf {\bibinfo {volume} {411}},\ \bibinfo
  {pages} {610} (\bibinfo {year} {1993})}\BibitemShut {NoStop}%
\bibitem [{\citenamefont {Chakrabarti}(1996)}]{Chakrabarti:1995dw}%
  \BibitemOpen
  \bibfield  {author} {\bibinfo {author} {\bibfnamefont {S.~K.}\ \bibnamefont
  {Chakrabarti}},\ }\href {\doibase 10.1103/PhysRevD.53.2901} {\bibfield
  {journal} {\bibinfo  {journal} {Phys. Rev.}\ }\textbf {\bibinfo {volume}
  {D53}},\ \bibinfo {pages} {2901} (\bibinfo {year} {1996})},\ \Eprint
  {http://arxiv.org/abs/astro-ph/9603117} {arXiv:astro-ph/9603117} \BibitemShut
  {NoStop}%
\bibitem [{\citenamefont {{Molteni}}\ \emph {et~al.}(1994)\citenamefont
  {{Molteni}}, \citenamefont {{Gerardi}},\ and\ \citenamefont
  {{Chakrabarti}}}]{1994ApJ...436..249M}%
  \BibitemOpen
  \bibfield  {author} {\bibinfo {author} {\bibfnamefont {D.}~\bibnamefont
  {{Molteni}}}, \bibinfo {author} {\bibfnamefont {G.}~\bibnamefont
  {{Gerardi}}}, \ and\ \bibinfo {author} {\bibfnamefont {S.~K.}\ \bibnamefont
  {{Chakrabarti}}},\ }\href {\doibase 10.1086/174897} {\bibfield  {journal}
  {\bibinfo  {journal} {\apj}\ }\textbf {\bibinfo {volume} {436}},\ \bibinfo
  {pages} {249} (\bibinfo {year} {1994})}\BibitemShut {NoStop}%
\bibitem [{\citenamefont {{Hoyle}}\ and\ \citenamefont
  {{Lyttleton}}(1939)}]{1939PCPS...35..405H}%
  \BibitemOpen
  \bibfield  {author} {\bibinfo {author} {\bibfnamefont {F.}~\bibnamefont
  {{Hoyle}}}\ and\ \bibinfo {author} {\bibfnamefont {R.~A.}\ \bibnamefont
  {{Lyttleton}}},\ }in\ \href {\doibase 10.1017/S0305004100021150} {\emph
  {\bibinfo {booktitle} {Proceedings of the Cambridge Philosophical
  Society}}},\ \bibinfo {series} {Proceedings of the Cambridge Philosophical
  Society}, Vol.~\bibinfo {volume} {35}\ (\bibinfo {year} {1939})\ pp.\
  \bibinfo {pages} {405--+}\BibitemShut {NoStop}%
\bibitem [{\citenamefont {{Bondi}}\ and\ \citenamefont
  {{Hoyle}}(1944)}]{1944MNRAS.104..273B}%
  \BibitemOpen
  \bibfield  {author} {\bibinfo {author} {\bibfnamefont {H.}~\bibnamefont
  {{Bondi}}}\ and\ \bibinfo {author} {\bibfnamefont {F.}~\bibnamefont
  {{Hoyle}}},\ }\href@noop {} {\bibfield  {journal} {\bibinfo  {journal}
  {Monthly Notices of the Royal Astronomical Society}\ }\textbf {\bibinfo
  {volume} {104}},\ \bibinfo {pages} {273} (\bibinfo {year}
  {1944})}\BibitemShut {NoStop}%
\bibitem [{\citenamefont {{Bondi}}(1952)}]{1952MNRAS.112..195B}%
  \BibitemOpen
  \bibfield  {author} {\bibinfo {author} {\bibfnamefont {H.}~\bibnamefont
  {{Bondi}}},\ }\href@noop {} {\bibfield  {journal} {\bibinfo  {journal}
  {Monthly Notices of the Royal Astronomical Society}\ }\textbf {\bibinfo
  {volume} {112}},\ \bibinfo {pages} {195} (\bibinfo {year}
  {1952})}\BibitemShut {NoStop}%
\bibitem [{\citenamefont {{Shapiro}}\ and\ \citenamefont
  {{Teukolsky}}(1986)}]{1986bhwd.book.....S}%
  \BibitemOpen
  \bibfield  {author} {\bibinfo {author} {\bibfnamefont {S.~L.}\ \bibnamefont
  {{Shapiro}}}\ and\ \bibinfo {author} {\bibfnamefont {S.~A.}\ \bibnamefont
  {{Teukolsky}}},\ }\href@noop {} {\emph {\bibinfo {title} {Black Holes, White
  Dwarfs and Neutron Stars: The Physics of Compact Objects, by Stuart
  L.~Shapiro, Saul A.~Teukolsky, pp.~672.~ISBN 0-471-87316-0.~Wiley-VCH , June
  1986.}}},\ edited by\ \bibinfo {editor} {\bibnamefont {{Shapiro, S.~L.~\&
  Teukolsky, S.~A.}}}\ (\bibinfo {year} {1986})\BibitemShut {NoStop}%
\bibitem [{\citenamefont {{Edgar}}(2004)}]{2004NewAR..48..843E}%
  \BibitemOpen
  \bibfield  {author} {\bibinfo {author} {\bibfnamefont {R.}~\bibnamefont
  {{Edgar}}},\ }\href {\doibase 10.1016/j.newar.2004.06.001} {\bibfield
  {journal} {\bibinfo  {journal} {\nar}\ }\textbf {\bibinfo {volume} {48}},\
  \bibinfo {pages} {843} (\bibinfo {year} {2004})},\ \Eprint
  {http://arxiv.org/abs/arXiv:astro-ph/0406166} {arXiv:astro-ph/0406166}
  \BibitemShut {NoStop}%
\bibitem [{\citenamefont {{Narayan}}(2000)}]{2000ApJ...536..663N}%
  \BibitemOpen
  \bibfield  {author} {\bibinfo {author} {\bibfnamefont {R.}~\bibnamefont
  {{Narayan}}},\ }\href {\doibase 10.1086/308956} {\bibfield  {journal}
  {\bibinfo  {journal} {The Astrophysical Journal}\ }\textbf {\bibinfo {volume}
  {536}},\ \bibinfo {pages} {663} (\bibinfo {year} {2000})},\ \Eprint
  {http://arxiv.org/abs/arXiv:astro-ph/9907328} {arXiv:astro-ph/9907328}
  \BibitemShut {NoStop}%
\bibitem [{\citenamefont {{{\v S}ubr}}\ and\ \citenamefont
  {{Karas}}(1999)}]{1999A&A...352..452S}%
  \BibitemOpen
  \bibfield  {author} {\bibinfo {author} {\bibfnamefont {L.}~\bibnamefont {{{\v
  S}ubr}}}\ and\ \bibinfo {author} {\bibfnamefont {V.}~\bibnamefont
  {{Karas}}},\ }\href@noop {} {\bibfield  {journal} {\bibinfo  {journal}
  {\aap}\ }\textbf {\bibinfo {volume} {352}},\ \bibinfo {pages} {452} (\bibinfo
  {year} {1999})},\ \Eprint {http://arxiv.org/abs/arXiv:astro-ph/9910401}
  {arXiv:astro-ph/9910401} \BibitemShut {NoStop}%
\bibitem [{\citenamefont {{Karas}}\ and\ \citenamefont {{{\v
  S}ubr}}(2001)}]{2001A&A...376..686K}%
  \BibitemOpen
  \bibfield  {author} {\bibinfo {author} {\bibfnamefont {V.}~\bibnamefont
  {{Karas}}}\ and\ \bibinfo {author} {\bibfnamefont {L.}~\bibnamefont {{{\v
  S}ubr}}},\ }\href {\doibase 10.1051/0004-6361:20011009} {\bibfield  {journal}
  {\bibinfo  {journal} {\aap}\ }\textbf {\bibinfo {volume} {376}},\ \bibinfo
  {pages} {686} (\bibinfo {year} {2001})},\ \Eprint
  {http://arxiv.org/abs/arXiv:astro-ph/0107232} {arXiv:astro-ph/0107232}
  \BibitemShut {NoStop}%
\bibitem [{\citenamefont {{Shakura}}\ and\ \citenamefont
  {{Sunyaev}}(1973)}]{1973A&A....24..337S}%
  \BibitemOpen
  \bibfield  {author} {\bibinfo {author} {\bibfnamefont {N.~I.}\ \bibnamefont
  {{Shakura}}}\ and\ \bibinfo {author} {\bibfnamefont {R.~A.}\ \bibnamefont
  {{Sunyaev}}},\ }\href@noop {} {\bibfield  {journal} {\bibinfo  {journal}
  {Astron. Astroph.}\ }\textbf {\bibinfo {volume} {24}},\ \bibinfo {pages}
  {337} (\bibinfo {year} {1973})}\BibitemShut {NoStop}%
\bibitem [{\citenamefont {{Gammie}}(2001)}]{2001ApJ...553..174G}%
  \BibitemOpen
  \bibfield  {author} {\bibinfo {author} {\bibfnamefont {C.~F.}\ \bibnamefont
  {{Gammie}}},\ }\href {\doibase 10.1086/320631} {\bibfield  {journal}
  {\bibinfo  {journal} {\apj}\ }\textbf {\bibinfo {volume} {553}},\ \bibinfo
  {pages} {174} (\bibinfo {year} {2001})},\ \Eprint
  {http://arxiv.org/abs/arXiv:astro-ph/0101501} {arXiv:astro-ph/0101501}
  \BibitemShut {NoStop}%
\bibitem [{\citenamefont {{Barausse}}\ and\ \citenamefont
  {{Rezzolla}}(2008)}]{2008PhRvD..77j4027B}%
  \BibitemOpen
  \bibfield  {author} {\bibinfo {author} {\bibfnamefont {E.}~\bibnamefont
  {{Barausse}}}\ and\ \bibinfo {author} {\bibfnamefont {L.}~\bibnamefont
  {{Rezzolla}}},\ }\href {\doibase 10.1103/PhysRevD.77.104027} {\bibfield
  {journal} {\bibinfo  {journal} {\prd}\ }\textbf {\bibinfo {volume} {77}},\
  \bibinfo {pages} {104027} (\bibinfo {year} {2008})},\ \Eprint
  {http://arxiv.org/abs/0711.4558} {arXiv:0711.4558 [gr-qc]} \BibitemShut
  {NoStop}%
\bibitem [{\citenamefont {{Ryan}}(1995)}]{1995PhRvD..52.3159R}%
  \BibitemOpen
  \bibfield  {author} {\bibinfo {author} {\bibfnamefont {F.~D.}\ \bibnamefont
  {{Ryan}}},\ }\href {\doibase 10.1103/PhysRevD.52.R3159} {\bibfield  {journal}
  {\bibinfo  {journal} {\prd}\ }\textbf {\bibinfo {volume} {52}},\ \bibinfo
  {pages} {3159} (\bibinfo {year} {1995})},\ \Eprint
  {http://arxiv.org/abs/arXiv:gr-qc/9506023} {arXiv:gr-qc/9506023} \BibitemShut
  {NoStop}%
\bibitem [{\citenamefont {{Papaloizou}}\ and\ \citenamefont
  {{Pringle}}(1984)}]{1984MNRAS.208..721P}%
  \BibitemOpen
  \bibfield  {author} {\bibinfo {author} {\bibfnamefont {J.~C.~B.}\
  \bibnamefont {{Papaloizou}}}\ and\ \bibinfo {author} {\bibfnamefont {J.~E.}\
  \bibnamefont {{Pringle}}},\ }\href@noop {} {\bibfield  {journal} {\bibinfo
  {journal} {Monthly Notices of the Royal Astronomical Society}\ }\textbf
  {\bibinfo {volume} {208}},\ \bibinfo {pages} {721} (\bibinfo {year}
  {1984})}\BibitemShut {NoStop}%
\bibitem [{\citenamefont {{Blaes}}(1985)}]{1985MNRAS.212P..37B}%
  \BibitemOpen
  \bibfield  {author} {\bibinfo {author} {\bibfnamefont {O.~M.}\ \bibnamefont
  {{Blaes}}},\ }\href@noop {} {\bibfield  {journal} {\bibinfo  {journal}
  {Monthly Notices of the Royal Astronomical Society}\ }\textbf {\bibinfo
  {volume} {212}},\ \bibinfo {pages} {37P} (\bibinfo {year}
  {1985})}\BibitemShut {NoStop}%
\bibitem [{\citenamefont {{Kojima}}(1986)}]{1986PThPh..75..251K}%
  \BibitemOpen
  \bibfield  {author} {\bibinfo {author} {\bibfnamefont {Y.}~\bibnamefont
  {{Kojima}}},\ }\href {\doibase 10.1143/PTP.75.251} {\bibfield  {journal}
  {\bibinfo  {journal} {Progress of Theoretical Physics}\ }\textbf {\bibinfo
  {volume} {75}},\ \bibinfo {pages} {251} (\bibinfo {year} {1986})}\BibitemShut
  {NoStop}%
\bibitem [{\citenamefont {{Liu}}\ and\ \citenamefont
  {{Shapiro}}(2010)}]{2010PhRvD..82l3011L}%
  \BibitemOpen
  \bibfield  {author} {\bibinfo {author} {\bibfnamefont {Y.~T.}\ \bibnamefont
  {{Liu}}}\ and\ \bibinfo {author} {\bibfnamefont {S.~L.}\ \bibnamefont
  {{Shapiro}}},\ }\href {\doibase 10.1103/PhysRevD.82.123011} {\bibfield
  {journal} {\bibinfo  {journal} {\prd}\ }\textbf {\bibinfo {volume} {82}},\
  \bibinfo {pages} {123011} (\bibinfo {year} {2010})},\ \Eprint
  {http://arxiv.org/abs/1011.0002} {arXiv:1011.0002 [astro-ph.HE]} \BibitemShut
  {NoStop}%
\bibitem [{\citenamefont {{Milosavljevi{\'c}}}\ and\ \citenamefont
  {{Phinney}}(2005)}]{2005ApJ...622L..93M}%
  \BibitemOpen
  \bibfield  {author} {\bibinfo {author} {\bibfnamefont {M.}~\bibnamefont
  {{Milosavljevi{\'c}}}}\ and\ \bibinfo {author} {\bibfnamefont {E.~S.}\
  \bibnamefont {{Phinney}}},\ }\href {\doibase 10.1086/429618} {\bibfield
  {journal} {\bibinfo  {journal} {\apjl}\ }\textbf {\bibinfo {volume} {622}},\
  \bibinfo {pages} {L93} (\bibinfo {year} {2005})},\ \Eprint
  {http://arxiv.org/abs/arXiv:astro-ph/0410343} {arXiv:astro-ph/0410343}
  \BibitemShut {NoStop}%
\bibitem [{\citenamefont {{Schnittman}}\ and\ \citenamefont
  {{Krolik}}(2008)}]{2008ApJ...684..835S}%
  \BibitemOpen
  \bibfield  {author} {\bibinfo {author} {\bibfnamefont {J.~D.}\ \bibnamefont
  {{Schnittman}}}\ and\ \bibinfo {author} {\bibfnamefont {J.~H.}\ \bibnamefont
  {{Krolik}}},\ }\href {\doibase 10.1086/590363} {\bibfield  {journal}
  {\bibinfo  {journal} {\apj}\ }\textbf {\bibinfo {volume} {684}},\ \bibinfo
  {pages} {835} (\bibinfo {year} {2008})},\ \Eprint
  {http://arxiv.org/abs/0802.3556} {arXiv:0802.3556} \BibitemShut {NoStop}%
\bibitem [{\citenamefont {{MacFadyen}}\ and\ \citenamefont
  {{Milosavljevi{\'c}}}(2008)}]{2008ApJ...672...83M}%
  \BibitemOpen
  \bibfield  {author} {\bibinfo {author} {\bibfnamefont {A.~I.}\ \bibnamefont
  {{MacFadyen}}}\ and\ \bibinfo {author} {\bibfnamefont {M.}~\bibnamefont
  {{Milosavljevi{\'c}}}},\ }\href {\doibase 10.1086/523869} {\bibfield
  {journal} {\bibinfo  {journal} {\apj}\ }\textbf {\bibinfo {volume} {672}},\
  \bibinfo {pages} {83} (\bibinfo {year} {2008})},\ \Eprint
  {http://arxiv.org/abs/arXiv:astro-ph/0607467} {arXiv:astro-ph/0607467}
  \BibitemShut {NoStop}%
\bibitem [{\citenamefont {{Cuadra}}\ \emph {et~al.}(2009)\citenamefont
  {{Cuadra}}, \citenamefont {{Armitage}}, \citenamefont {{Alexander}},\ and\
  \citenamefont {{Begelman}}}]{2009MNRAS.393.1423C}%
  \BibitemOpen
  \bibfield  {author} {\bibinfo {author} {\bibfnamefont {J.}~\bibnamefont
  {{Cuadra}}}, \bibinfo {author} {\bibfnamefont {P.~J.}\ \bibnamefont
  {{Armitage}}}, \bibinfo {author} {\bibfnamefont {R.~D.}\ \bibnamefont
  {{Alexander}}}, \ and\ \bibinfo {author} {\bibfnamefont {M.~C.}\ \bibnamefont
  {{Begelman}}},\ }\href {\doibase 10.1111/j.1365-2966.2008.14147.x} {\bibfield
   {journal} {\bibinfo  {journal} {\mnras}\ }\textbf {\bibinfo {volume}
  {393}},\ \bibinfo {pages} {1423} (\bibinfo {year} {2009})},\ \Eprint
  {http://arxiv.org/abs/0809.0311} {arXiv:0809.0311} \BibitemShut {NoStop}%
\bibitem [{\citenamefont {{Bogdanovi{\'c}}}\ \emph {et~al.}(2011)\citenamefont
  {{Bogdanovi{\'c}}}, \citenamefont {{Reynolds}},\ and\ \citenamefont
  {{Massey}}}]{2011ApJ...731....7B}%
  \BibitemOpen
  \bibfield  {author} {\bibinfo {author} {\bibfnamefont {T.}~\bibnamefont
  {{Bogdanovi{\'c}}}}, \bibinfo {author} {\bibfnamefont {C.~S.}\ \bibnamefont
  {{Reynolds}}}, \ and\ \bibinfo {author} {\bibfnamefont {R.}~\bibnamefont
  {{Massey}}},\ }\href {\doibase 10.1088/0004-637X/731/1/7} {\bibfield
  {journal} {\bibinfo  {journal} {\apj}\ }\textbf {\bibinfo {volume} {731}},\
  \bibinfo {pages} {7} (\bibinfo {year} {2011})},\ \Eprint
  {http://arxiv.org/abs/1005.2193} {arXiv:1005.2193 [astro-ph.CO]} \BibitemShut
  {NoStop}%
\bibitem [{\citenamefont {{Chang}}\ \emph {et~al.}(2010)\citenamefont
  {{Chang}}, \citenamefont {{Strubbe}}, \citenamefont {{Menou}},\ and\
  \citenamefont {{Quataert}}}]{2010MNRAS.407.2007C}%
  \BibitemOpen
  \bibfield  {author} {\bibinfo {author} {\bibfnamefont {P.}~\bibnamefont
  {{Chang}}}, \bibinfo {author} {\bibfnamefont {L.~E.}\ \bibnamefont
  {{Strubbe}}}, \bibinfo {author} {\bibfnamefont {K.}~\bibnamefont {{Menou}}},
  \ and\ \bibinfo {author} {\bibfnamefont {E.}~\bibnamefont {{Quataert}}},\
  }\href {\doibase 10.1111/j.1365-2966.2010.17056.x} {\bibfield  {journal}
  {\bibinfo  {journal} {\mnras}\ }\textbf {\bibinfo {volume} {407}},\ \bibinfo
  {pages} {2007} (\bibinfo {year} {2010})},\ \Eprint
  {http://arxiv.org/abs/0906.0825} {arXiv:0906.0825 [astro-ph.HE]} \BibitemShut
  {NoStop}%
\bibitem [{\citenamefont {{Yunes}}\ \emph {et~al.}(2011)\citenamefont
  {{Yunes}}, \citenamefont {{Kocsis}}, \citenamefont {{Loeb}},\ and\
  \citenamefont {{Haiman}}}]{2011arXiv1103.4609Y}%
  \BibitemOpen
  \bibfield  {author} {\bibinfo {author} {\bibfnamefont {N.}~\bibnamefont
  {{Yunes}}}, \bibinfo {author} {\bibfnamefont {B.}~\bibnamefont {{Kocsis}}},
  \bibinfo {author} {\bibfnamefont {A.}~\bibnamefont {{Loeb}}}, \ and\ \bibinfo
  {author} {\bibfnamefont {Z.}~\bibnamefont {{Haiman}}},\ }\href@noop {}
  {\bibfield  {journal} {\bibinfo  {journal} {ArXiv e-prints}\ } (\bibinfo
  {year} {2011})},\ \Eprint {http://arxiv.org/abs/1103.4609} {(Paper~I),
  arXiv:1103.4609 [astro-ph.CO]} \BibitemShut {NoStop}%
\bibitem [{\citenamefont {{Sakimoto}}\ and\ \citenamefont
  {{Coroniti}}(1981)}]{1981ApJ...247...19S}%
  \BibitemOpen
  \bibfield  {author} {\bibinfo {author} {\bibfnamefont {P.~J.}\ \bibnamefont
  {{Sakimoto}}}\ and\ \bibinfo {author} {\bibfnamefont {F.~V.}\ \bibnamefont
  {{Coroniti}}},\ }\href {\doibase 10.1086/159005} {\bibfield  {journal}
  {\bibinfo  {journal} {\apj}\ }\textbf {\bibinfo {volume} {247}},\ \bibinfo
  {pages} {19} (\bibinfo {year} {1981})}\BibitemShut {NoStop}%
\bibitem [{\citenamefont {{Frank}}\ \emph {et~al.}(2002)\citenamefont
  {{Frank}}, \citenamefont {{King}},\ and\ \citenamefont
  {{Raine}}}]{2002apa..book.....F}%
  \BibitemOpen
  \bibfield  {author} {\bibinfo {author} {\bibfnamefont {J.}~\bibnamefont
  {{Frank}}}, \bibinfo {author} {\bibfnamefont {A.}~\bibnamefont {{King}}}, \
  and\ \bibinfo {author} {\bibfnamefont {D.~J.}\ \bibnamefont {{Raine}}},\
  }\href@noop {} {\emph {\bibinfo {title} {Accretion Power in Astrophysics, by
  Juhan Frank and Andrew King and Derek Raine, pp.~398.~ISBN
  0521620538.~Cambridge, UK: Cambridge University Press, February 2002.}}},\
  edited by\ \bibinfo {editor} {\bibnamefont {{Frank, J., King, A., \& Raine,
  D.~J.}}}\ (\bibinfo {year} {2002})\BibitemShut {NoStop}%
\bibitem [{\citenamefont {{Armitage}}\ and\ \citenamefont
  {{Natarajan}}(2005)}]{2005ApJ...634..921A}%
  \BibitemOpen
  \bibfield  {author} {\bibinfo {author} {\bibfnamefont {P.~J.}\ \bibnamefont
  {{Armitage}}}\ and\ \bibinfo {author} {\bibfnamefont {P.}~\bibnamefont
  {{Natarajan}}},\ }\href {\doibase 10.1086/497108} {\bibfield  {journal}
  {\bibinfo  {journal} {\apj}\ }\textbf {\bibinfo {volume} {634}},\ \bibinfo
  {pages} {921} (\bibinfo {year} {2005})},\ \Eprint
  {http://arxiv.org/abs/arXiv:astro-ph/0508493} {arXiv:astro-ph/0508493}
  \BibitemShut {NoStop}%
\bibitem [{\citenamefont {{Hayasaki}}(2009)}]{2009PASJ...61...65H}%
  \BibitemOpen
  \bibfield  {author} {\bibinfo {author} {\bibfnamefont {K.}~\bibnamefont
  {{Hayasaki}}},\ }\href@noop {} {\bibfield  {journal} {\bibinfo  {journal}
  {\pasj}\ }\textbf {\bibinfo {volume} {61}},\ \bibinfo {pages} {65} (\bibinfo
  {year} {2009})},\ \Eprint {http://arxiv.org/abs/0805.3408} {arXiv:0805.3408}
  \BibitemShut {NoStop}%
\bibitem [{\citenamefont {{Nixon}}\ \emph {et~al.}(2011)\citenamefont
  {{Nixon}}, \citenamefont {{Cossins}}, \citenamefont {{King}},\ and\
  \citenamefont {{Pringle}}}]{2011MNRAS.tmp..363N}%
  \BibitemOpen
  \bibfield  {author} {\bibinfo {author} {\bibfnamefont {C.~J.}\ \bibnamefont
  {{Nixon}}}, \bibinfo {author} {\bibfnamefont {P.~J.}\ \bibnamefont
  {{Cossins}}}, \bibinfo {author} {\bibfnamefont {A.~R.}\ \bibnamefont
  {{King}}}, \ and\ \bibinfo {author} {\bibfnamefont {J.~E.}\ \bibnamefont
  {{Pringle}}},\ }\href {\doibase 10.1111/j.1365-2966.2010.17952.x} {\bibfield
  {journal} {\bibinfo  {journal} {\mnras}\ ,\ \bibinfo {pages} {363}} (\bibinfo
  {year} {2011})},\ \Eprint {http://arxiv.org/abs/1011.1914} {arXiv:1011.1914
  [astro-ph.HE]} \BibitemShut {NoStop}%
\bibitem [{\citenamefont {{Leung}}\ \emph {et~al.}(1997)\citenamefont
  {{Leung}}, \citenamefont {{Liu}}, \citenamefont {{Suen}}, \citenamefont
  {{Tam}},\ and\ \citenamefont {{Young}}}]{1997PhRvL..78.2894L}%
  \BibitemOpen
  \bibfield  {author} {\bibinfo {author} {\bibfnamefont {P.~T.}\ \bibnamefont
  {{Leung}}}, \bibinfo {author} {\bibfnamefont {Y.~T.}\ \bibnamefont {{Liu}}},
  \bibinfo {author} {\bibfnamefont {W.}~\bibnamefont {{Suen}}}, \bibinfo
  {author} {\bibfnamefont {C.~Y.}\ \bibnamefont {{Tam}}}, \ and\ \bibinfo
  {author} {\bibfnamefont {K.}~\bibnamefont {{Young}}},\ }\href {\doibase
  10.1103/PhysRevLett.78.2894} {\bibfield  {journal} {\bibinfo  {journal}
  {Physical Review Letters}\ }\textbf {\bibinfo {volume} {78}},\ \bibinfo
  {pages} {2894} (\bibinfo {year} {1997})},\ \Eprint
  {http://arxiv.org/abs/arXiv:gr-qc/9903031} {arXiv:gr-qc/9903031} \BibitemShut
  {NoStop}%
\bibitem [{\citenamefont {Papadopoulos}\ and\ \citenamefont
  {Font}(1999)}]{Papadopoulos:1998nc}%
  \BibitemOpen
  \bibfield  {author} {\bibinfo {author} {\bibfnamefont {P.}~\bibnamefont
  {Papadopoulos}}\ and\ \bibinfo {author} {\bibfnamefont {J.~A.}\ \bibnamefont
  {Font}},\ }\href {\doibase 10.1103/PhysRevD.59.044014} {\bibfield  {journal}
  {\bibinfo  {journal} {Phys.Rev.}\ }\textbf {\bibinfo {volume} {D59}},\
  \bibinfo {pages} {044014} (\bibinfo {year} {1999})},\ \Eprint
  {http://arxiv.org/abs/gr-qc/9808054} {arXiv:gr-qc/9808054 [gr-qc]}
  \BibitemShut {NoStop}%
\bibitem [{\citenamefont {Nagar}\ \emph {et~al.}(2004)\citenamefont {Nagar},
  \citenamefont {Diaz}, \citenamefont {Pons},\ and\ \citenamefont
  {Font}}]{Nagar:2004ns}%
  \BibitemOpen
  \bibfield  {author} {\bibinfo {author} {\bibfnamefont {A.}~\bibnamefont
  {Nagar}}, \bibinfo {author} {\bibfnamefont {G.}~\bibnamefont {Diaz}},
  \bibinfo {author} {\bibfnamefont {J.~A.}\ \bibnamefont {Pons}}, \ and\
  \bibinfo {author} {\bibfnamefont {J.~A.}\ \bibnamefont {Font}},\ }\href
  {\doibase 10.1103/PhysRevD.69.124028} {\bibfield  {journal} {\bibinfo
  {journal} {Phys.Rev.}\ }\textbf {\bibinfo {volume} {D69}},\ \bibinfo {pages}
  {124028} (\bibinfo {year} {2004})},\ \Eprint
  {http://arxiv.org/abs/gr-qc/0403077} {arXiv:gr-qc/0403077 [gr-qc]}
  \BibitemShut {NoStop}%
\bibitem [{\citenamefont {{Nagar}}\ \emph {et~al.}(2005)\citenamefont
  {{Nagar}}, \citenamefont {{Font}}, \citenamefont {{Zanotti}},\ and\
  \citenamefont {{de Pietri}}}]{2005PhRvD..72b4007N}%
  \BibitemOpen
  \bibfield  {author} {\bibinfo {author} {\bibfnamefont {A.}~\bibnamefont
  {{Nagar}}}, \bibinfo {author} {\bibfnamefont {J.~A.}\ \bibnamefont {{Font}}},
  \bibinfo {author} {\bibfnamefont {O.}~\bibnamefont {{Zanotti}}}, \ and\
  \bibinfo {author} {\bibfnamefont {R.}~\bibnamefont {{de Pietri}}},\ }\href
  {\doibase 10.1103/PhysRevD.72.024007} {\bibfield  {journal} {\bibinfo
  {journal} {\prd}\ }\textbf {\bibinfo {volume} {72}},\ \bibinfo {pages}
  {024007} (\bibinfo {year} {2005})},\ \Eprint
  {http://arxiv.org/abs/arXiv:gr-qc/0506070} {arXiv:gr-qc/0506070} \BibitemShut
  {NoStop}%
\bibitem [{\citenamefont {Nagar}\ \emph {et~al.}(2007)\citenamefont {Nagar},
  \citenamefont {Zanotti}, \citenamefont {Font},\ and\ \citenamefont
  {Rezzolla}}]{Nagar:2006eu}%
  \BibitemOpen
  \bibfield  {author} {\bibinfo {author} {\bibfnamefont {A.}~\bibnamefont
  {Nagar}}, \bibinfo {author} {\bibfnamefont {O.}~\bibnamefont {Zanotti}},
  \bibinfo {author} {\bibfnamefont {J.~A.}\ \bibnamefont {Font}}, \ and\
  \bibinfo {author} {\bibfnamefont {L.}~\bibnamefont {Rezzolla}},\ }\href
  {\doibase 10.1103/PhysRevD.75.044016} {\bibfield  {journal} {\bibinfo
  {journal} {Phys.Rev.}\ }\textbf {\bibinfo {volume} {D75}},\ \bibinfo {pages}
  {044016} (\bibinfo {year} {2007})},\ \Eprint
  {http://arxiv.org/abs/gr-qc/0610131} {arXiv:gr-qc/0610131 [gr-qc]}
  \BibitemShut {NoStop}%
\bibitem [{\citenamefont {{Nagar}}\ \emph {et~al.}(2007)\citenamefont
  {{Nagar}}, \citenamefont {{Zanotti}}, \citenamefont {{Font}},\ and\
  \citenamefont {{Rezzolla}}}]{2007PhRvD..75d4016N}%
  \BibitemOpen
  \bibfield  {author} {\bibinfo {author} {\bibfnamefont {A.}~\bibnamefont
  {{Nagar}}}, \bibinfo {author} {\bibfnamefont {O.}~\bibnamefont {{Zanotti}}},
  \bibinfo {author} {\bibfnamefont {J.~A.}\ \bibnamefont {{Font}}}, \ and\
  \bibinfo {author} {\bibfnamefont {L.}~\bibnamefont {{Rezzolla}}},\ }\href
  {\doibase 10.1103/PhysRevD.75.044016} {\bibfield  {journal} {\bibinfo
  {journal} {\prd}\ }\textbf {\bibinfo {volume} {75}},\ \bibinfo {pages}
  {044016} (\bibinfo {year} {2007})},\ \Eprint
  {http://arxiv.org/abs/arXiv:gr-qc/0610131} {arXiv:gr-qc/0610131} \BibitemShut
  {NoStop}%
\bibitem [{\citenamefont {{Esposito}}(1971)}]{1971ApJ...165..165E}%
  \BibitemOpen
  \bibfield  {author} {\bibinfo {author} {\bibfnamefont {F.~P.}\ \bibnamefont
  {{Esposito}}},\ }\href {\doibase 10.1086/150884} {\bibfield  {journal}
  {\bibinfo  {journal} {\apj}\ }\textbf {\bibinfo {volume} {165}},\ \bibinfo
  {pages} {165} (\bibinfo {year} {1971})}\BibitemShut {NoStop}%
\bibitem [{\citenamefont {{Kocsis}}\ and\ \citenamefont
  {{Loeb}}(2008)}]{2008PhRvL.101d1101K}%
  \BibitemOpen
  \bibfield  {author} {\bibinfo {author} {\bibfnamefont {B.}~\bibnamefont
  {{Kocsis}}}\ and\ \bibinfo {author} {\bibfnamefont {A.}~\bibnamefont
  {{Loeb}}},\ }\href {\doibase 10.1103/PhysRevLett.101.041101} {\bibfield
  {journal} {\bibinfo  {journal} {Physical Review Letters}\ }\textbf {\bibinfo
  {volume} {101}},\ \bibinfo {pages} {041101} (\bibinfo {year} {2008})},\
  \Eprint {http://arxiv.org/abs/0803.0003} {arXiv:0803.0003} \BibitemShut
  {NoStop}%
\bibitem [{\citenamefont {{Marklund}}\ \emph {et~al.}(2000)\citenamefont
  {{Marklund}}, \citenamefont {{Brodin}},\ and\ \citenamefont
  {{Dunsby}}}]{2000ApJ...536..875M}%
  \BibitemOpen
  \bibfield  {author} {\bibinfo {author} {\bibfnamefont {M.}~\bibnamefont
  {{Marklund}}}, \bibinfo {author} {\bibfnamefont {G.}~\bibnamefont
  {{Brodin}}}, \ and\ \bibinfo {author} {\bibfnamefont {P.~K.~S.}\ \bibnamefont
  {{Dunsby}}},\ }\href {\doibase 10.1086/308957} {\bibfield  {journal}
  {\bibinfo  {journal} {\apj}\ }\textbf {\bibinfo {volume} {536}},\ \bibinfo
  {pages} {875} (\bibinfo {year} {2000})},\ \Eprint
  {http://arxiv.org/abs/arXiv:astro-ph/9907350} {arXiv:astro-ph/9907350}
  \BibitemShut {NoStop}%
\bibitem [{\citenamefont {{Brodin}}\ \emph {et~al.}(2001)\citenamefont
  {{Brodin}}, \citenamefont {{Marklund}},\ and\ \citenamefont
  {{Servin}}}]{2001PhRvD..63l4003B}%
  \BibitemOpen
  \bibfield  {author} {\bibinfo {author} {\bibfnamefont {G.}~\bibnamefont
  {{Brodin}}}, \bibinfo {author} {\bibfnamefont {M.}~\bibnamefont
  {{Marklund}}}, \ and\ \bibinfo {author} {\bibfnamefont {M.}~\bibnamefont
  {{Servin}}},\ }\href {\doibase 10.1103/PhysRevD.63.124003} {\bibfield
  {journal} {\bibinfo  {journal} {\prd}\ }\textbf {\bibinfo {volume} {63}},\
  \bibinfo {pages} {124003} (\bibinfo {year} {2001})},\ \Eprint
  {http://arxiv.org/abs/arXiv:astro-ph/0004351} {arXiv:astro-ph/0004351}
  \BibitemShut {NoStop}%
\bibitem [{\citenamefont {{Papadopoulos}}\ \emph {et~al.}(2001)\citenamefont
  {{Papadopoulos}}, \citenamefont {{Stergioulas}}, \citenamefont {{Vlahos}},\
  and\ \citenamefont {{Kuijpers}}}]{2001A&A...377..701P}%
  \BibitemOpen
  \bibfield  {author} {\bibinfo {author} {\bibfnamefont {D.}~\bibnamefont
  {{Papadopoulos}}}, \bibinfo {author} {\bibfnamefont {N.}~\bibnamefont
  {{Stergioulas}}}, \bibinfo {author} {\bibfnamefont {L.}~\bibnamefont
  {{Vlahos}}}, \ and\ \bibinfo {author} {\bibfnamefont {J.}~\bibnamefont
  {{Kuijpers}}},\ }\href {\doibase 10.1051/0004-6361:20010839} {\bibfield
  {journal} {\bibinfo  {journal} {\aap}\ }\textbf {\bibinfo {volume} {377}},\
  \bibinfo {pages} {701} (\bibinfo {year} {2001})},\ \Eprint
  {http://arxiv.org/abs/arXiv:astro-ph/0107043} {arXiv:astro-ph/0107043}
  \BibitemShut {NoStop}%
\bibitem [{\citenamefont {{K{\"a}llberg}}\ \emph {et~al.}(2004)\citenamefont
  {{K{\"a}llberg}}, \citenamefont {{Brodin}},\ and\ \citenamefont
  {{Bradley}}}]{2004PhRvD..70d4014K}%
  \BibitemOpen
  \bibfield  {author} {\bibinfo {author} {\bibfnamefont {A.}~\bibnamefont
  {{K{\"a}llberg}}}, \bibinfo {author} {\bibfnamefont {G.}~\bibnamefont
  {{Brodin}}}, \ and\ \bibinfo {author} {\bibfnamefont {M.}~\bibnamefont
  {{Bradley}}},\ }\href {\doibase 10.1103/PhysRevD.70.044014} {\bibfield
  {journal} {\bibinfo  {journal} {\prd}\ }\textbf {\bibinfo {volume} {70}},\
  \bibinfo {pages} {044014} (\bibinfo {year} {2004})},\ \Eprint
  {http://arxiv.org/abs/arXiv:gr-qc/0312051} {arXiv:gr-qc/0312051} \BibitemShut
  {NoStop}%
\bibitem [{\citenamefont {{Clarkson}}\ \emph {et~al.}(2004)\citenamefont
  {{Clarkson}}, \citenamefont {{Marklund}}, \citenamefont {{Betschart}},\ and\
  \citenamefont {{Dunsby}}}]{2004ApJ...613..492C}%
  \BibitemOpen
  \bibfield  {author} {\bibinfo {author} {\bibfnamefont {C.~A.}\ \bibnamefont
  {{Clarkson}}}, \bibinfo {author} {\bibfnamefont {M.}~\bibnamefont
  {{Marklund}}}, \bibinfo {author} {\bibfnamefont {G.}~\bibnamefont
  {{Betschart}}}, \ and\ \bibinfo {author} {\bibfnamefont {P.~K.~S.}\
  \bibnamefont {{Dunsby}}},\ }\href {\doibase 10.1086/422497} {\bibfield
  {journal} {\bibinfo  {journal} {\apj}\ }\textbf {\bibinfo {volume} {613}},\
  \bibinfo {pages} {492} (\bibinfo {year} {2004})},\ \Eprint
  {http://arxiv.org/abs/arXiv:astro-ph/0310323} {arXiv:astro-ph/0310323}
  \BibitemShut {NoStop}%
\bibitem [{\citenamefont {{M{\"o}sta}}\ \emph {et~al.}(2010)\citenamefont
  {{M{\"o}sta}}, \citenamefont {{Palenzuela}}, \citenamefont {{Rezzolla}},
  \citenamefont {{Lehner}}, \citenamefont {{Yoshida}},\ and\ \citenamefont
  {{Pollney}}}]{2010PhRvD..81f4017M}%
  \BibitemOpen
  \bibfield  {author} {\bibinfo {author} {\bibfnamefont {P.}~\bibnamefont
  {{M{\"o}sta}}}, \bibinfo {author} {\bibfnamefont {C.}~\bibnamefont
  {{Palenzuela}}}, \bibinfo {author} {\bibfnamefont {L.}~\bibnamefont
  {{Rezzolla}}}, \bibinfo {author} {\bibfnamefont {L.}~\bibnamefont
  {{Lehner}}}, \bibinfo {author} {\bibfnamefont {S.}~\bibnamefont {{Yoshida}}},
  \ and\ \bibinfo {author} {\bibfnamefont {D.}~\bibnamefont {{Pollney}}},\
  }\href {\doibase 10.1103/PhysRevD.81.064017} {\bibfield  {journal} {\bibinfo
  {journal} {\prd}\ }\textbf {\bibinfo {volume} {81}},\ \bibinfo {pages}
  {064017} (\bibinfo {year} {2010})},\ \Eprint {http://arxiv.org/abs/0912.2330}
  {arXiv:0912.2330 [gr-qc]} \BibitemShut {NoStop}%
\bibitem [{\citenamefont {{Johnston}}\ \emph {et~al.}(1974)\citenamefont
  {{Johnston}}, \citenamefont {{Ruffini}},\ and\ \citenamefont
  {{Zerilli}}}]{1974PhLB...49..185J}%
  \BibitemOpen
  \bibfield  {author} {\bibinfo {author} {\bibfnamefont {M.}~\bibnamefont
  {{Johnston}}}, \bibinfo {author} {\bibfnamefont {R.}~\bibnamefont
  {{Ruffini}}}, \ and\ \bibinfo {author} {\bibfnamefont {F.}~\bibnamefont
  {{Zerilli}}},\ }\href {\doibase 10.1016/0370-2693(74)90505-X} {\bibfield
  {journal} {\bibinfo  {journal} {Physics Letters B}\ }\textbf {\bibinfo
  {volume} {49}},\ \bibinfo {pages} {185} (\bibinfo {year} {1974})}\BibitemShut
  {NoStop}%
\bibitem [{\citenamefont {{K{\"a}llberg}}\ \emph {et~al.}(2006)\citenamefont
  {{K{\"a}llberg}}, \citenamefont {{Brodin}},\ and\ \citenamefont
  {{Marklund}}}]{2006CQGra..23L...7K}%
  \BibitemOpen
  \bibfield  {author} {\bibinfo {author} {\bibfnamefont {A.}~\bibnamefont
  {{K{\"a}llberg}}}, \bibinfo {author} {\bibfnamefont {G.}~\bibnamefont
  {{Brodin}}}, \ and\ \bibinfo {author} {\bibfnamefont {M.}~\bibnamefont
  {{Marklund}}},\ }\href {\doibase 10.1088/0264-9381/23/2/L01} {\bibfield
  {journal} {\bibinfo  {journal} {Classical and Quantum Gravity}\ }\textbf
  {\bibinfo {volume} {23}},\ \bibinfo {pages} {L7} (\bibinfo {year} {2006})},\
  \Eprint {http://arxiv.org/abs/arXiv:gr-qc/0410005} {arXiv:gr-qc/0410005}
  \BibitemShut {NoStop}%
\bibitem [{\citenamefont {{Mosquera Cuesta}}(2002)}]{2002PhRvD..65f4009M}%
  \BibitemOpen
  \bibfield  {author} {\bibinfo {author} {\bibfnamefont {H.~J.}\ \bibnamefont
  {{Mosquera Cuesta}}},\ }\href {\doibase 10.1103/PhysRevD.65.064009}
  {\bibfield  {journal} {\bibinfo  {journal} {\prd}\ }\textbf {\bibinfo
  {volume} {65}},\ \bibinfo {pages} {064009} (\bibinfo {year}
  {2002})}\BibitemShut {NoStop}%
\bibitem [{\citenamefont {Yunes}\ \emph
  {et~al.}(2010{\natexlab{b}})\citenamefont {Yunes}, \citenamefont {Buonanno},
  \citenamefont {Hughes}, \citenamefont {Coleman~Miller},\ and\ \citenamefont
  {Pan}}]{Yunes:2009ef}%
  \BibitemOpen
  \bibfield  {author} {\bibinfo {author} {\bibfnamefont {N.}~\bibnamefont
  {Yunes}}, \bibinfo {author} {\bibfnamefont {A.}~\bibnamefont {Buonanno}},
  \bibinfo {author} {\bibfnamefont {S.~A.}\ \bibnamefont {Hughes}}, \bibinfo
  {author} {\bibfnamefont {M.}~\bibnamefont {Coleman~Miller}}, \ and\ \bibinfo
  {author} {\bibfnamefont {Y.}~\bibnamefont {Pan}},\ }\href {\doibase
  10.1103/PhysRevLett.104.091102} {\bibfield  {journal} {\bibinfo  {journal}
  {Phys.Rev.Lett.}\ }\textbf {\bibinfo {volume} {104}},\ \bibinfo {pages}
  {091102} (\bibinfo {year} {2010}{\natexlab{b}})},\ \Eprint
  {http://arxiv.org/abs/arXiv:0909.4263} {arXiv:arXiv:0909.4263 [gr-qc]}
  \BibitemShut {NoStop}%
\bibitem [{\citenamefont {{Yunes}}(2009)}]{2009GWN.....2....3Y}%
  \BibitemOpen
  \bibfield  {author} {\bibinfo {author} {\bibfnamefont {N.}~\bibnamefont
  {{Yunes}}},\ }\href@noop {} {\bibfield  {journal} {\bibinfo  {journal} {GW
  Notes, Vol.~2, p.~3-47}\ }\textbf {\bibinfo {volume} {2}},\ \bibinfo {pages}
  {3} (\bibinfo {year} {2009})}\BibitemShut {NoStop}%
\bibitem [{\citenamefont {Yunes}\ \emph
  {et~al.}(2010{\natexlab{c}})\citenamefont {Yunes}, \citenamefont {Buonanno},
  \citenamefont {Hughes}, \citenamefont {Pan}, \citenamefont {Barausse} \emph
  {et~al.}}]{Yunes:2010zj}%
  \BibitemOpen
  \bibfield  {author} {\bibinfo {author} {\bibfnamefont {N.}~\bibnamefont
  {Yunes}}, \bibinfo {author} {\bibfnamefont {A.}~\bibnamefont {Buonanno}},
  \bibinfo {author} {\bibfnamefont {S.~A.}\ \bibnamefont {Hughes}}, \bibinfo
  {author} {\bibfnamefont {Y.}~\bibnamefont {Pan}}, \bibinfo {author}
  {\bibfnamefont {E.}~\bibnamefont {Barausse}},  \emph {et~al.},\ }\href@noop
  {} {\  (\bibinfo {year} {2010}{\natexlab{c}})},\ \bibinfo {note} {* Temporary
  entry *},\ \Eprint {http://arxiv.org/abs/arXiv:1009.6013}
  {arXiv:arXiv:1009.6013 [gr-qc]} \BibitemShut {NoStop}%
\bibitem [{\citenamefont {{Yagi~et.~al.~}}()}]{Yagi:prep}%
  \BibitemOpen
  \bibfield  {author} {\bibinfo {author} {\bibfnamefont {K.}~\bibnamefont
  {{Yagi~et.~al.~}}},\ }\href@noop {} {}\bibinfo {note} {{in
  preparation}}\BibitemShut {NoStop}%
\bibitem [{\citenamefont {{Peters}}\ and\ \citenamefont
  {{Mathews}}(1963)}]{1963PhRv..131..435P}%
  \BibitemOpen
  \bibfield  {author} {\bibinfo {author} {\bibfnamefont {P.~C.}\ \bibnamefont
  {{Peters}}}\ and\ \bibinfo {author} {\bibfnamefont {J.}~\bibnamefont
  {{Mathews}}},\ }\href {\doibase 10.1103/PhysRev.131.435} {\bibfield
  {journal} {\bibinfo  {journal} {Physical Review}\ }\textbf {\bibinfo {volume}
  {131}},\ \bibinfo {pages} {435} (\bibinfo {year} {1963})}\BibitemShut
  {NoStop}%
\bibitem [{\citenamefont {{Peters}}(1964)}]{1964PhRv..136.1224P}%
  \BibitemOpen
  \bibfield  {author} {\bibinfo {author} {\bibfnamefont {P.~C.}\ \bibnamefont
  {{Peters}}},\ }\href {\doibase 10.1103/PhysRev.136.B1224} {\bibfield
  {journal} {\bibinfo  {journal} {Physical Review}\ }\textbf {\bibinfo {volume}
  {136}},\ \bibinfo {pages} {1224} (\bibinfo {year} {1964})}\BibitemShut
  {NoStop}%
\bibitem [{\citenamefont {Blanchet}(2006)}]{Blanchet:2002av}%
  \BibitemOpen
  \bibfield  {author} {\bibinfo {author} {\bibfnamefont {L.}~\bibnamefont
  {Blanchet}},\ }\href@noop {} {\bibfield  {journal} {\bibinfo  {journal}
  {Living Rev. Rel.}\ }\textbf {\bibinfo {volume} {9}},\ \bibinfo {pages} {4}
  (\bibinfo {year} {2006})},\ \bibinfo {note} {and references therein},\
  \Eprint {http://arxiv.org/abs/gr-qc/0202016} {gr-qc/0202016} \BibitemShut
  {NoStop}%
\bibitem [{\citenamefont {Lindblom}\ \emph {et~al.}(2008)\citenamefont
  {Lindblom}, \citenamefont {Owen},\ and\ \citenamefont
  {Brown}}]{Lindblom:2008cm}%
  \BibitemOpen
  \bibfield  {author} {\bibinfo {author} {\bibfnamefont {L.}~\bibnamefont
  {Lindblom}}, \bibinfo {author} {\bibfnamefont {B.~J.}\ \bibnamefont {Owen}},
  \ and\ \bibinfo {author} {\bibfnamefont {D.~A.}\ \bibnamefont {Brown}},\
  }\href {\doibase 10.1103/PhysRevD.78.124020} {\bibfield  {journal} {\bibinfo
  {journal} {Phys. Rev.}\ }\textbf {\bibinfo {volume} {D78}},\ \bibinfo {pages}
  {124020} (\bibinfo {year} {2008})},\ \Eprint {http://arxiv.org/abs/0809.3844}
  {arXiv:0809.3844 [gr-qc]} \BibitemShut {NoStop}%
\bibitem [{\citenamefont {{Lindblom}}(2009)}]{2009PhRvD..80f4019L}%
  \BibitemOpen
  \bibfield  {author} {\bibinfo {author} {\bibfnamefont {L.}~\bibnamefont
  {{Lindblom}}},\ }\href {\doibase 10.1103/PhysRevD.80.064019} {\bibfield
  {journal} {\bibinfo  {journal} {\prd}\ }\textbf {\bibinfo {volume} {80}},\
  \bibinfo {pages} {064019} (\bibinfo {year} {2009})},\ \Eprint
  {http://arxiv.org/abs/0907.0457} {arXiv:0907.0457 [gr-qc]} \BibitemShut
  {NoStop}%
\bibitem [{\citenamefont {{Lindblom}}\ \emph {et~al.}(2010)\citenamefont
  {{Lindblom}}, \citenamefont {{Baker}},\ and\ \citenamefont
  {{Owen}}}]{2010arXiv1008.1803L}%
  \BibitemOpen
  \bibfield  {author} {\bibinfo {author} {\bibfnamefont {L.}~\bibnamefont
  {{Lindblom}}}, \bibinfo {author} {\bibfnamefont {J.~G.}\ \bibnamefont
  {{Baker}}}, \ and\ \bibinfo {author} {\bibfnamefont {B.~J.}\ \bibnamefont
  {{Owen}}},\ }\href@noop {} {\bibfield  {journal} {\bibinfo  {journal} {ArXiv
  e-prints}\ } (\bibinfo {year} {2010})},\ \Eprint
  {http://arxiv.org/abs/1008.1803} {arXiv:1008.1803 [gr-qc]} \BibitemShut
  {NoStop}%
\bibitem [{\citenamefont {{Lightman}}\ and\ \citenamefont
  {{Eardley}}(1974)}]{1974ApJ...187L...1L}%
  \BibitemOpen
  \bibfield  {author} {\bibinfo {author} {\bibfnamefont {A.~P.}\ \bibnamefont
  {{Lightman}}}\ and\ \bibinfo {author} {\bibfnamefont {D.~M.}\ \bibnamefont
  {{Eardley}}},\ }\href {\doibase 10.1086/181377} {\bibfield  {journal}
  {\bibinfo  {journal} {\apjl}\ }\textbf {\bibinfo {volume} {187}},\ \bibinfo
  {pages} {L1+} (\bibinfo {year} {1974})}\BibitemShut {NoStop}%
\bibitem [{\citenamefont {{Shakura}}\ and\ \citenamefont
  {{Sunyaev}}(1976)}]{1976MNRAS.175..613S}%
  \BibitemOpen
  \bibfield  {author} {\bibinfo {author} {\bibfnamefont {N.~I.}\ \bibnamefont
  {{Shakura}}}\ and\ \bibinfo {author} {\bibfnamefont {R.~A.}\ \bibnamefont
  {{Sunyaev}}},\ }\href@noop {} {\bibfield  {journal} {\bibinfo  {journal}
  {\mnras}\ }\textbf {\bibinfo {volume} {175}},\ \bibinfo {pages} {613}
  (\bibinfo {year} {1976})}\BibitemShut {NoStop}%
\bibitem [{\citenamefont {{Bisnovatyi-Kogan}}\ and\ \citenamefont
  {{Blinnikov}}(1977)}]{1977A&A....59..111B}%
  \BibitemOpen
  \bibfield  {author} {\bibinfo {author} {\bibfnamefont {G.~S.}\ \bibnamefont
  {{Bisnovatyi-Kogan}}}\ and\ \bibinfo {author} {\bibfnamefont {S.~I.}\
  \bibnamefont {{Blinnikov}}},\ }\href@noop {} {\bibfield  {journal} {\bibinfo
  {journal} {\aap}\ }\textbf {\bibinfo {volume} {59}},\ \bibinfo {pages} {111}
  (\bibinfo {year} {1977})}\BibitemShut {NoStop}%
\bibitem [{\citenamefont {{Piran}}(1978)}]{1978ApJ...221..652P}%
  \BibitemOpen
  \bibfield  {author} {\bibinfo {author} {\bibfnamefont {T.}~\bibnamefont
  {{Piran}}},\ }\href {\doibase 10.1086/156069} {\bibfield  {journal} {\bibinfo
   {journal} {\apj}\ }\textbf {\bibinfo {volume} {221}},\ \bibinfo {pages}
  {652} (\bibinfo {year} {1978})}\BibitemShut {NoStop}%
\bibitem [{\citenamefont {{Ohsuga}}\ \emph {et~al.}(2009)\citenamefont
  {{Ohsuga}}, \citenamefont {{Mineshige}}, \citenamefont {{Mori}},\ and\
  \citenamefont {{Kato}}}]{2009PASJ...61L...7O}%
  \BibitemOpen
  \bibfield  {author} {\bibinfo {author} {\bibfnamefont {K.}~\bibnamefont
  {{Ohsuga}}}, \bibinfo {author} {\bibfnamefont {S.}~\bibnamefont
  {{Mineshige}}}, \bibinfo {author} {\bibfnamefont {M.}~\bibnamefont {{Mori}}},
  \ and\ \bibinfo {author} {\bibfnamefont {Y.}~\bibnamefont {{Kato}}},\
  }\href@noop {} {\bibfield  {journal} {\bibinfo  {journal} {\pasj}\ }\textbf
  {\bibinfo {volume} {61}},\ \bibinfo {pages} {L7+} (\bibinfo {year} {2009})},\
  \Eprint {http://arxiv.org/abs/0903.5364} {arXiv:0903.5364 [astro-ph.HE]}
  \BibitemShut {NoStop}%
\bibitem [{\citenamefont {{Hirose}}\ \emph
  {et~al.}(2009{\natexlab{a}})\citenamefont {{Hirose}}, \citenamefont
  {{Krolik}},\ and\ \citenamefont {{Blaes}}}]{2009ApJ...691...16H}%
  \BibitemOpen
  \bibfield  {author} {\bibinfo {author} {\bibfnamefont {S.}~\bibnamefont
  {{Hirose}}}, \bibinfo {author} {\bibfnamefont {J.~H.}\ \bibnamefont
  {{Krolik}}}, \ and\ \bibinfo {author} {\bibfnamefont {O.}~\bibnamefont
  {{Blaes}}},\ }\href {\doibase 10.1088/0004-637X/691/1/16} {\bibfield
  {journal} {\bibinfo  {journal} {\apj}\ }\textbf {\bibinfo {volume} {691}},\
  \bibinfo {pages} {16} (\bibinfo {year} {2009}{\natexlab{a}})},\ \Eprint
  {http://arxiv.org/abs/0809.1708} {arXiv:0809.1708} \BibitemShut {NoStop}%
\bibitem [{\citenamefont {{Hirose}}\ \emph
  {et~al.}(2009{\natexlab{b}})\citenamefont {{Hirose}}, \citenamefont
  {{Blaes}},\ and\ \citenamefont {{Krolik}}}]{2009ApJ...704..781H}%
  \BibitemOpen
  \bibfield  {author} {\bibinfo {author} {\bibfnamefont {S.}~\bibnamefont
  {{Hirose}}}, \bibinfo {author} {\bibfnamefont {O.}~\bibnamefont {{Blaes}}}, \
  and\ \bibinfo {author} {\bibfnamefont {J.~H.}\ \bibnamefont {{Krolik}}},\
  }\href {\doibase 10.1088/0004-637X/704/1/781} {\bibfield  {journal} {\bibinfo
   {journal} {\apj}\ }\textbf {\bibinfo {volume} {704}},\ \bibinfo {pages}
  {781} (\bibinfo {year} {2009}{\natexlab{b}})},\ \Eprint
  {http://arxiv.org/abs/0908.1117} {arXiv:0908.1117 [astro-ph.HE]} \BibitemShut
  {NoStop}%
\bibitem [{\citenamefont {{Turner}}\ \emph {et~al.}(2003)\citenamefont
  {{Turner}}, \citenamefont {{Stone}}, \citenamefont {{Krolik}},\ and\
  \citenamefont {{Sano}}}]{2003ApJ...593..992T}%
  \BibitemOpen
  \bibfield  {author} {\bibinfo {author} {\bibfnamefont {N.~J.}\ \bibnamefont
  {{Turner}}}, \bibinfo {author} {\bibfnamefont {J.~M.}\ \bibnamefont
  {{Stone}}}, \bibinfo {author} {\bibfnamefont {J.~H.}\ \bibnamefont
  {{Krolik}}}, \ and\ \bibinfo {author} {\bibfnamefont {T.}~\bibnamefont
  {{Sano}}},\ }\href {\doibase 10.1086/376615} {\bibfield  {journal} {\bibinfo
  {journal} {\apj}\ }\textbf {\bibinfo {volume} {593}},\ \bibinfo {pages} {992}
  (\bibinfo {year} {2003})},\ \Eprint
  {http://arxiv.org/abs/arXiv:astro-ph/0304511} {arXiv:astro-ph/0304511}
  \BibitemShut {NoStop}%
\bibitem [{\citenamefont {{Turner}}(2004)}]{2004ApJ...605L..45T}%
  \BibitemOpen
  \bibfield  {author} {\bibinfo {author} {\bibfnamefont {N.~J.}\ \bibnamefont
  {{Turner}}},\ }\href {\doibase 10.1086/386545} {\bibfield  {journal}
  {\bibinfo  {journal} {\apjl}\ }\textbf {\bibinfo {volume} {605}},\ \bibinfo
  {pages} {L45} (\bibinfo {year} {2004})},\ \Eprint
  {http://arxiv.org/abs/arXiv:astro-ph/0402539} {arXiv:astro-ph/0402539}
  \BibitemShut {NoStop}%
\bibitem [{\citenamefont {{Done}}\ and\ \citenamefont
  {{Davis}}(2008)}]{2008ApJ...683..389D}%
  \BibitemOpen
  \bibfield  {author} {\bibinfo {author} {\bibfnamefont {C.}~\bibnamefont
  {{Done}}}\ and\ \bibinfo {author} {\bibfnamefont {S.~W.}\ \bibnamefont
  {{Davis}}},\ }\href {\doibase 10.1086/589647} {\bibfield  {journal} {\bibinfo
   {journal} {\apj}\ }\textbf {\bibinfo {volume} {683}},\ \bibinfo {pages}
  {389} (\bibinfo {year} {2008})},\ \Eprint {http://arxiv.org/abs/0803.0584}
  {arXiv:0803.0584} \BibitemShut {NoStop}%
\bibitem [{\citenamefont {{Kollmeier}}\ \emph {et~al.}(2006)\citenamefont
  {{Kollmeier}}, \citenamefont {{Onken}}, \citenamefont {{Kochanek}},
  \citenamefont {{Gould}}, \citenamefont {{Weinberg}}, \citenamefont
  {{Dietrich}}, \citenamefont {{Cool}}, \citenamefont {{Dey}}, \citenamefont
  {{Eisenstein}}, \citenamefont {{Jannuzi}}, \citenamefont {{Le Floc'h}},\ and\
  \citenamefont {{Stern}}}]{2006ApJ...648..128K}%
  \BibitemOpen
  \bibfield  {author} {\bibinfo {author} {\bibfnamefont {J.~A.}\ \bibnamefont
  {{Kollmeier}}}, \bibinfo {author} {\bibfnamefont {C.~A.}\ \bibnamefont
  {{Onken}}}, \bibinfo {author} {\bibfnamefont {C.~S.}\ \bibnamefont
  {{Kochanek}}}, \bibinfo {author} {\bibfnamefont {A.}~\bibnamefont {{Gould}}},
  \bibinfo {author} {\bibfnamefont {D.~H.}\ \bibnamefont {{Weinberg}}},
  \bibinfo {author} {\bibfnamefont {M.}~\bibnamefont {{Dietrich}}}, \bibinfo
  {author} {\bibfnamefont {R.}~\bibnamefont {{Cool}}}, \bibinfo {author}
  {\bibfnamefont {A.}~\bibnamefont {{Dey}}}, \bibinfo {author} {\bibfnamefont
  {D.~J.}\ \bibnamefont {{Eisenstein}}}, \bibinfo {author} {\bibfnamefont
  {B.~T.}\ \bibnamefont {{Jannuzi}}}, \bibinfo {author} {\bibfnamefont
  {E.}~\bibnamefont {{Le Floc'h}}}, \ and\ \bibinfo {author} {\bibfnamefont
  {D.}~\bibnamefont {{Stern}}},\ }\href {\doibase 10.1086/505646} {\bibfield
  {journal} {\bibinfo  {journal} {\apj}\ }\textbf {\bibinfo {volume} {648}},\
  \bibinfo {pages} {128} (\bibinfo {year} {2006})},\ \Eprint
  {http://arxiv.org/abs/arXiv:astro-ph/0508657} {arXiv:astro-ph/0508657}
  \BibitemShut {NoStop}%
\bibitem [{\citenamefont {{Trump}}\ \emph {et~al.}(2009)\citenamefont
  {{Trump}}, \citenamefont {{Impey}}, \citenamefont {{Kelly}}, \citenamefont
  {{Elvis}}, \citenamefont {{Merloni}}, \citenamefont {{Bongiorno}},
  \citenamefont {{Gabor}}, \citenamefont {{Hao}}, \citenamefont {{McCarthy}},
  \citenamefont {{Huchra}}, \citenamefont {{Brusa}}, \citenamefont
  {{Cappelluti}}, \citenamefont {{Koekemoer}}, \citenamefont {{Nagao}},
  \citenamefont {{Salvato}},\ and\ \citenamefont
  {{Scoville}}}]{2009ApJ...700...49T}%
  \BibitemOpen
  \bibfield  {author} {\bibinfo {author} {\bibfnamefont {J.~R.}\ \bibnamefont
  {{Trump}}}, \bibinfo {author} {\bibfnamefont {C.~D.}\ \bibnamefont
  {{Impey}}}, \bibinfo {author} {\bibfnamefont {B.~C.}\ \bibnamefont
  {{Kelly}}}, \bibinfo {author} {\bibfnamefont {M.}~\bibnamefont {{Elvis}}},
  \bibinfo {author} {\bibfnamefont {A.}~\bibnamefont {{Merloni}}}, \bibinfo
  {author} {\bibfnamefont {A.}~\bibnamefont {{Bongiorno}}}, \bibinfo {author}
  {\bibfnamefont {J.}~\bibnamefont {{Gabor}}}, \bibinfo {author} {\bibfnamefont
  {H.}~\bibnamefont {{Hao}}}, \bibinfo {author} {\bibfnamefont {P.~J.}\
  \bibnamefont {{McCarthy}}}, \bibinfo {author} {\bibfnamefont {J.~P.}\
  \bibnamefont {{Huchra}}}, \bibinfo {author} {\bibfnamefont {M.}~\bibnamefont
  {{Brusa}}}, \bibinfo {author} {\bibfnamefont {N.}~\bibnamefont
  {{Cappelluti}}}, \bibinfo {author} {\bibfnamefont {A.}~\bibnamefont
  {{Koekemoer}}}, \bibinfo {author} {\bibfnamefont {T.}~\bibnamefont
  {{Nagao}}}, \bibinfo {author} {\bibfnamefont {M.}~\bibnamefont {{Salvato}}},
  \ and\ \bibinfo {author} {\bibfnamefont {N.~Z.}\ \bibnamefont {{Scoville}}},\
  }\href {\doibase 10.1088/0004-637X/700/1/49} {\bibfield  {journal} {\bibinfo
  {journal} {\apj}\ }\textbf {\bibinfo {volume} {700}},\ \bibinfo {pages} {49}
  (\bibinfo {year} {2009})},\ \Eprint {http://arxiv.org/abs/0905.1123}
  {arXiv:0905.1123 [astro-ph.CO]} \BibitemShut {NoStop}%
\bibitem [{\citenamefont {{Pessah}}\ \emph {et~al.}(2007)\citenamefont
  {{Pessah}}, \citenamefont {{Chan}},\ and\ \citenamefont
  {{Psaltis}}}]{2007ApJ...668L..51P}%
  \BibitemOpen
  \bibfield  {author} {\bibinfo {author} {\bibfnamefont {M.~E.}\ \bibnamefont
  {{Pessah}}}, \bibinfo {author} {\bibfnamefont {C.}~\bibnamefont {{Chan}}}, \
  and\ \bibinfo {author} {\bibfnamefont {D.}~\bibnamefont {{Psaltis}}},\ }\href
  {\doibase 10.1086/522585} {\bibfield  {journal} {\bibinfo  {journal} {\apjl}\
  }\textbf {\bibinfo {volume} {668}},\ \bibinfo {pages} {L51} (\bibinfo {year}
  {2007})},\ \Eprint {http://arxiv.org/abs/0705.0352} {arXiv:0705.0352}
  \BibitemShut {NoStop}%
\bibitem [{\citenamefont {{Lissauer}}\ \emph {et~al.}(2009)\citenamefont
  {{Lissauer}}, \citenamefont {{Hubickyj}}, \citenamefont {{D'Angelo}},\ and\
  \citenamefont {{Bodenheimer}}}]{2009Icar..199..338L}%
  \BibitemOpen
  \bibfield  {author} {\bibinfo {author} {\bibfnamefont {J.~J.}\ \bibnamefont
  {{Lissauer}}}, \bibinfo {author} {\bibfnamefont {O.}~\bibnamefont
  {{Hubickyj}}}, \bibinfo {author} {\bibfnamefont {G.}~\bibnamefont
  {{D'Angelo}}}, \ and\ \bibinfo {author} {\bibfnamefont {P.}~\bibnamefont
  {{Bodenheimer}}},\ }\href {\doibase 10.1016/j.icarus.2008.10.004} {\bibfield
  {journal} {\bibinfo  {journal} {\icarus}\ }\textbf {\bibinfo {volume}
  {199}},\ \bibinfo {pages} {338} (\bibinfo {year} {2009})},\ \Eprint
  {http://arxiv.org/abs/0810.5186} {arXiv:0810.5186} \BibitemShut {NoStop}%
\bibitem [{\citenamefont {{Dubus}}\ \emph {et~al.}(2001)\citenamefont
  {{Dubus}}, \citenamefont {{Hameury}},\ and\ \citenamefont
  {{Lasota}}}]{2001A&A...373..251D}%
  \BibitemOpen
  \bibfield  {author} {\bibinfo {author} {\bibfnamefont {G.}~\bibnamefont
  {{Dubus}}}, \bibinfo {author} {\bibfnamefont {J.}~\bibnamefont {{Hameury}}},
  \ and\ \bibinfo {author} {\bibfnamefont {J.}~\bibnamefont {{Lasota}}},\
  }\href {\doibase 10.1051/0004-6361:20010632} {\bibfield  {journal} {\bibinfo
  {journal} {\aap}\ }\textbf {\bibinfo {volume} {373}},\ \bibinfo {pages} {251}
  (\bibinfo {year} {2001})},\ \Eprint
  {http://arxiv.org/abs/arXiv:astro-ph/0102237} {arXiv:astro-ph/0102237}
  \BibitemShut {NoStop}%
\bibitem [{\citenamefont {{King}}\ \emph {et~al.}(2007)\citenamefont {{King}},
  \citenamefont {{Pringle}},\ and\ \citenamefont
  {{Livio}}}]{2007MNRAS.376.1740K}%
  \BibitemOpen
  \bibfield  {author} {\bibinfo {author} {\bibfnamefont {A.~R.}\ \bibnamefont
  {{King}}}, \bibinfo {author} {\bibfnamefont {J.~E.}\ \bibnamefont
  {{Pringle}}}, \ and\ \bibinfo {author} {\bibfnamefont {M.}~\bibnamefont
  {{Livio}}},\ }\href {\doibase 10.1111/j.1365-2966.2007.11556.x} {\bibfield
  {journal} {\bibinfo  {journal} {\mnras}\ }\textbf {\bibinfo {volume} {376}},\
  \bibinfo {pages} {1740} (\bibinfo {year} {2007})},\ \Eprint
  {http://arxiv.org/abs/arXiv:astro-ph/0701803} {arXiv:astro-ph/0701803}
  \BibitemShut {NoStop}%
\bibitem [{\citenamefont {{Haiman}}\ \emph {et~al.}(2009)\citenamefont
  {{Haiman}}, \citenamefont {{Kocsis}},\ and\ \citenamefont
  {{Menou}}}]{2009ApJ...700.1952H}%
  \BibitemOpen
  \bibfield  {author} {\bibinfo {author} {\bibfnamefont {Z.}~\bibnamefont
  {{Haiman}}}, \bibinfo {author} {\bibfnamefont {B.}~\bibnamefont {{Kocsis}}},
  \ and\ \bibinfo {author} {\bibfnamefont {K.}~\bibnamefont {{Menou}}},\ }\href
  {\doibase 10.1088/0004-637X/700/2/1952} {\bibfield  {journal} {\bibinfo
  {journal} {\apj}\ }\textbf {\bibinfo {volume} {700}},\ \bibinfo {pages}
  {1952} (\bibinfo {year} {2009})},\ \Eprint {http://arxiv.org/abs/0904.1383}
  {arXiv:0904.1383 [astro-ph.CO]} \BibitemShut {NoStop}%
\bibitem [{\citenamefont {{Goodman}}(2003)}]{2003MNRAS.339..937G}%
  \BibitemOpen
  \bibfield  {author} {\bibinfo {author} {\bibfnamefont {J.}~\bibnamefont
  {{Goodman}}},\ }\href {\doibase 10.1046/j.1365-8711.2003.06241.x} {\bibfield
  {journal} {\bibinfo  {journal} {\mnras}\ }\textbf {\bibinfo {volume} {339}},\
  \bibinfo {pages} {937} (\bibinfo {year} {2003})},\ \Eprint
  {http://arxiv.org/abs/arXiv:astro-ph/0201001} {arXiv:astro-ph/0201001}
  \BibitemShut {NoStop}%
\bibitem [{\citenamefont {{Novikov}}\ and\ \citenamefont
  {{Thorne}}(1973)}]{1973blho.conf..343N}%
  \BibitemOpen
  \bibfield  {author} {\bibinfo {author} {\bibfnamefont {I.~D.}\ \bibnamefont
  {{Novikov}}}\ and\ \bibinfo {author} {\bibfnamefont {K.~S.}\ \bibnamefont
  {{Thorne}}},\ }in\ \href@noop {} {\emph {\bibinfo {booktitle} {Black Holes
  (Les Astres Occlus)}}},\ \bibinfo {editor} {edited by\ \bibinfo {editor}
  {\bibnamefont {{A.~Giannaras}}}}\ (\bibinfo {year} {1973})\ pp.\ \bibinfo
  {pages} {343--450}\BibitemShut {NoStop}%
\bibitem [{\citenamefont {{Page}}\ and\ \citenamefont
  {{Thorne}}(1974)}]{1974ApJ...191..499P}%
  \BibitemOpen
  \bibfield  {author} {\bibinfo {author} {\bibfnamefont {D.~N.}\ \bibnamefont
  {{Page}}}\ and\ \bibinfo {author} {\bibfnamefont {K.~S.}\ \bibnamefont
  {{Thorne}}},\ }\href {\doibase 10.1086/152990} {\bibfield  {journal}
  {\bibinfo  {journal} {\apj}\ }\textbf {\bibinfo {volume} {191}},\ \bibinfo
  {pages} {499} (\bibinfo {year} {1974})}\BibitemShut {NoStop}%
\bibitem [{\citenamefont {{Artymowicz}}\ and\ \citenamefont
  {{Lubow}}(1996)}]{1996ApJ...467L..77A}%
  \BibitemOpen
  \bibfield  {author} {\bibinfo {author} {\bibfnamefont {P.}~\bibnamefont
  {{Artymowicz}}}\ and\ \bibinfo {author} {\bibfnamefont {S.~H.}\ \bibnamefont
  {{Lubow}}},\ }\href {\doibase 10.1086/310200} {\bibfield  {journal} {\bibinfo
   {journal} {\apjl}\ }\textbf {\bibinfo {volume} {467}},\ \bibinfo {pages}
  {L77+} (\bibinfo {year} {1996})}\BibitemShut {NoStop}%
\bibitem [{\citenamefont {{Ivanov}}\ \emph {et~al.}(1999)\citenamefont
  {{Ivanov}}, \citenamefont {{Papaloizou}},\ and\ \citenamefont
  {{Polnarev}}}]{1999MNRAS.307...79I}%
  \BibitemOpen
  \bibfield  {author} {\bibinfo {author} {\bibfnamefont {P.~B.}\ \bibnamefont
  {{Ivanov}}}, \bibinfo {author} {\bibfnamefont {J.~C.~B.}\ \bibnamefont
  {{Papaloizou}}}, \ and\ \bibinfo {author} {\bibfnamefont {A.~G.}\
  \bibnamefont {{Polnarev}}},\ }\href {\doibase
  10.1046/j.1365-8711.1999.02623.x} {\bibfield  {journal} {\bibinfo  {journal}
  {\mnras}\ }\textbf {\bibinfo {volume} {307}},\ \bibinfo {pages} {79}
  (\bibinfo {year} {1999})},\ \Eprint
  {http://arxiv.org/abs/arXiv:astro-ph/9812198} {arXiv:astro-ph/9812198}
  \BibitemShut {NoStop}%
\bibitem [{\citenamefont {{Armitage}}\ and\ \citenamefont
  {{Natarajan}}(2002)}]{2002ApJ...567L...9A}%
  \BibitemOpen
  \bibfield  {author} {\bibinfo {author} {\bibfnamefont {P.~J.}\ \bibnamefont
  {{Armitage}}}\ and\ \bibinfo {author} {\bibfnamefont {P.}~\bibnamefont
  {{Natarajan}}},\ }\href {\doibase 10.1086/339770} {\bibfield  {journal}
  {\bibinfo  {journal} {\apjl}\ }\textbf {\bibinfo {volume} {567}},\ \bibinfo
  {pages} {L9} (\bibinfo {year} {2002})},\ \Eprint
  {http://arxiv.org/abs/arXiv:astro-ph/0201318} {arXiv:astro-ph/0201318}
  \BibitemShut {NoStop}%
\bibitem [{\citenamefont {{Tanaka}}\ and\ \citenamefont
  {{Menou}}(2010)}]{2010ApJ...714..404T}%
  \BibitemOpen
  \bibfield  {author} {\bibinfo {author} {\bibfnamefont {T.}~\bibnamefont
  {{Tanaka}}}\ and\ \bibinfo {author} {\bibfnamefont {K.}~\bibnamefont
  {{Menou}}},\ }\href {\doibase 10.1088/0004-637X/714/1/404} {\bibfield
  {journal} {\bibinfo  {journal} {\apj}\ }\textbf {\bibinfo {volume} {714}},\
  \bibinfo {pages} {404} (\bibinfo {year} {2010})},\ \Eprint
  {http://arxiv.org/abs/0912.2054} {arXiv:0912.2054 [astro-ph.CO]} \BibitemShut
  {NoStop}%
\bibitem [{\citenamefont {{Lin}}\ and\ \citenamefont
  {{Papaloizou}}(1986)}]{1986ApJ...309..846L}%
  \BibitemOpen
  \bibfield  {author} {\bibinfo {author} {\bibfnamefont {D.~N.~C.}\
  \bibnamefont {{Lin}}}\ and\ \bibinfo {author} {\bibfnamefont
  {J.}~\bibnamefont {{Papaloizou}}},\ }\href {\doibase 10.1086/164653}
  {\bibfield  {journal} {\bibinfo  {journal} {\apj}\ }\textbf {\bibinfo
  {volume} {309}},\ \bibinfo {pages} {846} (\bibinfo {year}
  {1986})}\BibitemShut {NoStop}%
\bibitem [{\citenamefont {{Bryden}}\ \emph {et~al.}(1999)\citenamefont
  {{Bryden}}, \citenamefont {{Chen}}, \citenamefont {{Lin}}, \citenamefont
  {{Nelson}},\ and\ \citenamefont {{Papaloizou}}}]{1999ApJ...514..344B}%
  \BibitemOpen
  \bibfield  {author} {\bibinfo {author} {\bibfnamefont {G.}~\bibnamefont
  {{Bryden}}}, \bibinfo {author} {\bibfnamefont {X.}~\bibnamefont {{Chen}}},
  \bibinfo {author} {\bibfnamefont {D.~N.~C.}\ \bibnamefont {{Lin}}}, \bibinfo
  {author} {\bibfnamefont {R.~P.}\ \bibnamefont {{Nelson}}}, \ and\ \bibinfo
  {author} {\bibfnamefont {J.~C.~B.}\ \bibnamefont {{Papaloizou}}},\ }\href
  {\doibase 10.1086/306917} {\bibfield  {journal} {\bibinfo  {journal} {\apj}\
  }\textbf {\bibinfo {volume} {514}},\ \bibinfo {pages} {344} (\bibinfo {year}
  {1999})}\BibitemShut {NoStop}%
\bibitem [{\citenamefont {{Crida}}\ \emph {et~al.}(2006)\citenamefont
  {{Crida}}, \citenamefont {{Morbidelli}},\ and\ \citenamefont
  {{Masset}}}]{2006Icar..181..587C}%
  \BibitemOpen
  \bibfield  {author} {\bibinfo {author} {\bibfnamefont {A.}~\bibnamefont
  {{Crida}}}, \bibinfo {author} {\bibfnamefont {A.}~\bibnamefont
  {{Morbidelli}}}, \ and\ \bibinfo {author} {\bibfnamefont {F.}~\bibnamefont
  {{Masset}}},\ }\href {\doibase 10.1016/j.icarus.2005.10.007} {\bibfield
  {journal} {\bibinfo  {journal} {\icarus}\ }\textbf {\bibinfo {volume}
  {181}},\ \bibinfo {pages} {587} (\bibinfo {year} {2006})},\ \Eprint
  {http://arxiv.org/abs/arXiv:astro-ph/0511082} {arXiv:astro-ph/0511082}
  \BibitemShut {NoStop}%
\bibitem [{\citenamefont {{Goldreich}}\ and\ \citenamefont
  {{Tremaine}}(1980)}]{1980ApJ...241..425G}%
  \BibitemOpen
  \bibfield  {author} {\bibinfo {author} {\bibfnamefont {P.}~\bibnamefont
  {{Goldreich}}}\ and\ \bibinfo {author} {\bibfnamefont {S.}~\bibnamefont
  {{Tremaine}}},\ }\href {\doibase 10.1086/158356} {\bibfield  {journal}
  {\bibinfo  {journal} {\apj}\ }\textbf {\bibinfo {volume} {241}},\ \bibinfo
  {pages} {425} (\bibinfo {year} {1980})}\BibitemShut {NoStop}%
\bibitem [{\citenamefont {{Artymowicz}}\ and\ \citenamefont
  {{Lubow}}(1994)}]{1994ApJ...421..651A}%
  \BibitemOpen
  \bibfield  {author} {\bibinfo {author} {\bibfnamefont {P.}~\bibnamefont
  {{Artymowicz}}}\ and\ \bibinfo {author} {\bibfnamefont {S.~H.}\ \bibnamefont
  {{Lubow}}},\ }\href {\doibase 10.1086/173679} {\bibfield  {journal} {\bibinfo
   {journal} {\apj}\ }\textbf {\bibinfo {volume} {421}},\ \bibinfo {pages}
  {651} (\bibinfo {year} {1994})}\BibitemShut {NoStop}%
\bibitem [{\citenamefont {{Artymowicz}}(1993)}]{1993ApJ...419..155A}%
  \BibitemOpen
  \bibfield  {author} {\bibinfo {author} {\bibfnamefont {P.}~\bibnamefont
  {{Artymowicz}}},\ }\href {\doibase 10.1086/173469} {\bibfield  {journal}
  {\bibinfo  {journal} {\apj}\ }\textbf {\bibinfo {volume} {419}},\ \bibinfo
  {pages} {155} (\bibinfo {year} {1993})}\BibitemShut {NoStop}%
\bibitem [{\citenamefont {{Ward}}(1997)}]{1997Icar..126..261W}%
  \BibitemOpen
  \bibfield  {author} {\bibinfo {author} {\bibfnamefont {W.~R.}\ \bibnamefont
  {{Ward}}},\ }\href {\doibase 10.1006/icar.1996.5647} {\bibfield  {journal}
  {\bibinfo  {journal} {\icarus}\ }\textbf {\bibinfo {volume} {126}},\ \bibinfo
  {pages} {261} (\bibinfo {year} {1997})}\BibitemShut {NoStop}%
\bibitem [{\citenamefont {{Winters}}\ \emph {et~al.}(2003)\citenamefont
  {{Winters}}, \citenamefont {{Balbus}},\ and\ \citenamefont
  {{Hawley}}}]{2003ApJ...589..543W}%
  \BibitemOpen
  \bibfield  {author} {\bibinfo {author} {\bibfnamefont {W.~F.}\ \bibnamefont
  {{Winters}}}, \bibinfo {author} {\bibfnamefont {S.~A.}\ \bibnamefont
  {{Balbus}}}, \ and\ \bibinfo {author} {\bibfnamefont {J.~F.}\ \bibnamefont
  {{Hawley}}},\ }\href {\doibase 10.1086/374409} {\bibfield  {journal}
  {\bibinfo  {journal} {\apj}\ }\textbf {\bibinfo {volume} {589}},\ \bibinfo
  {pages} {543} (\bibinfo {year} {2003})},\ \Eprint
  {http://arxiv.org/abs/arXiv:astro-ph/0301589} {arXiv:astro-ph/0301589}
  \BibitemShut {NoStop}%
\bibitem [{\citenamefont {{Syer}}\ and\ \citenamefont
  {{Clarke}}(1995)}]{1995MNRAS.277..758S}%
  \BibitemOpen
  \bibfield  {author} {\bibinfo {author} {\bibfnamefont {D.}~\bibnamefont
  {{Syer}}}\ and\ \bibinfo {author} {\bibfnamefont {C.~J.}\ \bibnamefont
  {{Clarke}}},\ }\href@noop {} {\bibfield  {journal} {\bibinfo  {journal}
  {\mnras}\ }\textbf {\bibinfo {volume} {277}},\ \bibinfo {pages} {758}
  (\bibinfo {year} {1995})},\ \Eprint
  {http://arxiv.org/abs/arXiv:astro-ph/9505021} {arXiv:astro-ph/9505021}
  \BibitemShut {NoStop}%
\bibitem [{\citenamefont {{Hughes}}(2000)}]{2000PhRvD..61h4004H}%
  \BibitemOpen
  \bibfield  {author} {\bibinfo {author} {\bibfnamefont {S.~A.}\ \bibnamefont
  {{Hughes}}},\ }\href {\doibase 10.1103/PhysRevD.61.084004} {\bibfield
  {journal} {\bibinfo  {journal} {\prd}\ }\textbf {\bibinfo {volume} {61}},\
  \bibinfo {pages} {084004} (\bibinfo {year} {2000})},\ \Eprint
  {http://arxiv.org/abs/arXiv:gr-qc/9910091} {arXiv:gr-qc/9910091} \BibitemShut
  {NoStop}%
\bibitem [{\citenamefont {{Hughes}}(2001)}]{2001PhRvD..63d9902H}%
  \BibitemOpen
  \bibfield  {author} {\bibinfo {author} {\bibfnamefont {S.~A.}\ \bibnamefont
  {{Hughes}}},\ }\href {\doibase 10.1103/PhysRevD.63.049902} {\bibfield
  {journal} {\bibinfo  {journal} {\prd}\ }\textbf {\bibinfo {volume} {63}},\
  \bibinfo {pages} {049902} (\bibinfo {year} {2001})}\BibitemShut {NoStop}%
\bibitem [{\citenamefont {{Hughes}}(2002)}]{2002PhRvD..65f9902H}%
  \BibitemOpen
  \bibfield  {author} {\bibinfo {author} {\bibfnamefont {S.~A.}\ \bibnamefont
  {{Hughes}}},\ }\href {\doibase 10.1103/PhysRevD.65.069902} {\bibfield
  {journal} {\bibinfo  {journal} {\prd}\ }\textbf {\bibinfo {volume} {65}},\
  \bibinfo {pages} {069902} (\bibinfo {year} {2002})}\BibitemShut {NoStop}%
\bibitem [{\citenamefont {{Hughes}}(2003)}]{2003PhRvD..67h9901H}%
  \BibitemOpen
  \bibfield  {author} {\bibinfo {author} {\bibfnamefont {S.~A.}\ \bibnamefont
  {{Hughes}}},\ }\href {\doibase 10.1103/PhysRevD.67.089901} {\bibfield
  {journal} {\bibinfo  {journal} {\prd}\ }\textbf {\bibinfo {volume} {67}},\
  \bibinfo {pages} {089901} (\bibinfo {year} {2003})}\BibitemShut {NoStop}%
\bibitem [{\citenamefont {{Hughes}}(2008)}]{2008PhRvD..78j9902H}%
  \BibitemOpen
  \bibfield  {author} {\bibinfo {author} {\bibfnamefont {S.~A.}\ \bibnamefont
  {{Hughes}}},\ }\href {\doibase 10.1103/PhysRevD.78.109902} {\bibfield
  {journal} {\bibinfo  {journal} {\prd}\ }\textbf {\bibinfo {volume} {78}},\
  \bibinfo {pages} {109902} (\bibinfo {year} {2008})}\BibitemShut {NoStop}%
\bibitem [{\citenamefont {{Bardeen}}\ and\ \citenamefont
  {{Petterson}}(1975)}]{1975ApJ...195L..65B}%
  \BibitemOpen
  \bibfield  {author} {\bibinfo {author} {\bibfnamefont {J.~M.}\ \bibnamefont
  {{Bardeen}}}\ and\ \bibinfo {author} {\bibfnamefont {J.~A.}\ \bibnamefont
  {{Petterson}}},\ }\href {\doibase 10.1086/181711} {\bibfield  {journal}
  {\bibinfo  {journal} {Ap.~J.~Letters}\ }\textbf {\bibinfo {volume} {195}},\
  \bibinfo {pages} {L65+} (\bibinfo {year} {1975})}\BibitemShut {NoStop}%
\bibitem [{\citenamefont {{King}}\ \emph {et~al.}(2005)\citenamefont {{King}},
  \citenamefont {{Lubow}}, \citenamefont {{Ogilvie}},\ and\ \citenamefont
  {{Pringle}}}]{2005MNRAS.363...49K}%
  \BibitemOpen
  \bibfield  {author} {\bibinfo {author} {\bibfnamefont {A.~R.}\ \bibnamefont
  {{King}}}, \bibinfo {author} {\bibfnamefont {S.~H.}\ \bibnamefont {{Lubow}}},
  \bibinfo {author} {\bibfnamefont {G.~I.}\ \bibnamefont {{Ogilvie}}}, \ and\
  \bibinfo {author} {\bibfnamefont {J.~E.}\ \bibnamefont {{Pringle}}},\ }\href
  {\doibase 10.1111/j.1365-2966.2005.09378.x} {\bibfield  {journal} {\bibinfo
  {journal} {Monthly Notices of the Royal Astronomical Society}\ }\textbf
  {\bibinfo {volume} {363}},\ \bibinfo {pages} {49} (\bibinfo {year} {2005})},\
  \Eprint {http://arxiv.org/abs/arXiv:astro-ph/0507098}
  {arXiv:astro-ph/0507098} \BibitemShut {NoStop}%
\bibitem [{\citenamefont {{Lodato}}\ and\ \citenamefont
  {{Pringle}}(2006)}]{2006MNRAS.368.1196L}%
  \BibitemOpen
  \bibfield  {author} {\bibinfo {author} {\bibfnamefont {G.}~\bibnamefont
  {{Lodato}}}\ and\ \bibinfo {author} {\bibfnamefont {J.~E.}\ \bibnamefont
  {{Pringle}}},\ }\href {\doibase 10.1111/j.1365-2966.2006.10194.x} {\bibfield
  {journal} {\bibinfo  {journal} {\mnras}\ }\textbf {\bibinfo {volume} {368}},\
  \bibinfo {pages} {1196} (\bibinfo {year} {2006})},\ \Eprint
  {http://arxiv.org/abs/arXiv:astro-ph/0602306} {arXiv:astro-ph/0602306}
  \BibitemShut {NoStop}%
\bibitem [{\citenamefont {{Bregman}}\ and\ \citenamefont
  {{Alexander}}(2009)}]{2009ApJ...700L.192B}%
  \BibitemOpen
  \bibfield  {author} {\bibinfo {author} {\bibfnamefont {M.}~\bibnamefont
  {{Bregman}}}\ and\ \bibinfo {author} {\bibfnamefont {T.}~\bibnamefont
  {{Alexander}}},\ }\href {\doibase 10.1088/0004-637X/700/2/L192} {\bibfield
  {journal} {\bibinfo  {journal} {\apjl}\ }\textbf {\bibinfo {volume} {700}},\
  \bibinfo {pages} {L192} (\bibinfo {year} {2009})},\ \Eprint
  {http://arxiv.org/abs/0903.2051} {arXiv:0903.2051 [astro-ph.GA]} \BibitemShut
  {NoStop}%
\bibitem [{\citenamefont {{Yu}}\ \emph {et~al.}(2007)\citenamefont {{Yu}},
  \citenamefont {{Lu}},\ and\ \citenamefont {{Lin}}}]{2007ApJ...666..919Y}%
  \BibitemOpen
  \bibfield  {author} {\bibinfo {author} {\bibfnamefont {Q.}~\bibnamefont
  {{Yu}}}, \bibinfo {author} {\bibfnamefont {Y.}~\bibnamefont {{Lu}}}, \ and\
  \bibinfo {author} {\bibfnamefont {D.~N.~C.}\ \bibnamefont {{Lin}}},\ }\href
  {\doibase 10.1086/520622} {\bibfield  {journal} {\bibinfo  {journal} {\apj}\
  }\textbf {\bibinfo {volume} {666}},\ \bibinfo {pages} {919} (\bibinfo {year}
  {2007})},\ \Eprint {http://arxiv.org/abs/0705.3649} {arXiv:0705.3649}
  \BibitemShut {NoStop}%
\bibitem [{\citenamefont {{Ulubay-Siddiki}}\ \emph {et~al.}(2009)\citenamefont
  {{Ulubay-Siddiki}}, \citenamefont {{Gerhard}},\ and\ \citenamefont
  {{Arnaboldi}}}]{2009MNRAS.398..535U}%
  \BibitemOpen
  \bibfield  {author} {\bibinfo {author} {\bibfnamefont {A.}~\bibnamefont
  {{Ulubay-Siddiki}}}, \bibinfo {author} {\bibfnamefont {O.}~\bibnamefont
  {{Gerhard}}}, \ and\ \bibinfo {author} {\bibfnamefont {M.}~\bibnamefont
  {{Arnaboldi}}},\ }\href {\doibase 10.1111/j.1365-2966.2009.15089.x}
  {\bibfield  {journal} {\bibinfo  {journal} {\mnras}\ }\textbf {\bibinfo
  {volume} {398}},\ \bibinfo {pages} {535} (\bibinfo {year} {2009})},\ \Eprint
  {http://arxiv.org/abs/0909.5333} {arXiv:0909.5333 [astro-ph.CO]} \BibitemShut
  {NoStop}%
\bibitem [{\citenamefont {{Krumholz}}\ \emph {et~al.}(2005)\citenamefont
  {{Krumholz}}, \citenamefont {{McKee}},\ and\ \citenamefont
  {{Klein}}}]{2005ApJ...618..757K}%
  \BibitemOpen
  \bibfield  {author} {\bibinfo {author} {\bibfnamefont {M.~R.}\ \bibnamefont
  {{Krumholz}}}, \bibinfo {author} {\bibfnamefont {C.~F.}\ \bibnamefont
  {{McKee}}}, \ and\ \bibinfo {author} {\bibfnamefont {R.~I.}\ \bibnamefont
  {{Klein}}},\ }\href {\doibase 10.1086/426051} {\bibfield  {journal} {\bibinfo
   {journal} {\apj}\ }\textbf {\bibinfo {volume} {618}},\ \bibinfo {pages}
  {757} (\bibinfo {year} {2005})},\ \Eprint
  {http://arxiv.org/abs/arXiv:astro-ph/0409454} {arXiv:astro-ph/0409454}
  \BibitemShut {NoStop}%
\bibitem [{\citenamefont {{Krumholz}}\ \emph {et~al.}(2006)\citenamefont
  {{Krumholz}}, \citenamefont {{McKee}},\ and\ \citenamefont
  {{Klein}}}]{2006ApJ...638..369K}%
  \BibitemOpen
  \bibfield  {author} {\bibinfo {author} {\bibfnamefont {M.~R.}\ \bibnamefont
  {{Krumholz}}}, \bibinfo {author} {\bibfnamefont {C.~F.}\ \bibnamefont
  {{McKee}}}, \ and\ \bibinfo {author} {\bibfnamefont {R.~I.}\ \bibnamefont
  {{Klein}}},\ }\href {\doibase 10.1086/498844} {\bibfield  {journal} {\bibinfo
   {journal} {\apj}\ }\textbf {\bibinfo {volume} {638}},\ \bibinfo {pages}
  {369} (\bibinfo {year} {2006})},\ \Eprint
  {http://arxiv.org/abs/arXiv:astro-ph/0510410} {arXiv:astro-ph/0510410}
  \BibitemShut {NoStop}%
\bibitem [{\citenamefont {{Rees}}(1978)}]{1978PhyS...17..193R}%
  \BibitemOpen
  \bibfield  {author} {\bibinfo {author} {\bibfnamefont {M.~J.}\ \bibnamefont
  {{Rees}}},\ }\href {\doibase 10.1088/0031-8949/17/3/010} {\bibfield
  {journal} {\bibinfo  {journal} {\physscr}\ }\textbf {\bibinfo {volume}
  {17}},\ \bibinfo {pages} {193} (\bibinfo {year} {1978})}\BibitemShut
  {NoStop}%
\bibitem [{\citenamefont {{Begelman}}(1979)}]{1979MNRAS.187..237B}%
  \BibitemOpen
  \bibfield  {author} {\bibinfo {author} {\bibfnamefont {M.~C.}\ \bibnamefont
  {{Begelman}}},\ }\href@noop {} {\bibfield  {journal} {\bibinfo  {journal}
  {\mnras}\ }\textbf {\bibinfo {volume} {187}},\ \bibinfo {pages} {237}
  (\bibinfo {year} {1979})}\BibitemShut {NoStop}%
\bibitem [{\citenamefont {{Blondin}}(1986)}]{1986ApJ...308..755B}%
  \BibitemOpen
  \bibfield  {author} {\bibinfo {author} {\bibfnamefont {J.~M.}\ \bibnamefont
  {{Blondin}}},\ }\href {\doibase 10.1086/164548} {\bibfield  {journal}
  {\bibinfo  {journal} {\apj}\ }\textbf {\bibinfo {volume} {308}},\ \bibinfo
  {pages} {755} (\bibinfo {year} {1986})}\BibitemShut {NoStop}%
\bibitem [{\citenamefont {{Armitage}}(2007)}]{2007astro.ph..1485A}%
  \BibitemOpen
  \bibfield  {author} {\bibinfo {author} {\bibfnamefont {P.~J.}\ \bibnamefont
  {{Armitage}}},\ }\href@noop {} {\bibfield  {journal} {\bibinfo  {journal}
  {ArXiv Astrophysics e-prints}\ } (\bibinfo {year} {2007})},\ \Eprint
  {http://arxiv.org/abs/arXiv:astro-ph/0701485} {arXiv:astro-ph/0701485}
  \BibitemShut {NoStop}%
\bibitem [{\citenamefont {{Kratter}}\ \emph {et~al.}(2010)\citenamefont
  {{Kratter}}, \citenamefont {{Murray-Clay}},\ and\ \citenamefont
  {{Youdin}}}]{2010ApJ...710.1375K}%
  \BibitemOpen
  \bibfield  {author} {\bibinfo {author} {\bibfnamefont {K.~M.}\ \bibnamefont
  {{Kratter}}}, \bibinfo {author} {\bibfnamefont {R.~A.}\ \bibnamefont
  {{Murray-Clay}}}, \ and\ \bibinfo {author} {\bibfnamefont {A.~N.}\
  \bibnamefont {{Youdin}}},\ }\href {\doibase 10.1088/0004-637X/710/2/1375}
  {\bibfield  {journal} {\bibinfo  {journal} {\apj}\ }\textbf {\bibinfo
  {volume} {710}},\ \bibinfo {pages} {1375} (\bibinfo {year} {2010})},\ \Eprint
  {http://arxiv.org/abs/0909.2644} {arXiv:0909.2644 [astro-ph.EP]} \BibitemShut
  {NoStop}%
\bibitem [{\citenamefont {{Tanaka}}\ \emph {et~al.}(2002)\citenamefont
  {{Tanaka}}, \citenamefont {{Takeuchi}},\ and\ \citenamefont
  {{Ward}}}]{2002ApJ...565.1257T}%
  \BibitemOpen
  \bibfield  {author} {\bibinfo {author} {\bibfnamefont {H.}~\bibnamefont
  {{Tanaka}}}, \bibinfo {author} {\bibfnamefont {T.}~\bibnamefont
  {{Takeuchi}}}, \ and\ \bibinfo {author} {\bibfnamefont {W.~R.}\ \bibnamefont
  {{Ward}}},\ }\href {\doibase 10.1086/324713} {\bibfield  {journal} {\bibinfo
  {journal} {\apj}\ }\textbf {\bibinfo {volume} {565}},\ \bibinfo {pages}
  {1257} (\bibinfo {year} {2002})}\BibitemShut {NoStop}%
\bibitem [{\citenamefont {{Chandrasekhar}}(1943)}]{1943ApJ....97..255C}%
  \BibitemOpen
  \bibfield  {author} {\bibinfo {author} {\bibfnamefont {S.}~\bibnamefont
  {{Chandrasekhar}}},\ }\href {\doibase 10.1086/144517} {\bibfield  {journal}
  {\bibinfo  {journal} {\apj}\ }\textbf {\bibinfo {volume} {97}},\ \bibinfo
  {pages} {255} (\bibinfo {year} {1943})}\BibitemShut {NoStop}%
\bibitem [{\citenamefont {{Lynden-Bell}}(1962)}]{1962MNRAS.124..279L}%
  \BibitemOpen
  \bibfield  {author} {\bibinfo {author} {\bibfnamefont {D.}~\bibnamefont
  {{Lynden-Bell}}},\ }\href@noop {} {\bibfield  {journal} {\bibinfo  {journal}
  {\mnras}\ }\textbf {\bibinfo {volume} {124}},\ \bibinfo {pages} {279}
  (\bibinfo {year} {1962})}\BibitemShut {NoStop}%
\bibitem [{\citenamefont {{Binney}}\ and\ \citenamefont
  {{Tremaine}}(2008)}]{2008gady.book.....B}%
  \BibitemOpen
  \bibfield  {author} {\bibinfo {author} {\bibfnamefont {J.}~\bibnamefont
  {{Binney}}}\ and\ \bibinfo {author} {\bibfnamefont {S.}~\bibnamefont
  {{Tremaine}}},\ }\href@noop {} {\emph {\bibinfo {title} {Galactic Dynamics:
  Second Edition, by James Binney and Scott Tremaine.~ISBN 978-0-691-13026-2
  (HB).~Published by Princeton University Press, Princeton, NJ USA, 2008.}}},\
  edited by\ \bibinfo {editor} {\bibnamefont {{Binney, J.~\& Tremaine, S.}}}\
  (\bibinfo  {publisher} {Princeton University Press},\ \bibinfo {year}
  {2008})\BibitemShut {NoStop}%
\bibitem [{\citenamefont {{Ostriker}}(1999)}]{1999ApJ...513..252O}%
  \BibitemOpen
  \bibfield  {author} {\bibinfo {author} {\bibfnamefont {E.~C.}\ \bibnamefont
  {{Ostriker}}},\ }\href {\doibase 10.1086/306858} {\bibfield  {journal}
  {\bibinfo  {journal} {\apj}\ }\textbf {\bibinfo {volume} {513}},\ \bibinfo
  {pages} {252} (\bibinfo {year} {1999})},\ \Eprint
  {http://arxiv.org/abs/arXiv:astro-ph/9810324} {arXiv:astro-ph/9810324}
  \BibitemShut {NoStop}%
\bibitem [{\citenamefont {{Kim}}\ and\ \citenamefont
  {{Kim}}(2007)}]{2007ApJ...665..432K}%
  \BibitemOpen
  \bibfield  {author} {\bibinfo {author} {\bibfnamefont {H.}~\bibnamefont
  {{Kim}}}\ and\ \bibinfo {author} {\bibfnamefont {W.}~\bibnamefont {{Kim}}},\
  }\href {\doibase 10.1086/519302} {\bibfield  {journal} {\bibinfo  {journal}
  {\apj}\ }\textbf {\bibinfo {volume} {665}},\ \bibinfo {pages} {432} (\bibinfo
  {year} {2007})},\ \Eprint {http://arxiv.org/abs/0705.0084} {arXiv:0705.0084}
  \BibitemShut {NoStop}%
\bibitem [{\citenamefont {{Barausse}}(2007)}]{2007MNRAS.382..826B}%
  \BibitemOpen
  \bibfield  {author} {\bibinfo {author} {\bibfnamefont {E.}~\bibnamefont
  {{Barausse}}},\ }\href {\doibase 10.1111/j.1365-2966.2007.12408.x} {\bibfield
   {journal} {\bibinfo  {journal} {\mnras}\ }\textbf {\bibinfo {volume}
  {382}},\ \bibinfo {pages} {826} (\bibinfo {year} {2007})},\ \Eprint
  {http://arxiv.org/abs/0709.0211} {arXiv:0709.0211} \BibitemShut {NoStop}%
\bibitem [{\citenamefont {{Kocsis}}\ and\ \citenamefont
  {{Tremaine}}(2010)}]{2010arXiv1006.0001K}%
  \BibitemOpen
  \bibfield  {author} {\bibinfo {author} {\bibfnamefont {B.}~\bibnamefont
  {{Kocsis}}}\ and\ \bibinfo {author} {\bibfnamefont {S.}~\bibnamefont
  {{Tremaine}}},\ }\href@noop {} {\bibfield  {journal} {\bibinfo  {journal}
  {ArXiv e-prints}\ } (\bibinfo {year} {2010})},\ \Eprint
  {http://arxiv.org/abs/1006.0001} {arXiv:1006.0001 [astro-ph.GA]} \BibitemShut
  {NoStop}%
\bibitem [{\citenamefont {{D'Angelo}}\ \emph {et~al.}(2003)\citenamefont
  {{D'Angelo}}, \citenamefont {{Kley}},\ and\ \citenamefont
  {{Henning}}}]{2003ApJ...586..540D}%
  \BibitemOpen
  \bibfield  {author} {\bibinfo {author} {\bibfnamefont {G.}~\bibnamefont
  {{D'Angelo}}}, \bibinfo {author} {\bibfnamefont {W.}~\bibnamefont {{Kley}}},
  \ and\ \bibinfo {author} {\bibfnamefont {T.}~\bibnamefont {{Henning}}},\
  }\href {\doibase 10.1086/367555} {\bibfield  {journal} {\bibinfo  {journal}
  {\apj}\ }\textbf {\bibinfo {volume} {586}},\ \bibinfo {pages} {540} (\bibinfo
  {year} {2003})},\ \Eprint {http://arxiv.org/abs/arXiv:astro-ph/0308055}
  {arXiv:astro-ph/0308055} \BibitemShut {NoStop}%
\bibitem [{\citenamefont {{Paardekooper}}\ \emph {et~al.}(2010)\citenamefont
  {{Paardekooper}}, \citenamefont {{Baruteau}}, \citenamefont {{Crida}},\ and\
  \citenamefont {{Kley}}}]{2010MNRAS.401.1950P}%
  \BibitemOpen
  \bibfield  {author} {\bibinfo {author} {\bibfnamefont {S.}~\bibnamefont
  {{Paardekooper}}}, \bibinfo {author} {\bibfnamefont {C.}~\bibnamefont
  {{Baruteau}}}, \bibinfo {author} {\bibfnamefont {A.}~\bibnamefont {{Crida}}},
  \ and\ \bibinfo {author} {\bibfnamefont {W.}~\bibnamefont {{Kley}}},\ }\href
  {\doibase 10.1111/j.1365-2966.2009.15782.x} {\bibfield  {journal} {\bibinfo
  {journal} {\mnras}\ }\textbf {\bibinfo {volume} {401}},\ \bibinfo {pages}
  {1950} (\bibinfo {year} {2010})},\ \Eprint {http://arxiv.org/abs/0909.4552}
  {arXiv:0909.4552 [astro-ph.EP]} \BibitemShut {NoStop}%
\bibitem [{\citenamefont {{Nelson}}\ and\ \citenamefont
  {{Papaloizou}}(2004)}]{2004MNRAS.350..849N}%
  \BibitemOpen
  \bibfield  {author} {\bibinfo {author} {\bibfnamefont {R.~P.}\ \bibnamefont
  {{Nelson}}}\ and\ \bibinfo {author} {\bibfnamefont {J.~C.~B.}\ \bibnamefont
  {{Papaloizou}}},\ }\href {\doibase 10.1111/j.1365-2966.2004.07406.x}
  {\bibfield  {journal} {\bibinfo  {journal} {\mnras}\ }\textbf {\bibinfo
  {volume} {350}},\ \bibinfo {pages} {849} (\bibinfo {year} {2004})},\ \Eprint
  {http://arxiv.org/abs/arXiv:astro-ph/0308360} {arXiv:astro-ph/0308360}
  \BibitemShut {NoStop}%
\bibitem [{\citenamefont {{Laughlin}}\ \emph {et~al.}(2004)\citenamefont
  {{Laughlin}}, \citenamefont {{Steinacker}},\ and\ \citenamefont
  {{Adams}}}]{2004ApJ...608..489L}%
  \BibitemOpen
  \bibfield  {author} {\bibinfo {author} {\bibfnamefont {G.}~\bibnamefont
  {{Laughlin}}}, \bibinfo {author} {\bibfnamefont {A.}~\bibnamefont
  {{Steinacker}}}, \ and\ \bibinfo {author} {\bibfnamefont {F.~C.}\
  \bibnamefont {{Adams}}},\ }\href {\doibase 10.1086/386316} {\bibfield
  {journal} {\bibinfo  {journal} {\apj}\ }\textbf {\bibinfo {volume} {608}},\
  \bibinfo {pages} {489} (\bibinfo {year} {2004})},\ \Eprint
  {http://arxiv.org/abs/arXiv:astro-ph/0308406} {arXiv:astro-ph/0308406}
  \BibitemShut {NoStop}%
\bibitem [{\citenamefont {{Paardekooper}}\ and\ \citenamefont
  {{Mellema}}(2006)}]{2006A&A...459L..17P}%
  \BibitemOpen
  \bibfield  {author} {\bibinfo {author} {\bibfnamefont {S.}~\bibnamefont
  {{Paardekooper}}}\ and\ \bibinfo {author} {\bibfnamefont {G.}~\bibnamefont
  {{Mellema}}},\ }\href {\doibase 10.1051/0004-6361:20066304} {\bibfield
  {journal} {\bibinfo  {journal} {\aap}\ }\textbf {\bibinfo {volume} {459}},\
  \bibinfo {pages} {L17} (\bibinfo {year} {2006})},\ \Eprint
  {http://arxiv.org/abs/arXiv:astro-ph/0608658} {arXiv:astro-ph/0608658}
  \BibitemShut {NoStop}%
\bibitem [{\citenamefont {{Hirata}}(2010{\natexlab{a}})}]{2010arXiv1010.0758H}%
  \BibitemOpen
  \bibfield  {author} {\bibinfo {author} {\bibfnamefont {C.~M.}\ \bibnamefont
  {{Hirata}}},\ }\href@noop {} {\bibfield  {journal} {\bibinfo  {journal}
  {ArXiv e-prints}\ } (\bibinfo {year} {2010}{\natexlab{a}})},\ \Eprint
  {http://arxiv.org/abs/1010.0758} {arXiv:1010.0758 [astro-ph.HE]} \BibitemShut
  {NoStop}%
\bibitem [{\citenamefont {{Hirata}}(2010{\natexlab{b}})}]{2010arXiv1010.0759H}%
  \BibitemOpen
  \bibfield  {author} {\bibinfo {author} {\bibfnamefont {C.~M.}\ \bibnamefont
  {{Hirata}}},\ }\href@noop {} {\bibfield  {journal} {\bibinfo  {journal}
  {ArXiv e-prints}\ } (\bibinfo {year} {2010}{\natexlab{b}})},\ \Eprint
  {http://arxiv.org/abs/1010.0759} {arXiv:1010.0759 [astro-ph.HE]} \BibitemShut
  {NoStop}%
\bibitem [{\citenamefont {{Amin}}\ and\ \citenamefont
  {{Frolov}}(2006)}]{2006MNRAS.370L..42A}%
  \BibitemOpen
  \bibfield  {author} {\bibinfo {author} {\bibfnamefont {M.~A.}\ \bibnamefont
  {{Amin}}}\ and\ \bibinfo {author} {\bibfnamefont {A.~V.}\ \bibnamefont
  {{Frolov}}},\ }\href {\doibase 10.1111/j.1745-3933.2006.00185.x} {\bibfield
  {journal} {\bibinfo  {journal} {\mnras}\ }\textbf {\bibinfo {volume} {370}},\
  \bibinfo {pages} {L42} (\bibinfo {year} {2006})},\ \Eprint
  {http://arxiv.org/abs/arXiv:astro-ph/0603687} {arXiv:astro-ph/0603687}
  \BibitemShut {NoStop}%
\bibitem [{\citenamefont {{Lin}}\ and\ \citenamefont
  {{Papaloizou}}(2011{\natexlab{a}})}]{2011arXiv1103.5036L}%
  \BibitemOpen
  \bibfield  {author} {\bibinfo {author} {\bibfnamefont {M.}~\bibnamefont
  {{Lin}}}\ and\ \bibinfo {author} {\bibfnamefont {J.}~\bibnamefont
  {{Papaloizou}}},\ }\href@noop {} {\bibfield  {journal} {\bibinfo  {journal}
  {ArXiv e-prints}\ } (\bibinfo {year} {2011}{\natexlab{a}})},\ \Eprint
  {http://arxiv.org/abs/1103.5036} {arXiv:1103.5036 [astro-ph.EP]} \BibitemShut
  {NoStop}%
\bibitem [{\citenamefont {{Lin}}\ and\ \citenamefont
  {{Papaloizou}}(2011{\natexlab{b}})}]{2011arXiv1103.5025L}%
  \BibitemOpen
  \bibfield  {author} {\bibinfo {author} {\bibfnamefont {M.}~\bibnamefont
  {{Lin}}}\ and\ \bibinfo {author} {\bibfnamefont {J.}~\bibnamefont
  {{Papaloizou}}},\ }\href@noop {} {\bibfield  {journal} {\bibinfo  {journal}
  {ArXiv e-prints}\ } (\bibinfo {year} {2011}{\natexlab{b}})},\ \Eprint
  {http://arxiv.org/abs/1103.5025} {arXiv:1103.5025 [astro-ph.EP]} \BibitemShut
  {NoStop}%
\bibitem [{\citenamefont {{Kocsis}}\ and\ \citenamefont
  {{Sesana}}(2011)}]{2011MNRAS.411.1467K}%
  \BibitemOpen
  \bibfield  {author} {\bibinfo {author} {\bibfnamefont {B.}~\bibnamefont
  {{Kocsis}}}\ and\ \bibinfo {author} {\bibfnamefont {A.}~\bibnamefont
  {{Sesana}}},\ }\href {\doibase 10.1111/j.1365-2966.2010.17782.x} {\bibfield
  {journal} {\bibinfo  {journal} {\mnras}\ }\textbf {\bibinfo {volume} {411}},\
  \bibinfo {pages} {1467} (\bibinfo {year} {2011})},\ \Eprint
  {http://arxiv.org/abs/1002.0584} {arXiv:1002.0584 [astro-ph.CO]} \BibitemShut
  {NoStop}%
\bibitem [{\citenamefont {{Krolik}}(1999)}]{1999agnc.book.....K}%
  \BibitemOpen
  \bibfield  {author} {\bibinfo {author} {\bibfnamefont {J.~H.}\ \bibnamefont
  {{Krolik}}},\ }\href@noop {} {\emph {\bibinfo {title} {Active galactic nuclei
  : from the central black hole to the galactic environment /Julian
  H.~Krolik.~Princeton, N.~J.~: Princeton University Press, c1999.}}},\ edited
  by\ \bibinfo {editor} {\bibnamefont {{Krolik, J.~H.}}}\ (\bibinfo {year}
  {1999})\BibitemShut {NoStop}%
\bibitem [{\citenamefont {Buonanno}\ and\ \citenamefont
  {Damour}(1999)}]{Buonanno99}%
  \BibitemOpen
  \bibfield  {author} {\bibinfo {author} {\bibfnamefont {A.}~\bibnamefont
  {Buonanno}}\ and\ \bibinfo {author} {\bibfnamefont {T.}~\bibnamefont
  {Damour}},\ }\href@noop {} {\bibfield  {journal} {\bibinfo  {journal} {Phys.
  Rev.}\ }\textbf {\bibinfo {volume} {D59}},\ \bibinfo {pages} {084006}
  (\bibinfo {year} {1999})}\BibitemShut {NoStop}%
\bibitem [{\citenamefont {Buonanno}\ and\ \citenamefont
  {Damour}(2000)}]{Buonanno00}%
  \BibitemOpen
  \bibfield  {author} {\bibinfo {author} {\bibfnamefont {A.}~\bibnamefont
  {Buonanno}}\ and\ \bibinfo {author} {\bibfnamefont {T.}~\bibnamefont
  {Damour}},\ }\href@noop {} {\bibfield  {journal} {\bibinfo  {journal} {Phys.
  Rev.}\ }\textbf {\bibinfo {volume} {D62}},\ \bibinfo {pages} {064015}
  (\bibinfo {year} {2000})}\BibitemShut {NoStop}%
\bibitem [{\citenamefont {Damour}\ \emph {et~al.}(2000)\citenamefont {Damour},
  \citenamefont {Jaranowski},\ and\ \citenamefont {Schaefer}}]{Damour00}%
  \BibitemOpen
  \bibfield  {author} {\bibinfo {author} {\bibfnamefont {T.}~\bibnamefont
  {Damour}}, \bibinfo {author} {\bibfnamefont {P.}~\bibnamefont {Jaranowski}},
  \ and\ \bibinfo {author} {\bibfnamefont {G.}~\bibnamefont {Schaefer}},\
  }\href@noop {} {\bibfield  {journal} {\bibinfo  {journal} {Phys. Rev.}\
  }\textbf {\bibinfo {volume} {D62}},\ \bibinfo {pages} {084011} (\bibinfo
  {year} {2000})}\BibitemShut {NoStop}%
\bibitem [{\citenamefont {Damour}(2001)}]{Damour01}%
  \BibitemOpen
  \bibfield  {author} {\bibinfo {author} {\bibfnamefont {T.}~\bibnamefont
  {Damour}},\ }\href@noop {} {\bibfield  {journal} {\bibinfo  {journal} {Phys.
  Rev.}\ }\textbf {\bibinfo {volume} {D64}},\ \bibinfo {pages} {124013}
  (\bibinfo {year} {2001})}\BibitemShut {NoStop}%
\bibitem [{\citenamefont {Buonanno}\ \emph {et~al.}(2006)\citenamefont
  {Buonanno}, \citenamefont {Chen},\ and\ \citenamefont {Damour}}]{Buonanno06}%
  \BibitemOpen
  \bibfield  {author} {\bibinfo {author} {\bibfnamefont {A.}~\bibnamefont
  {Buonanno}}, \bibinfo {author} {\bibfnamefont {Y.}~\bibnamefont {Chen}}, \
  and\ \bibinfo {author} {\bibfnamefont {T.}~\bibnamefont {Damour}},\
  }\href@noop {} {\bibfield  {journal} {\bibinfo  {journal} {Phys. Rev.}\
  }\textbf {\bibinfo {volume} {D74}},\ \bibinfo {pages} {104005} (\bibinfo
  {year} {2006})}\BibitemShut {NoStop}%
\bibitem [{\citenamefont {Damour}\ \emph {et~al.}(2008)\citenamefont {Damour},
  \citenamefont {Jaranowski},\ and\ \citenamefont {Schaefer}}]{Damour:2008qf}%
  \BibitemOpen
  \bibfield  {author} {\bibinfo {author} {\bibfnamefont {T.}~\bibnamefont
  {Damour}}, \bibinfo {author} {\bibfnamefont {P.}~\bibnamefont {Jaranowski}},
  \ and\ \bibinfo {author} {\bibfnamefont {G.}~\bibnamefont {Schaefer}},\
  }\href {\doibase 10.1103/PhysRevD.78.024009} {\bibfield  {journal} {\bibinfo
  {journal} {Phys. Rev.}\ }\textbf {\bibinfo {volume} {D78}},\ \bibinfo {pages}
  {024009} (\bibinfo {year} {2008})},\ \Eprint {http://arxiv.org/abs/0803.0915}
  {arXiv:0803.0915 [gr-qc]} \BibitemShut {NoStop}%
\bibitem [{\citenamefont {Barausse}\ and\ \citenamefont
  {Buonanno}(2010)}]{Barausse:2009xi}%
  \BibitemOpen
  \bibfield  {author} {\bibinfo {author} {\bibfnamefont {E.}~\bibnamefont
  {Barausse}}\ and\ \bibinfo {author} {\bibfnamefont {A.}~\bibnamefont
  {Buonanno}},\ }\href {\doibase 10.1103/PhysRevD.81.084024} {\bibfield
  {journal} {\bibinfo  {journal} {Phys. Rev.}\ }\textbf {\bibinfo {volume}
  {D81}},\ \bibinfo {pages} {084024} (\bibinfo {year} {2010})},\ \Eprint
  {http://arxiv.org/abs/0912.3517} {arXiv:0912.3517 [gr-qc]} \BibitemShut
  {NoStop}%
\bibitem [{\citenamefont {Nagar}\ \emph {et~al.}(2007)\citenamefont {Nagar},
  \citenamefont {Damour},\ and\ \citenamefont {Tartaglia}}]{Nagar:2006xv}%
  \BibitemOpen
  \bibfield  {author} {\bibinfo {author} {\bibfnamefont {A.}~\bibnamefont
  {Nagar}}, \bibinfo {author} {\bibfnamefont {T.}~\bibnamefont {Damour}}, \
  and\ \bibinfo {author} {\bibfnamefont {A.}~\bibnamefont {Tartaglia}},\ }\href
  {\doibase 10.1088/0264-9381/24/12/S08} {\bibfield  {journal} {\bibinfo
  {journal} {Class. Quant. Grav.}\ }\textbf {\bibinfo {volume} {24}},\ \bibinfo
  {pages} {S109} (\bibinfo {year} {2007})},\ \Eprint
  {http://arxiv.org/abs/gr-qc/0612096} {arXiv:gr-qc/0612096} \BibitemShut
  {NoStop}%
\bibitem [{\citenamefont {Damour}\ and\ \citenamefont
  {Nagar}(2007)}]{Damour2007}%
  \BibitemOpen
  \bibfield  {author} {\bibinfo {author} {\bibfnamefont {T.}~\bibnamefont
  {Damour}}\ and\ \bibinfo {author} {\bibfnamefont {A.}~\bibnamefont {Nagar}},\
  }\href@noop {} {\bibfield  {journal} {\bibinfo  {journal} {Phys. Rev.}\
  }\textbf {\bibinfo {volume} {D76}},\ \bibinfo {pages} {064028} (\bibinfo
  {year} {2007})}\BibitemShut {NoStop}%
\bibitem [{\citenamefont {Bernuzzi}\ and\ \citenamefont
  {Nagar}(2010)}]{Bernuzzi:2010ty}%
  \BibitemOpen
  \bibfield  {author} {\bibinfo {author} {\bibfnamefont {S.}~\bibnamefont
  {Bernuzzi}}\ and\ \bibinfo {author} {\bibfnamefont {A.}~\bibnamefont
  {Nagar}},\ }\href {\doibase 10.1103/PhysRevD.81.084056} {\bibfield  {journal}
  {\bibinfo  {journal} {Phys. Rev.}\ }\textbf {\bibinfo {volume} {D81}},\
  \bibinfo {pages} {084056} (\bibinfo {year} {2010})},\ \Eprint
  {http://arxiv.org/abs/1003.0597} {arXiv:1003.0597 [gr-qc]} \BibitemShut
  {NoStop}%
\bibitem [{\citenamefont {Damour}\ \emph {et~al.}(1998)\citenamefont {Damour},
  \citenamefont {Iyer},\ and\ \citenamefont {Sathyaprakash}}]{Damour:1997ub}%
  \BibitemOpen
  \bibfield  {author} {\bibinfo {author} {\bibfnamefont {T.}~\bibnamefont
  {Damour}}, \bibinfo {author} {\bibfnamefont {B.~R.}\ \bibnamefont {Iyer}}, \
  and\ \bibinfo {author} {\bibfnamefont {B.~S.}\ \bibnamefont
  {Sathyaprakash}},\ }\href@noop {} {\bibfield  {journal} {\bibinfo  {journal}
  {Phys. Rev.}\ }\textbf {\bibinfo {volume} {D57}},\ \bibinfo {pages} {885}
  (\bibinfo {year} {1998})}\BibitemShut {NoStop}%
\bibitem [{\citenamefont {{Damour}}\ \emph {et~al.}(2009)\citenamefont
  {{Damour}}, \citenamefont {{Iyer}},\ and\ \citenamefont
  {{Nagar}}}]{Damour:2008gu}%
  \BibitemOpen
  \bibfield  {author} {\bibinfo {author} {\bibfnamefont {T.}~\bibnamefont
  {{Damour}}}, \bibinfo {author} {\bibfnamefont {B.~R.}\ \bibnamefont
  {{Iyer}}}, \ and\ \bibinfo {author} {\bibfnamefont {A.}~\bibnamefont
  {{Nagar}}},\ }\href {\doibase 10.1103/PhysRevD.79.064004} {\bibfield
  {journal} {\bibinfo  {journal} {\prd}\ }\textbf {\bibinfo {volume} {79}},\
  \bibinfo {pages} {064004} (\bibinfo {year} {2009})},\ \Eprint
  {http://arxiv.org/abs/0811.2069} {arXiv:0811.2069 [gr-qc]} \BibitemShut
  {NoStop}%
\bibitem [{\citenamefont {Pan}\ \emph {et~al.}(2010{\natexlab{a}})\citenamefont
  {Pan}, \citenamefont {Buonanno}, \citenamefont {Fujita}, \citenamefont
  {Racine},\ and\ \citenamefont {Tagoshi}}]{Pan:2010hz}%
  \BibitemOpen
  \bibfield  {author} {\bibinfo {author} {\bibfnamefont {Y.}~\bibnamefont
  {Pan}}, \bibinfo {author} {\bibfnamefont {A.}~\bibnamefont {Buonanno}},
  \bibinfo {author} {\bibfnamefont {R.}~\bibnamefont {Fujita}}, \bibinfo
  {author} {\bibfnamefont {E.}~\bibnamefont {Racine}}, \ and\ \bibinfo {author}
  {\bibfnamefont {H.}~\bibnamefont {Tagoshi}},\ }\href@noop {} {\  (\bibinfo
  {year} {2010}{\natexlab{a}})},\ \Eprint {http://arxiv.org/abs/1006.0431}
  {arXiv:1006.0431 [gr-qc]} \BibitemShut {NoStop}%
\bibitem [{\citenamefont {Fujita}\ and\ \citenamefont
  {Iyer}(2010)}]{Fujita:2010xj}%
  \BibitemOpen
  \bibfield  {author} {\bibinfo {author} {\bibfnamefont {R.}~\bibnamefont
  {Fujita}}\ and\ \bibinfo {author} {\bibfnamefont {B.~R.}\ \bibnamefont
  {Iyer}},\ }\href@noop {} {\  (\bibinfo {year} {2010})},\ \Eprint
  {http://arxiv.org/abs/1005.2266} {arXiv:1005.2266 [gr-qc]} \BibitemShut
  {NoStop}%
\bibitem [{\citenamefont {Damour}\ and\ \citenamefont
  {Nagar}(2009)}]{Damour:2009kr}%
  \BibitemOpen
  \bibfield  {author} {\bibinfo {author} {\bibfnamefont {T.}~\bibnamefont
  {Damour}}\ and\ \bibinfo {author} {\bibfnamefont {A.}~\bibnamefont {Nagar}},\
  }\href@noop {} {\bibfield  {journal} {\bibinfo  {journal} {Phys. Rev.}\
  }\textbf {\bibinfo {volume} {D79}},\ \bibinfo {pages} {081503} (\bibinfo
  {year} {2009})}\BibitemShut {NoStop}%
\bibitem [{\citenamefont {Buonanno}\ \emph {et~al.}(2009)\citenamefont
  {Buonanno} \emph {et~al.}}]{Buonanno:2009qa}%
  \BibitemOpen
  \bibfield  {author} {\bibinfo {author} {\bibfnamefont {A.}~\bibnamefont
  {Buonanno}} \emph {et~al.},\ }\href@noop {} {\bibfield  {journal} {\bibinfo
  {journal} {Phys. Rev.}\ }\textbf {\bibinfo {volume} {D79}},\ \bibinfo {pages}
  {124028} (\bibinfo {year} {2009})}\BibitemShut {NoStop}%
\bibitem [{\citenamefont {Pan}\ \emph {et~al.}(2010{\natexlab{b}})\citenamefont
  {Pan} \emph {et~al.}}]{Pan:2009wj}%
  \BibitemOpen
  \bibfield  {author} {\bibinfo {author} {\bibfnamefont {Y.}~\bibnamefont
  {Pan}} \emph {et~al.},\ }\href {\doibase 10.1103/PhysRevD.81.084041}
  {\bibfield  {journal} {\bibinfo  {journal} {Phys. Rev.}\ }\textbf {\bibinfo
  {volume} {D81}},\ \bibinfo {pages} {084041} (\bibinfo {year}
  {2010}{\natexlab{b}})},\ \Eprint {http://arxiv.org/abs/0912.3466}
  {arXiv:0912.3466 [gr-qc]} \BibitemShut {NoStop}%
\bibitem [{\citenamefont {Barack}\ and\ \citenamefont
  {Sago}(2009)}]{Barack:2009ey}%
  \BibitemOpen
  \bibfield  {author} {\bibinfo {author} {\bibfnamefont {L.}~\bibnamefont
  {Barack}}\ and\ \bibinfo {author} {\bibfnamefont {N.}~\bibnamefont {Sago}},\
  }\href {\doibase 10.1103/PhysRevLett.102.191101} {\bibfield  {journal}
  {\bibinfo  {journal} {Phys. Rev. Lett.}\ }\textbf {\bibinfo {volume} {102}},\
  \bibinfo {pages} {191101} (\bibinfo {year} {2009})},\ \Eprint
  {http://arxiv.org/abs/0902.0573} {arXiv:0902.0573 [gr-qc]} \BibitemShut
  {NoStop}%
\bibitem [{\citenamefont {Damour}(2010)}]{Damour:2009sm}%
  \BibitemOpen
  \bibfield  {author} {\bibinfo {author} {\bibfnamefont {T.}~\bibnamefont
  {Damour}},\ }\href {\doibase 10.1103/PhysRevD.81.024017} {\bibfield
  {journal} {\bibinfo  {journal} {Phys. Rev.}\ }\textbf {\bibinfo {volume}
  {D81}},\ \bibinfo {pages} {024017} (\bibinfo {year} {2010})},\ \Eprint
  {http://arxiv.org/abs/0910.5533} {arXiv:0910.5533 [gr-qc]} \BibitemShut
  {NoStop}%
\bibitem [{\citenamefont {Bardeen}\ \emph {et~al.}(1972)\citenamefont
  {Bardeen}, \citenamefont {Press},\ and\ \citenamefont
  {Teukolsky}}]{Bardeen:1972fi}%
  \BibitemOpen
  \bibfield  {author} {\bibinfo {author} {\bibfnamefont {J.~M.}\ \bibnamefont
  {Bardeen}}, \bibinfo {author} {\bibfnamefont {W.~H.}\ \bibnamefont {Press}},
  \ and\ \bibinfo {author} {\bibfnamefont {S.~A.}\ \bibnamefont {Teukolsky}},\
  }\href {\doibase 10.1086/151796} {\bibfield  {journal} {\bibinfo  {journal}
  {Astrophys. J.}\ }\textbf {\bibinfo {volume} {178}},\ \bibinfo {pages} {347}
  (\bibinfo {year} {1972})}\BibitemShut {NoStop}%
\bibitem [{\citenamefont {{Poisson}}\ and\ \citenamefont
  {{Will}}(1995)}]{1995PhRvD..52..848P}%
  \BibitemOpen
  \bibfield  {author} {\bibinfo {author} {\bibfnamefont {E.}~\bibnamefont
  {{Poisson}}}\ and\ \bibinfo {author} {\bibfnamefont {C.~M.}\ \bibnamefont
  {{Will}}},\ }\href {\doibase 10.1103/PhysRevD.52.848} {\bibfield  {journal}
  {\bibinfo  {journal} {\prd}\ }\textbf {\bibinfo {volume} {52}},\ \bibinfo
  {pages} {848} (\bibinfo {year} {1995})},\ \Eprint
  {http://arxiv.org/abs/arXiv:gr-qc/9502040} {arXiv:gr-qc/9502040} \BibitemShut
  {NoStop}%
\bibitem [{\citenamefont {Yunes}\ \emph {et~al.}(2009)\citenamefont {Yunes},
  \citenamefont {Arun}, \citenamefont {Berti},\ and\ \citenamefont
  {Will}}]{Yunes:2009yz}%
  \BibitemOpen
  \bibfield  {author} {\bibinfo {author} {\bibfnamefont {N.}~\bibnamefont
  {Yunes}}, \bibinfo {author} {\bibfnamefont {K.~G.}\ \bibnamefont {Arun}},
  \bibinfo {author} {\bibfnamefont {E.}~\bibnamefont {Berti}}, \ and\ \bibinfo
  {author} {\bibfnamefont {C.~M.}\ \bibnamefont {Will}},\ }\href@noop {} {\
  (\bibinfo {year} {2009})},\ \Eprint {http://arxiv.org/abs/0906.0313}
  {arXiv:0906.0313 [gr-qc]} \BibitemShut {NoStop}%
\bibitem [{\citenamefont {{Kocsis}}\ \emph {et~al.}(2007)\citenamefont
  {{Kocsis}}, \citenamefont {{Haiman}}, \citenamefont {{Menou}},\ and\
  \citenamefont {{Frei}}}]{2008kocsis-hmd}%
  \BibitemOpen
  \bibfield  {author} {\bibinfo {author} {\bibfnamefont {B.}~\bibnamefont
  {{Kocsis}}}, \bibinfo {author} {\bibfnamefont {Z.}~\bibnamefont {{Haiman}}},
  \bibinfo {author} {\bibfnamefont {K.}~\bibnamefont {{Menou}}}, \ and\
  \bibinfo {author} {\bibfnamefont {Z.}~\bibnamefont {{Frei}}},\ }\href@noop {}
  {\bibfield  {journal} {\bibinfo  {journal} {\prd}\ }\textbf {\bibinfo
  {volume} {76}},\ \bibinfo {pages} {022003} (\bibinfo {year}
  {2007})}\BibitemShut {NoStop}%
\end{thebibliography}%
\end{document}